\documentclass[11pt]{article}
\usepackage{eurosym}
\usepackage{amsfonts}
\usepackage{amssymb}
\usepackage{graphicx}
\usepackage{amsmath}
\usepackage{makeidx}
\usepackage{indentfirst}
\usepackage[T1]{fontenc}
\usepackage[utf8]{inputenc}

\newcounter{resultnum}[section]
\newcounter{conclusionnum}[section]
\newcounter{conditionnum}[section]
\newcounter{conjecturenum}[section]
\newcounter{examplenum}[section]
\newcounter{exercisenum}[section]
\newcounter{lemmanum}[section]
\newcounter{notationnum}[section]
\newcounter{theoremnum}[section]
\newcounter{definitionnum}[section]
\newcounter{corollarynum}[section]
\newcounter{remarknum}[section]
\newcounter{propositionnum}[section]
\newcounter{acknowledgementnum}[section]
\newcounter{algorithmnum}[section]
\newcounter{axiomnum}[section]
\newcounter{casenum}[section]
\newcounter{claimnum}[section]
\newcounter{summarynum}[section]
\newcounter{problemnum}[section]
\begin{document}

\title{Nonassociative geometric and quantum information flows and \\ R-flux deformations of wormhole solutions in string gravity}
\date{January 31, 2024; (version accepted to Fortschr. Physik)}
\author{{\textbf{Lauren\c{t}iu Bubuianu}\thanks{email: laurentiu.bubuianu@tvr.ro and laurfb@gmail.com}} \and {\small \textit{%
\ SRTV - Studioul TVR Ia\c{s}i} and \textit{University Appolonia}, 2 Muzicii street, Ia\c{s}i, 700399, Romania} \vspace{.1 in} \and {\textbf{Douglas
Singleton}} \thanks{email: dougs@mail.fresnostate.edu} \and {\small \textit{Department of Physics,\ California State University Fresno, Fresno, CA 93740-8031, USA;}} \and {\small \textit{Kavli Insitute for Theoretical Physics, University of
California Santa Barbara, Santa Barbara, CA 93106, USA} } \vspace{.1 in} \and 
\textbf{Sergiu I. Vacaru} \thanks{%
emails: sergiu.vacaru@fulbrightmail.org ; sergiu.vacaru@gmail.com ;  \textit{%
Address for correspondence in 2024 as a visiting fellow at CAS LMU, Munich, Germany and   YF CNU, Chernitvsi, Ukraine:\ } Vokzalna  street, 37-3, Chernivtsi, Ukraine, 58008} \\ {\small \textit{Department of Physics, California State
University at Fresno, Fresno, CA 93740, USA; }} \\ {\small \textit{Institute of Applied-Physics and Computer Sciences, Yu. Fedkovych
University, Chernivtsi, 58012, Ukraine;}} 
\\ {\small \textit{Center for Advanced Studies, Ludwig-Maximilians-Universit\"{a}t, Seestrasse 13, M\"{u}nchen, 80802, Germany}}
\vspace{.1 in} 
\and {\textbf{El\c{s}en Veli Veliev}} \thanks{email: elsen@kocaeli.edu.tr and elsenveli@hotmail.com} \\
{\small \textit{\ Department of Physics,\ Kocaeli University, 41380, Izmit, Turkey }}}
\maketitle

\begin{abstract}
This article consists of an introduction to the theory of nonassociative
geometric classical and quantum information flows defined by star products
with R-flux deformations in string gravity. Corresponding nonassociative
generalizations of the concepts of classical Shannon entropy, quantum von
Neumann entropy, R\'{e}nyi entropy are formulated. The fundamental geometric
and quantum information objects are computed following the Grigori Perelman
statistical thermodynamic approach to Ricci flows and gravity theories
generalized for phase spaces modelled as (co) tangent Lorentz bundles.
Nonassociative parametric deformations and nonholonomic thermo-geometric
versions of statistical generating functions, their quantum analogues as
density matrices are considered for deriving the entropy, energy and
fluctuation functionals. This allows us to define and compute respective
classical and quantum relative and conditional entropies, mutual information
and nonassociative entanglement and thermodynamic information variables. We
formulate the principles of nonassociative quantum geometric and information
flow theory, QGIF, and study the basic properties of such quasi-stationary
models related to modified gravity theories. Applications are considered for
nonassociative deformed and entangled couples of four-dimensional, 4-d,
wormholes (defined by respective spacetime and/or momentum type coordinates)
and nonassociative QGIFs of 8-d phase space generalized wormholes
configurations. Finally, we speculate on phase space black holes and
wormholes being transversable for nonassociative qubits, quantum channels
and entanglement witness; thought and laboratory experiments are discussed;
and perspectives for quantum computer modelling and tests of nonassociative
geometric flow and gravity theories are considered.

\vskip5pt \textbf{Keywords:}\ Nonassociative geometric flows and gravity;
quantum geometric and information flows; nonassociative qubits and
entanglement; exact solutions in modified gravity, nonassociative wormholes.
\end{abstract}

\tableofcontents


\section{Introduction}

\label{sec1}

The theory and applications of geometric, quantum and information methods in
classical and quantum field and gravity theories, gravity, particle physics,
condensed matter physics etc. consist nowadays a well-established and
perspective multi- and inter-disciplinary direction of research \cite%
{preskill,witten20,ryu16,vanraam10,jacobson16,
pastavski15,cover91,nielsen10,wilde13,hayashi17,watrous18,aolita15,nishioka18}%
. Such works involve various definitions and modifications of the concepts
of entropy and entanglement, study of quantum phase transitions and
topological phases, and applications in black hole physics, modified gravity
and modern cosmology.

\vskip5pt The main goal of this article is to provide an introduction to the
theory of nonassociative and noncommutative geometric and quantum
information flows, QGIFs, elaborated as a synthesis of theories of
non-Riemannian geometric flows \cite{svnc00,svnonh08,svmpnc09} and
developing nonholonomic geometric methods related to particle physics,
modified gravity, and quantum information theories \cite%
{sv20,lbdssv22,lbsvevv22}. This is also a review of geometric constructions
and thermodynamic methods and applications of the standard theory of Ricci
flows \cite{hamilton82,friedan80,perelman1}; on fundamental mathematical
results and methods we cite monographs \cite{kleiner06,morgan06,cao06} and,
for recent developments on high energy physics, we recommend \cite%
{kehagias19,biasio20,biasio21,lueben21,biasio22}.

\vskip5pt In this work, we also consider some important examples and
applications parametric solutions of the geometric flow and quantum
information theories in nonassociative gravity. Wormhole solutions encoding
nonassociative data from in string theory are constructed and analysed \cite%
{sv14a}.\footnote{%
We cite the results for nonholonomic associative and commutative theories
which will be extended in nonassociative forms in this paper.} Such a
program on nonassociative geometry and physics spans a wide area of
innovative research in nonassociative and noncommutative geometry with
R-flux deformations \cite%
{blumenhagen16,aschieri17,szabo19,partner01,partner02,partner03,lbdssv22,lbsvevv22}
and mathematical particle physics and gravity \cite%
{alvarez06,luest10,blumenhagen10,condeescu13,blumenhagen13,kupriyanov15,
gunaydin,kupriyanov19a}. In nonholonomic form, these works are related to a
series of papers devoted to relativistic geometric flows, emergent gravity,
and quantum methods in physics and information theory \cite%
{sv20,lbsvevv22,partner03,lbdssv22}. Here we note that nonassociative
geometric and quantum theories may encode various quasi-Hopf and exceptional
algebraic structures like octonions, quaternions, and Clifford (spinor)
configurations \cite{shafer95,baez02,jordan32,jordan34,okubo,mylonas13}. In
the unified nonassociative geometric and quantum information approach
formulated in this work, we concentrate on nonholonomic frame and
deformation of (non) linear connections methods with applications in string
gravity and M-theory. For simplicity, we do not consider octonion and/or
Clifford structures and limit our research activity to models of
nonassociativity and noncommutativity determined by star product
deformations involving non-geometric fluxes (R-fluxes). In theories with
small parameters (for instance, with a string small constant), real and
complex terms for geometric objects are computed for respective parametric
decompositions on the Planck and string constants.

\vskip5pt This multi-disciplinary review involves a number of advanced (non)
associative/ commutative geometric methods and various fundamental concepts
and results from the theory of geometric flows; mathematical particle
physics and quantum field theory, QFT; string/M-theory, and mathematical
relativity and cosmology concepts related to quantum information theory.
Readers are considered as top researchers in some such fields of activity
and possessing a wide theoretical background extended in the mentioned
directions. We assume that they are familiar with the results and research
methods outlined in standard monographs and reviews \cite%
{kleiner06,morgan06,cao06,weinberg95,becker06,blumenhagen12,
kramer03,misner,
hawking73,wald82,preskill,witten20,blumenhagen16,aschieri17,szabo19,partner01,partner02,partner03, lbdssv22, lbsvevv22}%
.

\subsection{Basic geometric ideas and motivations}

Let us explain and motivate our research program and show how the basic
ingredients (density matrices, quantum entropies and quantum thermodynamic
and information variables etc.) can be defined in statistical and geometric
forms which allows us to formulate the nonassociative QGIF theory:

\vskip5pt We begin with the main ideas which are necessary for constructing
the theory of (associative and commutative) geometric information flows,
GIFs, encoding information data into classical flows of fundamental
geometric objects on a positive real $\tau $-parameter, $0\leq \tau \leq
\tau _{0}$. In thermodynamic models, such a parameter is considered as a
temperature one \cite{perelman1}.\footnote{%
In our works, we use the terms Perelman's functionals, Pereman
thermodynamics etc. introduced in that famous preprint. They refer to the
author who published his works under the names Grigori Perelman, Grisha
Perelman and G. Perelman. The last name can be confused with other authors
being active in modern mathematics.} In the Ricci flow theory, the geometric
flow evolution of a (pseudo) Riemannian metric $\mathbf{g}(\tau
)=\{g_{\alpha \beta }(\tau )\}$ on a manifold $V$ is defined by the R.
Hamilton equations \cite{hamilton82,friedan80},\footnote{%
in our works, we consider manifolds of necessary smooth class which allow
the formulation of causal self-consistent and viable physical theories; such
a (pseudo) Riemannian manifold it is uniquely defined by a metric structure $%
\mathbf{g}$ following the condition that such a metric defines a unique
Levi-Civita, LC, connection $\nabla $, which is metric compatible, $\nabla 
\mathbf{g}=0,$ and with zero torsion;\ conventions on arbitrary frame/
coordinate indices and bases are standard ones as in \cite%
{perelman1,partner01,partner02,partner03,lbdssv22,lbsvevv22}} 
\begin{eqnarray}
\frac{\partial \mathbf{g}(\tau )}{\partial \tau } &=&-2Ric[\nabla ](\tau
)+\ldots \simeq -2\bigtriangleup \mathbf{g}(\tau ),  \label{heq} \\
&\simeq &-2\square \mathbf{g}(\tau ),\mbox{ for pseudo-Euclidean singnature }%
(+++-).  \label{heqps}
\end{eqnarray}%
In these formulas $\nabla \lbrack \mathbf{g}](\tau )$ denotes a family of
Levi-Civita, LC, connections; $Ric[\nabla ]$ and $\bigtriangleup =\nabla
^{2} $ are respectively the Ricci tensor and Laplace operator of a chosen $%
\nabla ,$ and dots are used for other possible nonlinear terms defined by
the Riemannian curvature in certain geometric/ physical models. In the
linear approximation for the metric tensors, the above formulas define some
diffusion type equations, which is stated by the operator $\bigtriangleup $.
For Lorentzian metrics and linearized equations, we obtain the wave type
d'Alambert operator $\square $. Here, we also note that for self-similar
configurations and a fixed parameter $\tau _{0},$ the equations (\ref{heq})
define Ricci solitons, which (for Lorentz signatures) contain as particular
cases the vacuum Einstein equations with cosmological constant $\Lambda
,Ric[\tau _{0},\nabla ]=\Lambda \mathbf{g}(\tau _{0}).$

\vskip5pt G. Perelman \cite{perelman1} introduced two Lyapunov type
functionals, $F[\tau ,R,\nabla ,f]$ and $W[\tau ,R,\nabla ,f]$ (we provide
explicit formulas in terms of fundamental geometric objects in the next
section). In such functional formulas, $Rs=R$ is the curvatures scalar
determined by geometric data $(\mathbf{g,}\nabla )$ and $f(\tau )$ defines
of $\tau $-family of normalization functions for defining integration
measures on $V$. The Ricci flow equations (\ref{heq}) can be derived
equivalently up to some normalizing terms following corresponding
variational procedures for the $F$- and/or $W$-functionals. It should be
noted that the second functional was defined as a "minus" entropy functional
and it is called also the W-entropy. The geometric technique with cutting
singularities and such functionals allowed G. Perelman to prove the famous
Poincar\'{e}-Thurston conjecture; on mathematical aspects of geometric flows
of Riemannian metrics and K\"{a}hler structures, see details in \cite%
{perelman1,kleiner06,morgan06,cao06}. To formulate and prove such
conjectures for non-Riemannian theories and study generalized topological,
relativistic and modified geometric constructions and mathematical physics
implications is a very difficult and unsolved problem in modern mathematics.

\vskip5pt In this article, we shall use and develop the results elaborated
in a series of our and co-authors' works \cite%
{sv20,lbsvevv22,partner03,lbdssv22} on generalized Ricci flows and
applications in mathematical particle physics, gravity, relativistic
thermodynamics and nonassociative MGTs.

\subsubsection{Perelman's functionals and geometric flows of nonholonomic
metric-affine structures}

The F- and W-functionals can be generalized in relativistic form, $\mathcal{F%
}[\tau ,\mathcal{R}s,\mathbf{D},f]$ and $\mathcal{W}[\tau ,\mathcal{R}s,%
\mathbf{D},f]$, for geometric evolution of metric-affine structures $(%
\mathbf{g}(\tau ),\mathbf{D}(\tau ))$ on nonholonomic (co) tangent Lorentz
bundles and/or Lorentz manifolds. Such a conventional spacetime $\mathbf{V}$
is usually modelled as a four-dimensional, 4-d, or higher dimension, Lorentz
manifold. A corresponding phase space $\mathcal{M}$ is modelled on tangent
Lorentz bundle, $T\mathbf{V},$ or cotangent bundle $T^{\ast }\mathbf{V}$
enabled with respective nonlinear connection, N-connection, structure $%
\mathbf{N}$ and adapted/ distinguished linear connection, d-connection $%
\mathbf{D}$.\footnote{\label{fndefncon}Such nonholonomic (equivalently
anholonomic, i.e. nonintegrable) spaces, modelled on Lorentz manifolds
and/or (co) tangent Lorentz bundles, are enabled with conventional and
necessary type $3+1,(3+1)+(3+1)+.....$ or $2(3)+2+2+...$ (dyadic) splitting
of total dimensions. For instance, for a tangent Lorentz bundle $TV,$ a
N-connection structure is defined as Withney direct sum, $\mathbf{N}:T%
\mathbf{V}=hTV\oplus vTV.$ In the next section, we shall provide all
necessary definitions and explain the conventions on abstract geometric
formalism and certain abstract and coordinate index decompositions of the
geometric objects considered in (non) associative/ commutative generalized
forms in \cite{partner01,partner02,partner03,lbdssv22,lbsvevv22}. Here note
that we use boldface symbols for geometric objects and spacetime/ phase
spaces enabled with N-connection structure and "cal" symbols for certain
functionals and curvature/ Ricci / torsion / entropy / energy type etc.
operators/ geometric objects which can be re-defined as thermodynamic
variables, subjected to certain quantization procedures etc.} For such
generalizations, $\mathcal{R}ic(\tau )=\mathcal{R}ic[\tau ,\mathbf{g},%
\mathbf{D}]$ and $\mathcal{R}s(\tau )=\mathcal{R}s[\tau ,\mathbf{g},\mathbf{D%
}]$ define respective families of Ricci tensors and scalar operators, when
for an N-adapted variational calculus, or following abstract geometric
formulations and proofs, the geometric flow equations (\ref{heq}) are
modified into 
\begin{eqnarray}
\frac{\partial \mathbf{g}(\tau )}{\partial \tau } &=&-2Ric[\mathbf{D}](\tau
)+\ldots ,\mbox{nonholonomic geometric flow equations};  \label{heqnonh} \\
&&Ric[\tau _{0},\mathbf{g},\mathbf{D}]=\Lambda \mathbf{g}(\tau _{0}),%
\mbox{for modified Ricci solitons/ Einstein equations}.
\label{ricciconfhonh}
\end{eqnarray}%
Such relativistic and modified systems of nonlinear PDEs with applications
in string theories, supergravity and noncommutative/ generalized Finsler
gravity and modified geometric flow scenarios were studied in a series of
our and co-authors previous works \cite%
{svnc00,svnonh08,svmpnc09,sv20,lbsvevv22} and in recent works by other
authors \cite{kehagias19,biasio20,biasio21,lueben21,biasio22}.

\subsubsection{Canonical nonholonomic spacetime and phase space geometric
flows}

It is not clear if and how nonholonomic F- and W-functionals and Ricci flow/
soliton equations of type (\ref{heqnonh}) and/or (\ref{ricciconfhonh}) could
be used for formulating and elaborating certain proofs of some generalized
Poincar\'{e}-Thurston conjectures for non-Riemannian geometric flows and
modified gravity theories, MGTs. Nevertheless, the relativistic and
non-Riemannian generalizations and applications are important for developing
renorm group, RG, flow models with theories with underlying nonlinear sigma
models and beta functions. Corresponding physical values determined by
solutions of geometric flow evolution and dynamical gravitational like field
equations, with additional nonholonomic constraints, are computed in the
framework of string gravity theory and various MGTs and quantum gravity, QG,
theories, swampland program etc. \cite%
{kehagias19,biasio20,biasio21,lueben21,biasio22,gomez19} with ultraviolet,
UV, completion and UV/ IR correspondence, i.e. from/ to infrared. Such
geometric and (quantum) physical models are defined by sophisticated systems
of nonlinear partial differential equations, PDEs. It is very difficult to
construct general classes of exact/ parametric solutions when, for instance,
the metric tensors can be generic off-diagonal and the coefficients depend
on some/all spacetime and phase space coordinates. In general, such
solutions of nonlinear PDEs are determined by generating functions and
generating (effective) sources, and integration functions/ constants and
can't be reduced to simplified diagonal ansatz of metric determined as
solutions of simplified systems of nonlinear ordinary differential
equations, ODEs.

\vskip5pt In a series of works on associative/ commutative phase spaces \cite%
{partner02,partner03,lbdssv22,lbsvevv22}), see also and references therein,
we considered the so-called canonical d-connection, $\widehat{\mathbf{D}}%
=\nabla \lbrack \mathbf{g]}+\widehat{\mathbf{Z}}[\mathbf{g,N]}$. Such a
linear connection preserves under parallelism the N-connection splitting
(i.e. it is N-adapted) being determined by a canonical distortion d-tensor $%
\widehat{\mathbf{Z}}[\mathbf{g]}$ in double connection metric compatible
forms, $\widehat{\mathbf{D}}\mathbf{g}=0$ and $\nabla \mathbf{g}=0.$ It
contains an induced canonical d-torsion structure $\widehat{\mathcal{T}}[%
\mathbf{g,N],}$ which is included via algebraic combinations of coefficients
in the d-tensor $\widehat{\mathbf{Z}}[\mathbf{g]}$.\footnote{%
In the next section, we shall provide all necessary definitions and formulas
for general cases with nonassociative star product generalizations. Appendix %
\ref{ass1} contains a summary of notations and conventions for abstract
geometric and N-/s-adapted index calculus.} Such a canonical d-torsion is
different from the torsion tensors in the Einstein-Cartan and/or string
theory because it is induced by the coefficients of metrics, which in
coordinate frames contain nontrivial off-diagonal terms determined by a
prescribed N-connection/ anholonomic frame structure. A canonical
d-connection $\widehat{\mathbf{D}}$ does not involve additional
algebraic/dynamical fields (subjected to related additional field equations)
for defining the corresponding d-torsion fields.

\vskip5pt The priority of the canonical d-connection compared to other types
of linear connection structures is that it allows to decouple and integrate
in certain general off-diagonal forms various "hat" geometric flow/ modified
Einstein equations of type (\ref{heqnonh}) and/or (\ref{ricciconfhonh}) for $%
\mathbf{D}=\widehat{\mathbf{D}}.$ For simplicity, we can consider
quasi-stationary type solutions with Killing symmetry on a time like
coordinate $t,$ with a Killing vector $\partial _{t},$ or inversely to
generate various classes of locally anisotropic cosmological solutions with
a Killing symmetry on a space like, or phase space like, coordinate. Here we
note that a $\widehat{\mathbf{D}}$ can be always reduced to a LC-connection $%
\nabla $ by imposing additional nonholonomic constraints when $\widehat{%
\mathbf{D}}_{\shortmid \widehat{\mathcal{T}}=0}=\nabla $ at the end after
some solutions of modified Ricci flow/ gravitational field equations have
been found in explicit form.

\vskip5pt New geometric methods involving canonical d-connections and
generalized Finsler d-connections, and nonholonomic dyadic variables, for
generating exact/parametric generic off-diagonal solutions of physically
important systems of nonlinear and nonholonomic systems of PDEs are reviewed
in a series of works \cite{partner02,partner03,lbsvevv22}. They summarize 25
years of research of activity and geometric applications in modern physics
and information theories and the so called the anholonomic frame and
connection deformation method, in brief, AFCDM. So, even though it is not
clear if and how mathematically well-defined and physically important
variants of generalized Poincar\'{e}-Thurston conjectures can be formulated
and proven for non-Riemannian geometric flows and MGTs, we can elaborate on
self-consistent causal nonholonomic geometric evolution scenarios of
modified gravitational and matter field equations using the AFCDM for the
canonical d-connection $\widehat{\mathbf{D}}$ and its various
supersymmetric/ nonassociative/ noncommutative/ fractional generalizations
etc.

\subsubsection{Statistical thermodynamics for canonical N-adapted geometric
flows}

The W-entropy was used for formulating a statistical thermodynamic theory of
Ricci flows of Riemannian metrics \cite{perelman1}. In such an approach, the
thermodynamic values are determined by fundamental geometric objects, for
instance, flows of metric, connections, related curvature scalar and Ricci
tensor, up to a normalizing function used for defining corresponding
integration measures. The constructions can be generalized in relativistic
form, for non-Riemannian theories, and distorted on spacetime/ phase spaces,
as we prove in \cite{sv20}. In abstract geometric form on a phase space $%
\mathcal{M}$ enabled with canonical d-connections and related geometric
d-objects, we can define functionals $\widehat{\mathcal{F}}(\tau )=\widehat{%
\mathcal{F}}[\tau ,\widehat{\mathcal{R}}s,\widehat{\mathbf{D}},\widehat{f}]$
and $\widehat{\mathcal{W}}(\tau)= \widehat{\mathcal{W}}[\tau ,\widehat{%
\mathcal{R}}s,\widehat{\mathbf{D}},\widehat{f}]$ and introduce a
corresponding statistical thermodynamic function (it is also called the
partition/ generating function) $\widehat{\mathcal{Z}}(\tau )=\widehat{%
\mathcal{Z}}[\tau ,\widehat{\mathcal{R}}s,\widehat{\mathbf{D}},\widehat{f}]$
for a canonical ensemble at temperature $T=\tau .$ Using $\widehat{\mathcal{Z%
}}(\tau ),$ we can compute in standard form the main thermodynamic variables
such as the canonical average flow energy, $\widehat{\mathcal{E}}(\tau )$,
flow entropy, $\widehat{\mathcal{S}}(\tau )$, and flow fluctuation
parameter, $\widehat{\sigma }(\tau ).$

\vskip5pt It should be noted here that G. Perelman speculated on possible
applications of his geometric and statistical thermodynamic models in string
theory and black hole, BH, physics \cite{perelman1}. He concluded that such
thermodynamic approach to gravity and geometric flows is very different from
the Bekenstein-Hawking thermodynamics for BHs \cite{bek1,bek2,haw1,haw2},
see the main concepts summaries GR and BHs in \cite{misner,hawking73,wald82}.

\vskip5pt We emphasize two outstanding properties of the F- and
W-functionals and related statistical/ geometric thermodynamic models:

\begin{enumerate}
\item Such functionals and thermodynamic variables can be computed in GR
and/or MGTs for any solution of the (modified) Einstein equations and not
only for configurations with conventional hypersurface horizons, holographic
and/or duality properties etc. What we need for formulating and
investigating physical properties of thermodynamic models for some general
classes of solutions is to know the Ricci tensor and scalar curvature of a
canonical d-connection computed for a found metric (in general, being
generic off-diagonal) and generalized for chosen (non) linear connections
data.

\item All functional and relativistic thermodynamic constructions for
geometric flows and be extended/ redefined for other types of MGTs, string
gravity models, generalized Finsler-Lagrange-Hamilton MGTs, supergravity,
nonassociative and noncommutative gravity. Such formulations and further
developments can be performed in the framework of (quantum) information
theories \cite{sv20,lbdssv22,lbsvevv22}.
\end{enumerate}

G. Perelman's thermodynamic paradigm for geometric flows and Ricci soliton
configurations can be extended/ modified to various classes of MGTs and
their solutions which, in general, do not have a standard interpretation
using the concept of Bekenstein-Hawing entropy. The W-entropy $\widehat{%
\mathcal{W}}(\tau)$ and the flow entropy $\widehat{\mathcal{S}}$ can be used
for formulating various types of classical and quantum thermodynamics models
which allow us to characterize locally anisotropic cosmological solutions,
generic off-diagonal quasi-stationary configurations, and various models of
geometric evolution, for instance, of nonholonomic Einstein systems.

\subsubsection{Geometric thermodynamics and information flows}

Geometric, quantum field and statistical thermodynamic methods are largely
used in classical and quantum information theory \cite%
{preskill,witten20,ryu16,vanraam10,jacobson16,aolita15,nishioka18}. Various
fundamental problems in BH physics and cosmology, holography and related
information theory are approached for generalizations of the
Bekenstein-Hawking entropy and thermodynamics \cite{bek1,bek2,haw1,haw2},
which can be considered only for some special classes of solutions involving
(effective) hyper-surfaces, duality conditions, and holographic
configurations.

\vskip5pt During the last four years, there were published a series of works 
\cite{sv20,lbsvevv22}, see also references therein, for elaborating a new
class of GIF and QGIF theories and applications in modern geometric
mechanics, particle physics, gravity and cosmology. The main innovative idea
exploited in such papers is to consider generalized Perelman geometric and
thermodynamic models when a statistical thermodynamic function $\mathcal{Z}%
(\tau )=\mathcal{Z}[\tau ,\mathcal{R}s,\mathbf{D},f]$ is transformed into a
quantum density matrix $\mathcal{\rho }(\tau,x^{i},p_{a}),$ where $%
(x^{i},p_{a})\in \ ^{\shortmid }\mathcal{M}=T^{\ast }\mathbf{V}$ for a
respective quantum thermodynamic model. In the quasi-classical
approximations, the nonlinear geometric evolution is defined by certain
modifications of the R. Hamilton equations and a respective classical GIF
interpretation, when the quantum fluctuations encode data for QGIF model.

\vskip5pt Using the GIF and QGIF formulation, we can define and compute all
basic ingredients for classical and quantum information theories:\ quantum
entropies/ channels, with nonholonomic information flow variants for Von
Neumann entropy and mutual information, etc. Such a new geometric and
quantum formalism was applied for an alternative study of entanglement of
quantum information flow systems, for various classes of locally anisotropic
BH and cosmological solutions when the concept of Bekenstein-Hawing entropy
is not applicable. In another turn, the generalized Perelman thermodynamic
variables can be always defined and computed in a self-consistent form for
various GIF and QGIF systems.

\subsubsection{Nonassociative gravity and geometric flows}

Nonassociative and noncommutative geometric deformation structures have been
studied in a series of important works on string gravity, quantum gravity,
QG, and quantum field theories, QFT, with magnetic and/or Kaluza-Klein
monopoles; original results and reviews are provided in \cite%
{luest10,blumenhagen10,condeescu13,blumenhagen13,kupriyanov15,
gunaydin,kupriyanov19a,blumenhagen16,aschieri17,szabo19,partner01,partner02,partner03,lbdssv22,lbsvevv22}%
. In string gravity and M-theory, nonassociative star product deformations, $%
\star $ deformations, are computed for a prescribed Moyal-Weyl tensor
product and determined by non-geometric fluxes (R-fluxes).\footnote{%
Necessary formulas will be provided in next section. In this introduction
section, we follow a geometric abstract formalism which is known for experts
in mathematical particle physics, geometry and physics, and mathematical
relativity.} Self-consistent approaches to nonassociative gravity were
elaborated \cite{blumenhagen16,aschieri17,szabo19} up to a level of
definition and parametric computation of nonassociative Ricci tensors $%
\mathcal{R}ic^{\star }[\mathbf{g}^{\star },\mathbf{\nabla }^{\star }]$
determined by $\star $ deformations of pseudo-Riemannian metrics. There are
considered R-flux deformations when $\star :\mathbf{g\rightarrow g}^{\star}=(%
\mathbf{\breve{g}}^{\star },\mathbf{\check{g}}^{\star })$ result in some
nonassociative symmetric, $\mathbf{\breve{g}}^{\star }$, and nonassociative
nonsymmeric, $\mathbf{\check{g}}^{\star }$, components;\ and respective star
product deformation of LC-connection, $\star :\ \mathbf{\nabla\rightarrow
\nabla }^{\star }.$ In such theories, the coefficients of fundamental
geometric objects and (effective) R-flux sources can be decomposed into real
and complex terms and parametric decompositions on the Planck constant, $%
\hbar ,$ and string constant, $\kappa ,$ Following a corresponding system of
abstract and index-free notations, the noncommutative configurations can be
generated in "pure" form for $\kappa =0.$ Nonassociative geometric and
vacuum gravity models were formulated in certain forms encoding quasi-Hopf 
\cite{blumenhagen16,aschieri17,szabo19} and/or exceptional algebraic
structures, for instance, octonionic and Clifford configurations $\kappa$
etc. \cite{shafer95,baez02,jordan32,jordan34,okubo,mylonas13}.

\vskip5pt To derive nonassociative deformations of the Einstein equations in
GR there were considered phase spaces constructed as star-deformations of
tangent and cotangent Lorentz bundles, $\mathcal{M}=T\mathbf{V}$ and $\
^{\shortmid }\mathcal{M}=T^{\ast }\mathbf{V,}$ on a base spacetime manifold $%
\mathbf{V,}$ when the nonassociative star product results in $\mathcal{M}
\rightarrow \ _{\star }\mathcal{M}$ and $\ ^{\shortmid }\mathcal{M}
\rightarrow \ _{\star}^{\shortmid }\mathcal{M}.$\footnote{%
In our partner works \cite{partner01,partner02,partner03,lbdssv22,lbsvevv22}%
, we elaborated a system of abstract/ index notations when left up/low $%
\star $-labels are used for emphasizing that certain geometric objects are
subjected to star product deformations. Another type of left up/low labels $%
\ "^{\shortmid }"$ are used for emphasizing that certain geometric objects
are defined on $TV,$ with local coordinates $u=(x,v)=\{u^{%
\alpha}=(x^{i},v^{a})\},$ but other ones on $T\ ^{\ast }V,$ with local
coordinates $\ ^{\shortmid }u=(x,p)=\{\ ^{\shortmid
}u^{\alpha}=(x^{i},p_{a})\}.$ Necessary details on such a system of
notations will be provided in the next sections, see also Appendix \ref%
{appendixa} on concepts, conventions, and notations.} In a commutative form,
a spacetime manifold $\mathbf{V}$ can be chosen as in GR, or another MGT
with metric-affine structure, and may encode as real parts certain
nontrivial R-flux contributions of star product deformations. The
nonassociative vacuum and non-vacuum gravitational equations consist of very
sophisticated systems of coupled nonlinear partial differential equations,
PDEs. Finding solutions of nonassocitaive physically important PDEs is more
difficult than in GR and associative/ commutative MDEs because the star
product and R-flux deformations involve more sophisticated coupling
properties and various off-diagonal dependencies on spacetime and phase
space local coordinates. Technically, it is not possible to decouple in
certain general forms all geometric and physical objects and find exact
solutions of such equations written in coordinate bases $\partial _{\alpha }$
as it was formulated in \cite{blumenhagen16,aschieri17}.

\vskip5pt It was a challenge to construct and analyse possible physical
implications, for instance, of some particular classes of exact/ parametric
solutions describing nonassociative black hole (BH) and/or cosmological
configurations; and such problems were solved in our partner works \cite%
{partner02,partner03,lbdssv22,lbsvevv22}. In those papers, an innovative
approach to nonassociative geometry and gravity was formulated for
star-products generalized for commutative nonholonomic distributions and
nonholonomic N-adapted frames $\mathbf{e}_{\alpha }$ (instead of coordinate
bases $\partial _{\alpha }$). Such a nonholonomic phase space geometry can
be constructed for a N-connection, structure, $\ ^{\shortmid }\mathbf{N}: T%
\mathbf{V}=hT\mathbf{V}\oplus cT\mathbf{V}$, where $\oplus $ is a Whithney
direct sum, stating a conventional horizontal and covertical (4+4) splitting
of dimensions for 4-d spacetimes $\mathbf{V.}$ We can also consider further
nonholonomic dyadic (2+2)+(2+2) decompositions with respective shell
connection splitting, $\ _{s}^{\shortmid }\mathbf{N}:T\mathbf{V}= \ ^{1}hT%
\mathbf{V}\oplus \ ^{2}vT\mathbf{V}\oplus \ ^{3}cT\mathbf{V\oplus }\ ^{4}cT%
\mathbf{\mathbf{V},}$ for shells $s=1,2,3,4$. This is necessary for
performing a general decoupling and integration strategy for physically
important systems of nonlinear partial differential equations, PDEs. The
AFCDM described above is generalized in such a way on associative and
commutative phase spaces as described in \cite{vbv18}.

\vskip5pt A ground-breaking result of \cite{partner02} consisted in a proof
that the nonassociative nonholonomic vacuum gravitational equations $%
\widehat{\mathcal{R}}ic^{\star }[\mathbf{g}^{\star },\widehat{\mathbf{D}}%
^{\star}]= \widehat{\Lambda }\mathbf{g}^{\star }$ can be decoupled and
integrated, i.e. solved, in very general forms for a so-called a canonical
d-connection structure $\widehat{\mathbf{D}}^{\star }$ constructed by star
product deformation of $\widehat{\mathbf{D}}$ described above. A matrix of
(effective) cosmological constants $\widehat{\Lambda }$ is related via
respective nonlinear symmetries to some (effective) sources encoding
nonassociative R-flux contributions and other terms describing matter
fields. For well-defined nonholonomic configurations, the symmetric and
nonsymmetric components of metrics decouple and can be computed in $\hbar $
and $\kappa $ parametric forms describing mixed nonassociative and
noncommutative star and R-flux deformations.

\vskip5pt Nonassociative vacuum gravitational equations MGTs and GR were
studied for star-product (non) holonomic structures and nonassociative
geometric models constructed on phase spaces modelled as (co) tangent
bundles, $T^{\ast }\mathbf{V}$ of a base Lorentz spacetime manifold, $%
\mathbf{V}$ \cite{partner02,partner03,lbdssv22,lbsvevv22}. In corresponding
commutative limits, we obtain theories with fundamental geometric objects
(for instance, $\mathbf{g}(x^{i},p_{a})$ and $\mathbf{D}(x^{i},p_{a})$)
determined by modified dispersion relations, MDRs, and depending both on
spacetime and momentum/ velocity like coordinates. Using star product
deformations of families of metrics $(\mathbf{g}(\tau ),\mathbf{D}%
(\tau))\rightarrow (\mathbf{g}^{\star }(\tau ),\mathbf{D}^{\star }(\tau )),$
we can generate various nonassociative variants of R. Hamilton equations, 
\begin{equation}
\frac{\partial \mathbf{g}^{\star }(\tau )}{\partial \tau }=-2\mathcal{R}%
ic^{\star }[\mathbf{g}^{\star },\mathbf{D}^{\star }](\tau )+\ldots ,
\label{nonassocham1}
\end{equation}%
where dots denote possible additional terms encoding contributions from
(non) associative geometric and MGTs theories. More details on such
equations will be provided in the next section for formulas (\ref%
{nonassocgeomfl}). A self-consistent formalism can be elaborated for
canonical nonholonomic configurations which are star-deformed into
nonassociative canonical nonholonomic configurations. For corresponding
systems of nonlinear PDEs, we can apply the AFCDM and construct exact and
parametric nonassociative solutions. This allows us to develop various
innovative geometric and analytic methods for investigating nonassociative
and nonholonomic evolution of geometric flows. We can apply a canonical
nonassociative variational procedure when modified geometric flow equations
are derived from the respective nonholonomic star (R-flux or other types)
deformed F- and W-functionals, $\widehat{\mathcal{F}}^{\star }(\tau)= 
\widehat{\mathcal{F}}^{\star }[\tau ,\ ^{\star }\widehat{\mathcal{R}}s, 
\widehat{\mathbf{D}}^{\star },\widehat{f}]$ and $\widehat{\mathcal{W}}%
^{\star }(\tau )=\widehat{\mathcal{W}}^{\star }[\tau ,\ ^{\star} \widehat{%
\mathcal{R}}s, \widehat{\mathbf{D}}^{\star },\widehat{f}]$ as it is
explained above.

\vskip5pt Constructing nonassociative deformations of commutative models of
information thermodynamics, we can formulate and elaborate on radically new
type theories of nonassociative quantum geometric and information flows
defined by nonlinear systems of PDEs, and respective functional equations,
which can be decoupled and solved in certain general exactly/ parametric
forms.

\subsection{The hypothesis, objectives, and structure of the work}

This work is the sixth one in a series of partner works devoted to
nonassociative geometry and exact and/or parametric solutions in such MGTs
and further developments to geometric and quantum information theories \cite%
{partner01,partner02,partner03,lbdssv22,lbsvevv22}. It is also a natural and
logical development of original nonassociative gravity theories \cite%
{blumenhagen16,aschieri17}. In a general context the nonassociative geometry
and physics were outlined in \cite{szabo19} but that had not included new
geometric methods of constructing exact/ parametric generic off-diagonal
solutions \cite{partner02,partner03} extending the AFCDM \cite{vbv18} for
integrating systems of nonlinear PDEs describing nonassociative geometric
flows and Ricci soliton configurations. This work provide a nonassociative
generalization and modification of our recent results and methods on the
theory geometric and (quantum) information flows with applications in
nonstandard particle physics, modified gravity and modern cosmological
theories.

\subsubsection{The main hypothesis and claims}

\label{sshypothesis}

The Hypotheses and Claims, and related Objectives, for performing a research
program on nonassociative geometry and physics, with off-diagonal solutions
and geometric and quantum information flow were stated and updated in the
partner works \cite{partner01,lbdssv22,lbsvevv22}. In this article, we
follow the

\vskip3pt \textbf{Hypothesis:} \emph{Star product and R-flux deformations
for string/ M-theory define nonassociative and noncommutative modifications
of classical gravity and QG theories, QM models and QFTs, which can be
unified as some thermodynamic information theories formulated in the
framework of respective GIF and QGIF theories.}

\vskip3pt The formulated Hypothesis is stated in more explicit form and
structured by such Claims C1-C6:

\begin{itemize}
\item C1:\ A general nonassociative and noncommutative geometric approach to
QG, and QGIF theories can be formulated using the nonholonomic geometric
formalism and methods elaborated for relativistic geometric flows and MGTs
and respective star product deformations.

\item C2:\ Nonassociative geometric flow evolution equations and
gravitational and matter field equations in MGTs and general relativity, GR,
can be derived in abstract nonholonomic geometric and/or variational forms
using nonassociative generalized Perelman functionals and non-integrable
constraints defining nonassociative Ricci solitons. Respective
nonassociative geometric spacetime and phase space theories can be
reformulated in generalized almost (symplectic) K\"{a}hler-Finsler variables
which allow to elaborate on geometric/ deformation quantization models and
develop other types of quantization methods in QG and QFT.

\item C3:\ Applying the AFCDM, we can decouple and integrate in general form
various physically important systems of nonlinear PDEs describing
nonassociative geometric flow models, MGTs, QFTs, and respective nonlinear
symmetries. This allows us to construct exact and parametric solutions for
nonassociative Einstein-Yang Mills-Dirac-Higgs systems and their
nonassociative geometric flow evolution equations. Such solutions defined by
generic off-diagonal metrics and (non) linear connections provide explicit
examples of (non) associative generalized Einstein-Eisenhart-Moffat
geometries, with symmetric and nonsymmetric metrics, and relativistic
Finsler-Lagrange-Hamilton geometries defined on phase spaces with spacetime
and momentum/velocity like coordinates.

\item C4:\ Various examples of generic off-diagonal metrics and nonlinear
and linear connections describe locally anisotropic generalizations of black
hole solutions, solitonic wave hierarchies, locally anisotropic wormhole and
cosmological configurations etc. Such solutions play an important role in
testing alternative gravity theories and elaborating scenarios with
anisotropic Universe acceleration and quasi-periodic cosmological structures
in dark energy/ matter physics. For corresponding classes of nonholonomic
constraints, we can extract standard particle models and solutions in GR. In
general forms, the classical and quantum solutions encode possible
nonassociative contributions from respective MGTs.

\item C5:\ Using nonassociative generalized relativistic Perelman
functionals, we can elaborate on thermodynamic models for GIF and QGIF
systems encoding nonassociative geometric structures and contributions from
various classes of classical and quantum theories. Nonassociative
generalized statistical and kinetic generated functions (with corresponding
density matrices for quantum phase space models) give rise to a new class of
nonassociative and noncommutative QG and QFT models and their GIF and QGIFs
formulations. We can define and compute all basic/ necessary ingredients for
QGIF theories, such as the Von Neuman entropy, W-entropy and quantum
channels, mutual nonassociative geometric information and entanglement
entropy.

\item C6: Nonassociative QGIFs allow the formulation of new transversable
wormhole protocols, their tests and creation of such configurations in the
lab and modelling on quantum computers. This provides new perspectives for
thought and lab tests of certain classes of (in general, nonassociative and
noncommutative) MGTs, string gravity and QG models involving advanced
mathematical methods from the Ricci flow and quantum information flows.
\end{itemize}

The Hypothesis and Claims C1-C6 summarize and update similar statements from
the partner works \cite{partner01,partner02,partner03,lbdssv22,lbsvevv22}.

\subsubsection{The objectives and structure}

\label{ssobjectives} To elaborate on the program stated by the above Claims
we outline the necessary results proved in the mentioned partner works
devoted to C1-C3 and provide new results for C4-C6 (when a series of
sub-objectives will be solved in future works related to C2-C4), we state
such Objectives for this article:

\vskip5pt \textbf{The first objective, Obj1}, for section \ref{sec2}, is to
review in brief the necessary results from \cite{lbsvevv22} on
nonassociative geometric flows and generalized Perelman thermodynamics and
then to formulate the classical nonassociative geometric information flow
theory, GIF. We define and compare basic concepts of nonassociative
generating functions and related geometric flow and Shannon entropies and
discuss properties of nonassociative conditional/ relative entropy.

\vskip5pt \textbf{The second objective, Obj2}, for section \ref{sec3}, is to
formulate and analyze the main properties of the basic ingredients of
nonassociative quantum geometric and quantum information flow theory, QGIF.
We define and study density matrices and quantum entropy for QGIFs and show
how to compute the respective conditional and relative entropy encoding
nonassociative data. We speculate also on GIF thermodynamics and
generalizations for quantum channels.

\vskip5pt \textbf{The third objective, Obj3}, for section \ref{sec4}, is to
show how QGIF models can be elaborated for physically important generic
off-diagonal solutions in nonassociative geometric flow and gravity.
Additionally to certain classes of nonassociative star deformations of BH
solutions studied in section 5 of \cite{lbsvevv22}, we apply the AFCDM to
construct certain variants of nonassociative wormhole solutions, compute GIF
thermodynamic variables and extend the constructions for QGIF entropic
values of such quasi-stationary configurations.

\vskip5pt \textbf{The fourth objective, Obj4}, motivated and commented in
subsection \ref{sscomments} is to establish a "wormhole bridge" between the
nonassociative GIF and QGIF theories toward quantum teleportation, thought
and lab tests of various models MGTs and QG and perspectives and
applications to quantum computer theory and modern cosmology. Such multi-
and inter-disciplinary research is supported by a unified background in the
frameworks of the AFCDM for constructing exact/ parametric generic
off-diagonal solutions of physically important systems of nonlinear PDEs
used in (nonassociative) geometric and information flow theories, GR and
MGTs, and related quantum (information) models.

\vskip5pt Finally, in section \ref{sec5}, we discuss and conclude the main
results obtained as solutions of objectives Obj1- Obj4 and supporting the 
\textbf{Hypothesis} of this work. Further perspectives, and queries on
providing more complete proofs of claims C1 - C6 (with generalizations,
examples and applications), in other partner works under elaboration are
considered.

To facilitate reading the text and understanding how the geometric methods
are applied for constructing parametric solutions of physically important
systems of nonlinear PDEs, we provide in the Appendix some Tables
summarizing notations and some important ansatz for generating
nonassociative wormhole solutions used in this and partner works.

\section{Statistical thermodynamics of classical nonassociative geometric
and information flows}

\label{sec2} In a series of works \cite{sv20,lbsvevv22,partner03,lbdssv22},
we formulated and studied possible applications in modern physics and
cosmology of the theory of associative and commutative geometric and
information flows, GIFs. The geometric and (quantum) information
constructions were performed as a synthesis of the main concepts and methods
of a relativistic Ricci flow theory and modified gravity formulated on
nonholonomic (co) tangent bundles \cite{svnonh08,svmpnc09} and classical/
quantum information theory \cite%
{preskill,witten20,ryu16,vanraam10,jacobson16,aolita15,nishioka18}. A
unified approach to nonassociative geometric flow theory and gravity models
determined by star product R-flux deformations in string/ M-theory is
elaborated in \cite{lbsvevv22}. This section aims to introduce the necessary
definitions for nonassociative geometric flow theory using respective
generalizations of Perelman's W-entropy and thermodynamics and then to
provide an introduction to the classical nonassociative GIF theory. The main
concepts and geometric/ physical objects are motivated and determined by
nonassociative generating functions and related geometric flow and Shannon
entropies, discussed properties of nonassociative conditional/ relative
entropy.

\subsection{Preliminaries: nonassociative geometry and nonholonomic dyadic
splitting}

We outline necessary concepts and results on nonassociative geometric flows
formulated in nonholonomic form in sections 2 and 3 and appendix A of \cite%
{lbsvevv22} following generalized abstract and coordinate index conventions
from \cite{partner01,partner02,partner03,lbdssv22}.

\subsubsection{Associative and commutative nonholonomic dyadic decompositions%
}

Originally, the nonassociative gravity theories defined by star products
determined by R-flux deformations \cite{blumenhagen16,aschieri17} were
constructed on respective phase spaces modeled as a cotangent bundle $\
^{\shortmid }\mathcal{M}=T^{\ast }V$ on a spacetime Lorentz manifold $V.$ To
generalize the AFCDM for constructing exact/parametric solutions encoding
nonassociative data we elaborated a nonholonomic shell decomposition
formalism which allow to decouple modified Einstein equations \cite%
{partner01,partner02}. In such an approach, $\ ^{\shortmid }\mathcal{M}$ is
enabled with conventional (2+2)+(2+2) splitting determined by a nonholonomic
dyadic, 2-d, decomposition into four oriented shells $s=1,2,3,4.$ In brief,
we shall use the term s-decomposition; and note that in literature there are
used equivalent terms like anholonomic, i.e. non-integrable. A s-splitting
is defined by a nonlinear connection, N-connection (equivalently,
s-connection), structure: 
\begin{eqnarray}
\ _{s}^{\shortmid }\mathbf{N}:\ \ _{s}T\mathbf{T}^{\ast }\mathbf{V} &=&\
^{1}hT^{\ast }V\oplus \ ^{2}vT^{\ast }V\oplus \ ^{3}cT^{\ast }V\oplus \
^{4}cT^{\ast }V,\mbox{ which is dual to }  \notag \\
\ _{s}\mathbf{N}:\ \ _{s}T\mathbf{TV} &=&\ ^{1}hTV\oplus \ ^{2}vTV\oplus \
^{3}vTV\oplus \ ^{4}vTV,\mbox{  for }s=1,2,3,4.  \label{ncon}
\end{eqnarray}%
In a local coordinate basis, a nonlinear s-connection defined by a Whitney
sum $\oplus $ as in (\ref{ncon}), for instance, is characterized by a
corresponding set of coefficients $\ _{s}^{\shortmid }\mathbf{N}=\{\
^{\shortmid }N_{\ i_{s}a_{s}}(\ ^{\shortmid }u)\},$ for any point $u=(x,p)=\
^{\shortmid }u=(\ _{1}x,\ _{2}y,\ _{3}p,\ _{4}p)\in \mathbf{T}^{\ast }%
\mathbf{V}$.\footnote{%
In our partner works, we follow a convention on indices and local
coordinates which is adapted dyadic to s-shell decompositions: For instance,
we use $\beta _{2}=(j_{1},b_{2}),$ where $j_{1}=1,2;b_{2}=3,4,$ for the
shell $s=2$ and the coordinate $u^{4}=y^{4}=t$ considered as a time like
one, $t.$ We write $u^{\beta _{2}}=(x^{i_{1}},y^{3},y^{4}),$ for $%
(x^{i_{1}},y^{3})$ being space like coordinates; this can be written also as 
$x^{i_{2}}=(x^{i_{1}},y^{a_{2}})$. On the first three shells, the
coordinates and indices are defined as $\ ^{\shortmid }u^{\beta
_{3}}=(x^{i_{1}},y^{b_{2}},p_{a_{3}}),$ for $a_{3}=5,6;$ and write
equivalently $\ ^{\shortmid }u^{i_{3}}=(x^{i_{1}},y^{b_{2}},p_{a_{3}}).$ In
a similar form, we split the indices and coordinates on the shell $s=4,$
when $\beta _{4}=(j_{3},b_{4}),$ for $j_{3}=1,2,...6$ and $b_{4}7,8;$ and
coordinates $\ _{4}^{\shortmid }u=\{\ ^{\shortmid }u^{\beta
_{4}}=(x^{i_{1}},y^{3},y^{4}=t,p_{5},p_{6},p_{7},p_{8}=E)\},$ for $E$ being
an energy type coordinate for a relativistic phase space $\ ^{\shortmid }%
\mathcal{M}.$ In similar form, we can label local coordinates and indices
for a relativistic phase space $\ \mathcal{M}=T\mathbf{V}$, when local
coordinates for $u=(x,v)$ are respective dual ones with velocity type
dependence on $v^{a},$ and the geometric objects do not involve a co-tangent
label "$^{\shortmid "}.$ Here, we note that only coordinate transforms and
diffeomorphisms are not efficient for our purposes to decouple and integrate
nonassociative modified Einstein equations. We need more general geometric
methods with N- and s-adapted frame transforms.} In s-coefficient form, we
can introduce N-elongated bases (N-/ s-adapted bases as linear N-operators): 
\begin{equation}
\ ^{\shortmid }\mathbf{e}_{\alpha _{s}}[\ ^{\shortmid }N_{\ i_{s}a_{s}}]=(\
^{\shortmid }\mathbf{e}_{i_{s}}=\ \frac{\partial }{\partial x^{i_{s}}}-\
^{\shortmid }N_{\ i_{s}a_{s}}\frac{\partial }{\partial p_{a_{s}}},\ \
^{\shortmid }e^{b_{s}}=\frac{\partial }{\partial p_{b_{s}}})\mbox{ on }\
_{s}T\mathbf{T}_{\shortmid }^{\ast }\mathbf{V,}  \label{nadapbdsc}
\end{equation}%
and, dual s-adapted bases/cobases,%
\begin{equation}
\ ^{\shortmid }\mathbf{e}^{\alpha _{s}}[\ ^{\shortmid }N_{\ i_{s}a_{s}}]=(\
^{\shortmid }\mathbf{e}^{i_{s}}=dx^{i_{s}},\ ^{\shortmid }\mathbf{e}%
_{a_{s}}=d\ p_{a_{s}}+\ ^{\shortmid }N_{\ i_{s}a_{s}}dx^{i_{s}})\mbox{ on }\
\ _{s}T^{\ast }\mathbf{T}_{\shortmid }^{\ast }\mathbf{V.}  \label{nadapbdss}
\end{equation}%
The terms of nonholonomic s-frames is motivated by the condition that, in
general, such s-bases satisfy certain anholonomy conditions, 
\begin{equation*}
\ ^{\shortmid }\mathbf{e}_{\beta _{s}}\ ^{\shortmid }\mathbf{e}_{\gamma
_{s}}-\ ^{\shortmid }\mathbf{e}_{\gamma _{s}}\ ^{\shortmid }\mathbf{e}%
_{\beta _{s}}=\ ^{\shortmid }w_{\beta _{s}\gamma _{s}}^{\tau _{s}}\
^{\shortmid }\mathbf{e}_{\tau _{s}},
\end{equation*}%
when details and computations are presented, for instance, in \cite%
{partner01,partner02} and references therein.\footnote{\label{fnshells}
General nonholonomic frames or, for instance, of type (\ref{nadapbdsc}) can
be related via frame transforms $\ ^{\shortmid }\mathbf{e}_{\gamma _{s}}=\
^{\shortmid }\mathbf{e}_{\ \gamma _{s}}^{\gamma _{s}^{\prime }}\ ^{\shortmid
}\mathbf{\partial }_{\gamma _{s}^{\prime }},$ using matrices $\ ^{\shortmid }%
\mathbf{e}_{\ \gamma _{s}}^{\gamma ;_{s}},$ to local coordinate bases $\
^{\shortmid }\mathbf{\partial }_{\gamma _{s}^{\prime }}.$ In our works, the
Einstein convention on repeating "up-low" indices is applied if a contrary
is not stated for some special cases. In a similar form, we can consider
s-splitting of dual frames $\ ^{\shortmid }\mathbf{e}^{\beta _{s}},$ when $\
^{\shortmid }\mathbf{e}^{\beta _{s}}\ ^{\shortmid }\mathbf{e}_{\gamma
_{s}}=\ \delta _{\ \ \gamma _{s}}^{\beta _{s}},$ with $\delta _{\ \ \gamma
_{s}}^{\beta _{s}}$ being the Kronecker symbol. In this work, there are used
(in the bulk) for explicit computations real co-fiber coordinates with
labels "$\ _{s}^{\shortmid }$"; in general, in nonassociative gravity \cite%
{aschieri17,partner01,partner02}, there are also considered complex
coordinates with labels $~$"$_{s}^{\shortparallel }".$} We can also consider
a conventional nonholonomic 4+4 splitting with N-connection coefficients $\
^{\shortmid }\mathbf{N}=\{\ ^{\shortmid }N_{\ ia}(x^{j},p_{b})\},$ when $%
i=1,...,4,$ and $b=5,...,8.$ We use bold letters if some geometric objects,
d-objects and/or s-objects, they are written in N-/ s-adapted form. We shall
put a left label $s$ for corresponding spaces and geometric objects if, for
instance, a phase space is enabled with a s-adapted dyadic structure as we
defined above, $\ _{s}\mathcal{M}$. Respectively, there be used terms like
d-tensor/ s-tensor, d-metric/ s-metric, d-connection/ s-connection etc.

A metric field in a phase space $\ ^{\shortmid }\mathcal{M}$ is defined as a
second rank symmetric tensor $\ ^{\shortmid }g=\{\ ^{\shortmid }g_{\alpha
\beta }\}\in TT^{\ast }V\otimes TT^{\ast }V$ with local signature $%
(+,+,+,-;+,+,+,-).$ Having prescribed a dyadic s-structure, we can express
it as s-metric $\ _{s}^{\shortmid }\mathbf{g}=\{\ ^{\shortmid }\mathbf{g}%
_{\alpha _{s}\beta _{s}}\},$ when the respective s-tensor components, i.e.
s-adapted coefficients, are parameterized 
\begin{eqnarray}
\ ^{\shortmid }g=\ _{s}^{\shortmid }\mathbf{g} &=&(h_{1}\ ^{\shortmid }%
\mathbf{g},~v_{2}\ ^{\shortmid }\mathbf{g},\ c_{3}\ ^{\shortmid }\mathbf{g,}%
c_{4}\ ^{\shortmid }\mathbf{g})\in T\mathbf{T}^{\ast }\mathbf{V}\otimes T%
\mathbf{T}^{\ast }\mathbf{V}  \label{sdm} \\
&=&\ ^{\shortmid }\mathbf{g}_{\alpha _{s}\beta _{s}}(\ _{s}^{\shortmid }u)\
\ ^{\shortmid }\mathbf{e}^{\alpha _{s}}\otimes _{s}\ ^{\shortmid }\mathbf{e}%
^{\beta _{s}}=\{\ \ ^{\shortmid }\mathbf{g}_{\alpha _{s}\beta _{s}}=(\ \
^{\shortmid }\mathbf{g}_{i_{1}j_{1}},\ \ ^{\shortmid }\mathbf{g}%
_{a_{2}b_{2}},\ \ ^{\shortmid }\mathbf{g}^{a_{3}b_{3}},\ \ ^{\shortmid }%
\mathbf{g}^{a_{4}b_{4}})\},  \notag
\end{eqnarray}%
where $\ ^{\shortmid }\mathbf{e}^{\alpha _{s}}$ can be chosen in s-adapted
form (\ref{nadapbdss}) and respective s-tensor products $\otimes _{s},$ see
geometric preliminaries in next section.

Another important nonholonomic geometric concept which is necessary for
elaborating the AFCDM is that of s-connection $\ _{s}^{\shortmid }\mathbf{D}$
with a (2+2)+(2+2) splitting (the term distinguished connection,
d-connection, can be used for a (4+4)-splitting), which is a linear
connection preserving under parallel transports a respective N-connection
splitting (\ref{ncon}). Usually we parameterize such a s-connection in the
form%
\begin{equation}
\ _{s}^{\shortmid }\mathbf{D}=(h_{1}\ ^{\shortmid }\mathbf{D},\ v_{2}\
^{\shortmid }\mathbf{D},\ c_{3}\ ^{\shortmid }\mathbf{D},\ c_{4}\
^{\shortmid }\mathbf{D})=\{\ ^{\shortmid }\mathbf{\Gamma }_{\ \ \beta
_{s}\gamma _{s}}^{\alpha _{s}}(\ _{s}^{\shortmid }u)\},  \label{sdc}
\end{equation}%
where indices split into respective dyadic components of a respective $%
h_{1},v_{2},c_{3},c_{4}$ decomposition. The geometric data $(\
_{s}^{\shortmid }\mathbf{N,}\ _{s}^{\shortmid }\mathbf{g,}\ _{s}^{\shortmid }%
\mathbf{D})$ define a N-/ s-adapted metric affine structure on $\
^{\shortmid}\mathcal{M}$ with independent $\ _{s}^{\shortmid }\mathbf{g}$
and $\ _{s}^{\shortmid }\mathbf{D.}$ In general coordinate frames/ abstract
frames, we obtain a "non-adapted" metric-affine structure which can be used
for formulating various classed of MGTs (for instance, of
Finsler-Lagrange-Hamilton type, or other type velocity/momentum depending
geometries, see review and discussions in \cite{vbv18}) on phase spaces $%
\mathcal{M}$ and $\ ^{\shortmid }\mathcal{M}$.

\subsubsection{Nonassociative canonical s-metric and s-connection variables}

The nonassociate geometric and physical models to be elaborated in this work
are derived for a nonholonomic s-adapted \textbf{star product} $\star _{s}$
which was defined in s-adapted form in our works \cite{partner01,partner02}
as a nonholonomic dyadic re-formulation of constructions with R-flux
deformations performed in from \cite{blumenhagen16,aschieri17}. Using
actions of $\ ^{\shortmid }\mathbf{e}_{i_{s}}$ (\ref{nadapbdsc}) on some
functions $\ f(x,p)$ and $\ q(x,p)$ defined on a phase space $\ _{s}\mathcal{%
M},$ we can compute always 
\begin{eqnarray}
f\star _{s}q&:=&\cdot \lbrack \mathcal{F}_{s}^{-1}(f,q)]  \label{starpn} \\
&=&\cdot \lbrack \exp (-\frac{1}{2}i\hbar (\ ^{\shortmid }\mathbf{e}%
_{i_{s}}\otimes \ ^{\shortmid }e^{i_{s}}-\ ^{\shortmid }e^{i_{s}}\otimes \
^{\shortmid }\mathbf{e}_{i_{s}})+\frac{i\mathit{\ell }_{s}^{4}}{12\hbar }%
R^{i_{s}j_{s}a_{s}}(p_{a_{s}}\ ^{\shortmid }\mathbf{e}_{i_{s}}\otimes \
^{\shortmid }\mathbf{e}_{j_{a}}-\ ^{\shortmid }\mathbf{e}_{j_{s}}\otimes
p_{a_{s}}\ ^{\shortmid }\mathbf{e}_{i_{s}}))]f\otimes q  \notag \\
&=&f\cdot q-\frac{i}{2}\hbar \lbrack (\ ^{\shortmid }\mathbf{e}_{i_{s}}f)(\
^{\shortmid }e^{i_{s}}q)-(\ ^{\shortmid }e^{i_{s}}f)(\ ^{\shortmid }\mathbf{e%
}_{i_{s}}q)]+\frac{i\mathit{\ell }_{s}^{4}}{6\hbar }%
R^{i_{s}j_{s}a_{s}}p_{a_{s}}(\ ^{\shortmid }\mathbf{e}_{i_{s}}f)(\
^{\shortmid }\mathbf{e}_{j_{s}}q)+\ldots .,  \notag
\end{eqnarray}%
where the constant $\mathit{\ell }$ characterizes the R-flux contributions
determined by an antisymmetric $R^{i_{s}j_{s}a_{s}}$ background in string
theory; the tensor product $\otimes $ can be written also in a s-adapted
form $\otimes _{s}.$ For explicit computations of star product deformations
s-adapted geometric objects and (physical) equations, there are considered
decompositions on small parameters $\hbar $ and $\kappa =\mathit{\ell }%
_{s}^{3}/6\hbar ,$ when the tensor products turn into usual multiplications
as in the third line of above formula. Such a star product (\ref{starpn})
transforms a nonholonomic dyadic phase space $\ _{s}\mathcal{M}$ into a
nonassociative one, $\ _{s}^{\star }\mathcal{M}$ enabled with s-adapted
2+2+... nonholonomic splitting.

For $_{s}\mathcal{M}\rightarrow \ _{s}^{\star }\mathcal{M},$ any symmetric
s-metric $\ _{s}^{\shortmid }\mathbf{g}$ (\ref{sdm}) transforms into a
general nonsymmetric one with respective symmetric, $\ _{\star
s}^{\shortparallel }\mathbf{g,}$ and nonsymmetric, $\ _{\star
s}^{\shortparallel }\mathfrak{g,}$ components, which can be re-defined in
respective $\ _{\star s}^{\shortmid }\mathbf{g}$ and $\ _{\star
s}^{\shortmid }\mathfrak{g}$ for respective nonholonomic frame and
coordinate systems.\footnote{%
In our works, we use labels $\ _{\star s}^{\shortparallel }$ instead of $\
_{s}^{\shortmid }\,\ $ when some geometric s-objects may contain complex
terms. For geometric constructions on (co) tangent bundles, it is always
possible to elaborate respective almost complex models when the basic
manifolds and (co) fibers are real. To study possible R-flux effects in
classical MGTs we can consider decompositions into pure real and imaginary
components and select terms with real variables. In a more general context,
we mentioned in \cite{partner01,partner02} the possibility to generalize the
constructions from \cite{aschieri17,drinf89} and define nonassociative (non)
symmetric and generalized connection s-structures on $\ _{s}^{\star }%
\mathcal{M}$ endowed also with quasi-Hopf s-structure determined by a
nonassociative algebra $\mathcal{A}_{s}^{\star }$. In abstract form, certain
labels $\ _{s}^{\shortmid }\,$ involve a procedure of transforming geometric
constructions into certain s-objects on real manifolds and bundle spaces.}
Using the star product symbol, the corresponding symmetric,$\ _{\star
s}^{\shortmid }\mathbf{g}$, and nonsymmetric , $\ _{\star }^{\shortmid }%
\mathfrak{g}_{\alpha _{s}\beta _{s}},$ components of a nonassociative metric
s-tensor on a phase space $\ _{s}\mathcal{M}$ can be parameterized in the
forms 
\begin{eqnarray}
\ _{\star s}^{\shortmid }\mathbf{g} &=&\ _{\star }^{\shortmid }\mathbf{g}%
_{\alpha _{s}\beta _{s}}\star _{s}(\ ^{\shortmid }\mathbf{e}^{\alpha
_{s}}\otimes \ ^{\shortmid }\mathbf{e}^{\beta _{s}}),\mbox{ where }\ _{\star
}^{\shortmid }\mathbf{g}(\ ^{\shortmid }\mathbf{e}_{\alpha _{s}},\
^{\shortmid }\mathbf{e}_{\beta _{s}})=\ _{\star }^{\shortmid }\mathbf{g}%
_{\alpha _{s}\beta _{s}}=\ _{\star }^{\shortmid }\mathbf{g}_{\beta
_{s}\alpha _{s}}\in \mathcal{A}_{s}^{\star }  \notag \\
\ \ _{\star }^{\shortmid }\mathfrak{g}_{\alpha _{s}\beta _{s}} &=&\ \
_{\star }^{\shortmid }\mathbf{g}_{\alpha _{s}\beta _{s}}-\kappa \mathcal{R}%
_{\quad \alpha _{s}}^{\tau _{s}\xi _{s}}\ \ ^{\shortmid }\mathbf{e}_{\xi
_{s}}\ \ _{\star }^{\shortmid }\mathbf{g}_{\beta _{s}\tau _{s}}=\ _{\star
}^{\shortmid }\mathfrak{g}_{\alpha _{s}\beta _{s}}^{[0]}+\ \ _{\star
}^{\shortmid }\mathfrak{g}_{\alpha _{s}\beta _{s}}^{[1]}(\kappa )=\ \
_{\star }^{\shortmid }\mathfrak{\check{g}}_{\alpha _{s}\beta _{s}}+\ \
_{\star }^{\shortmid }\mathfrak{a}_{\alpha _{s}\beta _{s}}.  \label{aux40a}
\end{eqnarray}%
In such formulas, the coefficients $\mathcal{R}_{\quad \alpha _{s}}^{\tau
_{s}\xi _{s}}$ are related to $R^{i_{s}j_{s}a_{s}}$ from\ (\ref{starpn}) via
algebraic relations which depend on they type of frame/ coordinated
transforms and multiplications on some real/complex coefficients. In (\ref%
{aux40a}), we distinguish $\ _{\star }^{\shortmid }\mathfrak{\check{g}}%
_{\alpha _{s}\beta _{s}}$ as the symmetric part and $\ _{\star }^{\shortmid }%
\mathfrak{a}_{\alpha _{s}\beta _{s}}$ as the anti-symmetric part, which are
defined and computed following formulas 
\begin{eqnarray}
\ _{\star }^{\shortmid }\mathfrak{\check{g}}_{\alpha _{s}\beta _{s}}:= &&%
\frac{1}{2}(\ _{\star }^{\shortmid }\mathfrak{g}_{\alpha _{s}\beta _{s}}+\
_{\star }^{\shortmid }\mathfrak{g}_{\beta _{s}\alpha _{s}})=\ _{\star
}^{\shortmid }\mathbf{g}_{\alpha _{s}\beta _{s}}-\frac{\kappa }{2}\left( 
\mathcal{R}_{\quad \beta _{s}}^{\tau _{s}\xi _{s}}\ ^{\shortmid }\mathbf{e}%
_{\xi _{s}}\ _{\star }^{\shortmid }\mathbf{g}_{\tau _{s}\alpha _{s}}+%
\mathcal{R}_{\quad \alpha _{s}}^{\tau _{s}\xi _{s}}\ ^{\shortmid }\mathbf{e}%
_{\xi _{s}}\ _{\star }^{\shortmid }\mathbf{g}_{\beta _{s}\tau _{s}}\right) 
\label{aux40b} \\
&=&\ _{\star }^{\shortmid }\mathfrak{\check{g}}_{\alpha _{s}\beta
_{s}}^{[0]}+\ _{\star }^{\shortmid }\mathfrak{\check{g}}_{\alpha _{s}\beta
_{s}}^{[1]}(\kappa ),  \notag \\
&&\mbox{ for }\ _{\star }^{\shortmid }\mathfrak{\check{g}}_{\alpha _{s}\beta
_{s}}^{[0]}=\ \ _{\star }^{\shortmid }\mathbf{g}_{\alpha _{s}\beta _{s}}%
\mbox{ and }\ \ _{\star }^{\shortmid }\mathfrak{\check{g}}_{\alpha _{s}\beta
_{s}}^{[1]}(\kappa )=-\frac{\kappa }{2}\left( \mathcal{R}_{\quad \beta
_{s}}^{\tau _{s}\xi _{s}}\ \ ^{\shortmid }\mathbf{e}_{\xi _{s}}\ \ _{\star
}^{\shortmid }\mathbf{g}_{\tau _{s}\alpha _{s}}+\mathcal{R}_{\quad \alpha
_{s}}^{\tau _{s}\xi _{s}}\ ^{\shortmid }\mathbf{e}_{\xi _{s}}\ _{\star
}^{\shortmid }\mathbf{g}_{\beta _{s}\tau _{s}}\right) ;  \notag \\
\ _{\star }^{\shortmid }\mathfrak{a}_{\alpha _{s}\beta _{s}}:= &&\frac{1}{2}%
(\ _{\star }^{\shortmid }\mathfrak{g}_{\alpha _{s}\beta _{s}}-\ _{\star
}^{\shortmid }\mathfrak{g}_{\beta _{s}\alpha _{s}})=\frac{\kappa }{2}\left( 
\mathcal{R}_{\quad \beta _{s}}^{\tau _{s}\xi _{s}}\ ^{\shortmid }\mathbf{e}%
_{\xi _{s}}\ _{\star }^{\shortmid }\mathbf{g}_{\tau _{s}\alpha _{s}}-%
\mathcal{R}_{\quad \alpha _{s}}^{\tau _{s}\xi _{s}}\ ^{\shortmid }\mathbf{e}%
_{\xi _{s}}\ _{\star }^{\shortmid }\mathbf{g}_{\beta _{s}\tau _{s}}\right)  
\notag \\
&=&\ _{\star }^{\shortmid }\mathfrak{a}_{\alpha _{s}\beta _{s}}^{[1]}(\kappa
)=\frac{1}{2}(\ _{\star }^{\shortmid }\mathfrak{g}_{\alpha _{s}\beta
_{s}}^{[1]}(\kappa )-\ _{\star }^{\shortmid }\mathfrak{g}_{\beta _{s}\alpha
_{s}}^{[1]}(\kappa )).  \label{aux40aa}
\end{eqnarray}%
Respective nonsymmetric inverse s-metrics can be also parameterized in
symmetric and nonsymmetric forms $\ _{\star }^{\shortmid }\mathfrak{g}%
^{\alpha _{s}\beta _{s}}=\ _{\star }^{\shortmid }\mathfrak{\check{g}}%
^{\alpha _{s}\beta _{s}}+\ _{\star }^{\shortmid }\mathfrak{a}^{\alpha
_{s}\beta _{s}}.$\footnote{%
Nevertheless, in such formulas, a matrix $\ _{\star }^{\shortmid }\mathfrak{%
\ \check{g}}^{\alpha _{s}\beta _{s}}$ is not the inverse to matrix $\
_{\star }^{\shortmid }\mathfrak{\check{g}}_{\alpha _{s}\beta _{s}}$ and a
matrix $\ _{\star }^{\shortmid }\mathfrak{a}^{\alpha _{s}\beta _{s}}$ is not
inverse to a matrix $\ _{\star }^{\shortmid }\mathfrak{a}_{\alpha _{s}\beta
_{s}}.$ Such values are computed not only inverting nondegenerate matrices
corresponding to some s-metrics but also using commutators and
anti-commutators, and contractions with s-tensors and s-metrics on $\
_{s}^{\star }\mathcal{M}.$ In nonassociative geometric models, we have to
apply more sophisticate procedures of star product deformations both of
differential/ integral and s-tenor calculus, see details in \cite%
{aschieri17,partner01,partner02}.}

A star product (\ref{starpn}) can be used for computing nonassociative
R-flux deformations of s-connection structures (\ref{sdc}),%
\begin{equation*}
\ _{s}^{\shortmid }\mathbf{D}=\{\ ^{\shortmid }\mathbf{\Gamma }_{\ \ \beta
_{s}\gamma _{s}}^{\alpha _{s}}(\ _{s}^{\shortmid }u)\}\rightarrow \
_{s}^{\shortmid }\mathbf{D}^{\star }=\{\ ^{\shortmid }\mathbf{\Gamma }%
_{\star \ \ \beta _{s}\gamma _{s}}^{\alpha _{s}}(\ _{s}^{\shortmid }u)\},
\end{equation*}%
and to construct $\star _{s}$-deforms of metric affine structures in
N-adapted forms, 
\begin{equation*}
(\ _{s}^{\shortmid }\mathbf{N,}\ _{s}^{\shortmid }\mathbf{g,}\
_{s}^{\shortmid }\mathbf{D})\rightarrow (\ _{s}^{\shortmid }\mathbf{N,}\ \
_{\star }^{\shortmid }\mathfrak{g}\mathbf{=}(\ _{\star }^{\shortmid }%
\mathfrak{\check{g}},\ \ _{\star }^{\shortmid }\mathfrak{a})\mathbf{,}\
_{s}^{\shortmid }\mathbf{D}^{\star }).
\end{equation*}%
For any $\ _{s}^{\shortmid }\mathbf{D}$ and $\ _{s}^{\shortmid }\mathbf{D}%
^{\star },$ we can define and compute in abstract/ coefficient forms the
corresponding star deformations of torsion s-tensors, $\ _{s}^{\shortmid }%
\mathcal{T}\rightarrow \ _{s}^{\shortmid }\widehat{\mathcal{T}}^{\star },$
and Riemann curvature s-tensors, $\ _{s}^{\shortmid }\mathcal{R}\rightarrow
\ _{s}^{\shortmid }\widehat{\mathcal{R}}^{\star },$ see details in \cite%
{partner01,partner02,lbsvevv22}. The nonholonomic structure can be
prescribed for a associative/commutative N-connection $\ _{s}^{\shortmid }%
\mathbf{N}$ not involving star deformations of type $\ _{s}^{\shortmid }%
\mathbf{N}\rightarrow \ _{s}^{\shortmid }\mathbf{N}$ $^{\star },$ which
allows to compute nonassociative R-flux deformations of fundamental
geometric s-objects (curvature/ torsion s-tensors etc.) in more simple
forms. Here we note that in \cite{blumenhagen16,aschieri17} there were
considered star product deformations without N-connection structure and for
(pseudo) Riemannian data, $(\ ^{\shortmid }g, \ ^{\shortmid
}\nabla)\rightarrow (\ \ _{\star }^{\shortmid }\mathfrak{g}= (\ \
_{\star}^{\shortmid }\mathfrak{\check{g}},\ \ _{\star }^{\shortmid }%
\mathfrak{a}), \ ^{\shortmid }\nabla ^{\star }),$ where $\ ^{\shortmid
}\nabla$ and $\ ^{\shortmid }\nabla ^{\star }$ are respective Levi-Civita,
LC, connections. Unfortunately, it is not possible to decouple nonholonomic/
nonassociative (modified) Einstein equations using $\ ^{\shortmid }\nabla $
and/or $\ ^{\shortmid }\nabla ^{\star }.$ The main result of \cite{partner02}
consisted in a proof that we can decouple and integrate such important
systems of nonlinear PDEs if we use some correspondingly defined auxiliary
s-connections, called respectively the canonical s-connections, $\
_{s}^{\shortmid }\widehat{\mathbf{D}}$ and $\ _{s}^{\shortmid }\widehat{%
\mathbf{D}}^{\star }.$ Let us explain how such a linear connection can be
defined in a nonassociative form which allow respective nonholonomic
constraints to $\ _{s}^{\shortmid }\widehat{\mathbf{D}}$ and/or $\
^{\shortmid }\nabla ^{\star },\ ^{\shortmid }\nabla :$

For any geometric data $(\ _{s}^{\shortmid }\mathbf{N,}\ _{\star s}^{\
\shortmid }\mathbf{g}, \ _{s}^{\shortmid }\mathbf{D}^{\star }),$ we can
define a canonical s-connection 
\begin{equation}
\ \ _{s}^{\shortmid }\widehat{\mathbf{D}}^{\star }=(h_{1}\ ^{\shortmid }%
\widehat{\mathbf{D}}^{\star },\ v_{2}\ ^{\shortmid }\widehat{\mathbf{D}}%
^{\star },\ c_{3}\ ^{\shortmid }\widehat{\mathbf{D}}^{\star },\ c_{4}\
^{\shortmid }\widehat{\mathbf{D}}^{\star })=\ ^{\shortmid }\nabla ^{\star
}+\ _{\star s}^{\shortmid }\widehat{\mathbf{Z}},  \label{candistrnas}
\end{equation}%
where the canonical distortion s-tensor $\ _{\star s}^{\shortmid }\widehat{%
\mathbf{Z}}[\ ^{\shortmid }\widehat{\mathcal{T}}^{\star }[\ _{s}^{\shortmid }%
\mathbf{N,}\ _{\star s}^{\ \shortmid }\mathbf{g}]]$ is an algebraic
functional of the canonical s-torsion $\ ^{\shortmid }\widehat{\mathcal{T}}%
^{\star },$ all defined by the conditions {\small 
\begin{equation}
(\ _{\star s}^{\shortmid }\mathbf{g,\ _{s}^{\shortmid }N})\rightarrow
\left\{ 
\begin{array}{cc}
\ _{\star }^{\shortmid }\mathbf{\nabla :}\  & 
\begin{array}{c}
\fbox{\ $\ \ _{\star }^{\shortmid }\mathbf{\nabla }\ _{\star s}^{\ \shortmid
}\mathbf{g}=0$;\ $_{\nabla }^{\shortmid }\mathcal{T}^{\star }=0$}%
\mbox{\ star
LC-connection}; \\ 
\end{array}
\\ 
\ _{s}^{\shortmid }\widehat{\mathbf{D}}^{\star }: & \fbox{$%
\begin{array}{c}
\ _{s}^{\shortmid }\widehat{\mathbf{D}}^{\star }\ _{\star s}^{\ \shortmid }%
\mathbf{g}=0;\ h_{1}\ ^{\shortmid }\widehat{\mathcal{T}}^{\star }=0,v_{2}\
^{\shortmid }\widehat{\mathcal{T}}^{\star }=0,c_{3}\ ^{\shortmid }\widehat{%
\mathcal{T}}^{\star }=0,c_{4}\ ^{\shortmid }\widehat{\mathcal{T}}^{\star }=0,
\\ 
h_{1}v_{2}\ ^{\shortmid }\widehat{\mathcal{T}}^{\star }\neq 0,h_{1}c_{s}\
^{\shortmid }\widehat{\mathcal{T}}^{\star }\neq 0,v_{2}c_{s}\ ^{\shortmid }%
\widehat{\mathcal{T}}^{\star }\neq 0,c_{3}c_{4}\ ^{\shortmid }\widehat{%
\mathcal{T}}^{\star }\neq 0,%
\end{array}%
$}\mbox{ canonical  s-connection }.%
\end{array}%
\right.  \label{twoconsstar}
\end{equation}%
} We note that in the definition both of $\ _{\star }^{\shortmid }\mathbf{%
\nabla }$ and $\ _{s}^{\shortmid }\widehat{\mathbf{D}}^{\star }$ we use the
s-tensor $_{\star s}^{\ \shortmid }\mathbf{g.}$\footnote{%
For defining linear connections, we can work directly with the nonsymmetric
metric $\ _{\star }^{\shortmid }\mathfrak{g}_{\alpha _{s}\beta _{s}}$ but
such a choice result in more strong coupling of tensor s-object which make
the procedure of constructing parametric solutions more sophisticate than
this variant (\ref{twoconsstar}) based on $_{\star s}^{\ \shortmid }\mathbf{%
g.}$ Working only up to $\kappa $-linear terms, all classes of canonical
s-connections are equivalent for those configurations when $\ _{\star
}^{\shortmid }\mathfrak{g}_{\alpha _{s}\beta _{s}}=\ _{\star }^{\shortmid }%
\mathfrak{\check{g}}_{\alpha _{s}\beta _{s}}+\ \ _{\star }^{\shortmid }%
\mathfrak{a}_{\alpha _{s}\beta _{s}},$ with $\ _{\star }^{\shortmid }%
\mathfrak{\check{g}}_{\alpha _{s}\beta _{s}}=\ _{\star }^{\ \shortmid }%
\mathbf{g}\ _{\alpha _{s}\beta _{s}}$ and $\ _{\star }^{\shortmid }\mathfrak{%
a}_{\alpha _{s}\beta _{s}|\kappa \rightarrow 0}\rightarrow 0.$ It should be
noted that there are non-vanishing terms of type $\ _{\star }^{\shortmid }%
\mathfrak{a}_{\alpha _{s}\beta _{s}}(\kappa \neq 0),$ but for the
nonholonomic structure can be chosen in such forms that for large classes of
exact/parametric solutions the nonsymmetric parts of s-metrics are decoupled
as induces and R-flux deformation terms. After certain classes of physically
important $\kappa $-parametric solutions are constructed in explicitly form,
we can consider general nonassociative frame and coordinate transforms. So,
prescribing certain nonholonomy conditions we do not violate any general
covariance in (non) associative/commutative forms but certain s-adapted and
coordinate systems are more important and convenient for decoupling and
solving nonlinear systems of PDEs.} The coefficient s-adapted formulas are
provided in \cite{partner01,partner02}. In this work, we shall provide only
certain coefficient formulas which are used for explicit computations in GIF
and QGIF re-definition of geometric s-objects.

Nonassociative and/or associative LC-configurations can be extracted if we
impose additional zero s-torsion conditions,%
\begin{eqnarray}
\ _{\star s}^{\shortmid }\widehat{\mathbf{Z}} &=&0,%
\mbox{ which is
equivalent to }\ _{s}^{\shortmid }\widehat{\mathbf{D}}_{\mid \
_{s}^{\shortmid }\widehat{\mathbf{T}}=0}^{\star }=\ \ _{\star }^{\shortmid
}\nabla ;  \label{lccondnonass} \\
\ _{s}^{\shortmid }\widehat{\mathbf{Z}} &=&0,\mbox{ which is
equivalent to }\ _{s}^{\shortmid }\widehat{\mathbf{D}}_{\mid \
_{s}^{\shortmid }\widehat{\mathbf{T}}=0}=\ \ ^{\shortmid }\nabla ,%
\mbox{ for
terms with }\kappa ^{0}.  \notag
\end{eqnarray}%
Such conditions result in additional nonholonomic constraints and PDEs which
can be solved for some special classes of generating/integration functions
and effective sources in already constructed solutions with a general $\
_{s}^{\shortmid }\widehat{\mathbf{D}}^{\star }.$ We note that all type of
metrics on $\ _{s}^{\star }\mathcal{M}$, and related s-metrics $\ _{\star
s}^{\ \shortmid }\mathbf{g}$, subjected/ or not to some conditions of type (%
\ref{lccondnonass}) contain certain nonzero anholonomy coefficients of frame
structures. In general, respective symmetric and nonsymmetric s-metrics can
be written in local coordinate forms as generic off-diagonal matrices.%
\footnote{%
Our (non) associative and nonholonomic geometric approach and the AFCDM can
be generalized for all spaces with nontrivial torsion structure including
nonassociative/ noncommutative and relativistic generalizations of the
Riemann-Cartan-Weyl geometry, anti-symmetric H-field torsion contributions
in (super) string theory, generalized Finsler-Lagrange-Hamilton theories
etc. We can formulate various models of gravitational and matter field
theories on (non) associative phase spaces enabled with star product R-flux
deformations. In such theories the canonical s-torsion $\ ^{\shortmid }%
\widehat{\mathcal{T}}^{\star }$ from (16) has to be generalized to
respective nontrivial torsion structure (we can also consider a nonzero
nonmetricity field, which results in more sophisticate systems of nonlinear
PDEs. This adds both technical difficulties in constructing exact/
parametric solutions and results in other types of physical effects etc..
Still, the AFCDM can be extended and applied to solving such nonlinear
systems. Such (non) commutative geometric and physical models are reviewed
and discussed in [65] (see Appendix B in the preprint version of that work).
For simplicity, in this paper, we consider only canonical s-connection
structures (when the nonholonomic structure and N-coefficients induce the
nontrivial torsion coefficients) which can be nonholonomically constrained
to LC configurations (17).}

To define and compute geometric and physical objects on a nonassociative
phase space $\ _{s}^{\star }\mathcal{M}$ we formulated (see details in \cite%
{partner01,partner02,partner03,lbsvevv22,lbdssv22} when we preserve the
number 2) this

\textbf{Convention 2 } :\ The commutative and nonassociative geometric data
derived for a star product $\star _{s}$ (\ref{starpn}), can be expressed in
such abstract/symbolic s-adapted forms: 
\begin{equation}
\begin{array}{ccc}
\fbox{$(\star _{N},\ \ \mathcal{A}_{N}^{\star },\ _{\star }^{\shortmid }%
\mathbf{g,\ _{\star }^{\shortmid }\mathfrak{g,}\ ^{\shortmid }N},\mathbf{\
^{\shortmid }e}_{\alpha }\mathbf{,\ ^{\shortmid }}\widehat{\mathbf{D}}%
^{\star })$} & \Leftrightarrow & \fbox{$(\star _{s},\ \ \mathcal{A}%
_{s}^{\star },\ _{\star s}^{\shortmid }\mathbf{g,}\ _{\star s}^{\shortmid }%
\mathbf{\mathfrak{g,}}\ \ _{s}^{\shortmid }\mathbf{N},\mathbf{\ ^{\shortmid
}e}_{\alpha _{s}}\mathbf{,}\ _{s}^{\shortmid }\widehat{\mathbf{D}}^{\star })$%
} \\ 
& \Uparrow &  \\ 
\fbox{$(\mathbf{\ ^{\shortmid }g,\ ^{\shortmid }N},\mathbf{\ ^{\shortmid }e}%
_{\alpha }\mathbf{,\ ^{\shortmid }}\widehat{\mathbf{D}})$} & \Leftrightarrow
& \fbox{$(\ \ _{s}^{\shortmid }\mathbf{g,\ }\ _{s}^{\shortmid }\mathbf{N},%
\mathbf{\ \ ^{\shortmid }e}_{\alpha _{s}}\mathbf{,}\ _{s}^{\shortmid }%
\widehat{\mathbf{D}}).$}%
\end{array}
\label{conv2s}
\end{equation}%
for certain canonical distortions $\ ^{\shortmid } \widehat{\mathbf{D}}%
^{\star }= \ _{\star }^{\shortmid }\nabla +\ ^{\shortmid }\widehat{\mathbf{Z}%
}^{\star },$ for a nonholonomic splitting 4+4, and $\ _{s}^{\shortmid }%
\widehat{\mathbf{D}}^{\star }=\ _{\star }^{\shortmid }\nabla + \
_{s}^{\shortmid }\widehat{\mathbf{Z}}^{\star },$ for a nonholonomic
s-splitting. Following the Convention 2, we can define and compute star
product deformations of fundamental geometric s-objects,%
\begin{eqnarray}
\ _{s}^{\shortmid }\mathcal{T} &\rightarrow &\ _{s}^{\shortmid }\widehat{%
\mathcal{T}}^{\star }=\{\ ^{\shortmid }\widehat{\mathbf{T}}_{\ \star \beta
_{s}\gamma _{s}}^{\alpha _{s}}\},%
\mbox{ nonassociative  canonical
s-torsion };\   \label{mafgeomobn} \\
\ _{s}^{\shortmid }\mathcal{R} &\rightarrow &\ _{s}^{\shortmid }\widehat{%
\mathcal{R}}^{\star }=\{\ ^{\shortmid }\widehat{\mathbf{R}}_{\ \beta
_{s}\gamma _{s}\delta _{s}}^{\star \alpha _{s}}\},%
\mbox{nonassociative
canonical Riemannian s-curvature };  \notag \\
\ _{s}^{\shortmid }\mathcal{R}ic &\rightarrow &\ _{s}^{\shortmid }\widehat{%
\mathcal{R}}ic^{\star }=\{\ ^{\shortmid }\widehat{\mathbf{R}}_{\ \beta
_{s}\gamma _{s}}^{\star }:=\ ^{\shortmid }\widehat{\mathbf{R}}_{\ \beta
_{s}\gamma _{s}\alpha _{s}}^{\star \alpha _{s}}\neq \ ^{\shortmid }\widehat{%
\mathbf{R}}_{\ \gamma _{s}\beta _{s}}^{\star }\},%
\mbox{ nonassociative
canonical Ricci s-tensor};  \notag \\
\ _{s}^{\shortmid }\mathcal{R}sc &\rightarrow &\ _{s}^{\shortmid }\widehat{%
\mathcal{R}}sc^{\star }=\{\ ^{\shortmid }\mathbf{g}^{\beta _{s}\gamma _{s}}\
^{\shortmid }\widehat{\mathbf{R}}_{\ \beta _{s}\gamma _{s}}^{\star }\},%
\mbox{  nonassociative  canonical  Riemannian scalar };  \notag \\
\ _{s}^{\shortmid }\mathcal{Q} &\rightarrow &\ _{s}^{\shortmid }\mathcal{Q}%
^{\star }=\{\ ^{\shortmid }\widehat{\mathbf{Q}}_{\gamma _{s}\alpha _{s}\beta
_{s}\ }^{\star }=\ ^{\shortmid }\widehat{\mathbf{D}}_{\gamma _{s}}^{\star }\
_{\star }^{\shortmid }\mathbf{g}_{\alpha _{s}\beta _{s}}\},%
\mbox{ zero
nonassociative  canonical nonmetricity s-tensor }.  \notag
\end{eqnarray}%
For instance, the nonassociative canonical Riemann s-tensor$\mathbf{\mathbf{%
\mathbf{\mathbf{\ ^{\shortmid }}}}}\widehat{\mathcal{\Re }}_{\quad }^{\star
}=\{\ ^{\shortmid }\widehat{\mathcal{\Re }}_{\quad \alpha _{s}\beta
_{s}\gamma _{s}}^{\star \mu _{s}}\}$ from (\ref{mafgeomobn}) can be defined
and computed for the data $(\ _{\star s}^{\shortmid }\mathfrak{g=\{\ \
_{\star }^{\shortmid }\mathbf{g}_{\alpha _{s}\beta _{s}}\}},\
_{s}^{\shortmid }\widehat{\mathbf{D}}^{\star }=\{\ ^{\shortmid }\widehat{%
\mathbf{\Gamma }}_{\star \alpha _{s}\beta _{s}}^{\gamma _{s}}\})$ and
written in a form with $\kappa $-linear decomposition, 
\begin{eqnarray}
\mathbf{\mathbf{\mathbf{\mathbf{\ ^{\shortmid }}}}}\widehat{\mathbf{R}}%
_{\quad \alpha _{s}\beta _{s}\gamma _{s}}^{\star \mu _{s}} &=&\mathbf{%
\mathbf{\mathbf{\mathbf{\ _{1}^{\shortmid }}}}}\widehat{\mathbf{R}}_{\quad
\alpha _{s}\beta _{s}\gamma _{s}}^{\star \mu _{s}}+\mathbf{\mathbf{\mathbf{%
\mathbf{\ _{2}^{\shortmid }}}}}\widehat{\mathbf{R}}_{\quad \alpha _{s}\beta
_{s}\gamma _{s}}^{\star \mu _{s}},\mbox{ where }  \label{nadriemhopfcan} \\
\mathbf{\mathbf{\mathbf{\mathbf{\ _{1}^{\ \shortmid }}}}}\widehat{\mathbf{R}}%
_{\quad \alpha _{s}\beta _{s}\gamma _{s}}^{\star \mu _{s}} &=&\ ^{\shortmid }%
\mathbf{e}_{\gamma _{s}}\ ^{\shortmid }\widehat{\Gamma }_{\star \alpha
_{s}\beta _{s}}^{\mu _{s}}-\ ^{\shortmid }\mathbf{e}_{\beta _{s}}\
^{\shortmid }\widehat{\Gamma }_{\star \alpha _{s}\gamma _{s}}^{\mu }+\
^{\shortmid }\widehat{\Gamma }_{\star \nu _{s}\tau _{s}}^{\mu _{s}}\star
_{s}(\delta _{\ \gamma _{s}}^{\tau _{s}}\ ^{\shortmid }\widehat{\Gamma }%
_{\star \alpha _{s}\beta _{s}}^{\nu _{s}}-\delta _{\ \beta _{s}}^{\tau
_{s}}\ ^{\shortmid }\widehat{\Gamma }_{\star \alpha _{s}\gamma _{s}}^{\nu
_{s}})+\ ^{\shortmid }w_{\beta _{s}\gamma _{s}}^{\tau _{s}}\star _{s}\
^{\shortmid }\widehat{\Gamma }_{\star \alpha _{s}\tau _{s}}^{\mu _{s}}, 
\notag \\
\ _{2}^{\shortmid }\widehat{\mathbf{R}}_{\quad \alpha _{s}\beta _{s}\gamma
_{s}}^{\star \mu _{s}} &=&i\kappa \ ^{\shortmid }\widehat{\Gamma }_{\star
\nu _{s}\tau _{s}}^{\mu _{s}}\star _{s}(\mathcal{R}_{\quad \gamma
_{s}}^{\tau _{s}\xi _{s}}\ ^{\shortmid }\mathbf{e}_{\xi _{s}}\ ^{\shortmid }%
\widehat{\Gamma }_{\star \alpha _{s}\beta _{s}}^{\nu _{s}}-\mathcal{R}%
_{\quad \beta _{s}}^{\tau _{s}\xi _{s}}\ ^{\shortmid }\mathbf{e}_{\xi _{s}}\
^{\shortmid }\widehat{\Gamma }_{\star \alpha _{s}\gamma _{s}}^{\nu _{s}}). 
\notag
\end{eqnarray}%
Such formulas are provided in abstract from for LC-configurations in \cite%
{blumenhagen16,aschieri17} and generalized for nonholonomic canonical
s-connections in \cite{partner01,partner02,partner03,lbsvevv22,lbdssv22}. We
use the results of those works in this article when the geometric and
quantum information objects depend additionally on a geometric/ information
flow $\tau $-parameter.\footnote{%
For developing the AFCDM and generating nonassociative solutions in certain
general forms, we have to work with nonassociative covector bundles enabled
with nonholonomic frames adapted to nonlinear connection and dyadic
structures etc. Only using the group of diffeomorphism is not possible to
decouple nonlinear systems of PDEs and construct off-diagonal solutions and
that is why the nonassociative diffeomorphic structure is not analyzed in
this work (it can be chosen to be the same as in \cite%
{blumenhagen16,aschieri17} but with frame/coordinate transforms adapted
N-connections). In this paper, we do not repeat all motivations and methods
from those works but concentrate on elaborating a new nonassociative quantum
geometric and information, QGIF, theory. The nonassociative BH and wormhole
solutions constructed in section \ref{sec5} are considered for possible
applications of nonassociative QGIF.}

\subsubsection{Parametric decomposition of fundamental nonassociative
geometric objects}

Let us explain how $\kappa $-parametric decompositions of fundamental
geometric s-objects can be derived from respective decompositions of
s-metrics and canonical s-connections. Considering a conventional parametric
decomposition of the star canonical s-connection in (\ref{nadriemhopfcan}), 
\begin{equation}
\ ^{\shortmid }\widehat{\mathbf{\Gamma }}_{\star \alpha _{s}\beta
_{s}}^{\gamma _{s}}=\ _{[0]}^{\shortmid }\widehat{\mathbf{\Gamma }}_{\star
\alpha _{s}\beta _{s}}^{\nu _{s}}+i\kappa \ _{[1]}^{\shortmid }\widehat{%
\mathbf{\Gamma }}_{\star \alpha _{s}\beta _{s}}^{\nu _{s}}=\
_{[00]}^{\shortmid }\widehat{\Gamma }_{\ast \alpha _{s}\beta _{s}}^{\nu
_{s}}+\ _{[01]}^{\shortmid }\widehat{\Gamma }_{\ast \alpha _{s}\beta
_{s}}^{\nu _{s}}(\hbar )+\ _{[10]}^{\shortmid }\widehat{\Gamma }_{\ast
\alpha _{s}\beta _{s}}^{\nu _{s}}(\kappa )+\ _{[11]}^{\shortmid }\widehat{%
\Gamma }_{\ast \alpha _{s}\beta _{s}}^{\nu _{s}}(\hbar \kappa )+O(\hbar
^{2},\kappa ^{2}...),  \label{paramscon}
\end{equation}%
we compute respective parametric decompositions of the nonassociative
canonical curvature tensor,%
\begin{equation*}
\ ^{\shortmid }\widehat{\mathbf{R}}_{\quad \alpha _{s}\beta _{s}\gamma
_{s}}^{\star \mu _{s}}=\mathbf{\mathbf{\mathbf{\mathbf{\ }}}}\
_{[00]}^{\shortmid }\widehat{\mathbf{R}}_{\quad \alpha _{s}\beta _{s}\gamma
_{s}}^{\star \mu _{s}}+\mathbf{\mathbf{\mathbf{\mathbf{\ }}}}\
_{[01]}^{\shortmid }\widehat{\mathbf{R}}_{\quad \alpha _{s}\beta _{s}\gamma
_{s}}^{\star \mu _{s}}(\hbar )+\mathbf{\mathbf{\mathbf{\mathbf{\ }}}}\
_{[10]}^{\shortmid }\widehat{\mathbf{R}}_{\quad \alpha _{s}\beta _{s}\gamma
_{s}}^{\star \mu _{s}}(\kappa )+\ _{[11]}^{\shortmid }\widehat{\mathbf{R}}%
_{\quad \alpha _{s}\beta _{s}\gamma _{s}}^{\star \mu _{s}}(\hbar \kappa
)+O(\hbar ^{2},\kappa ^{2},...).
\end{equation*}

Contracting the first and forth indices in s-adapted formulas (\ref%
{nadriemhopfcan}), we define the nonassociative canonical Ricci s-tensor, 
\begin{eqnarray*}
\mathbf{\mathbf{\mathbf{\mathbf{\ _{s}^{\shortmid }}}}}\widehat{\mathcal{\Re 
}}ic^{\star } &=&\mathbf{\mathbf{\mathbf{\mathbf{\ ^{\shortmid }}}}}\widehat{%
\mathbf{\mathbf{\mathbf{\mathbf{R}}}}}ic_{\alpha _{s}\beta _{s}}^{\star
}\star _{s}(\ \mathbf{^{\shortmid }e}^{\alpha _{s}}\otimes _{\star s}\ 
\mathbf{^{\shortmid }e}^{\beta _{s}}),\mbox{ where } \\
&&\mathbf{\mathbf{\mathbf{\mathbf{\ ^{\shortmid }}}}}\widehat{\mathbf{%
\mathbf{\mathbf{\mathbf{R}}}}}ic_{\alpha _{s}\beta _{s}}^{\star }:=\
_{s}^{\shortmid }\widehat{\mathcal{\Re }}ic^{\star }(\ ^{\shortmid }\mathbf{e%
}_{\alpha _{s}},\ ^{\shortmid }\mathbf{e}_{\beta _{s}})=\mathbf{\langle }\ 
\mathbf{\mathbf{\mathbf{\mathbf{\ ^{\shortmid }}}}}\widehat{\mathbf{\mathbf{%
\mathbf{\mathbf{R}}}}}ic_{\mu _{s}\nu _{s}}^{\star }\star _{s}(\ \mathbf{%
^{\shortmid }e}^{\mu _{s}}\otimes _{\star _{s}}\ ^{\shortmid }\mathbf{e}%
^{\nu _{s}}),\mathbf{\mathbf{\ }\ ^{\shortmid }\mathbf{e}}_{\alpha _{s}}%
\mathbf{\otimes _{\star s}\ ^{\shortmid }\mathbf{e}}_{\beta _{s}}\mathbf{%
\rangle }_{\star _{s}}.
\end{eqnarray*}%
Using (\ref{paramscon}), we express 
\begin{eqnarray}
&\ ^{\shortmid }\widehat{\mathbf{R}}ic_{\alpha _{s}\beta _{s}}^{\star }:= &%
\mathbf{\mathbf{\mathbf{\mathbf{\ ^{\shortmid }}}}}\widehat{\mathcal{\Re }}%
_{\quad \alpha _{s}\beta _{s}\mu _{s}}^{\star \mu _{s}}=\ _{[00]}^{\shortmid
}\widehat{\mathbf{\mathbf{\mathbf{\mathbf{R}}}}}ic_{\alpha _{s}\beta
_{s}}^{\star }+\mathbf{\mathbf{\mathbf{\mathbf{\ \ }}}}_{[01]}^{\shortmid }%
\widehat{\mathbf{\mathbf{\mathbf{\mathbf{R}}}}}ic_{\alpha _{s}\beta
_{s}}^{\star }(\hbar )+\mathbf{\mathbf{\mathbf{\mathbf{\ }}}}%
_{[10]}^{\shortmid }\widehat{\mathbf{\mathbf{\mathbf{\mathbf{R}}}}}%
ic_{\alpha _{s}\beta _{s}}^{\star }(\kappa )  \notag \\
&&+\mathbf{\mathbf{\mathbf{\mathbf{\ }}}}_{[11]}^{\shortmid }\widehat{%
\mathbf{\mathbf{\mathbf{\mathbf{R}}}}}ic_{\alpha _{s}\beta _{s}}^{\star
}(\hbar \kappa )+O(\hbar ^{2},\kappa ^{2},...),  \notag \\
&\mbox{where}&\ _{[00]}^{\shortmid }\widehat{\mathbf{R}}ic_{\alpha _{s}\beta
_{s}}^{\star }=\ _{[00]}^{\shortmid }\widehat{\mathcal{\Re }}_{\quad \alpha
_{s}\beta _{s}\mu _{s}}^{\star \mu _{s}}\mathbf{\mathbf{\mathbf{\mathbf{\ ,}}%
}}\ _{[01]}^{\shortmid }\mathbf{\mathbf{\mathbf{\mathbf{\widehat{\mathbf{%
\mathbf{\mathbf{\mathbf{R}}}}}}}}}ic_{\alpha _{s}\beta _{s}}^{\star }=\
_{[01]}^{\shortmid }\mathbf{\mathbf{\mathbf{\mathbf{\widehat{\mathcal{\Re }}}%
}}}_{\quad \alpha _{s}\beta _{s}\mu _{s}}^{\star \mu _{s}},
\label{driccicanonstar1} \\
&&\ _{[10]}^{\shortmid }\mathbf{\mathbf{\mathbf{\mathbf{\widehat{\mathbf{%
\mathbf{\mathbf{\mathbf{R}}}}}}}}}ic_{\alpha _{s}\beta _{s}}^{\star }=\
_{[10]}^{\shortmid }\mathbf{\mathbf{\mathbf{\mathbf{\widehat{\mathcal{\Re }}}%
}}}_{\quad \alpha _{s}\beta _{s}\mu _{s}}^{\star \mu _{s}},\
_{[11]}^{\shortmid }\widehat{\mathbf{\mathbf{\mathbf{\mathbf{R}}}}}%
ic_{\alpha _{s}\beta _{s}}^{\star }=\ _{[11]}^{\shortmid }\mathbf{\mathbf{%
\mathbf{\mathbf{\widehat{\mathcal{\Re }}}}}}_{\quad \alpha _{s}\beta _{s}\mu
_{s}}^{\star \mu _{s}}.  \notag
\end{eqnarray}%
It should be noted that such canonical Ricci s-tensors are not symmetric for
general (non) commutative and nonassociative cases even for terms
proportional to $\hbar ^{0}$ and/or $\kappa ^{0}$. This is a consequence of
nonholonomic structure.

We can compute abstract h1-v2-c3-c4 decompositions, or their s-adapted index
forms for formulas (\ref{nadriemhopfcan}) and (\ref{driccicanonstar1}) which
results in cumbersome formulas on tenths of pages. Fortunately, all results
can be found alternatively applying the Convention 2 to the formulas for the
nonasssociative LC-connection and respective fundamental geometric objects
provided in local coordinate bases in \cite{blumenhagen16,aschieri17} and 
\cite{partner01,partner02}.

Contracting the indices of (\ref{driccicanonstar1}) with $\
_{\star}^{\shortmid }\mathfrak{g}^{\mu _{s}\nu _{s}},$ we define and compute
the nonassociative nonholonomic canonical Ricci scalar curvature:%
\begin{eqnarray}
&\ _{s}^{\shortmid }\widehat{\mathbf{R}}sc^{\star }:= &\ _{\star
}^{\shortmid }\mathfrak{g}^{\mu _{s}\nu _{s}}\mathbf{\mathbf{\mathbf{\mathbf{%
\ ^{\shortmid }}}}}\widehat{\mathbf{\mathbf{\mathbf{\mathbf{R}}}}}ic_{\mu
_{s}\nu _{s}}^{\star }=\left( \ _{\star }^{\shortmid }\mathfrak{\check{g}}%
^{\mu _{s}\nu _{s}}+\ _{\star }^{\shortmid }\mathfrak{a}^{\mu _{s}\nu
_{s}}\right) \left( \mathbf{\mathbf{\mathbf{\mathbf{\ ^{\shortmid }}}}}%
\widehat{\mathbf{\mathbf{\mathbf{\mathbf{R}}}}}ic_{(\mu _{s}\nu
_{s})}^{\star }+\mathbf{\mathbf{\mathbf{\mathbf{\ ^{\shortmid }}}}}\widehat{%
\mathbf{\mathbf{\mathbf{\mathbf{R}}}}}ic_{[\mu _{s}\nu _{s}]}^{\star }\right)
\notag \\
&&=\ _{s}^{\shortmid }\widehat{\mathbf{\mathbf{\mathbf{\mathbf{R}}}}}%
ss^{\star }+\ _{s}^{\shortmid }\widehat{\mathbf{\mathbf{\mathbf{\mathbf{R}}}}%
}sa^{\star },  \notag \\
&\mbox{where}& \ _{s}^{\shortmid }\widehat{\mathbf{\mathbf{\mathbf{\mathbf{R}%
}}}}ss^{\star }=:\ _{\star }^{\shortmid }\mathfrak{\check{g}}^{\mu _{s}\nu
_{s}}\mathbf{\mathbf{\mathbf{\mathbf{\ ^{\shortmid }}}}}\widehat{\mathbf{%
\mathbf{\mathbf{\mathbf{R}}}}}ic_{(\mu _{s}\nu _{s})}^{\star }\mbox{ and }\
_{s}^{\shortmid }\widehat{\mathbf{\mathbf{\mathbf{\mathbf{R}}}}}sa^{\star
}:=\ _{\star }^{\shortmid }\mathfrak{a}^{\mu _{s}\nu _{s}}\mathbf{\mathbf{%
\mathbf{\mathbf{\ ^{\shortmid }}}}}\widehat{\mathbf{\mathbf{\mathbf{\mathbf{R%
}}}}}ic_{[\mu _{s}\nu _{s}]}^{\star }.  \label{ricciscsymnonsym}
\end{eqnarray}%
In such formulas, the respective symmetric $\left( ...\right) $ and
anti-symmetric $\left[ ...\right] $ operators are defined using the multiple 
$1/2$ when, for instance, $\ ^{\shortmid }\widehat{\mathbf{R}}ic_{\mu
_{s}\nu _{s}}^{\star }=\mathbf{\mathbf{\mathbf{\mathbf{\ ^{\shortmid }}}}}%
\widehat{\mathbf{\mathbf{\mathbf{\mathbf{R}}}}}ic_{(\mu _{s}\nu
_{s})}^{\star }+\mathbf{\mathbf{\mathbf{\mathbf{\ ^{\shortmid }}}}}\widehat{%
\mathbf{\mathbf{\mathbf{\mathbf{R}}}}}ic_{[\mu _{s}\nu _{s}]}^{\star }.$ We
note that the scalar functional \newline
$\ _{s}^{\shortmid }\widehat{\mathbf{R}}sc^{\star }[\ \ _{\star }^{\shortmid
}\mathfrak{\check{g}}_{\alpha _{s}\beta _{s}},\ \ _{\star }^{\shortmid }%
\mathfrak{a}_{\alpha _{s}\beta _{s}}]$ is different from the scalar
functional $\ _{s}^{\shortmid }\widehat{\mathbf{R}}sc_{\ _{\star
}^{\shortmid }\mathbf{\nabla }}^{\star }[\ \ _{\star }^{\shortmid }\mathfrak{%
\check{g}}_{\alpha _{s}\beta _{s}},\ \ _{\star }^{\shortmid }\mathfrak{a}%
_{\alpha _{s}\beta _{s}}]$ determined by $\ _{\star }^{\shortmid }\mathbf{%
\nabla }$. Such values are related by a distortion tensor $\ _{s}^{\shortmid
}\widehat{\mathbf{Z}}sc^{\star }[\ \ _{\star }^{\shortmid }\mathfrak{\check{g%
}}_{\alpha _{s}\beta _{s}},\ \ _{\star }^{\shortmid }\mathfrak{a}_{\alpha
_{s}\beta _{s}}]$ which can be computed by introducing $\ _{\star
}^{\shortmid }\nabla +\ _{s}^{\shortmid }\widehat{\mathbf{Z}}^{\star }$ into
(\ref{driccicanonstar1}) and (\ref{ricciscsymnonsym}).

Following abstract geometric principles formulated in \cite{misner}, we can
define for the canonical s-connection $\ _{s}^{\shortmid }\widehat{\mathbf{D}%
}^{\star }$ the nonassociatve and noncommutative modified vacuum Einstein
equations, 
\begin{equation}
\ ^{\shortmid }\widehat{\mathbf{R}}ic_{\alpha _{s}\beta _{s}}^{\star }-\frac{%
1}{2}\ _{\star }^{\shortmid }\mathfrak{g}_{\alpha _{s}\beta _{s}}\
_{s}^{\shortmid }\widehat{\mathbf{R}}sc^{\star }=\ _{s}^{\shortmid }\lambda
\ _{\star }^{\shortmid }\mathfrak{g}_{\alpha _{s}\beta _{s}}.
\label{nonassocdeinst1}
\end{equation}%
We note that such equations are for a nonassociative s-shell phase space $\
_{s}\mathcal{M}$ and involve at least a nontrivial at least one shell
cosmological constant ($\ _{s}^{\shortmid }\lambda \neq 0$ for any $s,$ or
some shells). They can be also derived by a canonical N-adapted star product
deformed variational calculus but such constructions are more cumbersome and
involve additional assumptions on the variational procedure and various
classes of nonholonomic constraints. Such equations can be considered also
as certain nonassociative canonical distortions of vacuum gravitational
equations in Finsler-Lagrange-Hamiton MGTs, when the nonassociative Ricci
s-tensor and scalar curvature are defined respectively by formulas (\ref%
{driccicanonstar1}) and (\ref{ricciscsymnonsym}) and respective distortions
and dual modifications of systems of nonlinear PDEs from \cite{vbv18}. Such
nonassociative vacuum gravitational field equations provide typical examples
of nonassociative Ricci solitons. Generalizing and applying the AFCDM we can
generate various classes of generic off-diagonal exact and parametric
solutions as we considered in \cite{partner02,partner03,lbsvevv22,lbdssv22}.
In section \ref{sec5}, we shall construct other classes of nonassociative
geometric flow and Ricci soliton solutions and study their associated GIF
and QGIF models.

\subsection{A brief review on nonassociative geometric flows and Perelman's
thermodynamics}

We outline the main results of sections 2 and 3 from \cite{lbdssv22} which
are important to elaborate and develop in next sections the nonassociatve
GIF and QGIF theories.

\subsubsection{Nonassociative generalizations of Perelman functionals and R.
Hamilton equations}

Our geometric arena consists from a nonassociative R-flux deformed phase
space$\ _{s}^{\star }\mathcal{M}$ enables with star product $\star _{s}$
structure (\ref{starpn}) adapted to a nonholonomic (2+2)+(2+2)
decomposition. Star product R-flux deformations of fundamental geometry
s-objects (\ref{mafgeomobn}) are encoded into canonical nonassociative
geometric data $[\ _{s}^{\shortmid }\mathfrak{g}^{\star },\ _{s}^{\shortmid }%
\widehat{\mathbf{D}}^{\star }].$ We follow the Convention 2 (\ref{conv2s})
with a $\kappa $-linear parametric decomposition when $\ _{\star}^{\shortmid
}\mathfrak{\check{g}}_{\alpha _{s}\beta _{s}}^{[0]}= \ _{\star}^{\shortmid }%
\mathbf{g}_{\alpha _{s}\beta _{s}}=\ ^{\shortmid }\mathbf{g}_{\alpha
_{s}\beta _{s}}.$

We can model geometric flows on a flow parameter $\tau $ of metric-affine
objects on $\ _{s}\mathcal{M}$ (we can consider in $[0]$-approximation, i.e.
zero power on $\kappa )$ using $\tau $-families of canonical data $(\
_{s}^{\shortmid }\mathbf{g}(\tau ),\ _{s}^{\shortmid }\widehat{\mathbf{D}}%
(\tau ))$ and flows of volume elements 
\begin{equation}
d\ \ ^{\shortmid }\mathcal{V}ol(\tau )=\sqrt{|\ \ ^{\shortmid }\mathbf{g}%
_{\alpha _{s}\beta _{s}}\ (\tau )|}\ \delta ^{8}\ ^{\shortmid }u^{\gamma
_{s}}(\tau ).  \label{volform}
\end{equation}%
Such a value is computed using N-elongated s-differentials $\delta ^{8}\
^{\shortmid }u^{\gamma _{s}}(\tau )$ which are linear on $\ ^{\shortmid}N_{\
i_{s}a_{s}}\ (\tau )$ as in (\ref{nadapbdss}). Corresponding nonassociative
flows are induced star product flows $\star _{s}(\tau )$ of (\ref{starpn})
as determined by flows of s-adapted frames $\ ^{\shortmid }\mathbf{e}%
_{i_{s}}(\tau ),$ see (\ref{nadapbdsc}).

For nonassociative geometric flows determined by flows of star products with
R-flux deformation and canonical data $[\ _{s}^{\shortmid }\mathbf{g}%
^{\star}(\tau ),\ _{s}^{\shortmid }\widehat{\mathbf{D}}^{\star }(\tau )],$
the Perelman type functionals where postulated (see formulas (12) and (13)
in \cite{lbdssv22}) in the form 
\begin{eqnarray}
\ _{s}^{\shortmid }\widehat{\mathcal{F}}^{\star }(\tau ) &=&\int_{\
_{s}^{\shortmid }\widehat{\Xi }}(\ _{s}^{\shortmid }\widehat{\mathbf{R}}%
sc^{\star }+|\ _{s}^{\shortmid }\widehat{\mathbf{D}}^{\star }\
_{s}^{\shortmid }\widehat{f}|^{2})\star e^{-\ \ _{s}^{\shortmid }\widehat{f}%
}\ d\ ^{\shortmid }\mathcal{V}ol(\tau ),\mbox{ and }  \label{naffunct} \\
\ _{s}^{\shortmid }\widehat{\mathcal{W}}^{\star }(\tau ) &=&\int_{\
_{s}^{\shortmid }\widehat{\Xi }}\left( 4\pi \tau \right) ^{-4}\ [\tau (\
_{s}^{\shortmid }\widehat{\mathbf{R}}sc^{\star }+\sum\nolimits_{s}|\
_{s}^{\shortmid }\widehat{\mathbf{D}}^{\star }\star \ _{s}^{\shortmid }%
\widehat{f}|)^{2}+\ _{s}^{\shortmid }\widehat{f}-8]\star e^{-\ \
_{s}^{\shortmid }\widehat{f}}\ d\ ^{\shortmid }\mathcal{V}ol(\tau ).
\label{nawfunct}
\end{eqnarray}%
The 8-d hypersurfrace integrals in such formulas are determined by a volume
element (\ref{volform}) and a set of shell normalizing functions $\
_{s}^{\shortmid }\widehat{f}(\tau ,\ _{s}^{\shortmid }u),$ which for
geometric constructions can be stated to satisfy the condition 
\begin{equation}
\int_{\ _{s}^{\shortmid }\widehat{\Xi }}\ _{s}^{\shortmid }\widehat{\nu }\ \
d\ ^{\shortmid }\mathcal{V}ol(\tau ):=\int_{t_{1}}^{t_{2}}\int_{\widehat{\Xi 
}_{t}}\ \int_{\ ^{\shortmid }\widehat{\Xi }_{E}}\ _{s}^{\shortmid }\widehat{%
\nu }\ \ d\ ^{\shortmid }\mathcal{V}ol(\tau )=1,  \label{normcond}
\end{equation}%
where the integration measures $\ _{s}^{\shortmid }\widehat{\nu }=\left(4\pi
\tau \right) ^{-4}e^{-\ _{s}^{\shortmid }\widehat{f}}$ are parameterized for
s-shells. For general topological considerations, such conditions may be not
considered. In other type commutative geometric models and applications in
high energy physics \cite{kehagias19,biasio20,biasio21,lueben21,biasio22}
the normalizing function is postulated as a dilaton field.\footnote{%
We can consider star-deformations of the volume form (\ref{volform}), $e^{-
\ _{s}^{\shortmid }\widehat{f}}d\ ^{\shortmid }\mathcal{V}ol(\tau)
\rightarrow e^{-\ \ _{s}^{\shortmid }\widehat{f}}d\ ^{\shortmid }\mathcal{V}%
ol^{\star }(\tau )=$ \newline
$e^{-\ \ _{s}^{\shortmid }\widehat{f}}\sqrt{|\ _{\star }^{\shortmid }\mathbf{%
g}_{\alpha _{s}\beta _{s}}\ (\tau )|} \delta \ ^{\shortmid }u^{\gamma
_{s}}(\tau ), $ for other types of adapted integration measures and
nonholonomic s-shells, for instance, involving $\ _{s}^{\shortmid }\mathfrak{%
g}^{\star }.$ Such transforms can be encoded into a normalizing function $\
_{s}^{\shortmid }\widehat{f}$ and adapted to a respective separation of
nonsymmetric components of s-metrics for $\kappa $--linear
parameterizations. Our approach involves the assumption that we simplify
further computations if we fix any convenient variant of star product and
s-frames, normalization functions and integration measures, which allow to
derive certain exact/parametric solutions in explicit forms. Finally, we can
change the frames of reference to some general ones and finally to state
normalizing conditions using functionals of type (\ref{naffunct}) and (\ref%
{nawfunct}).}

We can follow correspondingly s-adapted variational procedures for the
functional $\ _{s}^{\shortmid }\widehat{\mathcal{F}}^{\star }(\tau )$ (\ref%
{naffunct}) and/or $\ _{s}^{\shortmid }\widehat{\mathcal{W}}^{\star }(\tau )$
(\ref{nawfunct}) and derive nonassociative geometric flow evolution
equations,%
\begin{eqnarray}
\partial _{\tau }\ _{\star }^{\shortmid }\mathfrak{g}_{\alpha _{s}\beta
_{s}}(\tau ) &=&-2\ ^{\shortmid }\widehat{\mathbf{R}}_{\ \alpha _{s}\beta
_{s}}^{\star }(\tau ),  \label{nonassocgeomfl} \\
\partial _{\tau }\ _{s}^{\shortmid }\widehat{f}(\tau ) &=&\ _{s}^{\shortmid }%
\widehat{\mathbf{R}}sc^{\star }(\tau )-\ ^{\star }\widehat{\bigtriangleup }%
(\tau )\star \ \ _{s}^{\shortmid }\widehat{f}(\tau )+(\ _{s}^{\shortmid }%
\widehat{\mathbf{D}}^{\star }(\tau )\star \ \ _{s}^{\shortmid }\widehat{f}%
(\tau ))^{2}(\tau ).  \notag
\end{eqnarray}%
In these formulas, $\ ^{\star }\widehat{\bigtriangleup }(\tau )=[\
_{s}^{\shortmid }\widehat{\mathbf{D}}^{\star }(\tau )]^{2}$ are families of
the Laplace s-operators and the nonsymmetric components of $\ _{\star
}^{\shortmid }\mathfrak{g}_{\alpha _{s}\beta _{s}}(\tau )$ are computed
using $\kappa $-linear parameterizations (\ref{aux40a})--(\ref{aux40aa}). We
note that from different F- and W-functionals, there are obtained quite
different modified R. Hamilton equations \cite{hamilton82}, as we labeled by
dots in (\ref{nonassocham1}). Such equations can be equivalent for certain
conditions imposed on the nonholonomic frame structure and types of
normalization functions. We can relate two different procedures of deriving
generalized/ modified geometric flow evolution equations by re-defining the
nonholonomic structure and fixing corresponding type normalizing functions.
Such cumbersome proofs for associative and commutative Riemannian metrics
are provided in \cite{perelman1}, see details in monographs \cite%
{kleiner06,morgan06,cao06} and, for nonholonomic non-Riemannian
generalizations, in \cite{svnonh08}.\footnote{%
There is an alternative way to derive nonassociative geometric flow
equations (\ref{nonassocgeomfl}) when to begin with associative and
commutative s-adapted proofs for $\tau $-flows of families of s-metrics $\
_{s}^{\shortmid }\mathbf{g}(\tau )=\{\ ^{\shortmid }\mathbf{g}_{\alpha
_{s}\beta _{s}}(\tau )\}$ (\ref{sdm}) and canonical s-connections $\
_{s}^{\shortmid }\widehat{\mathbf{D}}(\tau )$ and then to apply the
Convention 2 (\ref{conv2s}). Here, we also note nonassociative geometric
flow equations can be also motivated as star product R-flux deformations of
a two-dimensional sigma model with beta functions and dilaton field as it
was stated by equations (79) and (80) in \cite{kehagias19}.}

Nonassociative Ricci solitons for the canonical s-connection $\
_{s}^{\shortmid }\widehat{\mathbf{D}}^{\star }$ are defined as self-similar
configurations of gradient geometric flows (\ref{nonassocgeomfl}) for a
fixed parameter $\tau _{0}.$ On $\ _{s}^{\star }\mathcal{M},$ the
corresponding Ricci soliton s-equations derived from a W-functional are of
type 
\begin{equation}
\ ^{\shortmid }\widehat{\mathbf{R}}_{\ \alpha _{s}\beta _{s}}^{\star }+\
^{\shortmid }\widehat{\mathbf{D}}_{\alpha _{s}}^{\star }\ ^{\shortmid }%
\widehat{\mathbf{D}}_{\beta _{s}}^{\star }\ ^{\shortmid }\varpi (\
_{s}^{\shortmid }u)=\lambda \ \ \ _{\star }^{\shortmid }\mathfrak{g}_{\alpha
_{s}\beta _{s}}.  \label{naricsol}
\end{equation}%
where $^{\shortmid }\varpi $ is a smooth potential function on every shell $%
s=1,2,3,4$ and $\lambda =const.$ The the nonassociatve and noncommutative
modified vacuum Einstein equations (\ref{nonassocdeinst1}) consist an
example of nonassociative Ricci soliton ones (\ref{naricsol}). For
Riemannian and Kaehler geometric flows, Ricci solitons geometries were
studied in details in \cite{kleiner06,morgan06,cao06}, see also \cite{sv20}
and references therein for nonholonomic s-adapted constructions. In
nonassociative form, the such equations were introduced and studied in \cite%
{lbdssv22}. We also note that systems of nonlinear PDEs can be derived as
deformed by star products and R-fluxes to nonassociative Ricci solitons
defined by standard Ricci soliton equations extend to nonholonomic phase
spaces. In abstract geometric form, we may follow the Convention 2 (\ref%
{conv2s}). The nonassociative geometric flow constructions performed in this
section for the canonical s-connection can be re-defined in terms of
respective LC-connections, $\ _{\star }^{\shortmid }\nabla $ and $\
^{\shortmid }\nabla .$ For this, we have to impose additional nonholonomic
constraints when $\ _{s}^{\shortmid }\widehat{\mathbf{D}}_{\mid\
_{s}^{\shortmid }\widehat{\mathbf{T}}=0}^{\star }= \ _{\star
}^{\shortmid}\nabla ,$ and then to consider $\kappa ^{0}$-terms an extract $%
\ ^{\shortmid }\nabla $-configurations. Such configurations can be generated
also following a procedure of type $\ _{s}^{\shortmid }\widehat{\mathbf{D}}%
^{\star }\rightarrow _{\mid \kappa ^{0}}\rightarrow \ _{s}^{\shortmid } 
\widehat{\mathbf{D}} \rightarrow _{\mid \ _{s}^{\shortmid }\widehat{\mathbf{T%
}}=0}\rightarrow \ ^{\shortmid }\nabla ,$ which is not equivalent to the
prevoious one. As a result, the nonassociative equations (\ref%
{nonassocgeomfl}) and (\ref{naricsol}) transform respectively into (\ref%
{heqnonh}) and (\ref{ricciconfhonh}) and/or (\ref{heq}) and (\ref{heqps})
postulated in the Introduction.

To elaborate on possible applications in modern gravity and cosmology, the
nonassociative F- and W-functionals and related geometric flow equations are
considered for a $\kappa $-linear parametric decomposition.

\subsubsection{Statistical models for nonassociative geometric flow
thermodynamics}

In statistical physics, a partition function $Z=\int \exp (-\beta E)d\omega
(E)$ for a canonical ensemble at temperature $\beta ^{-1}=\tau $ is defined
by a measure taken for a density of states $\omega (E).$ This allow to
construct a statistical thermodynamic model, when corresponding
thermodynamical variables are defined and computed respectively: the average
energy, $\ \left\langle E\right\rangle :=-\partial \log Z/\partial \beta ,$
the entropy $S:=\beta \left\langle E\right\rangle +\log Z$ and the
fluctuation parameter $\sigma :=\left\langle \left( E-\left\langle
E\right\rangle \right) ^{2}\right\rangle =\partial ^{2}\log Z/\partial \beta
^{2}.$ Perelman introduced in section 5.1 of \cite{perelman1} such a
partition function to describe Ricci flows of $n$-d Riemannian metrics on a
fixed closed manifold $M$ (with local coordinates $x^{\grave{\imath}},\grave{%
\imath}=1,2,3,...n),$%
\begin{equation}
\log Z=\int (-f+\frac{n}{2})(4\pi \tau )^{-n/2}e^{-f}d\mathcal{V}ol(\tau ),
\label{partitionf1}
\end{equation}%
where $d\mathcal{V}ol(\tau )=\sqrt{|g_{\grave{\imath}\grave{j}}(\tau ,x^{%
\grave{k}})|}d^{n}x^{\grave{k}}$ is determined by $g_{\grave{\imath}\grave{j}%
}(\tau )\simeq g_{\grave{\imath}\grave{j}}(\tau ,x^{\grave{k}})$ as a
solution of $\partial g_{\grave{\imath}\grave{j}}/\partial \tau =2(R_{\grave{%
\imath}\grave{j}}+ \nabla _{\grave{\imath}}\nabla _{\grave{j}}f).$ We
suppose that it is chosen such a solution which allows to define a
well-defined "density of states" mesures which allow integration on $M.$
There were provided certain important Remarks in sections 5.2 and 5.3$^{\ast
}$ about (possible) connections of such a statistical analogy to
renormalization group flows and black hole thermodynamics.

Considering a closed hypersurface in the nonassociative phase space $%
_{s}^{\star }M\subset \ _{s}^{\star }\mathcal{M}$ and the volume form $d \
^{\shortmid }\mathcal{V}ol(\tau )$ (\ref{volform}) is defined by the
associative component $\ \ ^{\shortmid }\mathbf{g}_{\alpha _{s}\beta _{s}}\
(\tau )$ of a solution of nonassociative geometric flow equations (\ref%
{nonassocham1}), we can generalize the partition function (\ref{partitionf1}%
) for relativistic phase spaces (we consider the dimension $n=8$), 
\begin{equation}
\log \ _{s}^{\star }\mathcal{Z}=\int (-\ _{s}^{\shortmid }f+4)(4\pi \tau
)^{-4}e^{-\ _{s}^{\shortmid }f}d\ ^{\shortmid }\mathcal{V}ol(\tau ).
\label{patitionfph}
\end{equation}%
For canonical s-flows with $\ ^{\shortmid }\widehat{\mathbf{D}}_{\alpha
_{s}}^{\star }$ determined by a solution of (\ref{nonassocgeomfl}), we
introduce 
\begin{equation}
\ _{s}\widehat{\mathcal{Z}}^{\star }(\tau )=\ _{s}^{\shortmid }\widehat{Z}%
(\tau )=\exp [\int_{\ _{s}^{\shortmid }\widehat{\Xi }}[-\ _{s}^{\shortmid }%
\widehat{f}+4]\ \left( 4\pi \tau \right) ^{-4}e^{-\ _{s}^{\shortmid }%
\widehat{f}}\ ^{\shortmid }\delta \ ^{\shortmid }\mathcal{V}(\tau ),
\label{spf}
\end{equation}%
as we defined by formulas (21) and (22) in \cite{lbdssv22}, where $%
_{s}^{\shortmid }\widehat{\Xi }$ labels a closed phase space hypersurface
defined for an additional (3+1)+(3+1) splitting; and when exact/pareamteric
solutions are supposed to be derived for a dyadic shell decomposition. Such
a double nonholonomic fibration $_{s}^{\star }\mathcal{M}$ allow to provide
a relativistic thermodynamic interpretation to such values: 
\begin{eqnarray}
\mbox{ average energy },\ _{s}^{\shortmid }\widehat{\mathcal{E}}^{\star
}(\tau )\ &=&-\tau ^{2}\int_{\ _{s}^{\shortmid }\widehat{\Xi }}\ \left( 4\pi
\tau \right) ^{-4}\left( \ _{s}^{\shortmid }\widehat{\mathbf{R}}sc^{\star
}+|\ _{s}^{\shortmid }\widehat{\mathbf{D}}^{\star }\ _{s}^{\shortmid }%
\widehat{f}|^{2}-\frac{4}{\tau }\right) \star e^{-\ _{s}^{\shortmid }%
\widehat{f}}\ ^{\shortmid }\delta \ ^{\shortmid }\mathcal{V}(\tau );
\label{nagthermodvalues} \\
\mbox{ entropy },\ \ _{s}^{\shortmid }\widehat{\mathcal{S}}^{\star }(\tau )
&=&-\int_{\ _{s}^{\shortmid }\widehat{\Xi }}\left( 4\pi \tau \right)
^{-4}\left( \tau (\ _{s}^{\shortmid }\widehat{\mathbf{R}}sc^{\star }+|\
_{s}^{\shortmid }\widehat{\mathbf{D}}^{\star }\ _{s}^{\shortmid }\widehat{f}%
|^{2})+\ _{s}^{\shortmid }\tilde{f}-8\right) \star e^{-\ _{s}^{\shortmid }%
\widehat{f}}\ ^{\shortmid }\delta \ ^{\shortmid }\mathcal{V}(\tau );  \notag
\\
\mbox{ fluctuation },\ _{s}^{\shortmid }\widehat{\sigma }^{\star }(\tau )
&=&2\ \tau ^{4}\int_{\ _{s}^{\shortmid }\widehat{\Xi }}\left( 4\pi \tau
\right) ^{-4}|\ \ ^{\shortmid }\widehat{\mathbf{R}}_{\alpha _{s}\beta
_{s}}^{\star }+\ ^{\shortmid }\widehat{\mathbf{D}}_{\alpha _{s}}^{\star }\
^{\shortmid }\widehat{\mathbf{D}}_{\beta _{s}}^{\star }\ \ _{s}^{\shortmid }%
\widehat{f}-\frac{1}{2\tau }\mathbf{g}_{\alpha _{s}\beta _{s}}^{\star
}|^{2}\star e^{-\ _{s}^{\shortmid }\widehat{f}}\ ^{\shortmid }\delta \
^{\shortmid }\mathcal{V}(\tau ).  \notag
\end{eqnarray}%
These formulas can be derived generalizing in s-adapted form the variational
procedure from \cite{perelman1,kleiner06,morgan06,cao06} using $\
_{s}^{\shortmid }\widehat{\mathcal{W}}^{\star }(\tau )$ (\ref{nawfunct}) and 
$\ \ _{s}\widehat{\mathcal{Z}}^{\star }(\tau )$ (\ref{spf}). In a geometric
symbolic form, we can construct such values for respective star product
deformations of s-adapted geometric/ thermodynamic objects from $_{s}%
\mathcal{M}$ to $_{s}^{\star }\mathcal{M}.$

\subsubsection{ Parametric canonical thermodynamic variables for
nonassociative geometric flows}

$\kappa $-linear parametric decompositions of nonassociative geometric
thermodynamic variables (\ref{nagthermodvalues}) are studied in \cite%
{lbdssv22} (see discussion of formulas (24)-(27) in that work). Using those
results, we can express such variables for the same (\ref{nawfunct}) in the
forms: 
\begin{eqnarray}
\ _{s}^{\shortmid }\widehat{\mathcal{F}}_{\kappa }^{\star }(\tau )
&=&\int_{\ _{s}^{\shortmid }\widehat{\Xi }}(\ \ _{s}^{\shortmid }\widehat{%
\mathbf{R}}sc+\ _{s}^{\shortmid }\widehat{\mathbf{K}}sc+|\ _{s}^{\shortmid }%
\widehat{\mathbf{D}}\ _{s}^{\shortmid }\widehat{f}|^{2})e^{-\ \
_{s}^{\shortmid }\widehat{f}}\ d\ ^{\shortmid }\mathcal{V}ol(\tau ),%
\mbox{
and }  \label{naffunctp} \\
\ _{s}^{\shortmid }\widehat{\mathcal{W}}_{\kappa }^{\star }(\tau )
&=&\int_{\ _{s}^{\shortmid }\widehat{\Xi }}\left( 4\pi \tau \right) ^{-4}\
[\tau (\ \ _{s}^{\shortmid }\widehat{\mathbf{R}}sc+\ _{s}^{\shortmid }%
\widehat{\mathbf{K}}sc+\sum\nolimits_{s}|\ _{s}^{\shortmid }\widehat{\mathbf{%
D}}\ _{s}^{\shortmid }\widehat{f}|)^{2}+\ _{s}^{\shortmid }\widehat{f}%
-8]e^{-\ \ _{s}^{\shortmid }\widehat{f}}\ d\ ^{\shortmid }\mathcal{V}ol(\tau
).  \label{nawfunctp}
\end{eqnarray}%
In these formulas, the Ricci s-scalar splits into two components, $%
_{s}^{\shortmid }\widehat{\mathbf{R}}sc^{\star }= \ _{s}^{\shortmid }%
\widehat{\mathbf{R}}sc+\ _{s}^{\shortmid }\widehat{\mathbf{K}}sc,$ where $\
_{s}^{\shortmid }\widehat{\mathbf{K}}sc:=\ _{\star }^{\shortmid }\mathfrak{g}%
^{\mu _{s}\nu _{s}}\ ^{\shortmid }\widehat{\mathbf{K}}_{\ \beta _{s}\gamma
_{s}}\left\lceil \hbar ,\kappa \right\rceil $ contains the coefficients
proportional to $\hbar $ and $\kappa .$ Explicit s-adapted coefficients for $%
\ _{s}^{\shortmid }\widehat{\mathbf{R}}sc$ and $\ _{s}^{\shortmid }\widehat{%
\mathbf{K}}sc$ are provided in formulas (48) and (49) from \cite{partner02}
with complex momentum coordinates (which can be re-defined in real ones).%
\footnote{%
We do no repeat details and formulas on such decompositions, and more
general ones for the Ricci s-tensor $\ ^{\shortmid }\widehat{\mathbf{R}}_{\
\alpha _{s}\beta _{s}}(\tau )+ \ ^{\shortmid }\widehat{\mathbf{K}}_{\ \alpha
_{s}\beta _{s}}(\tau ,\left\lceil \hbar ,\kappa \right\rceil ),$ in this
work because such terms can be encoded into certain efficient sources and
nonholonomic constraints. To generate solutions for geometric and
information flows in next sections it will be enough to use certain
parameterizations efficient sources.} The normalizing function $\ \
_{s}^{\shortmid }\widehat{f}$ is re-defined to include $\left\lceil
\hbar,\kappa \right\rceil $-terms from $\ ^{\shortmid }\widehat{\mathbf{D}}%
^{\star }\rightarrow \ _{s}^{\shortmid }\widehat{\mathbf{D}}$ and other
terms including $\kappa $-parametric decompositions.

We can also define nonassociative $\kappa $-linear R-flux generalizations of
the R. Hamilton equations. In the first case, we introduce R-flux
deformations of such equations using $\kappa $-parametric decompositions of (%
\ref{nonassocgeomfl}). An alternative and equivalent way, in the second
case, is to apply a s-adapted nonholonomic variational procedure to $\
_{s}^{\shortmid }\widehat{\mathcal{F}}_{\kappa }^{\star }(\tau )$ (\ref%
{naffunctp}), or $\ _{s}^{\shortmid }\widehat{\mathcal{W}}_{\kappa
}^{\star}(\tau )$ (\ref{nawfunctp}). In all cases, adapting corresponding
the nonholonomic structure, we obtain such phase geometric flow equations
with $\kappa $-terms encoding star product R-flux deformations, 
\begin{eqnarray}
\partial _{\tau }\ ^{\shortmid }\mathbf{g}_{\alpha _{s}\beta _{s}}(\tau )
&=&-2(\ ^{\shortmid }\widehat{\mathbf{R}}_{\ \alpha _{s}\beta _{s}}(\tau )+\
^{\shortmid }\widehat{\mathbf{K}}_{\ \alpha _{s}\beta _{s}}(\tau
,\left\lceil \hbar ,\kappa \right\rceil )),  \label{nonassocgeomflp} \\
\partial _{\tau }\ \ _{s}^{\shortmid }\widehat{f}(\tau ) &=&\
_{s}^{\shortmid }\widehat{\mathbf{R}}sc(\tau )+\ _{s}^{\shortmid }\widehat{%
\mathbf{K}}sc(\tau )-\ \widehat{\bigtriangleup }(\tau )\ \ _{s}^{\shortmid }%
\widehat{f}(\tau )+(\ _{s}^{\shortmid }\widehat{\mathbf{D}}(\tau )\ \
_{s}^{\shortmid }\widehat{f}(\tau ))^{2}(\tau ).  \notag
\end{eqnarray}%
In these formulas $\widehat{\bigtriangleup }$ is the Laplace operator
constructed from the canonical s-connection $\ _{s}^{\shortmid }\widehat{%
\mathbf{D}}.$\footnote{%
For self-similar configurations with $\tau =\tau _{0},$ the parametric
nonholonomic equations (\ref{nonassocgeomflp}) transform into a system of
nonlinear PDEs for $\kappa $--parametric canonical shell Ricci solitons.}

Introducing effective $\tau $-depending sources 
\begin{equation*}
\ ^{\shortmid }\Im _{\alpha _{s}\beta _{s}}^{\star }(\tau )=-(\ ^{\shortmid }%
\widehat{\mathbf{K}}_{\alpha _{s}\beta _{s}}(\tau )+\frac{1}{2}\partial
_{\tau }\ ^{\shortmid }\mathbf{g}_{\alpha _{s}\beta _{s}}(\tau )),
\end{equation*}%
we can write the $\kappa $-parametric nonassociative geometric flow
equations (\ref{nonassocgeomflp}) in the form 
\begin{equation}
\ ^{\shortmid }\widehat{\mathbf{R}}ic_{\alpha _{s}\beta _{s}}(\tau )=\
^{\shortmid }\Im _{\alpha _{s}\beta _{s}}^{\star }(\tau ),
\label{nonassocgeomflef}
\end{equation}%
which are similar to the modified Einstein equations (\ref{nonassocdeinst1})
with geometric objects and effective source running on $\tau $ and encoding
nonassociative contributions via some terms $\mathcal{R}_{\quad \alpha
_{s}}^{\tau _{s}\xi _{s}}(\tau ).$ On $\ _{s}^{\star }\mathcal{M},$ we can
consider certain frame transforms $\ ^{\shortmid }\widehat{\Im }_{\alpha
_{s}^{\prime }\beta _{s}^{\prime }}^{\star }= e_{\ \alpha
_{s}^{\prime}}^{\alpha _{s}}e_{\ \beta _{s}^{\prime }}^{\beta _{s}}\
^{\shortmid }\Im _{\alpha _{s}\beta _{s}}^{\star },$ when the effective
sources are parameterized in the form 
\begin{equation}
\ ^{\shortmid }\Im _{\star \ \beta _{s}}^{\alpha _{s}}~(\tau ,\ ^{\shortmid
}u^{\gamma _{s}})=[~_{1}^{\shortmid }\Im ^{\star }(\kappa ,\tau
,x^{k_{1}})\delta _{i_{1}}^{j_{1}},~_{2}^{\shortmid }\Im ^{\star }(\kappa
,\tau ,x^{k_{1}},y^{c_{2}})\delta _{b_{2}}^{a_{2}},~_{3}^{\shortmid }\Im
^{\star }(\kappa ,\tau ,x^{k_{2}},p_{c_{3}})\delta
_{a_{3}}^{b_{3}},~_{4}^{\shortmid }\Im ^{\star }(\kappa ,\tau ,~^{\shortmid
}x^{k_{3}},p_{c_{4}})\delta _{a_{4}}^{b_{4}}],  \label{cannonsymparamc2b}
\end{equation}%
i.e. $\ ^{\shortmid }\Im _{_{\beta _{s}\gamma _{s}}}^{\star }(\tau )=diag\{\
_{s}^{\shortmid }\Im ^{\star }(\tau )\}.$ A prescribed value $\
_{s}^{\shortmid }\Im ^{\star }~(\tau ,\ ^{\shortmid }u^{\gamma _{s}})$
imposes a s-shell nonholonomic constraint for $\tau $-derivatives of the
metrics s-coefficients $\partial _{\tau }\ ^{\shortmid }\mathbf{g}_{\alpha
_{s}\beta _{s}}(\tau ).$ For small parametric deformations and a fixed $%
\tau_{0}$, such constraints can be solved in explicit general forms, or
allow recurrent parametric computations of the coefficients of s-metrics and
s-connection for a corresponding class of solutions. Using effective sources
(\ref{cannonsymparamc2b}), The $\kappa $-linear parametric geometric flow
equations (\ref{nonassocgeomflp}) can be written equivalently as a $\tau $%
-family of R-flux deformed Einstein equations written in canonical
s-variables, 
\begin{equation}
\ ^{\shortmid }\widehat{\mathbf{R}}_{\ \ \gamma _{s}}^{\beta _{s}}(\tau )= {%
\delta }_{\ \ \gamma _{s}}^{\beta _{s}}\ _{s}^{\shortmid }\Im ^{\star
}(\tau).  \label{nonassocrf}
\end{equation}%
The data $\ _{s}^{\shortmid }\Im (\tau )^{\star }$ are considered as
generating sources in the AFCDM. Prescribing such values, we can decouple
(modified) geometric flow and/or gravitational field equations in some
general forms and then integrate them in terms of some classes of generating
and integration sources subjected to additional nonholonomic constraints,
see section 5 in \cite{partner02} (and section 4 in \cite{lbsvevv22}, for
nonassociative geometric flows). Certain general classes of solutions are
characterized by nonlinear symmetries transforming $\ _{s}^{\shortmid }\Im
^{\star }(\tau )$ into, running effective shell cosmological constants $\
_{s}^{\shortmid }\Lambda ^{\star }~(\tau ).$ We put a $\star $-label on
effective cosmological constants in order to emphasize that such values are
related via nonlinear symmetries to certain effective sources encoding
contributions from nonassociative star product R-flux deformations (at least
in $\kappa $-linear parametric form).

\subsubsection{Ansatz for generating off-diagonal solutions for
nonassociative geometric flows}

\label{ansatzss}We can generate generic off-diagonal quasi-stationary
solutions\footnote{%
such solutions can't be diagonalized in a finite phase space/ spacetime
region by coordinate frame transforms and posses symmetry on a time like
Killing vector, i.e. on $\partial _{4}=\partial _{t},$ at least on the
shells $s=1$ and 2, when the coefficients s-metrics and N-connections do not
depend on $y^{4}=t$} of nonassociative geometric flow equations (\ref%
{nonassocrf}) if we chose such an off-diagonal ansatz 
\begin{eqnarray}
d\widehat{s}^{2}(\tau ) &=&g_{i_{1}}(\tau
,x^{k_{1}})(dx^{i_{1}})^{2}+g_{a_{2}}(\tau ,x^{i_{1}},y^{3})(\mathbf{e}%
^{a_{2}}(\tau ))^{2}+  \notag \\
&&\ ^{\shortmid }g^{a_{3}}(\tau ,x^{i_{2}},p_{6})(\ ^{\shortmid }\mathbf{e}%
_{a_{3}}(\tau ))^{2}+\ ^{\shortmid }g^{a_{4}}(\tau ,\ ^{\shortmid
}x^{i_{3}},p_{7})(\ ^{\shortmid }\mathbf{e}_{a_{4}}(\tau ))^{2},\mbox{where }
\notag \\
\mathbf{e}^{a_{2}}(\tau ) &=&dy^{a_{2}}+N_{k_{1}}^{a_{2}}(\tau
,x^{i_{1}},y^{3})dx^{k_{1}},\ ^{\shortmid }\mathbf{e}_{a_{3}}(\tau
)=dp_{a_{3}}+\ ^{\shortmid }N_{a_{3}k_{2}}(\tau ,x^{i_{2}},p_{6})dx^{k_{2}},
\notag \\
\ ^{\shortmid }\mathbf{e}_{a_{4}}(\tau ) &=&dp_{a_{4}}+\ ^{\shortmid
}N_{a_{4}k_{3}}(\tau ,\ ^{\shortmid }x^{i_{3}},p_{7})d\ ^{\shortmid
}x^{k_{3}}.\   \label{ans1rf}
\end{eqnarray}%
Such a s-metric contains such Killing symmetries: on $\partial _{4}$ for the
shell $s=2;$ on $\partial _{5}$ for the shell $s=3;$ and on $\partial _{8}$
for the shell $s=4.$ For a fixed $\tau _{0}$ and self-similar
configurations, i.e. for the Ricci s-solitons (\ref{naricsol}), the
nonassociative geometric flow equations (\ref{nonassocgeomfl}) and/or (\ref%
{nonassocrf}), and modified Einstein equations (\ref{nonassocdeinst1}) can
be decoupled and solved in very general forms as we proved in \cite%
{partner02,partner03,lbdssv22} following the AFCDM. The method was extended
recently for nonassociative geometric flows in sections 4 and 5 of \cite%
{lbsvevv22}, when running of solutions on $\tau $ was investigated. In
general, we can generate various classes of type (\ref{ans1rf})
quasi-stationary, locally anisotropic cosmological solutions, so-called
rainbow metrics with Killing symmetry on $\partial _{7}$ and explicit
dependence on an energy type coordinate $p_{8}=E.$ Off-diagonal solutions of
(\ref{nonassocrf}) with a phase space Killing symmetry on a$\ ^{\shortmid
}\partial ^{a_{4}},$ for $a_{4}=7,$ or 8; and a spacetime Killing symmetry
on $\partial _{a_{2}}$, for $a_{2}=3,$ or 4, can be written in the form 
\begin{eqnarray}
d\widehat{s}^{2}(\tau ) &=&g_{i_{1}}(\tau )(dx^{i_{1}})^{2}+g_{a_{2}}(\tau )(%
\mathbf{e}^{a_{2}}(\tau ))^{2}+\ ^{\shortmid }g^{a_{3}}(\tau )(\ ^{\shortmid
}\mathbf{e}_{a_{3}}(\tau ))^{2}+\ ^{\shortmid }g^{a_{4}}(\tau )(\
^{\shortmid }\mathbf{e}_{a_{4}}(\tau ))^{2},\mbox{where }  \label{ssolutions}
\\
\mathbf{e}^{a_{2}}(\tau ) &=&dy^{a_{2}}+N_{k_{1}}^{a_{2}}(\tau )dx^{k_{1}},\
^{\shortmid }\mathbf{e}_{a_{3}}(\tau )=dp_{a_{3}}+\ ^{\shortmid
}N_{a_{3}k_{2}}(\tau )dx^{k_{2}},\ ^{\shortmid }\mathbf{e}_{a_{4}}(\tau
)=dp_{a_{4}}+\ ^{\shortmid }N_{a_{4}k_{3}}(\tau )d\ ^{\shortmid }x^{k_{3}},\ 
\notag
\end{eqnarray}%
where s-metric and N-connection coefficients are parameterized in the form 
\begin{equation*}
\frame{{\scriptsize $%
\begin{array}{ccccc}
\begin{array}{c}
{\ g}_{i_{1}}{\ (\tau ,x}^{k_{1}}) \\ 
{\ =e}^{\psi (\hbar ,\kappa ;\tau ,x^{k_{1}})}%
\end{array}
& 
\begin{array}{cc}
\begin{array}{c}
\begin{array}{c}
{g}_{a_{2}}{(\tau ,x}^{i_{1}},y^{3}) \\ 
{N}_{k_{1}}^{a_{2}}{(\tau ,x}^{i_{1}},y^{3})%
\end{array}
\\ 
\end{array}
& 
\begin{array}{c}
\mbox{quasi-} \\ 
\mbox{stationary}%
\end{array}
\\ 
\begin{array}{c}
\begin{array}{c}
{g}_{a_{2}}(\tau ,x^{i_{1}},t) \\ 
{N}_{k_{1}}^{a_{2}}(\tau ,x^{i_{1}},t)%
\end{array}
\\ 
\end{array}
& 
\begin{array}{c}
\mbox{locally anisotropic} \\ 
\mbox{cosmology}%
\end{array}%
\end{array}
& 
\begin{array}{c}
\begin{array}{c}
\ ^{\shortmid }{\ g}^{a_{3}}(\tau ,x^{i_{2}},p_{5}) \\ 
\ ^{\shortmid }N_{a_{3}k_{2}}(\tau ,x^{i_{2}},p_{5})%
\end{array}
\\ 
\\ 
\begin{array}{c}
\ ^{\shortmid }{\ g}^{a_{3}}(\tau ,x^{i_{2}},p_{6}) \\ 
\ ^{\shortmid }{N}_{a_{3}k_{2}}(\tau ,x^{i_{2}},p_{6})%
\end{array}
\\ 
\end{array}
& 
\begin{array}{c}
\ ^{\shortmid }{g}^{a_{4}}(\tau ,\ ^{\shortmid }x^{i_{3}},p_{7}) \\ 
\ ^{\shortmid }{N}_{a_{4}k_{3}}(\tau ,\ ^{\shortmid }{x}^{i_{3}},p_{7})%
\end{array}
& 
\begin{array}{c}
\mbox{fixed} \\ 
p_{8}=E_{0}%
\end{array}
\\ 
\begin{array}{c}
\tau \mbox{-flows of 2-d} \\ 
\mbox{Poisson eqs} \\ 
{\partial }_{11}^{2}{\psi +\partial }_{22}^{2}{\psi =} \\ 
{2\ }_{1}{\ \Im (\hbar ,\kappa ;\tau ,x}^{k_{1}})%
\end{array}
& 
\begin{array}{cc}
\begin{array}{c}
\begin{array}{c}
{g}_{a_{2}}(\tau ,x^{i_{1}},y^{3}) \\ 
{N}_{k_{1}}^{a_{2}}(\tau ,x^{i_{1}},y^{3})%
\end{array}
\\ 
\end{array}
& 
\begin{array}{c}
\mbox{quasi-} \\ 
\mbox{stationary}%
\end{array}
\\ 
\begin{array}{c}
\begin{array}{c}
{g}_{a_{2}}(\tau ,x^{i_{1}},t) \\ 
{N}_{k_{1}}^{a_{2}}(\tau ,x^{i_{1}},t)%
\end{array}
\\ 
\end{array}
& 
\begin{array}{c}
\mbox{locally anisotropic} \\ 
\mbox{cosmology}%
\end{array}%
\end{array}
& 
\begin{array}{c}
\begin{array}{c}
\\ 
\ ^{\shortmid }{g}^{a_{3}}(\tau ,x^{i_{2}},p_{5}) \\ 
\ ^{\shortmid }N_{a_{3}k_{2}}(\tau ,x^{i_{2}},p_{5})%
\end{array}
\\ 
\\ 
\begin{array}{c}
\ ^{\shortmid }{g}^{a_{3}}(\tau ,x^{i_{2}},p_{6}) \\ 
\ ^{\shortmid }{N}_{a_{3}k_{2}}(\tau ,x^{i_{2}},p_{6})%
\end{array}
\\ 
\end{array}
& 
\begin{array}{c}
\ ^{\shortmid }{g}^{a_{4}}(\tau ,\ ^{\shortmid }x^{i_{3}},E) \\ 
\ ^{\shortmid }{N}_{a_{4}k_{3}}(\tau ,\ ^{\shortmid }x^{i_{3}},E)%
\end{array}
& 
\begin{array}{c}
\mbox{rainbow} \\ 
\mbox{s-metrics} \\ 
\mbox{variable } \\ 
p_{8}=E%
\end{array}%
\end{array}%
$}}
\end{equation*}

The AFCDM allows us to construct more general classes of solutions with
dependencies on all spacetime/phase space coordinates but the formulas
became very cumbersome and we omit such considerations in this work. If a
Killing symmetry exists, the procedure of generating exact/parametric
solutions became more simple and various abstract geometric methods defined
by nonlinear symmetries can be applied. We shall construct and analyze
thermodynamic informational properties of explicit classes of
exact/parametric solutions of type (\ref{ans1rf}), i.e. for nonassociative
locally anisotropic wormhole configurations, in section \ref{sec5}.
Necessary generic off-diagonal ansatz and formulas are provided in Appendix %
\ref{asstables}. Nonassociative locally anisotropic cosmological solutions
of type (\ref{ssolutions}) and their geometric flow evolution will be
studied in further partner works. Geometric flows of nonassociative BH
spacetime and phase space configurations are studied in section 5 of \cite%
{lbsvevv22}; and geometric information flows of off-diagonal
quasi-stationary and/or locally anisotropic cosmological configurations in
associative and commutative MGTs are investigated in \cite{sv20}.

\subsection{Nonassociative GIF theory}

In this subsection, we introduce the main concepts and definitions and study
main properties of classical systems under evolution in nonassociative
geometric information flow, GIF, theory.

\subsubsection{Relating the Shannon entropy and Perelman entropy for
nonassociative GIFs:}

Let us consider a nonassociative canonical geometric flow system $\
_{s}^{\star }\widehat{A}(\tau )= \{\ _{s}^{\shortmid }\mathfrak{g}%
^{\star}(\tau ),\ _{s}^{\shortmid }\widehat{\mathbf{D}}^{\star }(\tau )\}$
defined by partition (statistical generating) function $\ _{s}^{\shortmid }%
\widehat{Z}(\tau )$ (\ref{spf}) when the star-entropy $\ _{s}^{\shortmid }%
\widehat{S}^{\star }(\tau )$ is computed as in (\ref{nagthermodvalues}). We
shall omit "hats" for symbols of some geometric/ information systems and
write $\ _{s}^{\star }A(\tau ),\ _{s}A(\tau ),\ _{s}^{\star }B,\ _{s}B$ etc.
if it will be not necessary to state that they are determined by canonical
s-variables even such geometric data can be introduced. For applications in
MGTs and to GIFs models, we shall use $\kappa $- parametric decompositions $%
\ _{s}^{\shortmid }\widehat{S}_{\kappa }^{\star }(\tau )$ determined by a $\
^{\shortmid }\mathbf{g}_{\alpha _{s}\beta _{s}}(\tau )$ as a solution of
nonassociative geometric flow equations (\ref{nonassocrf}) parameterized as
a s-metric (\ref{ssolutions}). We study flow evolution systems (or certain
nonassociative Ricci soliton configurations) when $\ _{s}^{\star }A$ and
respective geometric data are treated as random variables (i.e. take
randomly certain values for the respective metric-affine s-adapted geometric
objects) that takes values $\ _{s}^{\star }a_{1},...,\ _{s}^{\star }a_{r}$
with certain probabilities $q_{1},...,q_{r}.$ For instance, we can consider
some $\tau _{1},...,\tau _{r},$ when $\ _{s}^{\shortmid }\mathfrak{g}%
_{1}^{\star }=\ _{s}^{\shortmid }\mathfrak{g}^{\star }(\tau _{1}),...\
_{s}^{\shortmid }\mathfrak{g}_{r}^{\star }=\ _{s}^{\shortmid }\mathfrak{g}%
^{\star }(\tau _{r})$ and similarly for other types of geometric s-objects.

\vskip5pt

In the case of nonassociative Ricci solitons (\ref{naricsol}) and, in
particular, of nonassociative vacuum Einstein equations (\ref%
{nonassocdeinst1}), we may consider a GIF system $\ _{s}^{\star }A= \
_{s}^{\star }A(\tau _{0})= \{\ _{s}^{\shortmid }\mathfrak{g}^{\star },\
_{s}^{\shortmid }\widehat{\mathbf{D}}^{\star }\}$ determined by a solution
of type $^{\shortmid }\mathbf{g}_{\alpha _{s}\beta _{s}}(\tau )$ (\ref%
{ssolutions}) for $\tau =\tau _{0}.$ To elaborate on probabilistic models,
we can take $r$ such solutions generating some data $\ _{s}^{\shortmid } 
\mathfrak{g}_{1}^{\star },...,\ _{s}^{\shortmid }\mathfrak{g}_{1}^{\star }$
and respective nonassociative geometric s-objects and their $\kappa $-
parametric decompositions. The the data for such s-metrics are associated
random values $\ _{s}^{\star }a_{1},...,\ _{s}^{\star }a_{r}$ which are
taken with certain probabilities $q_{1},...,q_{r}.$ Any such solution is
already a statistical thermodynamic system characterized, for instance, by a 
$\ _{s}^{\shortmid }\widehat{S}_{\kappa }^{\star }(\tau _{0}).$

\vskip5pt In our approach, nonassociative GIF systems are modelled as
respective nonassociative geometric flow/ Ricci soliton systems determined
by certain classes of well-defined physical solutions and characterized by
respective nonassociative phase space entropy $\ _{s}^{\shortmid }\widehat{S}%
^{\star }$ and/or $\ _{s}^{\shortmid }\widehat{S}_{\kappa }^{\star }.$ Such
systems can be on necessary smooth class/ continuous on $\tau $-parameter,
and they can be modelled in discrete form, as we explained above. In next
subsections, we shall explain how $\ _{s}^{\shortmid }\widehat{S}^{\star }$
and/or $\ _{s}^{\shortmid }\widehat{S}_{\kappa }^{\star }$ can be related to
typical entropies used in classical information theories outlined in \cite%
{preskill,witten20} and references therein.

\vskip5pt Using the general definition of the Shannon entropy $S_{A}$ of a
probability distribution for such a random variable $\ _{s}^{\star }A$ on $\
_{s}^{\star }\mathcal{M},$ we write%
\begin{eqnarray}
S_{A}:= &&-\sum\nolimits_{z=1}^{z=r}p_{z}\log p_{z},(p_{z}%
\mbox{ are not
momentum variables but positive numbers related to probability})  \notag \\
&\hookrightarrow &  \label{shanonper} \\
\ ^{\shortmid }\widehat{S}_{A}^{\star } &=&\sum\nolimits_{z=1}^{z=r}\
_{s}^{\shortmid }\widehat{S}^{\star }(\tau _{z})\simeq
\sum\nolimits_{z=1}^{z=r}\ _{s}^{\shortmid }\widehat{S}_{\kappa }^{\star
}(\tau _{z}),%
\mbox{  nonassociative Ricci flows for a set of discrete values
of }\tau ;  \notag \\
&=&\ _{s}^{\shortmid }\widehat{S}^{\star }(\tau _{0})\simeq \
_{s}^{\shortmid }\widehat{S}_{\kappa }^{\star }(\tau _{0}),%
\mbox{ for a
solution of nonassociative Ricci soliton/ Einstein eqs. };  \notag \\
&=&\int\nolimits_{\tau _{r}}^{\tau _{r}}d\tau \ _{s}^{\shortmid }\widehat{S}%
^{\star }(\tau )\simeq \int\nolimits_{\tau _{r}}^{\tau _{r}}d\tau \
_{s}^{\shortmid }\widehat{S}_{\kappa }^{\star }(\tau ),%
\mbox{nonasociative
geometric flows with  continuous running of }\tau .  \notag
\end{eqnarray}%
The main assumption in the second line of these equations postulates is that
for such classical probabilistic and geometric flow systems encoding
nonassociative data the entropy is determined equivalently by the modified
Perelman entropy. Formulas (\ref{shanonper}) define variants of
Shannon--Perelman entropies for respective GIF systems which can be
continuous on $\tau ,$ and/or for discrete/fixed value for such a
temperature like parameter. For such constructions, we can consider
normalisation functions $\ _{s}^{\shortmid }\tilde{f}$ and use the method of
Lagrange multipliers in order to maximize $S_{A}:=-\sum%
\nolimits_{z=1}^{z=r}p_{z}\log p_{z}$ with the constraints $%
\sum\nolimits_{z=1}^{z=r}p_{z}=1.$ Certain solutions $\ ^{\shortmid }\mathbf{%
g}_{\alpha _{s}\beta _{s}}(\tau )$ (\ref{ssolutions}) are physical if the
corresponding $\ ^{\shortmid }\widehat{S}_{A}^{\star }\simeq
\sum\nolimits_{z=1}^{z=r}\ _{s}^{\shortmid }\widehat{S}_{\kappa}^{\star
}(\tau _{z})\geq 0.$\footnote{%
Hereafter a shell label will be omitted if it will be not important to
emphasize that it can be restricted for certain s-configurations, i.e. we
shall write $\ ^{\shortmid }\widehat{S}_{A}^{\star }$ instead of $\
_{s}^{\shortmid }\widehat{S}_{A}^{\star }.$} Such s-metrics should possess
well-defined relativistic Lorentz properties if we elaborate on
self-consistent nonassociative phase generalizations of the Einstein theory.

\vskip5pt In similar forms, we can define the Shannon--Perelman entropy $\ 
\widehat{S}_{A}^{\star }$ for random variables on $\ _{s}\mathcal{M}^{\star} 
$ on phase spaces with velocity type fiber coordinates. Respective values
for associative and commutative gravitational and geometric/ information
flow theory can be defined and computed in generalized
Finsler-Lagrange-Hamilton variables \cite{sv20}. In such cases, $\widehat{S}%
_{A}$ for $_{s}\mathcal{M}$ and $\ ^{\shortmid }\widehat{S}_{A}$ for $\
_{s}^{\shortmid }\mathcal{M}$ are related via certain Legendre transforms, $%
\widehat{S}_{A}\sim $ $\ ^{\shortmid }\widehat{S}_{A},$ which can be encoded
into certain canonical almost symplectic s-structures. If Legendre
transforms are not defined, the constructions with $(\ _{s}\widehat{Z}(\tau
),\ \widehat{S}_{A},\ _{s}\mathcal{M})$ can be independent/ different from $%
(\ _{s}^{\shortmid }\widehat{Z}(\tau ),\ ^{\shortmid }\widehat{S}_{A},\
_{s}^{\shortmid }\mathcal{M}),$ see (\ref{spf}). This holds true for star
product deformations.

\subsubsection{Nonassociative conditional entropy and relative entropy}

\paragraph{Conditional and relative entropy for nonassociative GIFs: \newline
}

Let us consider a toy model when Alice is trying to communicate with Bob
sending classical information using solutions for nonassociative geometric
flow equations and/or Ricci soliton equations like we explained above. Such
young researchers are skilled in Ricci flow theory, mathematical relativity,
and quantum information theory. For Alice (the emitter), it is associated a
random variable $\ _{s}^{\star }A,$ with possible values, i.e. letters
consisting from an alphabet: $\ _{s}^{\star }a_{1},...,\ _{s}^{\star
}a_{r_{A}};$ and for Bob there are associated random data $\ _{s}^{\star }B,$
with letters $\ _{s}^{\star }b_{1},...,\ _{s}^{\star }b_{q_{B}}$ (i.e. what
can see/hear/ understand the receiver). We analyze how many bits of
information sent by Alice (for instance, using a message of $N_{A}$ letters;
we should distinguish such a notation from that for N-connections when
indices are for some coordinate frames) will receive Bob? We introduce the
necessary concepts of probability theory considering respectively for $\
_{s}^{\star }A$ and $\ _{s}^{\star }B$ some discrete random data for the $%
\tau $-parameter $\tau _{1},...\tau _{r_{A}}$ and $\tau _{1},...\tau
_{q_{B}}.$ Then we show how certain additional assumptions on nonassociative
geometric flows with continuous parameter evolution, or for a nonassociative
Ricci soliton describing both in $\ _{s}^{\star }A$ and $\ _{s}^{\star }B,$
statistical thermodynamics result in a GIF model.

In the probability theory, a random variable is usually denoted $A,B,C,$
etc., when $P_{A}(a)$ is the probability to observe $A=a.$ For discrete sets
of variables (sent letters) $a_{\check{r}},$ where $\check{r}=1,...,N_{A},$
and (received letters) $b_{\check{q}},$ where $\check{q}=1,...,N_{B},$ we
have $\sum_{\check{r}}P_{A}(a_{\check{r}})=1$ and denote for two random
variables that $P_{A,B}(a_{\check{r}},b_{\check{q}})$ is the probability to
observe $A=a_{\check{r}}$ and $B=b_{\check{q}}.$ If there are not special
assumptions on interactions between Alice and Bob, we can write 
\begin{equation}
P_{B}(b_{\check{q}})=\sum_{\check{r}}P_{A,B}(a_{\check{r}},b_{\check{q}}),
\label{auxf01}
\end{equation}%
where $P_{B}(b_{\check{q}})$ is the probability that Bob get $B=b_{\check{q}%
} $ letters from a sum of all choices which Alice may send him.

The \textbf{conditional probability} is by definition 
\begin{equation}
P_{A|B}(a_{\check{r}}|b_{\check{q}}):=\frac{P_{A,B}(a_{\check{r}},b_{\check{q%
}})}{P_{B}(b_{\check{q}})}  \label{auxf02}
\end{equation}%
which means that when Bob does hear $B=b_{\check{q}}$ he estimates of the
probability that Alice sent $a_{\check{r}},$ where $P_{B}(b_{\check{q}})$ is
defined by (\ref{auxf01}). Respectively, we can define the Shannon entropy
of the conditional probability (\ref{auxf02}) when $B=b_{\check{q}},$%
\begin{equation*}
P_{A|B=b_{\check{q}}}=-\sum_{\check{r}}P_{A|B}(a_{\check{r}}|b_{\check{q}%
})\log P_{A|B}(a_{\check{r}}|b_{\check{q}}),
\end{equation*}%
which is the remaining entropy in Alice's letter (from Bob's point o view)
when he has read $B=b_{\check{q}}.$

We introduce 
\begin{eqnarray*}
S_{AB} &:=&\sum_{\check{r},\check{q}}P_{A,B}(a_{\check{r}},b_{\check{q}%
})\log P_{A,B}(a_{\check{r}},b_{\check{q}})%
\mbox{ is the entropy of joint
distiribution }P_{A,B}(a_{\check{r}},b_{\check{q}})\mbox{ for the pair }A,B;
\\
S_{B} &:=& \sum_{\check{r},\check{q}}P_{A,B}(a_{\check{r}},b_{\check{q}%
})\log P_{B}(b_{\check{q}})%
\mbox{ is the entropy of the probability
distribution }P_{B}(b_{\check{q}}) (\ref{auxf01});
\end{eqnarray*}%
and define the \textbf{conditional entropy,} 
\begin{equation}
S_{A|B}=S(A|B):=\sum_{\check{q}}P_{B}(b_{\check{q}})S_{A|B=b_{\check{q}}}.
\label{conde}
\end{equation}%
It states the entropy the entropy that remains in the probability
distribution $A$ once $B$ is known. Using above formulas, we can prove that 
\begin{equation*}
S_{A|B}=S_{AB}-S_{B}\geq 0.
\end{equation*}%
This follows from the fact that by definition $S_{A|B}$ is a sum of ordinary
entropies $S_{A|B=b_{\check{q}}}$ (which are nonnegative) with positive
coefficients, see formulas (\ref{conde}). Such inequality can be not true
for some quantum/ geometric flow models, but we can select only those
configurations which satisfy such conditions and use them for elaborating
well defined probabilistic models.

The \textbf{mutual information} between $A$ and $B$ is by definition%
\begin{equation}
I(A;B):=S_{A}-S_{AB}+S_{B}\geq 0.  \label{minf}
\end{equation}%
Let us explain the informational meaning of entropies in this formula: $%
S_{A} $ is the total information contained in Alice's message; $S_{A|B}$ is
the information content which Bob still does not have after he observed $B.$
In result, we can treat $I(A;B)$ (\ref{minf}) as the difference of
information about $A$ that Bob gains when he receives $B,$ and it measures
how much we learn about $A$ by observing $B.$

The proofs of inequality in (\ref{minf}) and all above formulas for some
discrete $\tau _{1},...\tau _{r_{A}}$ and $\tau _{1},...\tau _{q_{B}}$ with
respective sums on values are similar to those provided in section 2 of \cite%
{preskill,witten20}, when all concepts (on relative entropy and basic
properties as a classical information model) and basic results can be
extended for nonassociative geometric flows characterized by respective
nonassociative Ricci soliton stated for respective randomly fixed $\tau $%
-parameterized characterizing systems $A$ and $B.$ We omit details on
proving such algebraic formulas and provide only most important s-adapted
nonholonomic generalizations as in section 3 of \cite{lbsvevv22}, which in
this work are generalized for the nonassociative entropy$\ ^{\shortmid }%
\widehat{S}_{A}^{\star }$.

For nonassociative GIFs, we can define and compute: the canonical
conditional entropy 
\begin{eqnarray}
\ ^{\shortmid }\widehat{S}_{A|B}^{\star } &:=&\ ^{\shortmid }\widehat{S}%
_{AB}^{\star }-\ ^{\shortmid }\widehat{S}_{B}^{\star }\geq 0,%
\mbox{
nonassociative canonical conditional entropy };  \label{condes} \\
\ ^{\shortmid }I^{\star }(\ _{s}^{\star }A;\ _{s}^{\star }B) &:= &\
^{\shortmid }\widehat{S}_{A}^{\star }-\ ^{\shortmid }\widehat{S}_{AB}^{\star
}+\ ^{\shortmid }\widehat{S}_{B}^{\star }\geq 0,%
\mbox{nonassociative
canonical mutual information }.  \label{minfs}
\end{eqnarray}%
If such nonassociative thermodynamic information variables are nonnegative
for some classes of solutions, we consider that they are well-defined for a
respective GIF model. We shall extend such formulas in subsection \ref%
{ssnonassocgifes} for explicit classes of nonassociative s-metrics.

\paragraph{Monotonicity of relative entropy for nonassociative GIFs: \newline
}

The concept of relative entropy for nonassociative GIFs can be motivated as
follows. Let we observe propagation of nonassociative geometric flow for
some discrete values of $\tau $-parameter a random variable $\ _{s}^{\star}A 
$ when a chosen class of solutions (a model/theory) predicts a probability
distribution $Q_{A}.$ When we observe $A=a_{\check{r}},$ with a set of
outcomes $1,2,...,\check{r},$ we say that the prediction is $q_{\check{r}%
}=Q_{A}(a_{\check{r}}).$ It may be a variant when our class chosen solutions
does not describe exactly the GIF processes, i.e. our theory may be wrong,
and this process is actually described by some different probability
distribution $p_{\check{r}}=P_{A}(a_{\check{r}}).$ In classical probability
theory, the \textbf{relative entropy} (per observation, also called the
Kullback-Lieber difference, see formula (20) in \cite{witten20}) is defined
as 
\begin{equation}
S(P_{A}\parallel Q_{A}):=\sum_{\check{r}=1}^{N_{A}}p_{\check{r}}(\log p_{%
\check{r}}-\log q_{\check{r}})\geq 0.  \label{relentr}
\end{equation}%
If the initial hypothesis (theory) is correct, $P_{A}=Q_{A}$ and $%
S(P_{A}\parallel Q_{A})=0.$ For $S(P_{A}\parallel Q_{A})>0,$ the initial
hypothesis is not correct but we can be sure of this only if $N_{A}\
S(P_{A}\parallel Q_{A})\gg 1.$ Here we note that the relative entropy (\ref%
{relentr}) is an important measure of the difference between probability
distributions $P_{A}$ and $Q_{A}$ when the symmetry of arguments is broken
because by definition it is assumed that $Q_{A}$ was the initial hypothesis
and $P_{A}$ is the correct solution.

We can connect the relative entropy $S(P_{A}\parallel Q_{A})$ (\ref{relentr}%
) to the mutual information $I(A;B)$ (\ref{minf}). For instance, if we
consider 
\begin{equation*}
Q_{A,B}(a_{\check{r}},b_{\check{q}})=P_{A}(a_{\check{r}})P_{B}(b_{\check{q}%
}),
\end{equation*}%
we can prove 
\begin{equation}
S(P_{A,B}\parallel Q_{A,B})=S_{A}+S_{B}-S_{AB}=I(A;B)\geq 0.
\label{suadditivpropr}
\end{equation}%
The condition $I=0$ is possible if both distributions are the same which
mean that $A$ and $B$ to begin with. The property (\ref{suadditivpropr}) is
called \textbf{subadditivity} of entropy.

The formulas (\ref{relentr}) and (\ref{suadditivpropr}) can be considered
for nonassociative Ricc flows if we consider a partition function (\ref{spf}%
) 
\begin{equation*}
\ _{P}^{\shortmid }\widehat{Z}(\tau )=\exp [\int_{\ _{P}^{\shortmid }%
\widehat{\Xi }}[-\ _{P}^{\shortmid }\widehat{f}+4]\ \left( 4\pi \tau \right)
^{-4}e^{-\ _{P}^{\shortmid }\widehat{f}}\ ^{\shortmid }\delta \ ^{\shortmid }%
\mathcal{V[~}_{P}^{\shortmid }\mathbf{g}_{\alpha _{s}\beta _{s}}],
\end{equation*}%
for a solution of type $\ _{P}^{\shortmid }\mathbf{g}_{\alpha _{s}\beta
_{s}} $ (\ref{ssolutions}) considered for $\tau =\tau _{0};$ or a discrete
set $\tau _{\check{r}}.$ Such a partition function can be defined by a
partition function determined by another class of solutions $\
_{Q}^{\shortmid }\mathbf{g}_{\alpha _{s}\beta _{s}},$ when 
\begin{equation*}
\ _{Q}^{\shortmid }\widehat{Z}(\tau )=\exp [\int_{\ _{Q}^{\shortmid }%
\widehat{\Xi }}[-\ _{Q}^{\shortmid }\widehat{f}+4]\ \left( 4\pi \tau \right)
^{-4}e^{-\ _{Q}^{\shortmid }\widehat{f}}\ ^{\shortmid }\delta \ ^{\shortmid }%
\mathcal{V[~}_{P}^{\shortmid }\mathbf{g}_{\alpha _{s}\beta _{s}}].
\end{equation*}%
Such partition functions allow to compute an entropy of type $\ ^{\shortmid }%
\widehat{S}_{A}^{\star },\ ^{\shortmid }\widehat{S}_{B}^{\star }$ and $\
^{\shortmid }\widehat{S}_{AB}^{\star }$ (\ref{shanonper}) for nonassociative
GIFs as in (\ref{nagthermodvalues}). As a result, we state the condition of
subadditivity in nonassociative theories, which can proved for $\kappa $%
-parametric Ricci solitonic configurations: 
\begin{equation}
\ ^{\shortmid }\widehat{S}^{\star }(\ _{P}^{\shortmid }\widehat{Z}%
_{A,B}\parallel \ _{Q}^{\shortmid }\widehat{Z}_{A,B})=\ ^{\shortmid }%
\widehat{S}_{A}^{\star }+\ ^{\shortmid }\widehat{S}_{B}^{\star }-\
^{\shortmid }\widehat{S}_{AB}^{\star }=\ ^{\shortmid }I^{\star }(\
_{s}^{\star }A;\ _{s}^{\star }B)\geq 0.  \label{suadditivgif}
\end{equation}%
Finally we note that if the conditions (\ref{suadditivgif}) are not
satisfied for a class of GIF model this means that the corresponding
classical information model is not well-defined and we have to consider
other theories/ classes of solutions.

\subsubsection{Strong subadditivity and monotonicity of mutual information
for nonassociative GIFs}

Considering discrete values of $\tau $-parameter of two random GIF variables 
$\ _{s}^{\star }A$ and $\ _{s}^{\star }B$ and partition functions 
\begin{equation}
\ _{P}^{\shortmid }\widehat{Z}_{A,B}(a_{\check{r}},b_{\check{q}})\mbox{ and }%
\ _{Q}^{\shortmid }\widehat{Z}_{A,B}(a_{\check{r}},b_{\check{q}}),
\label{aux03}
\end{equation}%
we define the reduced distributions $\ _{P}^{\shortmid }\widehat{Z}_{A}$ and 
$\ _{Q}^{\shortmid }\widehat{Z}_{A},$ when%
\begin{equation*}
\ _{P}^{\shortmid }\widehat{Z}_{A}(a_{\check{r}})=\sum_{\check{q}}\
_{P}^{\shortmid }\widehat{Z}_{A,B}(a_{\check{r}},b_{\check{q}})\mbox{ and }\
_{Q}^{\shortmid }\widehat{Z}_{A}(a_{\check{r}})=\sum_{\check{q}}\
_{Q}^{\shortmid }\widehat{Z}_{A,B}(a_{\check{r}},b_{\check{q}}).
\end{equation*}%
If we observe only a nonassociative GIF evolution $\ _{s}^{\star }A,$ the
confidence after many trials that the initial model is controlled by $\
^{\shortmid }\widehat{S}^{\star }(\ _{P}^{\shortmid }\widehat{Z}%
_{A}\parallel \ _{Q}^{\shortmid }\widehat{Z}_{A})$ which is computed using
formulas (\ref{nagthermodvalues}) but for $\ _{P}^{\shortmid }\widehat{Z}%
_{A}(a_{\check{r}})$ and $\ _{Q}^{\shortmid }\widehat{Z}_{A}(a_{\check{r}}).$
In a similar form, we can compute $\ ^{\shortmid }\widehat{S}^{\star }(\
_{P}^{\shortmid }\widehat{Z}_{A,B}\parallel \ _{Q}^{\shortmid }\widehat{Z}%
_{A,B})$ for (\ref{aux03}). So, it is hard to disprove the initial
hypothesis if we observe only a nonassociative system $\ _{s}^{\star }A,$
but we can derive more correct conclusions if we use the condition of 
\textbf{monotonicity of relative entropy} for GIFs:%
\begin{equation}
\ ^{\shortmid }\widehat{S}^{\star }(\ _{P}^{\shortmid }\widehat{Z}%
_{A,B}\parallel \ _{Q}^{\shortmid }\widehat{Z}_{A,B})\geq \ ^{\shortmid }%
\widehat{S}^{\star }(\ _{P}^{\shortmid }\widehat{Z}_{A}\parallel \
_{Q}^{\shortmid }\widehat{Z}_{A}).  \label{aux04}
\end{equation}

Let us investigate an important special case for three random GIF variables $%
\ _{s}^{\star }A,\ _{s}^{\star }B$ and $\ _{s}^{\star }C.$ We can describe
such a combined system by a joint partition function (or probability
distribution for discrete nonassociative Ricci solitons) $\ _{P}^{\shortmid }%
\widehat{Z}_{A,B,C}(a_{\check{r}},b_{\check{q}},c_{\check{w}})$ extending
for 3 variables above constructions. In discrete $\tau $-parametric form,
for star-entropies, we can extend the formula (\ref{aux04}) for monotonicity
of relative entropy with three variables,%
\begin{equation}
\ ^{\shortmid }\widehat{S}^{\star }(\ _{P}^{\shortmid }\widehat{Z}%
_{A,B,C}\parallel \ _{Q}^{\shortmid }\widehat{Z}_{A,B,C})\geq \ ^{\shortmid }%
\widehat{S}^{\star }(\ _{P}^{\shortmid }\widehat{Z}_{A,B}\parallel \
_{Q}^{\shortmid }\widehat{Z}_{A,B})  \label{aux05a}
\end{equation}%
Using above studied relations between relative entropy and mutual
information, we can define and compute%
\begin{eqnarray}
\ ^{\shortmid }\widehat{S}^{\star }(\ _{P}^{\shortmid }\widehat{Z}_{A,B,C}
&\parallel &\ _{Q}^{\shortmid }\widehat{Z}_{A,B,C})=\ ^{\shortmid }I^{\star
}(\ _{s}^{\star }A;\ _{s}^{\star }B;\ _{s}^{\star }C)=\ ^{\shortmid }%
\widehat{S}_{A}^{\star }-\ ^{\shortmid }\widehat{S}_{ABC}^{\star }+\
^{\shortmid }\widehat{S}_{BC}^{\star }\mbox{ and }  \notag \\
\ ^{\shortmid }\widehat{S}^{\star }(\ _{P}^{\shortmid }\widehat{Z}_{A,B}
&\parallel &\ _{Q}^{\shortmid }\widehat{Z}_{A,B})=\ ^{\shortmid }I^{\star
}(\ _{s}^{\star }A;\ _{s}^{\star }B)=\ ^{\shortmid }\widehat{S}_{A}^{\star
}-\ ^{\shortmid }\widehat{S}_{AB}^{\star }+\ ^{\shortmid }\widehat{S}%
_{B}^{\star }.  \label{aux05}
\end{eqnarray}%
Formulas (\ref{aux04}), (\ref{aux05a}) and (\ref{aux05}) result in the
inequality 
\begin{equation*}
\ ^{\shortmid }\widehat{S}_{A}^{\star }-\ ^{\shortmid }\widehat{S}%
_{ABC}^{\star }+\ ^{\shortmid }\widehat{S}_{BC}^{\star }\geq \ ^{\shortmid }%
\widehat{S}_{A}^{\star }-\ ^{\shortmid }\widehat{S}_{AB}^{\star }+\
^{\shortmid }\widehat{S}_{B}^{\star }.
\end{equation*}%
It can be used to formulate the conditions: 
\begin{eqnarray}
\mbox{\bf strong subadditivity }:\ ^{\shortmid }\widehat{S}_{AB}^{\star }+\
^{\shortmid }\widehat{S}_{BC}^{\star } &\geq &\ ^{\shortmid }\widehat{S}%
_{B}^{\star }+\ ^{\shortmid }\widehat{S}_{ABC}^{\star };
\label{nonassocstrongsubadd} \\
\mbox{ \bf monotonicity of mutual information }:\ ^{\shortmid }I^{\star }(\
_{s}^{\star }A;\ _{s}^{\star }B\ _{s}^{\star }C) &\geq &\ ^{\shortmid
}I^{\star }(\ _{s}^{\star }A;\ _{s}^{\star }B).  \notag
\end{eqnarray}%
The intuition for (\ref{nonassocstrongsubadd}) is that the geometric flow
information about a nonassociative random system $\ _{s}^{\star }A$ by
observing both $\ _{s}^{\star }B$ and $\ _{s}^{\star }C$ is at least as much
as one could be observed only for $\ _{s}^{\star }B.$

We conclude this subsection with such remarks: Using concepts of classical
information theory, we can elaborate on a theory of nonassociative GIFs when
the modified Perelman thermodynamics is transformed into a model of
information thermodynamics. The entropy $\ ^{\shortmid }\widehat{S}%
_{A}^{\star }$ contains information on nonassociative star product R-flux
deformations of canonical geometric flow/ nonholonomic Ricci soliton
systems. For certain discrete values of the geometric evolution parameter $%
\tau ,$ we can define in standard classical information form respective
conditional probability distributions which are condition on some
observations. Such constructions have analogies for quantum mechanical cases
being generalized and studied for associative/ commutative QGIFs in \cite%
{sv20}. In section \ref{sec3}, we shall analyze how this property can be
extended for nonassociative QGIFs.

\subsection{The nonassociative GIF theory \& geometric thermodynamics of
nonholonomic Einstein systems}

\label{ssnonassocgifes}

In this subsection, we study classical nonassociative GIFs models defined by
solutions $\ ^{\shortmid }\mathbf{g}(\tau )=\{\ ^{\shortmid }\mathbf{g}%
_{\alpha _{s}\beta _{s}}(\tau )\}$ (\ref{ssolutions}) of effective
nonholonomic Einstein equations (\ref{nonassocrf})\ (in brief, NESs, i.e.
nonholonomic Einstein systems) encoding star product R-flux contributions
from string theory. In canonical s-adapted thermodynamic variables, such
systems are characterized by data $\left[ \ _{s}^{\shortmid }\widehat{%
\mathcal{W}}^{\star }(\tau );\ _{s}^{\shortmid }\widehat{\mathcal{Z}}^{\star
}(\tau ),\ _{s}^{\shortmid }\widehat{\mathcal{E}}^{\star }(\tau ),\
_{s}^{\shortmid }\widehat{\mathcal{S}}^{\star }(\tau ),\ _{s}^{\shortmid }%
\widehat{\sigma }^{\star }(\tau )\right] ,$ see formulas (\ref{nawfunct})
and (\ref{nagthermodvalues}). \ To elaborate on conditional GIF models
systems $\ _{s}^{\star }\widehat{A}[\ ^{\shortmid }\mathbf{g}(\tau )]$ and $%
\ _{s}^{\star }\widehat{B}[\ _{1}^{\shortmid }\mathbf{g}(\tau )]$ for two
exact/parametric solutions on $\ _{s}^{\shortmid }\mathcal{M}^{\star
}\otimes \ _{s}^{\shortmid }\mathcal{M}^{\star },$ when the local
coordinates are $(\ ^{\shortmid }u,\ _{1}^{\shortmid }u)$ and the
normalizing functions are of type $\ _{AB}\widehat{f}(\ ^{\shortmid }u,\
_{1}^{\shortmid }u).$ A s-metric structure on such tensor products of
nonholonomic phase spaces can be parameterized in the form$\
_{AB}^{\shortmid }\mathbf{g}(\tau )=\{\ ^{\shortmid }\mathbf{g}(\tau ),\
_{1}^{\shortmid }\mathbf{g}(\tau )\}.$ Respectively, we can define, for
instance, $\tau $-families of canonical s--connections $\ _{AB}^{\shortmid }%
\widehat{\mathbf{D}}^{\star }(\tau )=\ _{A}^{\shortmid }\widehat{\mathbf{D}}%
^{\star }(\tau )+\ _{B}^{\shortmid }\widehat{\mathbf{D}}^{\star }(\tau ),$
of type (\ref{candistrnas}), and corresponding canonical scalar curvatures
of type $\ _{s}^{\shortmid }\widehat{\mathbf{R}}sc(\tau )$ (\ref%
{ricciscsymnonsym}),%
\begin{equation*}
\ _{s}^{\shortmid }\widehat{\mathbf{R}}sc^{\star }(\tau )+\ _{s1}^{\shortmid
}\widehat{\mathbf{R}}sc^{\star }(\tau ),\mbox{ for }\kappa ^{0},%
\mbox{
results in }\ _{sAB}^{\shortmid }\widehat{R}(\tau )=\ _{s}^{\shortmid }%
\widehat{R}(\tau )+\ _{s1}^{\shortmid }\widehat{R}(\tau ).
\end{equation*}

To define continuous analogs of (\ref{suadditivgif}) for GIFs, we introduce
the thermodynamic generating function (as a generalization of partition
function (\ref{spf})) 
\begin{equation}
\ _{AB}^{\shortmid }\widehat{\mathcal{Z}}[\mathbf{g}(\tau ),\ _{1}\mathbf{g}%
(\tau )]=\ \int_{\ _{s}^{\shortmid }\widehat{\Xi }}\int_{\ _{s1}^{\shortmid }%
\widehat{\Xi }}(4\pi \tau )^{-8}e^{-\ _{AB}\widehat{f}}\sqrt{|\ ^{\shortmid }%
\mathbf{g}|}\sqrt{|\ _{1}^{\shortmid }\mathbf{g}|}d^{8}\ ^{\shortmid }u\
d^{8}\ _{1}^{\shortmid }u(-\ _{AB}\widehat{f}+8),\mbox{ for
}\ _{s}^{\shortmid }\mathcal{M}^{\star }\otimes \ _{s}^{\shortmid }\mathcal{M%
}^{\star }\mathbf{.}  \label{twogenf}
\end{equation}%
For nonassociative GIFs of NES systems, the canonical thermodynamic entropy
function can be computed using formulas 
\begin{eqnarray}
\ _{AB}^{\shortmid }\widehat{\mathcal{S}}^{\star }(\tau ) =\ ^{\shortmid }%
\widehat{\mathcal{S}}^{\star }\ [\ _{s}^{\star }\widehat{A},\ _{s}^{\star }%
\widehat{B}](\tau )&=& -\int_{\ _{s}^{\shortmid }\widehat{\Xi }}\int_{\
_{s1}^{\shortmid }\widehat{\Xi }}(4\pi \tau )^{-8}e^{-\ _{AB}\widehat{f}}%
\sqrt{|\ ^{\shortmid }\mathbf{g}|}\sqrt{|\ _{1}^{\shortmid }\mathbf{g}|}%
d^{8}\ ^{\shortmid }u\ d^{8}\ _{1}^{\shortmid }u  \label{twoentr} \\
&&\ \left[ \tau (\ _{s}^{\shortmid }\widehat{R}^{\star }+\ _{s1}^{\shortmid }%
\widehat{R}^{\star }+|\ \ _{AB}^{\shortmid }\widehat{\mathbf{D}}^{\star }%
\widehat{f}|^{2})+\ _{AB}\widehat{f}-16\right] .  \notag
\end{eqnarray}%
For $\tau $-continuous (nonassociative) geometric flow thermodynamic
variables, we shall use "calligraphic" symbols instead of "boldface" ones.

Using (\ref{twogenf}) and (\ref{twoentr}), we claim that the formulas for
the conditional entropy, (\ref{conde}) and (\ref{condes}), and mutual
information, (\ref{minf}) and (\ref{minfs}), are respectively generalized
for continuous GIFs of NES,\footnote{%
proofs can be performed in any point of respective causal curves on Lorentz
manifolds lifted on respective (co) tangent bundles used for modelling
nonassociative phase spaces and flow evolution of geometric/physical
s-objects} 
\begin{equation*}
\ ^{\shortmid }\widehat{\mathcal{S}}^{\star }\ [\ _{s}^{\star }\widehat{A}|\
_{s}^{\star }\widehat{B}]:=\ \ _{AB}^{\shortmid }\widehat{\mathcal{S}}%
^{\star }-\ ^{\shortmid }\widehat{\mathcal{S}}^{\star }[\ _{s}^{\star }%
\widehat{B}]\geq 0\mbox{
and }\ \ ^{\shortmid }\widehat{\mathcal{J}}^{\star }\ [\ _{s}^{\star }%
\widehat{A};\ _{s}^{\star }\widehat{B}]:=\ ^{\shortmid }\widehat{\mathcal{S}}%
^{\star }[\ _{s}^{\star }\widehat{A}]-\ \ _{AB}^{\shortmid }\widehat{%
\mathcal{S}}^{\star }+\ ^{\shortmid }\widehat{\mathcal{S}}^{\star }[\
_{s}^{\star }\widehat{B}]\geq 0.
\end{equation*}%
Similar definitions and formulas can be generalized respectively for the
relative entropy and mutual information (see (\ref{relentr}), (\ref%
{suadditivpropr}) and (\ref{suadditivgif})) $\tau $-continuous
nonassociative GIFs and NES.

Let us consider generalizations of above formulas for nonassociative GIFs of
three NES $\ _{s}^{\star }\widehat{A},\ _{s}^{\star }\widehat{B}$ and $\
_{s}^{\star }\widehat{C}.$ Using standard methods in any point of causal
curves lifted from Lorentz manifolds to (co) tangent bundles and applying
explicit integral N-adapted calculations, for instance, on $\
_{s}^{\shortmid }\mathcal{M}^{\star }\otimes \ _{s}^{\shortmid }\mathcal{M}%
^{\star }$ $\mathbf{\otimes }\ _{s}^{\shortmid }\mathcal{M}^{\star }$ that%
\begin{eqnarray*}
\ ^{\shortmid }\widehat{\mathcal{S}}^{\star }\ [\ _{AB}^{\shortmid }\widehat{%
\mathcal{Z}}||\ \ _{AB}^{\shortmid }\widehat{\mathcal{Z}}] &\geq &\
^{\shortmid }\widehat{\mathcal{S}}^{\star }\ [\ _{A}^{\shortmid }\widehat{%
\mathcal{Z}}||\ \ _{A}^{\shortmid }\widehat{\mathcal{Z}}]; \\
\ ^{\shortmid }\widehat{\mathcal{S}}^{\star }\ [\ _{ABC}^{\shortmid }%
\widehat{\mathcal{Z}}||\ \ _{ABC}^{\shortmid }\widehat{\mathcal{Z}}] &\geq
&\ ^{\shortmid }\widehat{\mathcal{S}}^{\star }\ [\ _{AB}^{\shortmid }%
\widehat{\mathcal{Z}}||\ \ _{AB}^{\shortmid }\widehat{\mathcal{Z}}],%
\mbox{monotonicity of canonical relative entropy}.
\end{eqnarray*}%
The conditions of strong subadditivity for $\tau $-continuous nonassociative
GIFs and NES entropies (see also formulas (\ref{nonassocstrongsubadd})) are
stated by formulas 
\begin{equation*}
\ _{A}^{\shortmid }\widehat{\mathcal{S}}^{\star }-\ \ _{ABC}^{\shortmid }%
\widehat{\mathcal{S}}^{\star }-\ \ _{BC}^{\shortmid }\widehat{\mathcal{S}}%
^{\star }\geq \ _{A}^{\shortmid }\widehat{\mathcal{S}}^{\star }-\ \
_{AB}^{\shortmid }\widehat{\mathcal{S}}^{\star }-\ \ _{B}^{\shortmid }%
\widehat{\mathcal{S}}^{\star },\mbox{ or }\ _{AB}\widehat{\mathcal{S}}+\
_{BC}\widehat{\mathcal{S}}\geq \ _{B}\widehat{\mathcal{S}}+\ _{ABC}\widehat{%
\mathcal{S}}.
\end{equation*}%
These formulas can be written in equivalent form as the condition of
monotonicity of $\tau $-continuous nonassociative GIFs and NES mutual
information, 
\begin{equation*}
\ ^{\shortmid }\widehat{\mathcal{J}}^{\star }\ [\ _{s}^{\star }\widehat{A};\
_{s}^{\star }\widehat{B}\ _{s}^{\star }\widehat{C}]\ \geq \ ^{\shortmid }%
\widehat{\mathcal{J}}^{\star }\ [\ _{s}^{\star }\widehat{A};\ _{s}^{\star }%
\widehat{B}].
\end{equation*}

Here we note that for three $\tau $-continuous nonassociative GIF systems,
there are involved thermodynamic generating functions generalizing (\ref%
{twogenf}) in the form, 
\begin{eqnarray}
\ _{ABC}\widehat{\mathcal{Z}}[\mathbf{g}(\tau ),\ _{1}\mathbf{g}(\tau ),\
_{2}\mathbf{g}(\tau )] &=&\ \int_{\ _{s}^{\shortmid }\widehat{\Xi }}\int_{\
_{s1}^{\shortmid }\widehat{\Xi }}\int_{\ _{s2}^{\shortmid }\widehat{\Xi }%
}(4\pi \tau )^{-12}e^{-\ _{ABC}\widehat{f}}\sqrt{|\ ^{\shortmid }\mathbf{g}|}%
\sqrt{|\ _{1}^{\shortmid }\mathbf{g}|}\sqrt{|\ _{2}^{\shortmid }\mathbf{g}|}%
d^{8}u\ d^{8}\ _{1}u\ d^{8}\ _{2}u  \notag \\
&&(-\ _{ABC}\widehat{f}+12),\mbox{ for }\ _{s}^{\shortmid }\mathcal{M}%
^{\star }\otimes \ _{s}^{\shortmid }\mathcal{M}^{\star }\otimes \
_{s}^{\shortmid }\mathcal{M}^{\star }\mathbf{,}  \label{threegenf}
\end{eqnarray}%
with a normalizing function $\ _{ABC}\widehat{f}(\ ^{\shortmid }u,\
_{1}^{\shortmid }u,\ _{2}^{\shortmid }u).$ In this formula, the $\tau $%
-running s-metric structure is defined as $\ _{ABC}^{\shortmid }\mathbf{g}%
(\tau )=\{\ ^{\shortmid }\mathbf{g}(\tau ),\ _{1}^{\shortmid }\mathbf{g}%
(\tau ),\ _{2}^{\shortmid }\mathbf{g}(\tau )\}.$ Using $\tau $-families of
respective canonical s--connections, we can define 
\begin{equation*}
\ _{ABC}^{\shortmid }\widehat{\mathbf{D}}^{\star }(\tau )=\ _{A}^{\shortmid }%
\widehat{\mathbf{D}}^{\star }(\tau )+\ _{B}^{\shortmid }\widehat{\mathbf{D}}%
^{\star }(\tau )+\ _{C}^{\shortmid }\widehat{\mathbf{D}}^{\star }(\tau ),
\end{equation*}%
when the corresponding canonical scalar curvatures of type $\
_{s}^{\shortmid }\widehat{\mathbf{R}}sc(\tau )$ (\ref{ricciscsymnonsym}),%
\begin{equation*}
\ _{s}^{\shortmid }\widehat{\mathbf{R}}sc^{\star }(\tau )+\ _{s1}^{\shortmid
}\widehat{\mathbf{R}}sc^{\star }(\tau )+\ _{s2}^{\shortmid }\widehat{\mathbf{%
R}}sc^{\star }(\tau ),\mbox{ for }\kappa ^{0},\mbox{
results in }\ _{sABC}^{\shortmid }\widehat{R}(\tau )=\ _{s}^{\shortmid }%
\widehat{R}(\tau )+\ _{s1}^{\shortmid }\widehat{R}(\tau )+\ _{s2}^{\shortmid
}\widehat{R}(\tau ).
\end{equation*}%
These formulas allow to define and compute the entropy function 
\begin{eqnarray}
\ _{ABC}\widehat{\mathcal{S}}^{\star }(\tau ) &=&\ \widehat{\mathcal{S}}%
^{\star }[\widehat{A},\widehat{B},\widehat{C}](\tau ):=-\ \int_{\
_{s}^{\shortmid }\widehat{\Xi }}\int_{\ _{s1}^{\shortmid }\widehat{\Xi }%
}\int_{\ _{s2}^{\shortmid }\widehat{\Xi }}(4\pi \tau )^{-24}e^{-\ _{ABC}%
\widehat{f}}\sqrt{|\mathbf{g}|}\sqrt{|\ _{1}\mathbf{g}|}\sqrt{|\ _{2}\mathbf{%
g}|}d^{4}u\ d^{4}\ _{1}u\ d^{4}\ _{2}u  \notag \\
&&\left[ \tau \left( \ _{s}\widehat{R}^{\star }+\ _{s1}\widehat{R}^{\star
}+\ _{s2}\widehat{R}^{\star }+|\widehat{\mathbf{D}}_{ABC}^{\star }\widehat{f}%
+\ _{1}\widehat{\mathbf{D}}_{ABC}^{\star }\widehat{f}+\ _{2}\widehat{\mathbf{%
D}}_{ABC}^{\star }\widehat{f}|^{2}\right) +\ _{ABC}\widehat{f}-24\right] .
\label{threeentr}
\end{eqnarray}

Finally, we note that we can work, for instance, with the W-entropy $\
^{\shortmid }\widehat{\mathcal{W}}^{\star }[\ _{s}^{\star }\widehat{A}]$ (as
in (\ref{nawfunct}) because it has the meaning of "minus-entropy") instead
of the S-entropy $\ ^{\shortmid }\widehat{\mathcal{S}}^{\star }[\
_{s}^{\star }\widehat{A}]$ and formulate a theory of classical
nonassociative GIFs. Nevertheless, the second variant has certain priorities
because it allows to formulate a model of nonassociative information
thermodynamics with variables (\ref{nagthermodvalues}).

\section{Nonassociative geometric and quantum information flow theory}

\label{sec3} In this section, we elaborate on the theory of nonassociative
quantum information flows, QGIFs, for nonholonomic Einstein systems, NESs.
Such nonassociative geometric and physical models can be constructed as star
product R-flux deformations of the constructions performed in subsections
3.2, 3.3 and 3.4 of \cite{lbsvevv22}. The main assumption is that at least
for small nonholonomic deformations and $\kappa $--parametric decompositions
the geometric/ physical s-objects and fundamental geometric/ physical
equations are described by certain linearized quantum evolution models. In
nonlinear forms, the geometric structures are defined by classical nonlinear
GIF systems, which can be star product deformed and quantized in the
vicinity of certain configurations for NESs. In such an approach, the basic
concepts of quantum information theory \cite%
{preskill,witten20,ryu16,vanraam10,jacobson16,aolita15,nishioka18} can be
extended in nonholonomic form to associative/ commutative geometric flows 
\cite{sv20} and then completed with $\kappa $--parametric terms encoding
nonassociative data.

\subsection{Basic ingredients and main properties of QGIF theories}

Let us show how basic concepts of quantum mechanics and information theory
can be generalized and used for formulating models of QGIF theories. A
classical nonassociative GIFs is defined by a canonical partition function $%
\ _{s}^{\shortmid }\widehat{Z}[\ ^{\shortmid }\mathbf{g}(\tau )]$ (\ref{spf}%
). Such a partition function is determined by a solution $\ ^{\shortmid }%
\mathbf{g}(\tau )$ of a $\kappa $--parametric nonassociative geometric flows
equations (\ref{nonassocrf}). As a statistical thermodynamic model, a
thermo-field theory is modified for geometric flows in some forms which
allow to construct a thermo-geometric flow model with $\tau$ considered as a
temperature like parameter following Perelman's ideas \cite{perelman1}. The
nonassociative geometric thermodynamic constructions are characterized by
respective star entropy and energy variables, $\ _{s}^{\shortmid }\widehat{%
\mathcal{S}}^{\star }(\tau )$ and $\ _{s}^{\shortmid }\widehat{\mathcal{E}}%
^{\star}(\tau ),$ see formulas (\ref{nagthermodvalues}).\footnote{%
Such nonassociative geometric flow configurations are also characterized by
respective nonassociative symmetric, $\ _{\star s}^{\shortmid }\mathbf{g}%
\mathfrak{(\tau )}$, and nonsymmetric metric, $\ _{\star }^{\shortmid }%
\mathfrak{g}_{\alpha _{s}\beta _{s}}\mathfrak{(\tau )},$ with such star and
s-tensor products decompositions, 
\begin{eqnarray*}
\ _{\star s}^{\shortmid }\mathbf{g}\mathfrak{(\tau )} &=&\ _{\star
}^{\shortmid }\mathbf{g}_{\alpha _{s}\beta _{s}}\mathfrak{(\tau )}\star _{s}%
\mathfrak{(\tau )}[\ ^{\shortmid }\mathbf{e}^{\alpha _{s}}\mathfrak{(\tau )}%
\otimes _{\star s}\ ^{\shortmid }\mathbf{e}^{\beta _{s}\mathfrak{(\tau )}}],%
\mbox{ where }\ _{\star }^{\shortmid }\mathbf{g}(\ ^{\shortmid }\mathbf{e}%
_{\alpha _{s}},\ ^{\shortmid }\mathbf{e}_{\beta _{s}})=\ _{\star
}^{\shortmid }\mathbf{g}_{\alpha _{s}\beta _{s}}=\ _{\star }^{\shortmid }%
\mathbf{g}_{\beta _{s}\alpha _{s}} \\
\ \ _{\star }^{\shortmid }\mathfrak{g}_{\alpha _{s}\beta _{s}}\mathfrak{%
(\tau )} &=&\ \ _{\star }^{\shortmid }\mathbf{g}_{\alpha _{s}\beta _{s}}%
\mathfrak{(\tau )}-\kappa \mathcal{R}_{\quad \alpha _{s}}^{\tau _{s}\xi
_{s}}\ \mathfrak{(\tau )}\ ^{\shortmid }\mathbf{e}_{\xi _{s}}\mathfrak{(\tau
)}\ \ _{\star }^{\shortmid }\mathbf{g}_{\beta _{s}\tau _{s}}\mathfrak{(\tau )%
}=\ _{\star }^{\shortmid }\mathfrak{g}_{\alpha _{s}\beta _{s}}^{[0]}%
\mathfrak{(\tau )}+\ \ _{\star }^{\shortmid }\mathfrak{g}_{\alpha _{s}\beta
_{s}}^{[1]}(\kappa ,\mathfrak{\tau })=\ \ _{\star }^{\shortmid }\mathfrak{%
\check{g}}_{\alpha _{s}\beta _{s}}\mathfrak{(\tau )}+\ \ _{\star
}^{\shortmid }\mathfrak{a}_{\alpha _{s}\beta _{s}}\mathfrak{(\tau )},
\end{eqnarray*}%
where $\mathcal{R}_{\quad \alpha _{s}}^{\tau _{s}\xi _{s}}$ are related to $%
R^{i_{s}j_{s}a_{s}}$ from\ (\ref{starpn}) via frame transforms and
multiplications on some real/complex coefficients. In holonomic form,
respective details are presented in \cite{aschieri17} and, for nonholonomic
s-adapted generalizations, in \cite{partner01,partner02}. In these formulas
(not writing dependencies on $\mathfrak{\tau }$), $\ _{\star }^{\shortmid }%
\mathfrak{\check{g}}_{\alpha _{s}\beta _{s}}$ is the symmetric part and $\
_{\star }^{\shortmid }\mathfrak{a}_{\alpha _{s}\beta _{s}}$ is the
anti-symmetric part, 
\begin{eqnarray*}
\ _{\star }^{\shortmid }\mathfrak{\check{g}}_{\alpha _{s}\beta _{s}} &:=&%
\frac{1}{2}(\ _{\star }^{\shortmid }\mathfrak{g}_{\alpha _{s}\beta _{s}}+\
_{\star }^{\shortmid }\mathfrak{g}_{\beta _{s}\alpha _{s}})=\ _{\star
}^{\shortmid }\mathbf{g}_{\alpha _{s}\beta _{s}}-\frac{\kappa }{2}\left( 
\mathcal{R}_{\quad \beta _{s}}^{\tau _{s}\xi _{s}}\ \ ^{\shortmid }\mathbf{e}%
_{\xi _{s}}\ _{\star }^{\shortmid }\mathbf{g}_{\tau _{s}\alpha _{s}}+%
\mathcal{R}_{\quad \alpha _{s}}^{\tau _{s}\xi _{s}}\ \ ^{\shortmid }\mathbf{e%
}_{\xi _{s}}\ _{\star }^{\shortmid }\mathbf{g}_{\beta _{s}\tau _{s}}\right)
=\ _{\star }^{\shortmid }\mathfrak{\check{g}}_{\alpha _{s}\beta
_{s}}^{[0]}+\ _{\star }^{\shortmid }\mathfrak{\check{g}}_{\alpha _{s}\beta
_{s}}^{[1]}(\kappa ), \\
&&\mbox{ for }\ _{\star }^{\shortmid }\mathfrak{\check{g}}_{\alpha _{s}\beta
_{s}}^{[0]}=\ \ _{\star }^{\shortmid }\mathbf{g}_{\alpha _{s}\beta _{s}}%
\mbox{ and }\ \ _{\star }^{\shortmid }\mathfrak{\check{g}}_{\alpha _{s}\beta
_{s}}^{[1]}(\kappa )=-\frac{\kappa }{2}\left( \mathcal{R}_{\quad \beta
_{s}}^{\tau _{s}\xi _{s}}\ \ ^{\shortmid }\mathbf{e}_{\xi _{s}}\ \ _{\star
}^{\shortmid }\mathbf{g}_{\tau _{s}\alpha _{s}}+\mathcal{R}_{\quad \alpha
_{s}}^{\tau _{s}\xi _{s}}\ ^{\shortmid }\mathbf{e}_{\xi _{s}}\ \ _{\star
}^{\shortmid }\mathbf{g}_{\beta _{s}\tau _{s}}\right) ; \\
\ _{\star }^{\shortmid }\mathfrak{a}_{\alpha _{s}\beta _{s}}&:=&\frac{1}{2}%
(\ _{\star }^{\shortmid }\mathfrak{g}_{\alpha _{s}\beta _{s}}-\ _{\star
}^{\shortmid }\mathfrak{g}_{\beta _{s}\alpha _{s}})=\frac{\kappa }{2}\left( 
\mathcal{R}_{\quad \beta _{s}}^{\tau _{s}\xi _{s}}\ ^{\shortmid }\mathbf{e}%
_{\xi _{s}}\ \ _{\star }^{\shortmid }\mathbf{g}_{\tau _{s}\alpha _{s}}-%
\mathcal{R}_{\quad \alpha _{s}}^{\tau _{s}\xi _{s}}\ ^{\shortmid }\mathbf{e}%
_{\xi _{s}}\ _{\star }^{\shortmid }\mathbf{g}_{\beta _{s}\tau _{s}}\right)
=\ _{\star }^{\shortmid }\mathfrak{a}_{\alpha _{s}\beta _{s}}^{[1]}(\kappa )=%
\frac{1}{2}(\ _{\star }^{\shortmid }\mathfrak{g}_{\alpha _{s}\beta
_{s}}^{[1]}(\kappa )-\ _{\star }^{\shortmid }\mathfrak{g}_{\beta _{s}\alpha
_{s}}^{[1]}(\kappa )).
\end{eqnarray*}%
For solutions constructed using the AFCDM, such nonassociative geometric
s-objects consist from $\kappa $-parametric R-deformations of some
associative and commutative symmetric s-metrics. So, to elaborate on GIF or
QGIF models we define and compute thermodynamic variables as in GR, or MGTs,
and then deform the nonholonomic geometric structures using respective $%
\kappa $-parametric terms.} For reviews of thermo-field models with emphasis
on the applications in quantum information theory, we cite \cite%
{preskill,witten20,nishioka18}. We also assume that the reader has a
background knowledge about the basics of QFT in curved spaces, GR and
nonholonomic geometric models in modern physics \cite%
{birrell82,misner,hawking73,wald82,vbv18}.

\subsubsection{Quantum state vectors and density matrices associated to
models of GIFs and QGIFs}

\vskip5pt To extend the nonlinear geometric approach elaborated for GIF
theories in a form to describe QGIFs we consider an exact/parametric
solution of modified classical (nonassociative) geometric flow equations
with $\kappa $--parametric terms. For such a configuration, we formulate a
procedure for computing linear quantum fluctuations and defining quantum
information variables. We suppose that in any point $\ ^{\shortmid }u\in
U^{\star }\subset \ _{s}\mathcal{M}^{\star }$ along a causal curve covering
an open region $U^{\star }$ a Hilbert space $\mathcal{H}_{\mathcal{A}}$ is
defined. A state vector $\psi _{\mathcal{A}}\in \mathcal{H}_{\mathcal{A}}$
is an infinite dimensional complex vector function. For various computations
in quantum information theory, such a value can be approximated to a vector
in complex spaces of finite dimension.\footnote{%
In this section, we shall write in brief $\mathcal{A}$ instead of $\
^{\shortmid }\widehat{\mathcal{A}}^{\star }$ and put respective labels
without $\star $ and "$\ ^{\shortmid }"$ in order to simplify the system of
notations but only if that will not result in ambiguities. Here we note that
in this work the geometric and quantum model constructions with quantum
density matrices encode certain $\kappa $--parametric nonassociative
geometric data.} Locally on $U^{\star },$ a state vector $\psi _{\mathcal{A}%
} $ is considered as a solution of the Schr\"{o}dinger equation with as a
well-defined quantum Hamiltonian $\boldsymbol{H}_{\mathcal{A}}$, see details
in \cite{preskill,witten20}. In a self-consistent QGIF model a $\boldsymbol{H%
}_{\mathcal{A}}$ (modular Hamiltonian) is related to a $\ _{s}^{\shortmid }%
\widehat{\mathcal{E}}^{\star }(\tau )$ (\ref{nagthermodvalues}).

\vskip5pt We shall work also with combined Hilbert spaces defined as tensor
products, $\mathcal{H}_{\mathcal{AB}}=\mathcal{H}_{\mathcal{A}}\otimes 
\mathcal{H}_{\mathcal{B}}.$ In such spaces, $\mathcal{H}_{\mathcal{A}}$ is
an associate Hilbert space associated to a complementary system $\mathcal{A}%
, $ when $\psi _{\mathcal{AB}}=\psi _{\mathcal{A}}\otimes \psi _{\mathcal{B}%
}\in \mathcal{H}_{\mathcal{AB}}$ for $\psi _{\mathcal{A}}=1_{\mathcal{A}}$
taken as the unity state vector. A quantum system is represented by matrices
of dimension $\underline{N}\times \underline{M}$ if $\dim \widehat{\mathcal{H%
}}_{\mathcal{A}}=\underline{N}$ and $\dim \mathcal{H}_{\mathcal{B}}=%
\underline{M}.$ For quantum data, we shall underline symbols for certain
indices, dimensions and symbols if it will be necessary to avoid ambiguities
with the N-connection symbol $\mathbf{N}$ and other s-shell notations. The
Schmidt decomposition can be performed for a pure state function, 
\begin{equation}
\ \psi _{\mathcal{AB}}=\sum_{\underline{i}}\sqrt{p_{\underline{i}}}\psi _{%
\mathcal{A}}^{\underline{i}}\otimes \psi _{\mathcal{B}}^{\underline{i}},
\label{schmidt}
\end{equation}%
for any index $\underline{i}=1,2,....$ (up to a finite value). A state
vector $\psi _{\mathcal{A}}^{\underline{i}}$ is orthonormal if $<\psi _{%
\mathcal{A}}^{\underline{i}},\psi _{\mathcal{A}}^{\underline{j}}>= <\psi _{%
\mathcal{B}}^{\underline{i}},\psi _{\mathcal{B}}^{\underline{j}}>=\delta ^{%
\underline{i}\underline{j}},$ where $\delta ^{\underline{i}\underline{j}}$
is the Kronecker symbol. Considering $p_{\underline{i}}>0$ and $\sum_{%
\underline{i}}p_{\underline{i}}=1,$ we treat $p_{\underline{i}}$ as
probabilities. In general, such $\psi _{\mathcal{A}}^{\underline{i}}$ and/or 
$\psi _{\mathcal{B}}^{\underline{i}}$ do not define bases of $\mathcal{H}_{%
\mathcal{A}}$ and/or $\mathcal{H}_{\mathcal{B}}.$

\vskip5pt The quantum density matrix of a system $\mathcal{A},$ 
\begin{equation}
\ \rho _{\mathcal{A}}:=\sum_{\underline{a}}p_{\underline{a}}|\psi _{\mathcal{%
A}}^{\underline{a}}><\otimes \psi _{\mathcal{A}}^{\underline{a}}|,
\label{densmcan}
\end{equation}%
is defined as a Hermitian and positive semi-definite operator with trace $%
Tr_{\mathcal{H}_{\mathcal{A}}}\rho _{\mathcal{A}}=1.$ In this work, the hat
symbol will be used if necessary to emphasize that the constructions are
associated to canonical nonholonomic variables and respective
(nonassocitive) geometric flow and gravitational systems. Having defined an
operator $\rho _{\mathcal{A}}$ (\ref{densmcan}) we can compute the
expectation value of any operator $\widehat{\mathcal{O}}_{\mathcal{A}}$
characterizing additionally such a system,%
\begin{eqnarray}
<\ \widehat{\mathcal{O}}>_{\mathcal{AB}} &=&<\psi _{\mathcal{AB}}|\ \widehat{%
\mathcal{O}}\otimes 1_{\mathcal{B}}|\psi _{\mathcal{AB}}>=\sum_{\underline{i}%
}p_{\underline{i}}<\psi _{\mathcal{A}}^{\underline{i}}|\widehat{\mathcal{O}}%
|\psi _{\mathcal{A}}^{\underline{i}}><\psi _{\mathcal{B}}^{\underline{i}}|1_{%
\mathcal{B}}|\psi _{\mathcal{B}}^{\underline{i}}>,  \notag \\
<\widehat{\mathcal{O}}>_{\mathcal{A}} &=&\sum_{\underline{i}}p_{\underline{i}%
}<\psi _{\mathcal{A}}^{\underline{i}}|\ \widehat{\mathcal{O}}_{\mathcal{A}%
}|\psi _{\mathcal{A}}^{\underline{i}}>=Tr_{\mathcal{H}_{\mathcal{A}}}\ \rho
_{\mathcal{A}}\ \widehat{\mathcal{O}}_{\mathcal{A}}.  \label{expectvalues}
\end{eqnarray}%
Here we note that for arbitrary nonholonomic frame transforms and
deformations of d-connections we can elaborate general covariant models when 
$<\mathcal{O}>_{\mathcal{A}}=Tr_{\mathcal{H}_{\mathcal{A}}}\rho _{\mathcal{A}%
}\mathcal{O}_{\mathcal{A}},$ or other type nonholonomic variables with, or
without, "caps"; for nonholonomic mechanical like variables, such
constructions with "tilde" are elaborated in \cite{sv20}.

\subsubsection{Thermo-geometric double states and nonassociative QGIF models}

\label{ssthgdsnonassoc}Synthesizing main concepts and methods of
thermo-field dynamics and QM, (nonassociative) geometric thermodynamics we
can construct models of \textbf{thermo-geometric and information flows}. A
similar approach but not involving geometric flows is outlined for
thermo-field double states with entanglement is developed in section II.B 2b
from \cite{nishioka18} and \cite{cottrell19}.

\paragraph{Thermo-field double states and reduced density matrix: \newline
}

For a modular Hamiltonian $\boldsymbol{H}_{\mathcal{A}},$ when $\boldsymbol{H%
}_{\mathcal{A}}|\psi _{\mathcal{A}}^{\underline{a}}>=E_{\underline{a}}|\psi
_{\mathcal{A}}^{\underline{a}}>,$ we can write 
\begin{eqnarray}
|\Psi >&=& Z^{-1/2}\sum_{\underline{a}}e^{-E_{\underline{a}}/2\tau }|\psi _{%
\mathcal{A}}^{\underline{a}}> \otimes |\psi _{\mathcal{B}}^{\underline{a}}>, %
\mbox{ thermofield double state};  \label{thfield} \\
\rho _{\mathcal{A}} &=&Z^{-1}\sum_{\underline{a}}e^{-E_{\underline{a}}/\tau
}|\psi _{\mathcal{A}}^{\underline{a}}><\otimes \psi _{\mathcal{A}}^{%
\underline{a}}|=Z^{-1}e^{-\tau ^{-1}\boldsymbol{H}_{\mathcal{A}}},%
\mbox{
reduced density matrix}.  \notag
\end{eqnarray}%
In these formulas, the quantum states are normalized with the partition
function $Z=\sum_{\underline{a}}e^{-E_{\underline{a}}/\tau }.$ The
thermo-field double state is emphasized when the partial trace is taken over
states of the subsystem $\mathcal{B},$ when the subsystem $\mathcal{A}$
becomes a Gibbs state of temperature $\tau $.

\paragraph{Thermo-geometric flow double states encoding nonassociative
QGIFs: \newline
}

\label{ssthgeomds}We can consider a discrete model for a set of data $\
^{\shortmid }\widehat{\mathcal{E}}_{\underline{a}}^{\star }:=\
_{s}^{\shortmid }\widehat{\mathcal{E}}^{\star }(\tau _{\underline{a}}),$ for 
$\underline{a}=1,2,...,\check{r},$ using the energy variable in (\ref%
{nagthermodvalues}). The values $\ ^{\shortmid }\widehat{\mathcal{E}}_{%
\underline{a}}^{\star }$ are determined by certain volume forms and
integration on respective closed regions $U^{\star }\subset \ _{s}\mathcal{M}%
^{\star }$ and they can be transformed into certain constant data for
respective nonlinear transforms relating effective sources to effective
shell cosmological constants, see details in \cite%
{partner02,partner03,lbsvevv22} (we shall consider some new examples in
section \ref{sec4}). Using the canonical partition function $\
_{s}^{\shortmid }\widehat{Z}[\ ^{\shortmid }\mathbf{g}(\tau )]$ (\ref{spf})
determined by a solution $\ ^{\shortmid }\mathbf{g}(\tau )$ for $\kappa $%
--parametric nonassociative geometric flows equations (\ref{nonassocrf}), we
can define as above a canonical density matrix 
\begin{eqnarray}
|\widehat{\Psi } > &=&\ _{s}^{\shortmid }\widehat{Z}^{-1/2}\sum_{\underline{a%
}}e^{-\ ^{\shortmid }\widehat{\mathcal{E}}_{\underline{a}}^{\star }/2\tau
}|\psi _{\mathcal{A}}^{\underline{a}}>\otimes |\psi _{\mathcal{B}}^{%
\underline{a}}>,  \label{dmq1} \\
\ ^{\shortmid }\widehat{\rho }_{\mathcal{A}} &=&\ _{s}^{\shortmid }\widehat{%
\rho }=\ _{s}^{\shortmid }\widehat{Z}^{-1}\sum_{\underline{a}}e^{-\
^{\shortmid }\widehat{\mathcal{E}}_{\underline{a}}^{\star }/\tau }|\widehat{%
\psi }_{\mathcal{A}}^{\underline{a}}>\otimes <\widehat{\psi }_{\mathcal{A}}^{%
\underline{a}}|=\ _{s}^{\shortmid }\widehat{Z}^{-1}e^{-\widehat{\boldsymbol{H%
}}_{\mathcal{A}}/\tau },  \notag
\end{eqnarray}%
where $\widehat{\boldsymbol{H}}_{\mathcal{A}}|\widehat{\psi }_{\mathcal{A}}^{%
\underline{a}}>=\ ^{\shortmid }\widehat{\mathcal{E}}_{\underline{a}}^{\star
}|\widehat{\psi }_{\mathcal{A}}^{\underline{a}}>.$

The thermo-field double states (\ref{thfield}) are important to study the
thermal nature of BHs considered as models of QFT on a background geometry
with horizon, see section IV.E in \cite{nishioka18} and references therein.
Geometric flows on a temperature like parameter $\tau $ can be considered as
also statistical thermodynamic theory \cite{perelman1}. Respective QINF
models formulated with thermo-geometric canonical double states and/or
density matries (\ref{dmq1}) consist a different class of thermo-geometric
theories. In $\kappa $--parametric form, they define nonassociative
geometric flow and/or gravitational configurations.

\paragraph{Nonassociative QGIFs and entanglementent purification: \newline
}

Given a discrete nonassociative QGIF model or a nonholonomic Ricci solitonic
configuration, suppose the system is in a pure canonical state $|\widehat{%
\Psi }>$ (\ref{dmq1}). For a total Hilbert space $\ ^{tot}\mathcal{H},$ we
can define 
\begin{equation}
\ _{tot}^{\shortmid }\widehat{\rho }:=|\widehat{\Psi }><\widehat{\Psi }|,%
\mbox{ when }<\widehat{\Psi }|\widehat{\Psi }>=1\mbox{ and }tr_{tot}(\
_{tot}^{\shortmid }\widehat{\rho })=1.  \label{puredm}
\end{equation}%
We suppose that we can divide the total QGIF system into two subsystems $%
\mathcal{A}$ and $\mathcal{B=}\overline{\mathcal{A}}$ complementary to each
other. To model nonassociatiative QGIFs this means that defining and
computing thermodynamic variables (\ref{nagthermodvalues}) we have to
consider that such conditions are imposed for 
\begin{equation}
\mathcal{A}=\{\ _{s}^{\star }A=(\ _{s}^{\shortmid }\mathfrak{g}^{\star },\
_{s}^{\shortmid }\widehat{\mathbf{D}}^{\star };\tau _{1},...\tau _{r_{A}};\
_{s}^{\shortmid }\widehat{\Xi })\}\mbox{ and }\mathcal{B}=\{\ _{s}^{\star
}B=(\ _{s}^{\shortmid }\mathfrak{g}^{\star },\ _{s}^{\shortmid }\widehat{%
\mathbf{D}}^{\star };\tau _{1},...\tau _{q_{B}};\ _{s}^{\shortmid }\widehat{%
\Xi })\}  \label{nonassocqgif}
\end{equation}%
determined by a solution $^{\shortmid }\mathbf{g}_{\alpha _{s}\beta
_{s}}(\tau )$ (\ref{ssolutions}). In what follows, we assume that $\ ^{tot}%
\mathcal{H}=\mathcal{H}_{\mathcal{A}}\otimes \mathcal{H}_{\mathcal{B}}.$
Such an assumption is not necessarily valid for general QFTs (for instance,
in certain models with gauge symmetries and/or QG), see discussions and
references to footnote 3 in \cite{nishioka18}. To elaborate nonassociative
QGIF theories we consider that at least for the models with two/ three etc.
geometric flows such an assumption is used by analogy with two/three spin
chain systems even a rigorous study of GIF and QGIF evolution of NESs and
nonholonomic Yang Mills, YM, and/or Einstein-Dirac, ED, NED, systems with
star product deformations request other additional claims and conditions.

Let us denote an orthonormal basis in $\mathcal{H}_{\mathcal{B}}$ as $\psi _{%
\mathcal{B}}^{\underline{i}}$ following the conventions for formulas (\ref%
{schmidt}) and (\ref{densmcan}). The \textbf{reduced canonical density matrix%
} of system $\mathcal{A}$ is defined%
\begin{equation}
\ ^{\shortmid }\widehat{\rho }_{\mathcal{A}}:=tr_{\mathcal{B}}(\
_{tot}^{\shortmid }\widehat{\rho })= \sum_{\underline{i}}<\psi _{\mathcal{B}%
}^{\underline{i}}|\ _{tot}^{\shortmid }\widehat{\rho }|\psi _{\mathcal{B}}^{%
\underline{i}}>.  \label{candm}
\end{equation}%
Here we not that if $\ _{tot}^{\shortmid }\widehat{\rho }=\ ^{\shortmid }%
\widehat{\rho }_{\mathcal{A}}\otimes \ ^{\shortmid }\widehat{\rho }_{%
\mathcal{B}},$ the partial trace recovers both density matrices of
subsystems, $^{\shortmid }\widehat{\rho }_{\mathcal{A}}:=tr_{\mathcal{B}}(\
_{tot}^{\shortmid }\widehat{\rho })$ and $^{\shortmid }\widehat{\rho }_{%
\mathcal{B}}:=tr_{\mathcal{A}}(\ _{tot}^{\shortmid }\widehat{\rho }).$ Such
density matrices can be used for computing as in (\ref{expectvalues}) the $%
tr(\ _{tot}^{\shortmid }\widehat{\rho }\ \mathcal{O})=tr_{\mathcal{A}}(\
^{\shortmid }\widehat{\rho }_{\mathcal{A}}\mathcal{O}_{\mathcal{A}})$ for
any operator $\mathcal{O}=\mathcal{O}_{\mathcal{A}}\otimes I_{\mathcal{B}},$
where the identity operator $I_{\mathcal{B}}\in \mathcal{H}_{\mathcal{B}}.$
A reduced $\ _{tot}^{\shortmid }\widehat{\rho }$ needs not be pure to allow
to reconstruct correlation functions for QGIFs in $\mathcal{A}$. Inversely,
if the complete data for QGIF system $\mathcal{A}$ is given by $^{\shortmid }%
\widehat{\rho }_{\mathcal{A}},$ it is possible to construct a pure canonical
density matrix by enlarging the $\mathcal{H}_{\mathcal{A}}$ in a form that
the partial trace (of enlarged $^{\shortmid }\widehat{\rho },$ which can be
taken in pure form (\ref{schmidt})) recovers $^{\shortmid }\widehat{\rho}_{%
\mathcal{A}}.$ In quantum information, this procedure is called \textbf{%
entanglement purification} which is performed by enlarging the Hilbert
space. In our approach, it enlarge also the phase space $\ _{s}\mathcal{M}%
^{\star }$ using respective classes of $\kappa $--parametric solutions as we
explain below:

A general canonical density matrix can be written in the form (\ref{densmcan}%
) when $|\psi _{\mathcal{A}}^{\underline{a}}>$ is an orthonormal basis in $%
\mathcal{H}_{\mathcal{A}}.$ For QGIFs, the entanglement purification has to
be performed for a thermal state with the Boltzmann weight $\widehat{p}_{%
\underline{a}}=\ _{s}^{\shortmid }\widehat{Z}^{-1}e^{-\ ^{\shortmid }%
\widehat{\mathcal{E}}_{\underline{a}}^{\star }/2\tau }$ as in (\ref{dmq1}).
We copy $\mathcal{H}_{\mathcal{A}}$ into another Hilbert space $\mathcal{H}_{%
\mathcal{B}},$ with a basis $|\psi _{\mathcal{B}}^{\underline{a}}>,$ where $%
\mathcal{B=}\overline{\mathcal{A}}.$ Then, it is possible to define a pure
density matrix of type (\ref{puredm}),%
\begin{equation}
\ ^{\shortmid }\widehat{\rho }=|\widehat{\chi }><\widehat{\chi }|,%
\mbox{ for
}|\widehat{\chi }>:=\sum_{\underline{a}}\sqrt{\widehat{p}_{\underline{a}}}|%
\widehat{\psi }_{\mathcal{A}}^{\underline{a}}><\otimes \widehat{\psi }_{%
\mathcal{B}}^{\underline{a}}|,  \label{purific}
\end{equation}%
considered in an enlarged Hilbert space $\mathcal{H}_{\mathcal{A}}\otimes 
\mathcal{H}_{\mathcal{B}}.$ This procedure is important for understanding
what an observer restricted to nonassociative QGIF system $\mathcal{A}$
modelled both as a thermal nonassocitaive geometric flow at temperature $%
\tau $ and QM model on respective Hilbert spaces and their tensor products.

\paragraph{Entanglement QGIF entropy encoding nonassociative star product
data at finite temperature: \newline
}

Now we can introduce a measure of both GIF and QM entanglement by defining
the entanglement canonical entropy of a nonassociative QGIF system $\mathcal{%
A}=\{\ _{s}^{\star }A\}=\mathcal{A}^{\star },$ 
\begin{equation}
\ ^{\shortmid }\widehat{\mathcal{S}}_{\mathcal{A}}^{\star }\ =-tr_{\mathcal{A%
}}[\ ^{\shortmid }\widehat{\rho }_{\mathcal{A}}\log \ ^{\shortmid }\widehat{%
\rho }_{\mathcal{A}}].  \label{entanglentrcan}
\end{equation}%
This variable is defined as a von Neumann entropy for $\ ^{\shortmid }%
\widehat{\rho }_{\mathcal{A}}$ (\ref{candm}) (for similar motivations and
details we refer to formula (7) in \cite{nishioka18}. For $\ ^{\shortmid }%
\widehat{\rho }_{\mathcal{A}}=\ _{s}^{\shortmid }\widehat{\rho }$ from (\ref%
{densmcan}), such an entropy is determined by a canonical partition function 
$\ _{s}^{\shortmid }\widehat{Z}[\ ^{\shortmid }\mathbf{g}(\tau )]$ (\ref{spf}%
) corresponding a solution $\ ^{\shortmid }\mathbf{g}(\tau )$ for $\kappa $%
--parametric nonassociative geometric flows equations (\ref{nonassocrf}). It
also is a quantum variant of the thermodynamic entropy $\ _{s}^{\shortmid }%
\widehat{\mathcal{S}}^{\star }(\tau )$ from (\ref{nagthermodvalues}). For a
pure state (\ref{puredm}), $\ ^{\shortmid }\widehat{\mathcal{S}}_{\mathcal{A}%
}^{\star }$ (\ref{entanglentrcan}) vanishes. Here we note that in QFT the
entanglement entropy remains finite in a finite-dimensional system but
suffers, in general, from UV divergences. In the standard theory of Ricci
flows of Riemannian metrics, Perelman elaborated a procedure for cutting
singularities under geometric flow evolution \cite{perelman1}. For
relativistic geometric flows and nonassociative star product deformations 
\cite{sv20,lbdssv22}, we may construct and select well-defined physically
important solutions using the AFCDM.

We can purify a QGIF system $\mathcal{A}$ also as a thermal system in the
extended Hilbert space $\mathcal{H}_{\mathcal{A}}\otimes \mathcal{H}_{%
\mathcal{B}}$ as we described above in (\ref{purific}). Then every
expectation value for a local operator in $\mathcal{A}$ is representable in
the thermo-geometric double state (\ref{dmq1}) of the total system $\mathcal{%
A}\cup \mathcal{B}$. For such systems, the QGIF entanglement entropy
measures the thermo-dynamic entropy both as a GIF and quantum subsystem $%
\mathcal{A}$, when 
\begin{equation}
\ ^{\shortmid }\widehat{\mathcal{S}}_{\mathcal{A}}^{\star }\ =-tr_{\mathcal{A%
}}[\ ^{\shortmid }\widehat{\rho }_{\mathcal{A}}(-\ ^{\shortmid }\widehat{%
\mathcal{E}}^{\star }/\tau -\log \ _{s}^{\shortmid }\widehat{Z})]=(<\widehat{%
\boldsymbol{H}}_{\mathcal{A}}>-\ ^{\shortmid }\widehat{\mathcal{F}}_{%
\mathcal{A}}^{\star })/\tau ,  \label{entanglentrcan1}
\end{equation}%
where the canonical thermal free energy is $\ ^{\shortmid }\widehat{\mathcal{%
F}}_{\mathcal{A}}^{\star }:=\tau \log \ _{s}^{\shortmid }\widehat{Z}.$

\subsubsection{Separable and entangled QGIFs and Bell states}

\paragraph{Separable and entangled QGIFs: \newline
}

Let the QGIF entanglement entropy $\ ^{\shortmid }\widehat{\mathcal{S}}_{%
\mathcal{A}}^{\star }$ (\ref{entanglentrcan}) is defined. We want to
understand what are the physical meaning and measures for such a
nonassociative geometric flow and quantum subsystem $\mathcal{A}$. A pure
state $|\widehat{\Psi }>$ (\ref{dmq1}) can be represented in the form%
\begin{equation}
|\widehat{\Psi }>=\sum_{\underline{a},\underline{b}}\widehat{c}_{\underline{a%
}\underline{b}}|\underline{a}>_{\mathcal{A}}\otimes |\underline{b}>_{%
\mathcal{B}},  \label{dmq1a}
\end{equation}%
where the orthornormal bases are defined respectively $|\underline{a}>_{%
\mathcal{A}}\subset \mathcal{H}_{\mathcal{A}},$ for $\underline{a}=1,...,d_{%
\mathcal{A}};$ and $|\underline{b}>_{\mathcal{B}}\subset \mathcal{H}_{%
\mathcal{B}},$ for $\underline{b}=1,...,d_{\mathcal{B}}.$ The complex
coefficients $\widehat{c}_{\underline{a}\underline{b}}$ define a matrix of
dimension $d_{\mathcal{A}}\times d_{\mathcal{B}}.$ There two different types
of QGIFs determined by a (\ref{dmq1a}) depending on factorization, or not,
of $\widehat{c}_{\underline{a}\underline{b}}:$ {\small 
\begin{equation*}
\begin{tabular}{lllll}
& $\widehat{c}_{\underline{a}\underline{b}}$ & $|\widehat{\Psi }>=$ & $\
^{\shortmid }\widehat{\mathcal{S}}_{\mathcal{A}}^{\star }=%
\begin{array}{c}
\\ 
\end{array}%
$ & $\ ^{\shortmid }\widehat{\rho }_{\mathcal{A}}$ \\ 
separable, & $=\widehat{c}_{\underline{a}}^{\mathcal{A}}\widehat{c}_{%
\underline{b}}^{\mathcal{B}}$ & $%
\begin{vmatrix}
\begin{array}{c}
|\widehat{\Psi }_{\mathcal{A}}>\otimes |\widehat{\Psi }_{\mathcal{B}}>,%
\mbox{ where } \\ 
|\widehat{\Psi }_{\mathcal{A}}>:=\sum_{\underline{a}}\widehat{c}_{\underline{%
a}}^{\mathcal{A}}|\underline{a}>_{\mathcal{A}} \\ 
|\widehat{\Psi }_{\mathcal{B}}>:=\sum_{\underline{b}}\widehat{c}_{\underline{%
b}}^{\mathcal{B}}|\underline{a}>_{\mathcal{B}}%
\end{array}
&  \\ 
& 
\end{vmatrix}%
$ & $%
\begin{vmatrix}
\begin{array}{c}
\ _{s}^{\shortmid }\widehat{\mathcal{S}}^{\star }(\tau )(\ref%
{nagthermodvalues}) \\ 
\mbox{background} \\ 
\mbox{Perelman entropy}, \\ 
\ ^{\shortmid }\widehat{\mathcal{S}}_{\mathcal{A}}^{\star }=0, \\ 
\mbox{ trivial Ricci flows},%
\end{array}%
\end{vmatrix}%
$ & $%
\begin{tabular}{l}
$\ ^{\shortmid }\widehat{\rho }_{\mathcal{A}}=|\widehat{\Psi }_{\mathcal{A}%
}><\widehat{\Psi }_{\mathcal{A}}|,$ \\ 
pure state;%
\end{tabular}%
$ \\ 
entangled, & $\neq \widehat{c}_{\underline{a}}^{\mathcal{A}}\widehat{c}_{%
\underline{b}}^{\mathcal{B}}$ & $%
\begin{vmatrix}
\begin{array}{c}
\\ 
\mbox{Schmidt decomp. }(\ref{schmidt})\  \\ 
=\sum\limits_{_{\underline{i}}}^{\min (d_{\mathcal{A}},d_{\mathcal{B}})}%
\sqrt{p_{\underline{i}}}\psi _{\mathcal{A}}^{\underline{i}}\otimes \psi _{%
\mathcal{B}}^{\underline{i}},%
\end{array}%
\end{vmatrix}%
$ & $\ ^{\shortmid }\widehat{\mathcal{S}}_{\mathcal{A}}^{\star }(\tau )$\ (%
\ref{entanglentrcan}) & $%
\begin{array}{c}
\begin{array}{c}
\ ^{\shortmid }\widehat{\rho }_{\mathcal{A}}=\  \\ 
\sum\limits_{\underline{i}}^{d_{\mathcal{B}}}\ _{\mathcal{B}}<\psi ^{%
\underline{i}}|\widehat{\Psi }><\widehat{\Psi }|\psi ^{\underline{i}}>_{%
\mathcal{B}} \\ 
=\sum\limits_{_{\underline{i}}}^{\min (d_{\mathcal{A}},d_{\mathcal{B}})}p_{%
\underline{i}}\psi _{\mathcal{A}}^{\underline{i}}\otimes \psi _{\mathcal{B}%
}^{\underline{i}}%
\end{array}
\\ 
\mbox{mixed state}.%
\end{array}%
$%
\end{tabular}%
\end{equation*}
}

In summary, an entangled QGIF state is a coupled superposition of several
such sub-states. An observer being under such a nonassociative quantum
geometric flow is in a mixed state when the pure ground state $|\widehat{%
\Psi }>$ in the total system in entangled as we discussed above. The
entanglement entropy measures how much a given QGIF system differs from a
separable system. It reaches the maximum value when a given QGIF is a
superposition of all possible quantum states with an equal weight.
Entanglement can be both of mixed nonassociative geometric and quantum
origin for some separate type stated with respective Perelman, Shannon,
and/or von Neumann entropy.

\paragraph{Two QGIF systems: \newline
}

In quantum information theory, typical examples of entanglement and Bell
states are studied for so-called two spin systems (see, for instance,
formulas (18)-(23) from \cite{nishioka18}). Similar physical constructions
can be modeled for two QGIF systems $\mathcal{A}=\{\ _{s}^{\star }A\}=%
\mathcal{A}^{\star }$ and $\mathcal{B}=\{\ _{s}^{\star }B\}=\mathcal{B}%
^{\star}$ of type (\ref{nonassocqgif}) with respective spanned Hilbert
spaces $\mathcal{H}_{\mathcal{A}}$ and $\mathcal{H}_{\mathcal{B}},$ when $%
\mathcal{H}_{\mathcal{A},\mathcal{B}}=\{|\hat{1}>_{\mathcal{A},\mathcal{B}},|%
\hat{2}>_{\mathcal{A},\mathcal{B}}\}.$ To define such two bases we use the
conditions for orthonormal bases when $\ _{\mathcal{A},\mathcal{B}}<\hat{%
\imath}|\hat{\jmath}>\ _{\mathcal{A},\mathcal{B}}= \delta _{\hat{\imath}\hat{%
\jmath}}$ for $\hat{\imath},\hat{\jmath}=\hat{1},\hat{2}.$ We use hats on
symbols and numbers in order to emphasize that the constructions are
performed for canonical geometric and quantum thermodynamic models with (\ref%
{dmq1}). The total Hilbert space for modelling two QGIF systems consists
from the tensor product $\ ^{tot}\mathcal{H}=\mathcal{H}_{\mathcal{A}%
}\otimes \mathcal{H}_{\mathcal{B}}.$ It includes a 4-d orthonormal basis $\{|%
\hat{1}\hat{1}>,|\hat{1}\hat{2}>,|\hat{2}\hat{1}>,|\hat{2}\hat{2}>\}\subset
\ ^{tot}\mathcal{H},$ where $|\hat{\imath}\hat{\jmath}>=|\hat{\imath}>_{%
\mathcal{A}}\otimes |\hat{\jmath}>_{\mathcal{B}}.$

Let us consider explicit examples of ground state and reduced density matrix
for a $\mathcal{A}$ obtained by taking the partial trace over $\mathcal{H}_{%
\mathcal{B}}$ as in (\ref{puredm}):%
\begin{eqnarray*}
|\widehat{\Psi }^{\star }> &=&\frac{e^{-(\ ^{\shortmid }\widehat{\mathcal{E}}%
_{\underline{1}}^{\star }+\ ^{\shortmid }\widehat{\mathcal{E}}_{\underline{2}%
}^{\star })/2\tau }}{\sqrt{2}\ \ ^{\shortmid }\widehat{Z}_{\underline{1}%
}^{1/2}\ ^{\shortmid }\widehat{Z}_{\underline{2}}^{1/2}}(|\hat{1}\hat{2}>-|%
\hat{2}\hat{1}>)\mbox{ and } \\
\ ^{\shortmid }\widehat{\rho }_{\mathcal{A}}^{\star } &=&\frac{1}{2}(\frac{%
e^{-\ ^{\shortmid }\widehat{\mathcal{E}}_{\underline{1}}^{\star }/\tau }}{\
^{\shortmid }\widehat{Z}_{\underline{1}}}|\hat{1}>_{\mathcal{A}}\ _{\mathcal{%
A}}<\hat{1}|+\frac{e^{-\ ^{\shortmid }\widehat{\mathcal{E}}_{\underline{2}%
}^{\star }/\tau }}{\ ^{\shortmid }\widehat{Z}_{\underline{2}}}|\hat{2}>_{%
\mathcal{A}}\ _{\mathcal{A}}<\hat{2}|)=\left( 
\begin{array}{cc}
e^{-\ ^{\shortmid }\widehat{\mathcal{E}}_{\underline{1}}^{\star }/\tau }/2\
^{\shortmid }\widehat{Z}_{\underline{1}} & 0 \\ 
0 & e^{-\ ^{\shortmid }\widehat{\mathcal{E}}_{\underline{2}}^{\star }/\tau
}/2\ ^{\shortmid }\widehat{Z}_{\underline{2}}%
\end{array}%
\right) ,
\end{eqnarray*}%
where $\ ^{\shortmid }\widehat{Z}_{\underline{1}}=\ _{s}^{\shortmid }%
\widehat{Z}[\ ^{\shortmid }\mathbf{g}_{\underline{1}}(\tau )]$ and $\
^{\shortmid }\widehat{Z}_{\underline{2}}=\ _{s}^{\shortmid }\widehat{Z}[\
^{\shortmid }\mathbf{g}_{\underline{2}}(\tau )]$, see formula (\ref{spf}).
Using formula $\ ^{\shortmid }\widehat{\mathcal{S}}_{\mathcal{A}}^{\star }$ (%
\ref{entanglentrcan}) for this $\ ^{\shortmid }\widehat{\rho }_{\mathcal{A}%
}^{\star }$, we compute 
\begin{equation*}
\ ^{\shortmid }\mathcal{S}_{\mathcal{A}}^{\star }\ =-tr_{\mathcal{A}}\left[
\left( 
\begin{array}{cc}
e^{-\ ^{\shortmid }\widehat{\mathcal{E}}_{\underline{1}}^{\star }/\tau }/2\
^{\shortmid }\widehat{Z}_{\underline{1}} & 0 \\ 
0 & e^{-\ ^{\shortmid }\widehat{\mathcal{E}}_{\underline{2}}^{\star }/\tau
}/2\ ^{\shortmid }\widehat{Z}_{\underline{2}}%
\end{array}%
\right) \left( 
\begin{array}{cc}
\log [e^{-\ ^{\shortmid }\widehat{\mathcal{E}}_{\underline{1}}^{\star }/\tau
}/2\ ^{\shortmid }\widehat{Z}_{\underline{1}}] & 0 \\ 
0 & \log [e^{-\ ^{\shortmid }\widehat{\mathcal{E}}_{\underline{2}}^{\star
}/\tau }/2\ ^{\shortmid }\widehat{Z}_{\underline{2}}]%
\end{array}%
\right) \right] .
\end{equation*}%
If we do not consider contributions of nonassociative geometric flows (when $%
\ ^{\shortmid }\widehat{\mathcal{E}}_{\underline{1}}^{\star }=0$ and $\
^{\shortmid }\widehat{\mathcal{E}}_{\underline{2}}^{\star }=0,$ for
normalized $\ ^{\shortmid }\widehat{Z}_{\underline{1}}=1$ and $\ ^{\shortmid
}\widehat{Z}_{\underline{2}}=1$) , we obtain $\ ^{\shortmid }\mathcal{S}_{%
\mathcal{A}}^{\star }\ \approx \mathcal{S}_{\mathcal{A}}=\log 2$ as in
formula (21) from \cite{nishioka18}. Such a maximally entanglement state
with not pure $\ ^{\shortmid }\widehat{\rho }_{\mathcal{A}}^{\star }$ is
defined for two QGIFs.

\paragraph{Examples of Bell type of QGIF entanglement and nonassociative
canonical qubit systems: \newline
}

Nonassociative two qubit systems can be defined by such totally four
independent maximally entangled QGIFs,%
\begin{eqnarray}
|\widehat{B}_{1}^{\star } &>&=\frac{1}{\sqrt{2}\ }(\frac{e^{-\ ^{\shortmid }%
\widehat{\mathcal{E}}_{\underline{1}}^{\star }/\tau }}{\ \ ^{\shortmid }%
\widehat{Z}_{\underline{1}}}|\hat{1}\hat{1}>+\frac{e^{-\ ^{\shortmid }%
\widehat{\mathcal{E}}_{\underline{2}}^{\star }/\tau }}{\ \ ^{\shortmid }%
\widehat{Z}_{\underline{2}}}|\hat{2}\hat{2}>),\ |\widehat{B}_{2}^{\star }>=%
\frac{1}{\sqrt{2}\ }(\frac{e^{-\ ^{\shortmid }\widehat{\mathcal{E}}_{%
\underline{1}}^{\star }/\tau }}{\ \ ^{\shortmid }\widehat{Z}_{\underline{1}}}%
|\hat{1}\hat{1}>-\frac{e^{-\ ^{\shortmid }\widehat{\mathcal{E}}_{\underline{2%
}}^{\star }/\tau }}{\ \ ^{\shortmid }\widehat{Z}_{\underline{2}}}|\hat{2}%
\hat{2}>),  \label{qgifbell} \\
|\widehat{B}_{3}^{\star } &>&=\frac{e^{-(\ ^{\shortmid }\widehat{\mathcal{E}}%
_{\underline{1}}^{\star }+\ ^{\shortmid }\widehat{\mathcal{E}}_{\underline{2}%
}^{\star })/2\tau }}{\sqrt{2}\ \ ^{\shortmid }\widehat{Z}_{\underline{1}%
}^{1/2}\ ^{\shortmid }\widehat{Z}_{\underline{2}}^{1/2}}(|\hat{1}\hat{2}>+|%
\hat{2}\hat{1}>),\ |\widehat{B}_{4}^{\star }>=\frac{e^{-(\ ^{\shortmid }%
\widehat{\mathcal{E}}_{\underline{1}}^{\star }+\ ^{\shortmid }\widehat{%
\mathcal{E}}_{\underline{2}}^{\star })/2\tau }}{\sqrt{2}\ \ ^{\shortmid }%
\widehat{Z}_{\underline{1}}^{1/2}\ ^{\shortmid }\widehat{Z}_{\underline{2}%
}^{1/2}}(|\hat{1}\hat{2}>-|\hat{2}\hat{1}>).  \notag
\end{eqnarray}%
These QGIF systems are analogs of Bell states (Einstein-Podolsky-Rosen)
pairs in quantum information theory. As quantum systems such states manifest
QM aspects in the sense of violating the Bell's inequalities (they can be
considered in a local hidden variable theory and account for the
probabilistic features of QG with a hidden variables and a probability
density). But they also include nonassociative geometric flow data via $\
^{\shortmid }\widehat{\mathcal{E}}_{\underline{1}}^{\star },\ ^{\shortmid }%
\widehat{\mathcal{E}}_{\underline{2}}^{\star }$ and $\ ^{\shortmid }\widehat{%
Z}_{\underline{1}}^{1/2},\ ^{\shortmid }\widehat{Z}_{\underline{2}}^{1/2}.$

We can generalize (\ref{qgifbell}) for systems of $\grave{n}$ qubits. For
instance, the QGIF generalizations of Greenberger-Horne-Zelinger, GHZ,
states \cite{greenberger},%
\begin{equation*}
|\widehat{G}HZ^{\star }>=\frac{1}{\sqrt{2}\ }\left( [\frac{e^{-\ ^{\shortmid
}\widehat{\mathcal{E}}_{\underline{1}}^{\star }/\tau }}{\ \ ^{\shortmid }%
\widehat{Z}_{\underline{1}}}|\hat{1}>]^{\grave{n}}+[\frac{e^{-\ ^{\shortmid }%
\widehat{\mathcal{E}}_{\underline{2}}^{\star }/\tau }}{\ \ ^{\shortmid }%
\widehat{Z}_{\underline{2}}}|\hat{2}>]^{\grave{n}}\right) .
\end{equation*}%
For trivial geometric flows with $\ ^{\shortmid }\widehat{\mathcal{E}}_{%
\underline{1}}^{\star }=\ ^{\shortmid }\widehat{\mathcal{E}}_{\underline{2}%
}^{\star }=0$ and $\ ^{\shortmid }\widehat{Z}_{\underline{1}}=\ ^{\shortmid }%
\widehat{Z}_{\underline{2}}=1,$ we obtain entangled states of $\grave{n}$
qubits. In general, such states can be defined in non-equivalent forms, when
certain sub-systems are not fully separable see details in section C of \cite%
{nishioka18}.

\subsection{Inequalities for entropies of nonassociative Ricci soliton and
QGIFs}

We outline most important inequalities and properties of the QGIF
entanglement entropy $\ ^{\shortmid }\widehat{\mathcal{S}}_{\mathcal{A}%
}^{\star }$ (\ref{entanglentrcan}) for the density matrix $\ ^{\shortmid }%
\widehat{\rho }_{\mathcal{A}}$ (\ref{candm}), when r $\ ^{\shortmid }%
\widehat{\rho }_{\mathcal{A}}=\ _{s}^{\shortmid }\widehat{\rho }$ \ from (%
\ref{densmcan}). Proofs consist from star product R-flux deformations,
following Convention 2 (\ref{conv2s}), of respective definitions and
formulas commutative QGIFs. For thermo-field systems, such constructions are
reviewed in \cite{nishioka18}.

\subsubsection{(Strong) subadditivite of entangled nonassociative Einstein
systems}

In this subsection, we discuss important properties of nonassociative QGIFs
and nonholonomic Einstein systeme related to the strong subadditivity
property of nonassociative entanglement and generalized Perelman's entropies.

\paragraph{Entanglement entropy for complementary nonassociative QGIFs and
Ricci soliotns: \newline
}

If $\ \widehat{\mathcal{B}}=\overline{\widehat{\mathcal{A}}},$ we have such
a condition for entropies:$\ \ ^{\shortmid }\widehat{\mathcal{S}}_{\mathcal{A%
}}^{\star }=\ ^{\shortmid }\widehat{\mathcal{S}}_{\overline{\mathcal{A}}%
}^{\star }.$ It can be proven using formulas with $\rho _{\mathcal{A}}$ (\ref%
{densmcan}) for a pure ground state wave function and then extending the
formulas for the thermodynamic GIF entropy $\ _{s}^{\shortmid }\widehat{%
\mathcal{S}}^{\star }(\tau )$ from (\ref{nagthermodvalues}) related to a $\
^{\shortmid }\widehat{\mathcal{S}}_{\mathcal{A}}^{\star }$ (\ref%
{entanglentrcan}) for the same s-metric $\underline{\mathbf{g}}$ and
respective normalizations on $\widehat{\mathcal{A}}$ and $\overline{\widehat{%
\mathcal{A}}}.$ For quantum models of GIF thermodynamic systems, $\ \ \
^{\shortmid }\widehat{\mathcal{S}}_{\mathcal{A}}^{\star }\neq \ \
^{\shortmid }\widehat{\mathcal{S}}_{\mathcal{B}}^{\star }$ if $\widehat{%
\mathcal{A}}\mathcal{\cup }\widehat{\mathcal{B}}$ is a mixed state. In
result, we have general inequalities, 
\begin{equation*}
\ ^{\shortmid }\widehat{\mathcal{S}}_{\mathcal{A}}^{\star }\neq \
^{\shortmid }\widehat{\mathcal{S}}_{\mathcal{B}}^{\star }.
\end{equation*}%
Such which can be also proven in any quasi-classical limit, for instance, in
a model with WKB approximation. For some special subclasses of nonholonomic
deformations and certain classes of normalizing functions such conditions
may transform in equalities.

\paragraph{Subadditivity conditions: \newline
}

Such inequalities are satisfied for \ disjoint nonassociative subsystems $%
\widehat{\mathcal{A}}$ and $\widehat{\mathcal{B}},$ 
\begin{equation}
\ ^{\shortmid }\widehat{\mathcal{S}}_{\mathcal{A\cup B}}^{\star }\leq \
^{\shortmid }\widehat{\mathcal{S}}_{\mathcal{A}}^{\star }+\ ^{\shortmid }%
\widehat{\mathcal{S}}_{\mathcal{B}}^{\star }\mbox{
and }|\ ^{\shortmid }\widehat{\mathcal{S}}_{\mathcal{A}}^{\star }-\
^{\shortmid }\widehat{\mathcal{S}}_{\mathcal{B}}^{\star }|\leq \ ^{\shortmid
}\widehat{\mathcal{S}}_{\mathcal{A\cup B}}^{\star }.  \label{subaditcond}
\end{equation}%
They can be verified at least in $\kappa $-linear parametric form including
nonassociative data.

\paragraph{Strong subadditivity: \newline
}

Let us consider three disjointed nonassociative QGIF gravitational
subsystems $\widehat{\mathcal{A}},\widehat{\mathcal{B}}$ and $\widehat{%
\mathcal{C}}$ and conditions of convexity of a function built from
respective density matrices and unitarity of systems \cite%
{preskill,witten20,nishioka18}. As in quantum information theory, 
\begin{equation*}
\ ^{\shortmid }\widehat{\mathcal{S}}_{\mathcal{A\cup B\cup C}}^{\star }+\
^{\shortmid }\widehat{\mathcal{S}}_{\mathcal{B}}^{\star }\leq \ ^{\shortmid }%
\widehat{\mathcal{S}}_{\mathcal{A\cup B}}^{\star }+\ ^{\shortmid }\widehat{%
\mathcal{S}}_{B\mathcal{\cup C}}^{\star }\mbox{ and }\ \ ^{\shortmid }%
\widehat{\mathcal{S}}_{\mathcal{A}}^{\star }+\ ^{\shortmid }\widehat{%
\mathcal{S}}_{\mathcal{C}}^{\star }\leq \ ^{\shortmid }\widehat{\mathcal{S}}%
_{\mathcal{A\cup B}}^{\star }+\ ^{\shortmid }\widehat{\mathcal{S}}_{B%
\mathcal{\cup C}}^{\star }.
\end{equation*}%
These strong subadditivity conditions contain as particular classes the
inequalities (\ref{subaditcond}).

\subsubsection{Relative entropy and mutual information of nonassociative
gravity and QGIFs}

\label{ssrelentrna}The concept of \textit{relative entropy} of two
nonassociative QGIFs described by density matrices $\ ^{\shortmid }\widehat{%
\rho }_{\mathcal{A}}$ and $\ ^{\shortmid }\widehat{\sigma }_{\mathcal{A}}$
can be defined and computed in canonical nonholonomic variables as in
standard quantum information theories, 
\begin{equation}
\ ^{\shortmid }\widehat{\mathcal{S}}^{\star }(\ ^{\shortmid }\widehat{\rho }%
_{\mathcal{A}}\shortparallel \ ^{\shortmid }\widehat{\sigma }_{\mathcal{A}%
})=Tr_{\mathcal{H}_{\mathcal{B}}}[\ ^{\shortmid }\widehat{\rho }_{\mathcal{A}%
}(\log \ ^{\shortmid }\widehat{\rho }_{\mathcal{A}}-\log \ ^{\shortmid }%
\widehat{\sigma }_{\mathcal{A}})],  \label{relativentr}
\end{equation}%
where $\ ^{\shortmid }\widehat{\mathcal{S}}^{\star }(\ ^{\shortmid }\widehat{%
\rho }_{\mathcal{A}}\shortparallel \ ^{\shortmid }\widehat{\rho }_{\mathcal{A%
}})=0.$ Using relative entropy, we can define a measure of "distance"
between two nonassociative QGIFs, or nonholonomic Ricci solitons, with a
norm $||\ ^{\shortmid }\widehat{\rho }_{\mathcal{A}}||=tr(\sqrt{(\
^{\shortmid }\widehat{\rho }_{\mathcal{A}})(\ ^{\shortmid }\widehat{\rho }_{%
\mathcal{A}}^{\dag })}).$

There are such important formulas and conditions for the relative entropy of
two nonassociative QGIFs:

\begin{enumerate}
\item for tensor products of density matrices of type $\ _{1}\widehat{\rho }%
_{\mathcal{A}}\otimes \ _{2}\widehat{\rho }_{\mathcal{A}}$ and $\ _{1}%
\widehat{\sigma }_{\mathcal{A}}\otimes \ _{2}\widehat{\sigma }_{\mathcal{A}}$%
, 
\begin{equation*}
\ \ ^{\shortmid }\widehat{\mathcal{S}}^{\star }(\ _{1}^{\shortmid }\widehat{%
\rho }_{\mathcal{A}}\otimes \ _{2}^{\shortmid }\widehat{\rho }_{\mathcal{A}%
}\shortparallel \ _{1}^{\shortmid }\widehat{\sigma }_{\mathcal{A}}\otimes \
_{2}^{\shortmid }\widehat{\sigma }_{\mathcal{A}})=\ ^{\shortmid }\widehat{%
\mathcal{S}}^{\star }(\ _{1}^{\shortmid }\widehat{\rho }_{\mathcal{A}%
}\shortparallel \ _{1}^{\shortmid }\widehat{\sigma }_{\mathcal{A}})+\
^{\shortmid }\widehat{\mathcal{S}}^{\star }(\ _{2}^{\shortmid }\widehat{\rho 
}_{\mathcal{A}}\shortparallel \ _{2}^{\shortmid }\widehat{\sigma }_{\mathcal{%
A}});
\end{equation*}

\item positivity conditions, $\ ^{\shortmid }\widehat{\mathcal{S}}^{\star
}(\ ^{\shortmid }\widehat{\rho }_{\mathcal{A}}\shortparallel \ ^{\shortmid }%
\widehat{\sigma }_{\mathcal{A}})\geq \frac{1}{2}||\ ^{\shortmid }\widehat{%
\rho }_{\mathcal{A}}-\ ^{\shortmid }\widehat{\sigma }_{\mathcal{A}}||^{2},$
i.e.$\ \ ^{\shortmid }\widehat{\mathcal{S}}^{\star }(\ ^{\shortmid }\widehat{%
\rho }_{\mathcal{A}}\shortparallel \ ^{\shortmid }\widehat{\sigma }_{%
\mathcal{A}})\geq 0;$

\item monotonicity property, $\ ^{\shortmid }\widehat{\mathcal{S}}^{\star
}(\ ^{\shortmid }\widehat{\rho }_{\mathcal{A}}\shortparallel \ ^{\shortmid }%
\widehat{\sigma }_{\mathcal{A}})\geq \ ^{\shortmid }\widehat{\mathcal{S}}%
^{\star }(tr_{s}\ ^{\shortmid }\widehat{\rho }_{\mathcal{A}}|tr_{s}\
^{\shortmid }\widehat{\sigma }_{\mathcal{A}}).$
\end{enumerate}

The relative entropy $\ ^{\shortmid }\widehat{\mathcal{S}}^{\star }(\
^{\shortmid }\widehat{\rho }_{\mathcal{A}}\shortparallel \ ^{\shortmid }%
\widehat{\sigma }_{\mathcal{A}})$ (\ref{relativentr}) is related to the
entanglement entropy $\ ^{\shortmid }\widehat{\mathcal{S}}_{\mathcal{A}%
}^{\star }$ (\ref{entanglentrcan}) following the$\ $ formula 
\begin{equation*}
\ ^{\shortmid }\widehat{\mathcal{S}}_{\mathcal{A}}^{\star }(\ ^{\shortmid }%
\widehat{\rho }_{\mathcal{A}}\shortparallel 1_{\mathcal{A}}/k_{\mathcal{A}%
})=\log k_{\mathcal{A}}-\ \ ^{\shortmid }\widehat{\mathcal{S}}_{\mathcal{A}%
}^{\star }(\ ^{\shortmid }\widehat{\rho }_{\mathcal{A}}),
\end{equation*}
where $1_{\mathcal{A}}$ is the $k_{\mathcal{A}}\times k_{\mathcal{A}}$ unit
matrix for a $k_{\mathcal{A}}$-dimensional Hilbert space associated to the
region $\widehat{\mathcal{A}}.$

Considering $\ ^{\shortmid }\widehat{\rho }_{\mathcal{A\cup B\cup C}}$ as a
density matrix on $\widehat{\mathcal{A}}\mathcal{\cup }\widehat{\mathcal{%
\mathcal{B}}}\mathcal{\cup }\widehat{\mathcal{C}},$ when $\ \ ^{\shortmid }%
\widehat{\rho }_{\mathcal{A\cup B}}$ is written for its restriction on $%
\widehat{\mathcal{A}}\mathcal{\cup }\widehat{\mathcal{B}}$ and $\
^{\shortmid }\widehat{\rho }_{\mathcal{B}}$ is stated for its restriction on 
$\widehat{\mathcal{B}};$ and using the formula $\ $%
\begin{equation*}
tr_{\mathcal{A\cup B\cup C}}[\ \ ^{\shortmid }\widehat{\rho }_{\mathcal{%
A\cup B\cup C}}(\mathcal{O}_{\mathcal{A\cup B}}\otimes 1_{\mathcal{C}}/k_{%
\mathcal{C}})]=tr_{\mathcal{A\cup B}}(\ ^{\shortmid }\ \widehat{\rho }_{%
\mathcal{A\cup B}}\mathcal{O}_{\mathcal{A\cup B}}),
\end{equation*}
we can prove as in standard quantum information theory such identities 
\begin{eqnarray*}
\ ^{\shortmid }\widehat{\mathcal{S}}^{\star }(\ ^{\shortmid }\widehat{\rho }%
_{\mathcal{A\cup B\cup C}}\shortparallel 1_{\mathcal{A\cup B\cup C}}/k_{%
\mathcal{A\cup B\cup C}}) &=&\ ^{\shortmid }\widehat{\mathcal{S}}^{\star }(\
^{\shortmid }\widehat{\rho }_{\mathcal{A\cup B}}\shortparallel 1_{\mathcal{%
A\cup B}}/k_{\mathcal{A\cup B}})+\ ^{\shortmid }\widehat{\mathcal{S}}^{\star
}(\ ^{\shortmid }\widehat{\rho }_{\mathcal{A\cup B\cup C}}\shortparallel \
^{\shortmid }\widehat{\rho }_{\mathcal{A\cup B}}\otimes 1_{\mathcal{C}}/k_{%
\mathcal{C}}), \\
\ ^{\shortmid }\widehat{\mathcal{S}}^{\star }(\ ^{\shortmid }\widehat{\rho }%
_{\mathcal{B\cup C}}\shortparallel 1_{\mathcal{B\cup C}}/k_{\mathcal{B\cup C}%
}) &=&\ ^{\shortmid }\widehat{\mathcal{S}}^{\star }(\ ^{\shortmid }\widehat{%
\rho }_{\mathcal{B}}\shortparallel 1_{\mathcal{B}}/k_{\mathcal{B}})+\
^{\shortmid }\widehat{\mathcal{S}}^{\star }(\ ^{\shortmid }\widehat{\rho }_{%
\mathcal{B\cup C}}\shortparallel \ ^{\shortmid }\widehat{\rho }_{\mathcal{B}%
}\otimes 1_{\mathcal{C}}/k_{\mathcal{C}});
\end{eqnarray*}%
\begin{eqnarray*}
&&\mbox{ and inequalities }\ ^{\shortmid }\widehat{\mathcal{S}}^{\star }(\
^{\shortmid }\widehat{\rho }_{\mathcal{A\cup B\cup C}}\shortparallel \
^{\shortmid }\widehat{\rho }_{\mathcal{A\cup B}}\otimes 1_{\mathcal{C}}/k_{%
\mathcal{C}})\geq \ \ ^{\shortmid }\widehat{\mathcal{S}}^{\star }(\
^{\shortmid }\widehat{\rho }_{\mathcal{B\cup C}}\shortparallel \ ^{\shortmid
}\widehat{\rho }_{\mathcal{B}}\otimes 1_{\mathcal{C}}/k_{\mathcal{C}}), \\
\ ^{\shortmid }\widehat{\mathcal{S}}^{\star }(\ ^{\shortmid }\widehat{\rho }%
_{\mathcal{A\cup B\cup C}} &\shortparallel &1_{\mathcal{A\cup B\cup C}}/k_{%
\mathcal{A\cup B\cup C}})+\ \ \ ^{\shortmid }\widehat{\mathcal{S}}^{\star
}(\ ^{\shortmid }\widehat{\rho }_{\mathcal{B}}\shortparallel 1_{\mathcal{B}%
}/k_{\mathcal{B}})\geq \\
\ \ \ ^{\shortmid }\widehat{\mathcal{S}}^{\star }(\ ^{\shortmid }\widehat{%
\rho }_{\mathcal{A\cup B}} &\shortparallel &1_{\mathcal{A\cup B}}/k_{%
\mathcal{A\cup B}})+\ ^{\shortmid }\widehat{\mathcal{S}}^{\star }(\
^{\shortmid }\widehat{\rho }_{\mathcal{B\cup C}}\shortparallel 1_{\mathcal{%
B\cup C}}/k_{\mathcal{B\cup C}}).
\end{eqnarray*}

We introduce the \textit{mutual information, }$\ ^{\shortmid }\widehat{%
\mathcal{J}}^{\star },$ as a measure of correlation between two
nonassociative QGIFs $\widehat{\mathcal{A}}$ and $\widehat{\mathcal{B}}$
involving also a third system $\widehat{\mathcal{C}},$ 
\begin{equation*}
\ ^{\shortmid }\widehat{\mathcal{J}}^{\star }(\widehat{\mathcal{A}},\widehat{%
\mathcal{B}}):=\ ^{\shortmid }\widehat{\mathcal{S}}_{\mathcal{A}}^{\star }+\
^{\shortmid }\widehat{\mathcal{S}}_{\mathcal{B}}^{\star }-\ ^{\shortmid }%
\widehat{\mathcal{S}}^{\star }{}_{\mathcal{A\cup B}}\geq 0\mbox{ and }\ \
^{\shortmid }\widehat{\mathcal{J}}^{\star }(\widehat{\mathcal{A}},\widehat{%
\mathcal{B}}\mathcal{\cup }\widehat{\mathcal{C}})\leq \ ^{\shortmid }%
\widehat{\mathcal{J}}^{\star }(\widehat{\mathcal{A}},\widehat{\mathcal{B}}).
\end{equation*}%
Using such inequalities and the formula $\ ^{\shortmid }\widehat{\mathcal{J}}%
^{\star }(\widehat{\mathcal{A}},\widehat{\mathcal{B}})=\ ^{\shortmid }%
\widehat{\mathcal{S}}^{\star }(\ ^{\shortmid }\widehat{\rho }_{\mathcal{%
A\cup B}}\shortparallel \ ^{\shortmid }\widehat{\rho }_{\mathcal{A}}\otimes
\ ^{\shortmid }\widehat{\rho }_{\mathcal{B}}),$ we can prove important
inequalities for the entanglement of nonassociative QGIFs,%
\begin{equation*}
\ \ ^{\shortmid }\widehat{\mathcal{J}}^{\star }(\widehat{\mathcal{A}},%
\widehat{\mathcal{B}})\geq 0\mbox{
and }\ \ ^{\shortmid }\widehat{\mathcal{J}}^{\star }(\widehat{\mathcal{A}},%
\widehat{\mathcal{B}}\mathcal{\cup }\widehat{\mathcal{C}})\leq \ \
^{\shortmid }\widehat{\mathcal{J}}^{\star }(\widehat{\mathcal{A}},\widehat{%
\mathcal{B}}),\mbox{ for }\ \ ^{\shortmid }\widehat{\mathcal{J}}^{\star }(%
\widehat{\mathcal{A}},\widehat{\mathcal{B}})=\ ^{\shortmid }\widehat{%
\mathcal{S}}^{\star }(\ ^{\shortmid }\widehat{\rho }_{\mathcal{A\cup B}%
}\shortparallel \ ^{\shortmid }\widehat{\rho }_{\mathcal{A}}\otimes \
^{\shortmid }\widehat{\rho }_{\mathcal{B}}).
\end{equation*}

The mutual information $\ ^{\shortmid }\widehat{\mathcal{J}}^{\star }$
between two nonassociative QGIFs is a measure how much the density matrix $\
^{\shortmid }\widehat{\rho }_{\mathcal{A\cup B}}$ differs from a separable
state $\ \ ^{\shortmid }\widehat{\rho }_{\mathcal{A}}\otimes \ ^{\shortmid }%
\widehat{\rho }_{\mathcal{B}}.$ Quantum correlations entangle even spacetime
disconnected regions of the phase space under geometric flow evolution. Star
product and R-flux deformations contribute in nonassociative forms to such
entanglement.

\subsubsection{The R\'{e}nyi entropy for nonassociative QGIFs}

\label{ssrenyientr}For applications in modern quantum information, it is
used concept of R\'{e}nyi entropy \cite{renyi61}. It allows to compute the
entanglement entropy of QFTs by developing the so-called replica method
reviewed in section IV of \cite{nishioka18}, see also further
generalizations in \cite{bao19}. The approach can be generalized for
nonassociative QGIF using for classical nonassociative GIFs is defined by a
canonical partition function $\ _{s}^{\shortmid }\widehat{Z}[\ ^{\shortmid }%
\mathbf{g}(\tau )]$ (\ref{spf}) as a statistical density $\widehat{\rho }%
(\beta ,\widehat{\mathcal{E}}\ ,\mathbf{g})$ used for defining density
matrices $\ ^{\shortmid }\widehat{\rho }_{\mathcal{A}}$ of type (\ref%
{thfield})\ and or (\ref{dmq1}). In \cite{sv20}, we extended the replica
method to the Perelman thermodynamical model and related classical and
quantum information theories for NES. Similar constructions can be performed
on nonassociative deformed phase spaces.

Let us define for a nonassociative QGIF density matrix$\ ^{\shortmid }%
\widehat{\rho }_{\mathcal{A}}$ the R\'{e}nyi entropy: 
\begin{equation}
\ _{r}^{\shortmid }\widehat{\mathcal{S}}^{\star }(\widehat{\mathcal{A}}):=%
\frac{1}{1-r}\log [tr_{\mathcal{A}}(\ ^{\shortmid }\widehat{\rho }_{\mathcal{%
A}})^{r}],  \label{renentr}
\end{equation}%
with an integer replica parameter $r.$ A corresponding replica computational
formalism can be elaborated for an analytic continuation of $r$ to a real
number with a defined limit $\ ^{\shortmid }\widehat{\mathcal{S}}_{\mathcal{A%
}}^{\star }(\ ^{\shortmid }\widehat{\rho }_{\mathcal{A}})=\lim_{r\rightarrow
1}\ _{r}^{\shortmid }\widehat{\mathcal{S}}^{\star }(\widehat{\mathcal{A}})$
and normalization $tr_{\mathcal{A}}(\ ^{\shortmid }\widehat{\rho }_{\mathcal{%
A}})$ for $r\rightarrow 1.$ It assumed that for such conditions (\ref%
{renentr}) reduces to the nonassociative entanglement entropy $\ ^{\shortmid
}\widehat{\mathcal{S}}_{\mathcal{A}}^{\star }$ (\ref{entanglentrcan}).

Following Convention 2 (\ref{conv2s}) and similar (associative and
commutative) formulas proven in \cite{zycz03}, we can introduce important
inequalities for the derivative on the replica parameter, $\partial _{r},$ 
\begin{equation}
\partial _{r}(\ _{r}^{\shortmid }\widehat{\mathcal{S}}^{\star }\mathcal{)}%
\leq 0,\ \partial _{r}\left( \frac{r-1}{r}\ \ _{r}^{\shortmid }\widehat{%
\mathcal{S}}^{\star }\right) \geq 0,\ \partial _{r}[(r-1)\ \ _{r}^{\shortmid
}\widehat{\mathcal{S}}^{\star }\mathcal{]}\geq 0,\ \partial
_{rr}^{2}[(r-1)](\ _{r}^{\shortmid }\widehat{\mathcal{S}}^{\star }\mathcal{)}%
\leq 0.  \label{aux07}
\end{equation}%
A usual thermodynamical interpretation of such formulas is possible for GIF
and NES with a conventional modular Hamiltonian $\widehat{H}_{\mathcal{A}}=\
^{\shortmid }\widehat{\mathcal{E}}^{\star }$ and effective statistical
density $\ ^{\shortmid }\widehat{\rho }_{\mathcal{A}}:=e^{-2\pi \widehat{H}_{%
\mathcal{A}}},$ see assumptions for nonassociative geometric thermofield
theories (\ref{thfield}) and (\ref{dmq1}). \ For such generalizations, the
value $\beta _{r}=2\pi r$ is considered as the inverse temperature when the
effective "thermal" statistical generation (partition) function is defined $%
\ _{r}^{\shortmid }\widehat{\mathcal{Z}}(\beta _{r}):=tr_{\mathcal{A}}(\
^{\shortmid }\widehat{\rho }_{\mathcal{A}})^{r}=tr_{\mathcal{A}}(e^{-\beta
_{r}\widehat{H}_{\mathcal{A}}})$ similarly to $\ _{s}^{\shortmid }\widehat{Z}%
[\ ^{\shortmid }\mathbf{g}(\tau )]$ (\ref{spf}). In canonical forms, there
are computed such statistical mechanics values and conditions:%
\begin{eqnarray*}
\mbox{modular energy} &:&\ _{r}^{\shortmid }\widehat{\mathcal{E}}^{\star
}(\beta _{r}):=-\partial _{\beta _{r}}\log [\ _{r}^{\shortmid }\widehat{%
\mathcal{Z}}(\beta _{r})]\geq 0; \\
\mbox{modular entropy} &:&\ \ _{r}^{\shortmid }\mathcal{\breve{S}}^{\star
}(\beta _{r}):=\left( 1-\beta _{r}\partial _{\beta _{r}}\right) \log [\
_{r}^{\shortmid }\widehat{\mathcal{Z}}(\beta _{r})]\geq 0; \\
\mbox{modular capacity} &:&\ _{r}^{\shortmid }\widehat{\mathcal{C}}^{\star
}(\beta _{r}):=\beta _{r}^{2}\partial _{\beta _{r}}^{2}\log [\
_{r}^{\shortmid }\widehat{\mathcal{Z}}(\beta _{r})]\geq 0.
\end{eqnarray*}%
These inequalities invlove some conditions from (\ref{aux07}) and
characterize the stability of nonassociative GIFs considered as thermal
systems when the replica parameter is used as the inverse temperature for a
respective modular Hamiltonian. \footnote{%
The nonassociative QGIF theories with the modular entropy $\ _{r}^{\shortmid
}\widehat{\mathcal{S}}^{\star }$ can be transformed into models derived for
GIFs and associated thermodynamic models and with the R\'{e}nyi entropy and
inversely. The encode nonassociative data which can be computed for $\kappa $%
-linear parametric decompositions. Such transforms in canonical s-adapted
forms, for respective classes of parametric solutions, can be performed
using formulas $\ _{r}^{\shortmid }\mathcal{\breve{S}}^{\star
}:=r^{2}\partial _{r}\left( \frac{r-1}{r}\ \ _{r}^{\shortmid }\widehat{%
\mathcal{S}}^{\star }\right) $ and (for inverse transforms) $\ \ \
_{r}^{\shortmid }\widehat{\mathcal{S}}^{\star }\mathcal{=}\frac{r}{r-1}%
\int_{1}^{r}dr^{\prime }\ \ _{r}^{\shortmid }\mathcal{\breve{S}}^{\star
}/(r^{\prime })^{2}.$}

The concept of relative entropy $\ ^{\shortmid }\widehat{\mathcal{S}}%
^{\star}(\ ^{\shortmid }\widehat{\rho }_{\mathcal{A}}\shortparallel \
^{\shortmid }\widehat{\sigma }_{\mathcal{A}})$ (\ref{relativentr}) can be
extended to that of relative R\'{e}nyi entropy (for a review and similar
constructions, see section II.E.3b in \cite{nishioka18}). Considering a
nonassociative QGIF system involving density matrices $\ ^{\shortmid }%
\widehat{\rho }_{\mathcal{A}}$ and $\ ^{\shortmid }\widehat{\sigma }_{%
\mathcal{A}},$ we compute 
\begin{eqnarray}
\ _{r}^{\shortmid }\widehat{\mathcal{S}}^{\star }(\ ^{\shortmid }\widehat{%
\rho }_{\mathcal{A}}\shortparallel \ ^{\shortmid }\widehat{\sigma }_{%
\mathcal{A}})\ &=&\frac{1}{r-1}\log \left[ tr\left( (\ ^{\shortmid }\widehat{%
\sigma }_{\mathcal{A}})^{(1-r)/2r}\ \ ^{\shortmid }\ \widehat{\rho }_{%
\mathcal{A}}(\ \ ^{\shortmid }\widehat{\sigma }_{\mathcal{A}%
})^{(1-r)/2r}\right) ^{r}\right] ,\mbox{ for }r\in (0,1)\cup (1,\infty );
\label{relatrenyi} \\
\mbox{ or }\ \ _{1}^{\shortmid }\widehat{\mathcal{S}}^{\star }(\ ^{\shortmid
}\widehat{\rho }_{\mathcal{A}}\shortparallel \ ^{\shortmid }\widehat{\sigma }%
_{\mathcal{A}}) &=&\ ^{\shortmid }\widehat{\mathcal{S}}^{\star }(\
^{\shortmid }\ \widehat{\rho }_{\mathcal{A}}\shortparallel \ ^{\shortmid }%
\widehat{\sigma }_{\mathcal{A}})\mbox{ and }  \notag \\
\ _{\infty }^{\shortmid }\widehat{\mathcal{S}}^{\star }(\ ^{\shortmid }\ 
\widehat{\rho }_{\mathcal{A}}\shortparallel \ ^{\shortmid }\widehat{\sigma }%
_{\mathcal{A}}) &=&\log ||(\ \ ^{\shortmid }\widehat{\sigma }_{\mathcal{A}%
})^{-1/2}\ \ ^{\shortmid }\widehat{\rho }_{\mathcal{A}}(\ ^{\shortmid }%
\widehat{\sigma }_{\mathcal{A}})^{-1/2}||_{\infty }.  \notag
\end{eqnarray}%
In any point of causal curves on respective nonholononomic Lorentz
manifolds/ bundles / co-bundles, we can prove monotonic properties, 
\begin{equation*}
\ _{r}^{\shortmid }\widehat{\mathcal{S}}^{\star }(\ ^{\shortmid }\widehat{%
\rho }_{\mathcal{A}}\shortparallel \ ^{\shortmid }\widehat{\sigma }_{%
\mathcal{A}})\geq \ \ _{r}^{\shortmid }\widehat{\mathcal{S}}^{\star
}(tr_{s}\ \ ^{\shortmid }\widehat{\rho }_{\mathcal{A}}|tr_{s}\ \ ^{\shortmid
}\widehat{\sigma }_{\mathcal{A}})\mbox{ and }\partial _{r}[\ _{r}^{\shortmid
}\widehat{\mathcal{S}}^{\star }(\ ^{\shortmid }\widehat{\rho }_{\mathcal{A}%
}\shortparallel \ ^{\shortmid }\widehat{\sigma }_{\mathcal{A}})]\geq 0.
\end{equation*}%
This reduces the nonassociative relative R\'{e}nyi entropy to the
nonassociative R\'{e}nyi entropy for QGIFs using formula $\ _{r}^{\shortmid }%
\widehat{\mathcal{S}}^{\star }(\ ^{\shortmid }\widehat{\rho}_{\mathcal{A}%
}\shortparallel 1_{\mathcal{A}}/k_{\mathcal{A}})=\log k_{\mathcal{A}}-\
_{r}^{\shortmid }\widehat{\mathcal{S}}^{\star }(\widehat{\mathcal{A}}).$

\section{Nonassociative entanglement \& transversability of phase space
wormholes}

\label{sec4} The goal of this section is to construct explicit s-forms and
analyze the physical properties of two classes of nonassociative modified
wormhole solutions on phase spaces $\ _{s}\mathcal{M}^{\star }$. Such
off-diagonal s-metrics describe: 1) double 4-d wormholes involving distinct
spacetime and momentum type coordinates; and 2) higher dimension
nonassociative wormholes with warped momentum coordinates, which can be
connected to nonassociative higher dimension BHs, for instance, of
off-diagonal and generalized Tangherlini type \cite{lbdssv22,lbsvevv22}.
Such solutions are with nontrivial nonholonomic structure and/or with warped
factors \cite{vwh3,vbv18,warpwormh}. We note that in associative and
commutative form various classes of nonholonomic wormhole and BH solutions
were constructed applying the AFCDM to GR and MGTs, see reviews of results
in \cite{vbv18,wormh21a}. In this work, nonassociative star product R-flux
deformations of prime phase space wormhole metrics, and their
theoretical-informational properties are investigated. Necessary technical
results and methods are summarized in appendix \ref{asstables}.

\vskip5pt It is well known that transversable wormholes are forbidden in GR
in the sense that it is not possible to send causal signals through its
throat faster than we can send it through outside. A number of theoretic
constructions and modifications were exploited with the aim of solving this
problem; for reviews of results and methods of constructing wormhole
solutions, we cite \cite{morris88,kar94,roy20,souza22,sv14a}. A series of
recent works exploit the idea that wormholes seem to be transversable for
qubits. Such innovative ideas, proposals for thought experiments and
preliminary experimental data on running experiments on Google's quantum
computer are provided in \cite{wormh22}. That team of researchers set out to
create a wormhole in the lab based on transversable wormhole protocol \cite%
{wormh19} (using negative energy shock waves allowing the qubit to traverse
it), see also \cite{wormh19a,wormh21}. Their approach is based on J.
Wheeler, G. 't Hooft and L. Susskind ideas that our Universe is a
hologram-like projection of the quantum world (discussions of early ideas,
original works and references can be found in \cite%
{misner,thooft93,susskind95}). In such models, the spacetime arises from
some quantum information theories for QFTs in lower dimensions and when in
the semi-classical region the geometry and effective thermodynamics are
controlled by lower dimensions. Then, the holographic duality was involved
in exploring models of QG, which mathematically connects gravitational
systems to quantum systems, and when spacetime properties are described by
two different theories, a recent review on wormholes and holography can be
found in \cite{wormh21a}. We cite here two important papers on diving into
traversable wormholes and teleportation through the wormholes \cite%
{wormh17,wormh18}. Such constructions originate from two famous works \cite%
{er35}, on Einstein-Rosen bridges/ wormholes, EP; and \cite{epr35}, on the
Einstein-Podolsky-Rosen paradox, EPR, and ideas on entanglement. In \cite%
{wormh13}, it was conjectured that ER =EPR and suggested that the concept of
wormhole configurations predicted by GR is equivalent to quantum
entanglement.

\vskip5pt If the above-described constructions with wormholes and
entanglement, when ER=EPR, (all elaborated in the framework of the
holographic paradigm), are true, it could help us to solve the biggest
incompatibility problem in physics: How the gravity can be formulated in the
language of quantum physics using quantum information concepts and how this
can be related to QG? In this approach, to explore QG, physicists use a
technique called holographic duality which, in its turn, is elaborated in
the framework of the Bekenstein-Hawking BH entropy paradigm \cite%
{bek1,bek2,haw1,haw2}. It works always when certain high symmetry solutions
of some (modified) gravitational equations involve hypersurface/ holographic
configurations. For more general classes of generic off-diagonal solutions
in GR and in MGTs, for instance, with nonassociative star product R-flux
deformations, and relativistic geometric flow models on Lorentz manifolds or
(co) tangent Lorentz bundles, we have to elaborate on more advanced
theoretic constructions based on the concept of Perelman thermodynamics \cite%
{perelman1}. Nonassociativity and noncommutativity arise naturally both in
quantum physics and string theory which motivated us to formulate in
sections \ref{sec3} and \ref{sec4} respective models of statistical
thermodynamics for nonassociative GIFs and nonassociative QGIF theories.
They involve nonassociative modifications of the concept of entanglement
which are investigated in this section considering applications for more
general classes of exact/parametric solutions in nonassociative geometric
and information flow theory and nonassociative gravity \cite{sv20}.

\subsection{Nonassociative geometric flow deformations of double 4-d phase
space wormholes}

\label{ssdoublenworm} We construct a new class of parametric solutions of
nonassociative geometric flow equations (\ref{nonassocrf}) defined by ansatz
for quasi-stationary s-metrics of type (\ref{ans1rf}). For small $\kappa $%
-parametric and running $\tau $-dependencies, such solutions define two
types of nonholonomic 4-d wormholes:\ one configuration on a Lorentzian base
spacetime manifold and another one on the co-fibers space with momentum like
coordinates. Both geometric/ physical objects encode nonassociative star
product R-flux deformations and can be modelled to possess nonlinear
symmetries for respective running cosmological constants. The nonholonomic
structure can be prescribed in some forms generating nonassociative
wormholes with ellipsoidal symmetries or, in certain associative and
commutative limits, describe couples of nonholonomic 4-d wormholes like in 
\cite{vwh3,sv14a,vbv18}. The thermodynamic properties of such double
wormhole solutions can not be characterized, in general, using
generalizations of the concept of Bekenstein-Hawking entropy but we can
always compute corresponding modified Pereman's thermodynamics variables (%
\ref{nagthermodvalues}). In analytic form, this is always possible using $%
\kappa $-linear $F$- and/or $W$-functionals $\ _{s}^{\shortmid }\widehat{%
\mathcal{F}}_{\kappa }^{\star}(\tau )$ (\ref{naffunctp}), or $\
_{s}^{\shortmid }\widehat{\mathcal{W}}_{\kappa }^{\star }(\tau )$ (\ref%
{nawfunctp}). We show how corresponding quantum information basic
ingredients for nonassociative wormhole GIF and QGIF models can be computed
in explicit form using formulas from sections \ref{sec2} and \ref{sec3}.

\subsubsection{Prime metrics as Morris-Thorne and generalized
Ellis-Bronnikov wormholes}

Let us consider a 8-d phase space diagonal metric 
\begin{eqnarray}
d\mathring{s}^{2} &=&\ ^{\shortmid }\mathring{g}_{\alpha _{s}^{\prime }}(d\
^{\shortmid }u^{\alpha _{s}^{\prime }})^{2}=\ ^{\shortmid }\mathring{g}%
_{i_{2}^{\prime }}(d\ x^{i_{2}^{\prime }})^{2}+\ ^{\shortmid }\mathring{g}%
^{a_{s}^{\prime }}(d\ p_{a_{s}^{\prime }})^{2}  \label{pm1} \\
&=&(1-\frac{b(r)}{r})^{-1}dr^{2}+r^{2}d\theta ^{2}+r^{2}\sin ^{2}\theta
d\varphi ^{2}-e^{2\Theta (r)}dt^{2}+  \notag \\
&&(1-\frac{\ ^{\shortmid }b(p_{r})}{p_{r}}%
)^{-1}d(p_{r})^{2}+(p_{r})^{2}(dp_{\theta })^{2}+(p_{r})^{2}\sin
^{2}(p_{\theta })(dp_{\varphi })^{2}-e^{2\ ^{\shortmid }\Theta
(p_{r})}dE^{2}.  \notag
\end{eqnarray}%
In this formula, we use spherical coordinates both on base and cofiber
spaces, $\ ^{\shortmid }u^{\alpha _{s}^{\prime }}=(r,\theta ,\varphi
,t,p_{r},p_{\theta },p_{\varphi },E)=(x^{i_{2}^{\prime
}},p_{a_{s}^{\prime}}) $ for $\alpha _{s}^{\prime }=1,2,...8;$ and,
respectively, $x^{i_{2}^{\prime}}=(x^{i_{1}^{\prime }},y^{a_{2}^{\prime }}),
p_{a_{s}^{\prime}}=(p_{a_{3}^{\prime }},p_{a_{3}^{\prime }}),$ with $%
i_{1}=1,2;a_{2}^{\prime}=3,4;a_{3}^{\prime }=5,6$ and $a_{4}^{\prime }=7,8.$
Such a prime s-metric (with zero N-connection coefficients) describes two
generic Morris-Thorne wormhole configurations \cite{morris88}, when the
Lorentzian spacetime base contains a standard wormhole and the co-fiber part%
\footnote{%
with momentum like coordinates; for well-defined physical models, we should
consider corresponding unities when the units $[x^{i_{2}^{\prime }}]$ are
the same as $[p_{a_{s}^{\prime }}]$ up to multiplication on some constant
coefficients}contains another wormhole. Let us explain which conditions must
satisfy the coefficients of (\ref{pm1}) for generating respective wormhole
configurations. For simplicity, we provide details on the spacetime part
when the conditions for momentum variables and coefficients can be stated by
a dual analogy. In explicit form, $e^{2\Theta(r)}$ is a red-shift function
and $b(r)$ is the shape function defined in spherically polar coordinates $%
u^{i_{2}}=(r,\theta ,\varphi ,t).$ We can define usual Ellis-Bronnikov, EB,
wormholes for $\Theta (r)=0$ and when $b(r)=\ _{0}b^{2}/r$ taken for a zero
tidal wormhole with $\ _{0}b$ being the throat radius. The papers \cite%
{kar94,roy20,souza22} can be considered for details and recent reviews of
results on wormhole physics. Here we note that a generalized EB is
characterized additionally by even integers $2k$ (with $k=1,2,...$) and when 
$r(l)=(l^{2k}+\ _{0}b^{2k})^{1/2k}$ is taken as a proper radial distance
coordinate (tortoise) and the cylindrical angular coordinate $\phi \in
\lbrack 0,2\pi )$ is called parallel. In such new coordinates, $-\infty
<l<\infty $, which is different from the cylindrical radial coordinate $%
\rho, $ when $0\leq \rho <\infty .$ In result, the spacetime component of
the prime metric (\ref{pm1}) can be written in the form 
\begin{eqnarray*}
d\mathring{s}_{[h]}^{2}&=&dl^{2}+r^{2}(l)d\theta ^{2}+r^{2}(l)\sin
^{2}\theta d\varphi ^{2}-dt^{2}, \mbox{when } \\
dl^{2}&=&(1-\frac{b(r)}{r})^{-1}dr^{2}\mbox{ and }b(r)=r-r^{3(1-k)}(r^{2k}-\
_{0}b^{2k})^{(2-1/k))}.
\end{eqnarray*}

To apply the AFCDM for generating new classes of generic off-diagonal
solutions is important to avoid off-diagonal deformations with coordinate
and frame coefficient singularities. For instance, we can consider frame
transforms to a parametrization with some trivial N-connection coefficients $%
\check{N}_{i}^{a}=\check{N}_{i}^{a}(u^{\alpha }(l,\theta ,\varphi ,t))$ and $%
\check{g}_{\beta }(u^{j}(l,\theta ,\varphi ),u^{3}(l,\theta ,\varphi ))$
(triviality means that such coefficients are generated by coordinate
transforms for integrable distributions). This is possible introducing new
coordinates $u^{1}=x^{1}=l,u^{2}=\theta ,$ and $u^{3}=y^{3}=\varphi +\
^{3}B(l,\theta),u^{4}=y^{4}=t+\ ^{4}B(l,\theta ),$ when 
\begin{eqnarray*}
\mathbf{\check{e}}^{3} &=&d\varphi =du^{3}+\check{N}_{i}^{3}(l,%
\theta)dx^{i}=du^{3}+\check{N}_{1}^{3}(l,\theta)dl +\check{N}%
_{2}^{3}(l,\theta)d\theta , \\
\mathbf{\check{e}}^{4} &=&dt=du^{4}+\check{N}_{i}^{4}(l,\theta)dx^{i}=du^{4}+%
\check{N}_{1}^{4}(l,\theta )dl+ \check{N}_{2}^{4}(l,\theta) d\theta ,
\end{eqnarray*}%
for $\mathring{N}_{i}^{3}=-\partial \ ^{3}B/\partial x^{i}$ and $\mathring{N}%
_{i}^{4}=-\partial \ ^{4}B/\partial x^{i}.$ Using such nonlinear
coordinates, the quadratic elements for above wormhole solutions can be
parameterized as a prime d-metric, 
\begin{equation}
d\mathring{s}_{[h]}^{2}=\check{g}_{i_{2}}(l,\theta ,\varphi )[\mathbf{\check{%
e}}^{i_{2}}(l,\theta ,\varphi )]^{2},  \label{pmwh}
\end{equation}%
where $\check{g}_{1}=1,\check{g}_{2}=r^{2}(l),\check{g}_{3}=r^{2}(l)\sin
^{2}\theta $ and $\check{g}_{4}=-1.$

Considering similar constructions in the co-fiber (momentum) space, we
construct%
\begin{eqnarray}
d\ ^{\shortmid }\mathring{s}_{[c]}^{2} &=&\ ^{\shortmid }\check{g}^{a_{s}}(\
^{\shortmid }l,p_{\theta },p_{\varphi })[\ ^{\shortmid }\mathbf{\check{e}}%
_{a_{s}}(\ ^{\shortmid }l,p_{\theta },p_{\varphi })]^{2},\mbox {where }
\label{pmwc} \\
\ ^{\shortmid }\mathbf{\check{e}}_{5} &=&\delta p_{l}=d\ ^{\shortmid
}u^{5}=d\ ^{\shortmid }l,\mbox{ for }\ ^{\shortmid }\check{N}_{5i_{3}}(\
^{\shortmid }l,p_{\theta },p_{\varphi })=0;\ ^{\shortmid }\mathbf{\check{e}}%
_{6}=\delta p_{\theta }=d\ ^{\shortmid }u^{6}=d\ p_{\theta },\mbox{ for }\
^{\shortmid }\check{N}_{6i_{3}}(\ ^{\shortmid }l,p_{\theta },p_{\varphi })=0;
\notag \\
\ ^{\shortmid }\mathbf{\check{e}}_{7} &=&\delta p_{\varphi }=\delta \
^{\shortmid }u^{7}=d\ ^{\shortmid }u^{7}+\ ^{\shortmid }\check{N}_{7i_{3}}(\
^{\shortmid }l,p_{\theta })d\ ^{\shortmid }x^{i_{3}};\ ^{\shortmid }\mathbf{%
\check{e}}_{8}=\delta E=\delta \ ^{\shortmid }u^{8}=dE+\ ^{\shortmid }\check{%
N}_{8i_{3}}(\ ^{\shortmid }l,p_{\theta })d\ ^{\shortmid }x^{i_{3}},  \notag
\end{eqnarray}%
where $i_{3}=1,2,3,4,5,6;\ ^{\shortmid }\check{g}^{5}=1,\ ^{\shortmid }%
\check{g}^{6}=[p_{r}(\ ^{\shortmid }l)]^{2},\ ^{\shortmid }\check{g}%
^{7}=[p_{r}(\ ^{\shortmid }l)]^{2}\sin ^{2}p_{\theta }$ and $\ ^{\shortmid }%
\check{g}^{8}=-1.$

Summarizing (\ref{pmwh}) and (\ref{pmwc}), we express (\ref{pm1}) in an
equivalent s-adapted form (it is not generic, because such a quadratic line
element was generated by coordinate transforms on the total phase space), 
\begin{equation}
d\mathring{s}^{2}=\ ^{\shortmid }\check{g}_{\alpha _{s}}(d\ ^{\shortmid
}u^{\alpha _{s}})^{2}=\ ^{\shortmid }\check{g}_{i_{2}}(d\ x^{i_{2}})^{2}+\
^{\shortmid }\check{g}^{a_{3}}(d\ ^{\shortmid }u^{a_{3}})^{2}+\ ^{\shortmid }%
\check{g}^{a_{4}}(d\ \ ^{\shortmid }\mathbf{\check{e}}_{a_{4}})^{2}.
\label{pm1a}
\end{equation}%
In this formula, we omitted priming indices for the coordinate system
following the notations that 
\begin{equation*}
\ ^{\shortmid }u^{\alpha _{s}}=\ ^{\shortmid }u^{\alpha _{s}}(\ ^{\shortmid
}u^{\alpha _{s}^{\prime }})=\ ^{\shortmid }u^{\alpha _{s}}(l,\theta ,\varphi
,t,p_{l},p_{\theta },p_{\varphi },E)=(x^{i_{2}},p_{a_{s}})=(\ ^{\shortmid
}x^{i_{3}},p_{a_{4}}).
\end{equation*}%
The diagonal metric (\ref{pm1}) and s-metric (\ref{pm1a})\ (because of
curved coordinates, it is off-diagonal) define in equivalent forms a matter
deformed solution of the Ricci soliton equations%
\begin{equation}
\ ^{\shortmid }Ric[\tau _{0},\ \ ^{\shortmid }\mathring{g},\ ^{\shortmid }%
\mathbf{\mathring{\nabla}}]=\ ^{\shortmid }Ric[\tau _{0}, \ ^{\shortmid }%
\mathbf{\check{g}},\ ^{\shortmid }\mathbf{\check{D}}]=\ _{s}^{\shortmid }\Im
,  \label{twowhricci}
\end{equation}%
see (\ref{ricciconfhonh}) for $\ ^{\shortmid }\mathbf{\check{D}}\simeq \
^{\shortmid }\check{\nabla}.$ In such cases, the system of nonlinear PDE (%
\ref{twowhricci}) can be considered as a variant of Einstein equations in
8-d gravity with momentum variables stated on associative and commutative
phase space $\ _{s}^{\shortmid }\mathcal{M}.$ The effective source $\
_{s}^{\shortmid }\Im $ can be determined in explicit form using
corresponding energy-momentum tensors for matter as in GR, or MGTs
considered as 4-d and/or 8-d gravity theories. Such double 4-d wormhole
phase space configurations deforms substantially (for certain well-defined
conditions, we can consider small parametric deformations) if $\ _{s}%
\mathcal{M}\rightarrow \ _{s}^{\shortmid }\mathcal{M}^{\star },$ with $\
_{s}^{\shortmid }\Im \rightarrow \ _{s}^{\shortmid }\Im ^{\star },$ when we
consider nonassociative geometric flows and star product R-flux
deformations. The associative and commutative modified Einstein equations (%
\ref{twowhricci}) have to be extended, for instance, to nonassociative
geometric flow equations (\ref{nonassocrf}). The s-metric (\ref{pm1a}) can
be transformed (for respective nonholonomic configurations) into certain
variants of nonassociative $\tau $-running double 4-d phase space wormhole
configurations. Such star deformed wormholes are entangled both at the
classical level, as some GIF systems (because of geometric flows on phase
space) and as some QGIF systems, if there are involved scenarios with
nonassociative quantum and thermo-geometric theories.

\subsubsection{Quasi-stationary gravitational polarizations for
nonassociative deformed wormholes}

Using the AFCDM, we can generate parametric solutions describing
nonassociative geometric flow deformations of the double 4-d wormhole
s-metric (\ref{pm1a}). This geometric and analytic method for generating
off-diagonal solutions is outlined in Appendix \ref{asstables}, see Table 3
and respective $\tau $-families of quasi-stationary s-metrics (\ref{qst8}).
In terms of gravitational $\eta $-polarizations, we can model R-flux $\tau $%
-running deformations of type 
\begin{equation}
\ _{s}^{\shortmid }\mathbf{\check{g}}=[\ ^{\shortmid }\check{g}_{\alpha
_{s}},\ ^{\shortmid }\check{N}_{i_{s-1}}^{a_{s}}]\rightarrow \
_{s}^{\shortmid }\mathbf{g}(\tau )=[\ ^{\shortmid }\eta _{\alpha _{s}}(\tau
)\ ^{\shortmid }\check{g}_{\alpha _{s}}(\tau ),\ ^{\shortmid }\eta
_{i_{s-1}}^{a_{s}}(\tau )\ ^{\shortmid }\check{N}_{i_{s-1}}^{a_{s}}(\tau )],
\label{taufluxdef1}
\end{equation}%
substituting respective data, $\ _{s}^{\shortmid }\mathbf{\mathring{g}}%
(\tau)\rightarrow \ _{s}^{\shortmid }\mathbf{\check{g}}$ and $\
_{s}^{\shortmid}\Im \rightarrow \ _{s}^{\shortmid }\Im ^{\star }(\tau )$ in (%
\ref{offdiagpolfr1}).\footnote{%
Here we note that $\ _{s}^{\shortmid }\Im $ and $\ _{s}^{\shortmid }\Im
^{\star }(\tau )$ are stated for certain type of wormhole energy-momentum
tensors, and respective nonholonomic, $\tau $- and $\kappa $-linear
deformations of such sources which, for target configurations encode
nonassociative data.} Such $\tau $-families of target metrics are determined
by prime associative and commutative wormhole phase s-metrics with
coefficients $\ ^{\shortmid }\mathbf{\check{g}}_{\alpha _{s}\beta _{s}}$ not
depending on the flow parameter $\tau $, when $\ _{s}^{\shortmid }\mathbf{g}%
_{\alpha _{s}\beta _{s}}(\tau )$ are s-metrics solving the nonassociative
geometric flow equations (\ref{nonassocrf}).

In phase space coordinates adapted to prime double wormhole solutions (\ref%
{pm1a}), we parameterize the $\tau $-running generating functions (\ref%
{etapolgen}) in the form:%
\begin{eqnarray}
\psi (\tau ) &\simeq &\psi (\hbar ,\kappa ;\tau ,l,\theta ;\ _{1}\Lambda
^{\star }(\tau );\ \ _{1}^{\shortmid }\Im ^{\star }(\tau )),\eta _{4}(\tau
)\ \simeq \eta _{4}(\tau ,x^{k_{1}},l,\theta ,\varphi ;\ _{2}\Lambda ^{\star
}(\tau );\ \ _{2}^{\shortmid }\Im ^{\star }(\tau )),  \label{etaplegen1} \\
\ ^{\shortmid }\eta ^{6}(\tau ) &\simeq &\ ^{\shortmid }\eta ^{6}(\tau
,l,\theta ,\varphi ,p_{l};\ _{3}\Lambda ^{\star }(\tau );\ \ _{3}^{\shortmid
}\Im ^{\star }(\tau )),\ ^{\shortmid }\eta ^{8}(\tau )\simeq \ ^{\shortmid
}\eta ^{8}(\tau ,l,\theta ,\varphi ,p_{l},p_{\varphi };\ _{4}\Lambda ^{\star
}(\tau );\ \ _{4}^{\shortmid }\Im ^{\star }(\tau )).  \notag
\end{eqnarray}%
Such $\eta (\tau )$-deformations for (\ref{etaplegen1}) result in a $\tau $%
-families of quasi-stationary s-metrics: 
\begin{eqnarray}
d\ \ ^{\shortmid }\widehat{s}^{2}(\tau ) &=& \ ^{\shortmid }g_{\alpha
_{s}\beta _{s}}(\hbar ,\kappa ,\tau ,l,\theta ,\varphi ,p_{l},p_{\varphi };\
^{\shortmid }\check{g}_{\alpha _{s}};\eta _{4}(\tau ),\ ^{\shortmid }\eta
^{6}(\tau ),\ ^{\shortmid }\eta ^{8}(\tau ),\ _{s}\Lambda ^{\star }(\tau );\
\ _{s}^{\shortmid }\Im ^{\star }(\tau ))d~\ ^{\shortmid }u^{\alpha _{s}}d~\
^{\shortmid }u^{\beta _{s}}  \notag \\
&=&e^{\psi (\tau )}[(dx^{1}(l,\theta ))^{2}+(dx^{2}(l,\theta ))^{2}]-
\label{doublwnonassocwh} \\
&&\frac{[\partial _{3}(\eta _{4}(\tau )\ \check{g}_{4}(\tau ))]^{2}}{|\int
d\varphi \ _{2}^{\shortmid }\Im ^{\star }(\tau )\partial _{3}(\eta _{4}(\tau
)\ \check{g}_{4}(\tau ))|\ (\eta _{4}(\tau )\check{g}_{4}(\tau ))}\{dy^{3}+%
\frac{\partial _{i_{1}}[\int d\varphi \ _{2}\Im (\tau )\ \partial _{3}(\eta
_{4}(\tau )\check{g}_{4}(\tau ))]}{\ _{2}^{\shortmid }\Im ^{\star }(\tau
)\partial _{3}(\eta _{4}(\tau )\check{g}_{4}(\tau ))}dx^{i_{1}}\}^{2}+ 
\notag \\
&&\eta _{4}(\tau )\check{g}_{4}(\tau ))\{dt+[\ _{1}n_{k_{1}}(\tau )+\
_{2}n_{k_{1}}(\tau )\int \frac{d\varphi \lbrack \partial _{3}(\eta _{4}(\tau
)\check{g}_{4}(\tau ))]^{2}}{|\int dy^{3}\ _{2}^{\shortmid }\Im ^{\star
}(\tau )\partial _{3}(\eta _{4}(\tau )\check{g}_{4}(\tau ))|\ [\eta
_{4}(\tau )\check{g}_{4}(\tau )]^{5/2}}]dx^{k_{1}}\}-  \notag
\end{eqnarray}%
\begin{eqnarray*}
&&\frac{[\ ^{\shortmid }\partial ^{5}(\ ^{\shortmid }\eta ^{6}(\tau )\
^{\shortmid }\check{g}^{6}(\tau ))]^{2}}{|\int dp_{l}~\ _{3}^{\shortmid }\Im
^{\star }(\tau )\ \ ^{\shortmid }\partial ^{5}(\ ^{\shortmid }\eta ^{6}(\tau
)\ \ ^{\shortmid }\check{g}^{6}(\tau ))\ |\ (\ ^{\shortmid }\eta ^{6}(\tau
)\ ^{\shortmid }\check{g}^{6}(\tau ))}\{dp_{l}+\frac{\ ^{\shortmid }\partial
_{i_{2}}[\int dp_{l}\ _{3}^{\shortmid }\Im (\tau )\ ^{\shortmid }\partial
^{5}(\ ^{\shortmid }\eta ^{6}(\tau )\ ^{\shortmid }\check{g}^{6}(\tau ))]}{%
~\ _{3}^{\shortmid }\Im ^{\star }(\tau )\ ^{\shortmid }\partial ^{5}(\
^{\shortmid }\eta ^{6}(\tau )\ ^{\shortmid }\check{g}^{6}(\tau ))}%
dx^{i_{2}}\}^{2}+ \\
&&(\ ^{\shortmid }\eta ^{6}(\tau )\ ^{\shortmid }\check{g}^{6}(\tau
))\{dp_{\theta }+[\ _{1}^{\shortmid }n_{k_{2}}(\tau )+\ _{2}^{\shortmid
}n_{k_{2}}(\tau )\int \frac{dp_{l}[\ ^{\shortmid }\partial ^{5}(\
^{\shortmid }\eta ^{6}(\tau )\ ^{\shortmid }\check{g}^{6}(\tau ))]^{2}}{%
|\int dp_{l}\ _{3}^{\shortmid }\Im ^{\star }(\tau )\ \partial ^{5}(\
^{\shortmid }\eta ^{6}(\tau )\ ^{\shortmid }\check{g}^{6}(\tau ))|\ [\
^{\shortmid }\eta ^{6}(\tau )\ ^{\shortmid }\check{g}^{6}(\tau )]^{5/2}}%
]dx^{k_{2}}\}-
\end{eqnarray*}%
\begin{eqnarray*}
&&\frac{[\ ^{\shortmid }\partial ^{7}(\ ^{\shortmid }\eta ^{8}(\tau)\
^{\shortmid}\check{g}^{8}(\tau))]^{2}}{|\int dp_{7}\ _{4}^{\shortmid }\Im
^{\star}(\tau)\ ^{\shortmid }\partial ^{8}(\ ^{\shortmid }\eta ^{7}(\tau)\
^{\shortmid }\check{g}^{7}(\tau))\ |\ (\ ^{\shortmid }\eta ^{7}(\tau)\
^{\shortmid}\check{g}^{7}(\tau))}\{dp_{\varphi }+\frac{\
^{\shortmid}\partial _{i_{3}}[\int dp_{\varphi}\ _{4}^{\shortmid}\Im
^{\star}(\tau)\ ^{\shortmid}\partial ^{7}(\ ^{\shortmid}\eta ^{8}(\tau) \
^{\shortmid}\check{g}^{8}(\tau))]}{\ _{4}^{\shortmid}\Im ^{\star}(\tau) \
^{\shortmid}\partial ^{7}(\ ^{\shortmid}\eta ^{8}(\tau)\ ^{\shortmid}\check{g%
}^{8}(\tau))}d\ ^{\shortmid}x^{i_{3}}\}^{2}+ \\
&&(\ ^{\shortmid}\eta ^{8}(\tau)\ ^{\shortmid}\check{g}^{8}(\tau))\{dE+[\
_{1}n_{k_{3}}(\tau )+\ _{2}n_{k_{3}}(\tau )\int \frac{dp_{7} [\
^{\shortmid}\partial ^{7}(\ ^{\shortmid}\eta ^{8}(\tau)\ ^{\shortmid}\check{g%
}^{8}(\tau))]^{2}}{|\int dp_{7}\ _{4}^{\shortmid}\Im ^{\star }(\tau)[\
^{\shortmid}\partial ^{7}(\ ^{\shortmid }\eta ^{8}(\tau)\ ^{\shortmid }%
\check{g}^{8}(\tau))]|\ [\ ^{\shortmid}\eta ^{8}(\tau)\ ^{\shortmid }\check{g%
}^{8}(\tau)]^{5/2}}]d\ ^{\shortmid }x^{k_{3}}\}.
\end{eqnarray*}%
Such s-metrics preserve a quasi-stationary character under nonassociative
geometric flow evolution and possess certain nonlinear symmetries relating
effective nonassociative sources $\ _{s}^{\shortmid }\Im ^{\star }(\tau )$
to certain nonzero running effective cosmological constants $\
_{s}^{\shortmid }\Lambda ^{\star }(\tau )$.\footnote{%
Formulas for nonlinear symmetries involving generating functions and sources
are presented in Table 3 in Appendix and details and proofs can be found in
partner works \cite{partner02,partner03}} For self-similar configurations
with $\tau =\tau _{0},$ they generate parametric solutions of vacuum
nonassociative gravitational equations with effective cosmological constants 
$_{s}^{\shortmid }\Lambda _{0}^{\star }=\ _{s}^{\shortmid }\Lambda
^{\star}(\tau _{0}).$ The target s-metrics $\ _{\check{\eta}}^{\shortmid }%
\mathbf{g}(\tau )= \ _{s}^{\shortmid }\mathbf{g}(\tau )$ (\ref%
{doublwnonassocwh}) (for geometric symbolic computations, we shall use such
notations) are defined by a $\tau $-family $\psi (\tau ,l,\theta )$ of
solutions of 2-d Poisson equations 
\begin{equation*}
\partial _{11}^{2}\psi (\tau ,l,\theta )+\partial _{22}^{2}\psi (\tau
,l,\theta )=2\ \ _{1}^{\shortmid }\Im ^{\star }(\tau ,l,\theta ),
\end{equation*}%
which, correspondingly, are related via 2-d conformal transforms to some
classes of solutions of $\partial _{11}^{2}\psi (\tau )+\partial
_{22}^{2}\psi (\tau )=2 \ _{1}\Lambda ^{\star }(\tau ).$ The general form of
such generic off-diagonal metrics encode also, for instance, $\tau $%
-families of integration functions $\ _{1}n_{k_{1}}(\tau ,l,\theta )$ and $\
_{2}n_{k_{1}}(\tau ,l,\theta );$ $\ _{1}n_{k_{2}}(\tau ,l,\theta ;t,p_{l})$
and $\ _{2}n_{k_{2}}(\tau ,l,\theta ;t,p_{l});$ and $\
_{1}n_{k_{3}}(\tau,l,\theta ;t,p_{l})$ and $\ _{2}n_{k_{3}}(\tau ,l,\theta ;
t,p_{l}; p_{\theta},p_{\varphi }).$

The solutions (\ref{doublwnonassocwh}) are, in general, with nontrivial
canonical s-torsion $\ ^{\shortmid }\widehat{\mathcal{T}}^{\star }$. We have
to impose additional nonholonomic constraints (on the respective classes of
generating and integration functions) of type (\ref{lccondnonass}) if we
wont to extract LC-configurations and model their nonassociative geometric
flow evolution. There are also nontrivial $\kappa $-linear terms for the
symmetric, $\ _{\star }^{\shortmid }\mathfrak{\check{g}}_{\alpha
_{s}\beta_{s}},$ and nonsymmetric, $\ _{\star }^{\shortmid }\mathfrak{a}%
_{\alpha _{s}\beta _{s}},$ components of nonassociative s-metrics $\
_{\star}^{\shortmid }\mathfrak{g}_{\alpha _{s}\beta _{s}}$ which can be
computed using formals (\ref{aux40a}) - (\ref{aux40aa}). Following the
AFCDM, such nonsymmetric metric coefficients are decoupled from the
symmetric ones and generated as some nonholonomic induced variables and
functionals of the coefficients of the symmetric s-metrics and N-connections
from the prime and target solutions. General $\tau $- and $\kappa $%
-nonassociative deformations of prime wormhole metrics do not result in
other families of wormhole s-metrics. The wormhole character can be modified
and transformed, for instance, into certain solitonic configurations etc. We
can not provide explicit physical important realizations for such solutions.
Nevertheless, such target s-metrics can be characterized by respective
nonlinear symmetries and Perelman variables as we shall consider in next
subsections.

\subsubsection{Parametric nonassociative geometric flows of double wormhole
s-metrics}

\label{ssdwhsmall}We can impose certain nonholonomic conditions when
s-metrics of type (\ref{doublwnonassocwh}) describe nonsymmetric $\tau $%
-evolution of double 4-d wormholes on phase space $\ _{s}^{\shortmid }%
\mathcal{M}^{\star }$ and the wormhole character of solutions is preserved
but with some small parametric $\kappa $-linear terms, for instance, with
small ellipsoid deformation of throats and effective polarizations of
physical constants. For such purposes, we consider small deformations of the
polarization functions from (\ref{taufluxdef1}) (we can use $\kappa $ as a
small parameter and/or consider any other one, $\epsilon ,$ even for
associative and commutative deformations). The $\eta $-polarizations are
transformed into small $\chi $-polarizations with some nonholonomic
modifications by some families of $\zeta $-functions when $\ _{s}^{\shortmid
}\eta (\tau )\ ^{\shortmid }\check{g}_{\alpha _{s}}(\tau )\sim \ ^{\shortmid
}\zeta _{\alpha _{s}}(\tau) (1+\kappa \ ^{\shortmid }\chi _{\alpha
_{s}}(\tau ))\ \ ^{\shortmid }\check{g}_{\alpha _{s}}(\tau ).$ Introducing
such values into respective s-metrics (\ref{doublwnonassocwh}), we construct 
$\tau $-families of small wormhole deformations. We provide respective
formulas in Appendix, see formulas (\ref{whpolf1}).

We can model elliptic wormhole deformations as a particular case of
s-metrics of type (\ref{whpolf1}) if we chose generating functions of type%
\begin{eqnarray}
\chi _{4}(\tau ,l,\theta ,\varphi ) &=&\underline{\chi }(\tau ,l,\theta)\sin
(\omega _{0}\varphi +\varphi _{0}),\ |\ \zeta _{4}(\tau ,l,\theta ,\varphi
)|\simeq 1;  \notag \\
\ ^{\shortmid }\chi ^{6}(\tau ,l,\theta ,\varphi ,t,p_{l},p_{\theta }) &=&0,
\ |\ ^{\shortmid }\zeta ^{6}(\tau ,l,\theta ,\varphi ,t,p_{l},p_{\theta
})|=1;  \label{elipsoid1} \\
\ ^{\shortmid }\chi ^{8}(\tau ,l,\theta ,\varphi ,t,p_{l},p_{\varphi
},p_{\theta },p_{\varphi }) &=&\ \ ^{\shortmid }\underline{\chi }(\tau
,l,\theta ,\varphi ,t,p_{l},p_{\varphi },p_{\theta })\sin (\omega
_{0}p_{\varphi }+p_{\varphi \lbrack 0]}),\ |\ ^{\shortmid }\zeta ^{8}(\tau
,l,\theta ,\varphi ,t,p_{l},p_{\theta },p_{\varphi })|\simeq 1;  \notag
\end{eqnarray}%
as for quasi-stationary cylindric configurations with $\varphi $-anisotropic
and/or deformations. Here we note that different $\tau $-families of
s-solutions of type (\ref{whpolf1}) can be constructed if we change the
order of angular coordinates in the primary and target d-metrics, $\theta
\leftrightarrow \varphi $ and/or $p_{\theta }\leftrightarrow p_{\varphi }.$
We omit details on such applications of the AFCDM which can be considered
for (non) associative/commutative wormholes, see \cite{vwh3,sv14a,vbv18} and
Appendix \ref{asstables}. Such phase space double wormholes relates 4-d BE
configurations (one on the base spacetime and another one in the cofiber)
with nonassociative solutions constructed and studied in \cite%
{partner03,lbdssv22,lbsvevv22}. The $\kappa $-parameter plays the role of
eccentricity of corresponding rotoid configurations being under $\tau $%
-evolution. For BHs and BEs derived for a fixed $\tau _{0}$, we can apply
the concept of Bekenstein-Hawking entropy \cite{bek1,bek2,haw1,haw2} and
elaborate on respective thermodynamic and quantum information theories as in 
\cite{wormh22,wormh19,wormh21,wormh17,wormh18,wormh13}. But such
constructions and respective interpretations are possible only for very
special classes of diagonal phase space wormhole configurations. If we
consider solutions with variable temperature like parameter $\tau ,$ certain
generic off-diagonal terms with locall anisotropy, with nonassociative star
product R-flux deformations etc., we have to elaborate on nonassociative
generalizations of the GIF and QGIF theories from \cite{sv20}. This is the
goal of the next two subsections.

\subsubsection{GIFs thermodynamic variables for nonassociative double 4-d
wormholes}

We can define $\kappa $-linear parametric decompositions of the
nonassociative generalized Perelman variables (\ref{nagthermodvalues}), see
also respective F- and W-functionals (\ref{naffunct}) and/or (\ref{nawfunct}%
), for $\tau $-evolving double wormhole configurations (\ref%
{doublwnonassocwh}). The computations are similar to those presented in \cite%
{partner03,lbdssv22,lbsvevv22} for double nonassociative BHs and BEs; and
simplify substantially for classes of solutions with nonlinear symmetries $\
_{s}^{\shortmid }\Im ^{\star }(\tau )\rightarrow \ _{s}^{\shortmid }\Lambda
^{\star }(\tau ).$ We omit such technical details for s-adapted coefficient
constructions because all formulas can be derived in geometric symbolic form
for respective volume functionals $\ ^{\shortmid }\delta \ ^{\shortmid }%
\mathcal{V}(\tau )$ in phase spaces, see the formula for the statistical
generating function $_{s}\widehat{\mathcal{Z}}^{\star }(\tau )$ (\ref{spf}).
Such volume forms depend on the type of prime and target s-metrics of
respective solutions, but all thermodynamic, GIF and QGIF variables can be
normalized in certain ways for physically viable nonholonomic configurations
subjected to obey a series of important inequalities considered in sections %
\ref{sec2} and \ref{sec3}.

The nonassociative geometric flow thermodynamic variables (\ref%
{nagthermodvalues}) for quasi-stationary $\tau $-parametric s-metrics (\ref%
{offdiagpolfr1}) elements are computed for such general data for the prime
s-metrics $\ _{s}^{\shortmid }\mathbf{\mathring{g},}\eta $-polarization
functions and $\tau $-running cosmological constants $\
_{s}^{\shortmid}\Lambda ^{\star }(\tau ),$%
\begin{eqnarray}
\ _{s}^{\shortmid }\widehat{\mathcal{W}}_{\kappa }^{\star }(\tau )
&=&\int\nolimits_{\tau ^{\prime }}^{\tau }\frac{d\tau }{(4\pi \tau )^{4}}%
\frac{\tau \lbrack \ _{1}\Lambda ^{\star }(\tau )+\ _{2}\Lambda ^{\star
}(\tau )+\ _{3}^{\shortmid }\Lambda ^{\star }(\tau )+\ _{4}^{\shortmid
}\Lambda ^{\star }(\tau )]^{2}-8}{\sqrt{|\ _{1}\Lambda ^{\star }(\tau )\
_{2}\Lambda ^{\star }(\tau )\ _{3}^{\shortmid }\Lambda ^{\star }(\tau )\
_{4}^{\shortmid }\Lambda ^{\star }(\tau )|}}\ _{\eta }^{\shortmid }\mathcal{%
\mathring{V}}_{\kappa }(\tau ),  \label{thvcannpd} \\
\ _{s}^{\shortmid }\widehat{\mathcal{Z}}_{\kappa }^{\star }(\tau ) &=& \exp
[ \int\nolimits_{\tau ^{\prime }}^{\tau }\frac{d\tau }{(2\pi \tau )^{4}}%
\frac{1}{\sqrt{|\ _{1}\Lambda ^{\star }(\tau )\ _{2}\Lambda ^{\star }(\tau
)\ _{3}^{\shortmid }\Lambda ^{\star }(\tau )\ _{4}^{\shortmid }\Lambda
^{\star }(\tau )|}}\ _{\eta }^{\shortmid }\mathcal{\mathring{V}}(\tau )], 
\notag \\
\ _{s}^{\shortmid }\widehat{\mathcal{E}}_{\kappa }^{\star }(\tau )
&=&-\int\nolimits_{\tau ^{\prime }}^{\tau }\frac{d\tau }{(4\pi )^{4}\tau ^{3}%
}\frac{\tau \lbrack \ _{1}\Lambda ^{\star }(\tau )+\ _{2}\Lambda ^{\star
}(\tau )+\ _{3}^{\shortmid }\Lambda ^{\star }(\tau )+\ _{4}^{\shortmid
}\Lambda ^{\star }(\tau )]-4}{\sqrt{|\ _{1}\Lambda ^{\star }(\tau )\
_{2}\Lambda ^{\star }(\tau )\ _{3}^{\shortmid }\Lambda ^{\star }(\tau )\
_{4}^{\shortmid }\Lambda ^{\star }(\tau )|}}\ _{\eta }^{\shortmid }\mathcal{%
\mathring{V}}_{\kappa }(\tau ),  \notag \\
\ _{s}^{\shortmid }\widehat{\mathcal{S}}_{\kappa }^{\star }(\tau )
&=&-\int\nolimits_{\tau ^{\prime }}^{\tau }\frac{d\tau }{(4\pi \tau )^{4}}%
\frac{\tau \lbrack \ _{1}\Lambda ^{\star }(\tau )+\ _{2}\Lambda ^{\star
}(\tau )+\ _{3}^{\shortmid }\Lambda ^{\star }(\tau )+\ _{4}^{\shortmid
}\Lambda ^{\star }(\tau )]-8}{\sqrt{|\ _{1}\Lambda ^{\star }(\tau )\
_{2}\Lambda ^{\star }(\tau )\ _{3}^{\shortmid }\Lambda ^{\star }(\tau )\
_{4}^{\shortmid }\Lambda ^{\star }(\tau )|}}\ _{\eta }^{\shortmid }\mathcal{%
\mathring{V}}_{\kappa }(\tau ).  \notag
\end{eqnarray}%
In these formulas, we use the running phase space volume functional 
\begin{equation}
\ _{\eta }^{\shortmid }\mathcal{\mathring{V}}_{\kappa }(\tau )=\int_{\
_{s}^{\shortmid }\widehat{\Xi }}\ ^{\shortmid }\delta \ _{\eta }^{\shortmid }%
\mathcal{V}_{\kappa }(\ _{s}^{\shortmid }\Im ^{\star }(\tau ),\ ~^{\shortmid
}\mathring{g}_{\alpha _{s}}),  \label{volumfpsp}
\end{equation}%
which can be computed in explicit forms if we prescribe certain classes of
generating $\eta $-functions, effective generating sources $\
_{s}^{\shortmid }\Im (\tau ),$ coefficients of a prime s-metric $\
^{\shortmid }\mathring{g}_{\alpha _{s}}$ and nonholonomic distributions
defined on a closed hyper-surface $\ _{s}^{\shortmid }\widehat{\Xi }\subset
\ _{s}^{\shortmid }\mathcal{M}^{\star }.$

Changing the nonassociative geometric data for s-metrics (\ref{offdiagpolfr1}%
) into respective ones for double wormhole $\tau $-running configurations $\
_{\check{\eta}}^{\shortmid }\mathbf{g}(\tau )$ (\ref{doublwnonassocwh}) with 
$\ _{1}\Lambda ^{\star }(\tau )= \ _{2}\Lambda ^{\star }(\tau )=\check{%
\Lambda}^{\star }(\tau )$ and $\ _{3}^{\shortmid }\Lambda ^{\star }(\tau )=
\ _{4}^{\shortmid }\Lambda ^{\star }(\tau )=\ ^{\shortmid }\check{\Lambda}%
^{\star }(\tau )$, we obtain: 
\begin{eqnarray}
\ _{\check{\eta}}^{\shortmid }\widehat{\mathcal{Z}}_{\kappa }^{\star }(\tau)
&=& \exp [\int\nolimits_{\tau ^{\prime }}^{\tau }\frac{d\tau }{(2\pi \tau
)^{4}}\frac{1}{|\check{\Lambda}(\tau )\ ^{\shortmid }\check{\Lambda}(\tau )|}%
\ _{\check{\eta}}^{\shortmid }\mathcal{\check{V}}^{\star }(\tau )] ,
\label{statistgenf2wh} \\
\ _{\check{\eta}}^{\shortmid }\widehat{\mathcal{E}}_{\kappa }^{\star }(\tau
) &=&-\int\nolimits_{\tau ^{\prime }}^{\tau }\frac{d\tau }{128\pi ^{4}\tau
^{3}}\frac{\tau (\check{\Lambda}(\tau )+\ ^{\shortmid }\check{\Lambda}(\tau
))-2}{|\check{\Lambda}(\tau )\ ^{\shortmid }\check{\Lambda}(\tau )|}\ _{%
\check{\eta}}^{\shortmid }\mathcal{\check{V}}^{\star }(\tau ),
\label{energy2wh} \\
\ _{\check{\eta}}^{\shortmid }\widehat{\mathcal{S}}_{\kappa }^{\star }(\tau
) &=&-\int\nolimits_{\tau ^{\prime }}^{\tau }\frac{d\tau }{128(\pi \tau )^{4}%
}\frac{\tau (\check{\Lambda}(\tau )+\ ^{\shortmid }\check{\Lambda}(\tau ))-4%
}{|\check{\Lambda}(\tau )\ ^{\shortmid }\check{\Lambda}(\tau )|}\ _{\check{%
\eta}}^{\shortmid }\mathcal{\check{V}}^{\star }(\tau ).  \label{perentr2wh}
\end{eqnarray}%
Such thermodynamic variables values are determined by the volume functionals 
\begin{equation}
\ _{\check{\eta}}^{\shortmid }\mathcal{\check{V}}^{\star }(\tau )=\int_{\
_{s}^{\shortmid }\widehat{\Xi }}\ ^{\shortmid }\delta \ _{\check{\eta}%
}^{\shortmid }\mathcal{V}(\ _{s}^{\shortmid }\Im ^{\star }(\tau ),\ \check{g}%
_{\alpha _{s}}).  \label{volumfunct2wh}
\end{equation}%
Such functionals include primary data $\check{g}_{\alpha _{s}}$ (\ref{pm1a})
for two 4-d phase space wormhole configurations being $\eta $-deformed by
respective $\tau$-running nonassociative data. We use the symbol $\check{\eta%
}$ to emphasize that corresponding nonassociative geometric/ thermodynamic
objects are $\eta $-generated from some primary data $\check{g}_{\alpha
_{s}} $ and $\check{N}_{i_{s-1}}^{a_{s}}(\tau )$ as in (\ref{taufluxdef1}).
The thermodynamic variables $[ \ _{\check{\eta}}^{\shortmid }\widehat{%
\mathcal{Z}}_{\kappa }^{\star }(\tau ),\ _{\check{\eta}}^{\shortmid }%
\widehat{\mathcal{E}}_{\kappa }^{\star }(\tau ),\ _{\check{\eta}}^{\shortmid
}\widehat{\mathcal{S}}_{\kappa }^{\star }(\tau )] $ are well-defined for $%
\tau $-running nonholonomic configurations with respective generating
functions, generating sources and for certain classes of integration
functions. Formulas (\ref{statistgenf2wh}), (\ref{energy2wh}) and (\ref%
{perentr2wh}) can be computed also in terms of $\chi $-generating functions
if we consider small parametric quasi-stationary solutions of type (\ref%
{whpolf1} and write in geometric symbolic form $[\ _{\check{\chi}%
}^{\shortmid }\widehat{\mathcal{Z}}_{\kappa }^{\star }(\tau ),\ _{\check{\chi%
}}^{\shortmid }\widehat{\mathcal{E}}_{\kappa }^{\star }(\tau ), \ _{\check{%
\chi}}^{\shortmid }\widehat{\mathcal{S}}_{\kappa }^{\star }(\tau )].$ Both
for the $\eta $- or $\chi $-generating functions, or other type ones, the
functionals $\ _{s}^{\shortmid }\widehat{\mathcal{E}}_{\kappa}^{\star }(\tau
)$ and $\ _{s}^{\shortmid }\widehat{\mathcal{S}}_{\kappa}^{\star }(\tau )$
can be treated as respective effective energy and entropy flow transports in
a phase space media evolving in the interval $\tau ^{\prime }<\tau .$ Only
for certain special 4-d BE configurations of type (\ref{elipsoid1}) with
fixed $\tau _{0}$ we can compute hypersurface values like in
Bekenstein-Hawking thermodynamics. For more general classes of solutions,
the (nonassociative) wormhole deformations are characterized by modified
Perelman thermodynamic models.

The Shannon--Perelman entropy (\ref{shanonper}) for nonassociative
quasi-stationary double wormhole $\tau $-running solutions $\ _{\check{\eta}%
}^{\shortmid }\mathbf{g}(\tau )$ (\ref{doublwnonassocwh}) can be associated
respectively with a probability distribution for such a random variable $\
_{s}^{\star }A$ on $\ _{s}^{\star }\mathcal{M},$ when%
\begin{eqnarray*}
\ _{\check{\eta}}^{\shortmid }\widehat{S}_{A}^{\star } &\simeq
&\sum\nolimits_{z=1}^{z=r} \ _{\check{\eta}}^{\shortmid }\widehat{\mathcal{S}%
}_{\kappa }^{\star }(\tau _{z}), 
\mbox{  nonassociative Ricci flows for a
set of discrete values of }\tau ; \\
&=&\ _{\check{\eta}}^{\shortmid }\widehat{\mathcal{S}}_{\kappa
}^{\star}(\tau _{0}), 
\mbox{ for a solution of nonassociative Ricci soliton/
Einstein eqs. }; \\
&\simeq &\int\nolimits_{\tau _{r}}^{\tau _{r}}d\tau \ \ _{\check{\eta}%
}^{\shortmid }\widehat{\mathcal{S}}_{\kappa }^{\star }(\tau ),%
\mbox{nonasociative geometric flows with  continuous running of }\tau .
\end{eqnarray*}%
For explicit computations, we can consider $\ _{\check{\eta}}^{\shortmid }%
\widehat{\mathcal{S}}_{\kappa }^{\star }(\tau )$ (\ref{perentr2wh}), or a
variant as $\ _{\check{\chi}}^{\shortmid }\widehat{\mathcal{S}}_{\kappa
}^{\star }(\tau ).$ Such nonassociative GIF models are physical if the
corresponding $\ ^{\shortmid }\widehat{S}_{A}^{\star }\simeq
\sum\nolimits_{z=1}^{z=r}\ \ _{\check{\eta}}^{\shortmid }\widehat{\mathcal{S}%
}_{\kappa }^{\star }(\tau _{z})\geq 0.$ Similar conditions on the
nonholonomic structure of such quasi-stationary configurations are imposed
to result in inequalities $\ _{\check{\eta}}^{\shortmid }\widehat{\mathcal{S}%
}_{\kappa }^{\star }(\tau _{0})$ $\geq 0,$ or $\int\nolimits_{\tau
_{r}}^{\tau _{r}}d\tau \ \ _{\check{\eta}}^{\shortmid }\widehat{\mathcal{S}}%
_{\kappa }^{\star }(\tau )\geq 0.$

Let us sketch how we can define and compute continuous analogs of (\ref%
{suadditivgif}) for nonassociative GIFs of double wormhole solutions $\ _{%
\check{\eta}}^{\shortmid }\mathbf{g}(\tau )$. We introduce the thermodynamic
generating function (as a generalization of the partition function (\ref%
{statistgenf2wh}) on $\ _{s}^{\shortmid }\mathcal{M}^{\star }\otimes \
_{s}^{\shortmid }\mathcal{M}^{\star }$) 
\begin{equation*}
\ _{AB}^{\shortmid }\widehat{\mathcal{Z}}[\ _{\check{\eta}}^{\shortmid }%
\mathbf{g}(\tau ),\ _{1\check{\eta}}^{\shortmid }\mathbf{g}\ (\tau
)]=\int_{\ _{s}^{\shortmid }\widehat{\Xi }}\int_{\ _{1s}^{\shortmid }%
\widehat{\Xi }}(4\pi \tau )^{-8}e^{-\ _{AB}\widehat{f}}\sqrt{|\ _{\check{\eta%
}}^{\shortmid }\mathbf{g}(\tau )|}\sqrt{|\ _{1\check{\eta}}^{\shortmid }%
\mathbf{g}(\tau )|}d^{8}\ ^{\shortmid }u\ d^{8}\ _{1}^{\shortmid }u(-\ _{AB}(%
\hat{f}+8).
\end{equation*}%
The corresponding generalization for the nonassociative GIFs of NES systems
of the canonical thermodynamic entropy function (\ref{perentr2wh}) can be
computed using formulas 
\begin{eqnarray*}
\ _{AB}^{\shortmid }\widehat{\mathcal{S}}_{\check{\eta}}^{\star }(\tau )=\
^{\shortmid }\widehat{\mathcal{S}}_{\check{\eta}}^{\star }\ [\ _{s}^{\star }%
\widehat{A},\ _{s}^{\star }\widehat{B}](\tau )&=&-\int_{\ _{s}^{\shortmid }%
\widehat{\Xi }}\int_{\ _{s1}^{\shortmid }\widehat{\Xi }}(4\pi \tau )^{-8}\
e^{-\ _{AB}\widehat{f}}\sqrt{|\ _{\check{\eta}}^{\shortmid }\mathbf{g}(\tau
)|}\sqrt{|\ _{1\check{\eta}}^{\shortmid }\mathbf{g}(\tau )|}d^{8}\
^{\shortmid }u\ d^{8}\ _{1}^{\shortmid }u \\
&& [\tau (\ _{s}^{\shortmid }\widehat{R}_{\check{\eta}}^{\star }(\tau )+\
_{s1}^{\shortmid }\widehat{R}_{\check{\eta}}^{\star }(\tau )+|\
_{AB}^{\shortmid }\widehat{\mathbf{D}}^{\star }\widehat{f}|^{2})+\ _{AB}%
\widehat{f}-16 ] .
\end{eqnarray*}%
In such formulas, respective scalar s-curvatures can be transformed into $%
\tau $-running cosmological constants $\check{\Lambda}^{\star }(\tau )$ and $%
\ ^{\shortmid }\check{\Lambda}^{\star }(\tau ).$ Using such computations for 
$\ _{\check{\eta}}^{\shortmid }\mathbf{g}(\tau )$ and above thermodynamic
information variables, we claim that the formulas for the conditional
entropy, (\ref{conde}) and (\ref{condes}), and mutual information, (\ref%
{minf}) and (\ref{minfs}), are respectively generalized for continuous GIFs
of nonassociative $\tau $-running quasi-stationary deformations of wormhole
solutions, 
\begin{eqnarray*}
\ ^{\shortmid }\widehat{\mathcal{S}}_{\check{\eta}}^{\star }\ [\ _{s}^{\star
}\widehat{A}|\ _{s}^{\star }\widehat{B}] &:=&\ _{AB}^{\shortmid }\widehat{%
\mathcal{S}}_{\check{\eta}}^{\star }-\ ^{\shortmid }\widehat{\mathcal{S}}_{%
\check{\eta}}^{\star }[\ _{s}^{\star }\widehat{B}]\geq 0\mbox{ and }\  \\
\ ^{\shortmid }\widehat{\mathcal{J}}_{\check{\eta}}^{\star }\ [\ _{s}^{\star
}\widehat{A};\ _{s}^{\star }\widehat{B}] &:=&\ ^{\shortmid }\widehat{%
\mathcal{S}}_{\check{\eta}}^{\star }[\ _{s}^{\star }\widehat{A}]-\
_{AB}^{\shortmid }\widehat{\mathcal{S}}_{\check{\eta}}^{\star }+\
^{\shortmid }\widehat{\mathcal{S}}_{\check{\eta}}^{\star }[\ _{s}^{\star }%
\widehat{B}]\geq 0.
\end{eqnarray*}%
It is not possible to prove that such conditions are satisfied for arbitrary 
$\eta $-polarizations, general classes of integration functions and
arbitrary data for some effective $\ _{1}^{\shortmid }\Im ^{\star }$ and $%
\check{\Lambda}^{\star }(\tau )$ and $^{\shortmid }\check{\Lambda}%
^{\star}(\tau ).$ We claim that prescribing such physical important
inequalities for double wormhole GIFs we impose certain criteria of
viability on a corresponding class of nonassociative deformations and
related nonlinear symmetries. Such conditions can be always proven for any
point along causal curves in $\ _{s}^{\star }\mathcal{M}$ if we consider
parametric solutions with small with $\chi $-polarizations (of type (\ref%
{whpolf1}). This approach does not results in non-physical/ -causal models
if, at least, we consider $\kappa $-parametric data $[\ _{\check{\chi}%
}^{\shortmid}\widehat{\mathcal{Z}}_{\kappa }^{\star }(\tau ), \ _{\check{\chi%
}}^{\shortmid}\widehat{\mathcal{E}}_{\kappa }^{\star }(\tau ), \ _{\check{%
\chi}}^{\shortmid }\widehat{\mathcal{S}}_{\kappa }^{\star }(\tau )] $ and
their two information systems $AB,$ or three systems $ABC,$ as for formulas (%
\ref{threegenf}) and (\ref{threeentr}).

Finally, we note that above formulas can be defined and computed for more
special classes of solutions with classical geometric flow entanglement of a
4-d wormhole on the base manifold and a 4-d wormhole in co-fiber space (with
momentum like variables). Such solutions can be constructed to be
independent as primary metrics but $\tau $-evolution and/or nonassociative
star product R-flux deformations result in generic off-diagonal terms on $\
_{s}^{\star }\mathcal{M}$ then above star labeled GIF variables are not
trivial.

\subsubsection{QGIFs entanglement entropy and Bell states for nonassociative
double 4-d wormholes}

Nonassociative QGIF models can be elaborated in quasi-stationary form for
thermo-geometric double wormhole states on nonassociative phase space $\
_{s}^{\star }\mathcal{M}.$ The abstract geometric and quantum formulas from
subsection \ref{ssthgdsnonassoc} can be computed in explicit forms for
s-metrics $\ _{\check{\eta}}^{\shortmid }\mathbf{g}(\tau )$ (\ref%
{doublwnonassocwh}). Using energy functionals (\ref{energy2wh}), we define a
discrete model with effective energy $\ _{\check{\eta}}^{\shortmid }\widehat{%
\mathcal{E}}_{\underline{a}}^{\star }:=\ _{\check{\eta}}^{\shortmid }%
\widehat{\mathcal{E}}_{\kappa }^{\star }(\tau _{\underline{a}}),$ for $%
\underline{a}=1,2,...,\check{r}.$ Such thermodynamic variables are derived
for the canonical partition function $\ _{\check{\eta}}^{\shortmid }\widehat{%
\mathcal{Z}}_{\kappa }^{\star }= \sum_{\underline{a}}\ _{\check{\eta}%
}^{\shortmid }\widehat{\mathcal{Z}}_{\kappa }^{\star }(\tau _{\underline{a}%
}),$ when (\ref{statistgenf2wh}) is considering for discrete $\tau _{%
\underline{a}}.$ For any data $(\ _{\check{\eta}}^{\shortmid }\widehat{%
\mathcal{Z}}_{\kappa }^{\star }, \ _{\check{\eta}}^{\shortmid }\widehat{%
\mathcal{E}}_{\underline{a}}^{\star }),$ we can define canonical quantum
variables (\ref{dmq1}), 
\begin{eqnarray}
|\widehat{\Psi }_{\check{\eta}}^{\star }> &= &(\ _{\check{\eta}}^{\shortmid}%
\widehat{\mathcal{Z}}_{\kappa }^{\star })^{-1/2}\sum_{\underline{a}}e^{-\ _{%
\check{\eta}}^{\shortmid }\widehat{\mathcal{E}}_{\underline{a}}^{\star
}/2\tau }|\psi _{\mathcal{A}}^{\underline{a}}>\otimes |\psi _{\mathcal{B}}^{%
\underline{a}}>,  \label{dmq1wh2} \\
\ _{\check{\eta}}^{\shortmid }\widehat{\rho }_{\mathcal{A}}^{\star } &=&(\ _{%
\check{\eta}}^{\shortmid }\widehat{\mathcal{Z}}_{\kappa }^{\star
})^{-1}\sum_{\underline{a}}e^{-\ _{\check{\eta}}^{\shortmid }\widehat{%
\mathcal{E}}_{\underline{a}}^{\star }/\tau }|\widehat{\psi }_{\mathcal{A}}^{%
\underline{a}}>\otimes <\widehat{\psi }_{\mathcal{A}}^{\underline{a}}|=(\ _{%
\check{\eta}}^{\shortmid }\widehat{\mathcal{Z}}_{\kappa }^{\star
})^{-1}e^{-\ _{\check{\eta}}^{\shortmid }\widehat{\boldsymbol{H}}_{\mathcal{A%
}}^{\star }/\tau },  \notag
\end{eqnarray}%
where $\ _{\check{\eta}}^{\shortmid }\widehat{\boldsymbol{H}}_{\mathcal{A}%
}^{\star }|\widehat{\psi }_{\mathcal{A}}^{\underline{a}}>= \ _{\check{\eta}%
}^{\shortmid }\widehat{\mathcal{E}}_{\underline{a}}^{\star }|\widehat{\psi }%
_{\mathcal{A}}^{\underline{a}}>.$ The thermodynamic and quantum s-objects in
(\ref{dmq1wh2}) contain additional abstract labels "$\check{\eta}$ " and "$%
\star $" stating that the constructions are for nonassociative double
wormhole configurations with $\check{\eta}$-generating functions. Explicit
computations of such variables can be performed for $\kappa $-linear
decompositions when, for $\chi $-generating functions of type (\ref%
{elipsoid1}), we can consider $\ _{\chi }^{\shortmid }\widehat{\boldsymbol{H}%
}_{\mathcal{A}}^{\star }|\widehat{\psi }_{\mathcal{A}}^{\underline{a}}>= \
_{\chi }^{\shortmid }\widehat{\mathcal{E}}_{\underline{a}}^{\star }|\widehat{%
\psi }_{\mathcal{A}}^{\underline{a}}>$.

The density matrix $\ _{\check{\eta}}^{\shortmid }\widehat{\rho }_{\mathcal{A%
}}^{\star }$ can be used for introducing a measure of both nonassociative
GIF and QM entanglement. The von Neumann--Perelman entropy can be defined as
the entanglement canonical entropy of a nonassociative QGIF system $\mathcal{%
A}=\{\ _{\check{\eta}}^{\star }A\}=\mathcal{A}^{\star },$ when 
\begin{equation}
\ _{\check{\eta}}^{\shortmid }\widehat{\mathcal{S}}_{\mathcal{A}}^{\star }\
=-tr_{\mathcal{A}}[\ _{\check{\eta}}^{\shortmid }\widehat{\rho }_{\mathcal{A}%
}^{\star }\log \ _{\check{\eta}}^{\shortmid }\widehat{\rho }_{\mathcal{A}%
}^{\star }].  \label{entanglentrcanwh1}
\end{equation}%
A quantum system $\mathcal{A}$ associated to a corresponding QGIF system can
be purified as a thermal system in the extended Hilbert space $\mathcal{H}_{%
\mathcal{A}}\otimes \mathcal{H}_{\mathcal{B}}$ as we described above in (\ref%
{purific}). For such configurations, every expectation value for a local
operator in $\mathcal{A}$ can be represented in a thermo-geometric double
state (\ref{dmq1wh2}) of the total system $\mathcal{A}\cup \mathcal{B}$.
This means that for the nonassociative geometric evolution on double
wormhole configurations the nonassociative QGIF entanglement entropy
measures the thermodynamic entropy both as a nonassociative GIF model and as
a quantum subsystem $\mathcal{A}$. The formula (\ref{entanglentrcanwh1}) can
be written in different forms using corresponding nonassociative quantum
information thermodynamic relations, when 
\begin{equation*}
\ _{\check{\eta}}^{\shortmid }\widehat{\mathcal{S}}_{\mathcal{A}}^{\star }\
=-tr_{\mathcal{A}}[\ _{\check{\eta}}^{\shortmid }\widehat{\rho }_{\mathcal{A}%
}(-\ _{\check{\eta}}^{\shortmid }\widehat{\mathcal{E}}^{\star }/\tau -\log \
\ _{\check{\eta}}^{\shortmid }\widehat{\mathcal{Z}}_{\kappa }^{\star })]=(<\
_{\check{\eta}}^{\shortmid }\widehat{\boldsymbol{H}}_{\mathcal{A}}^{\star
}>-\ _{\check{\eta}}^{\shortmid }\widehat{\mathcal{F}}_{\mathcal{A}}^{\star
})/\tau ,
\end{equation*}%
for the canonical thermal free energy $\ _{\check{\eta}}^{\shortmid }%
\widehat{\mathcal{F}}_{\mathcal{A}}^{\star }:= \tau \log \ \ _{\check{\eta}%
}^{\shortmid }\widehat{\mathcal{Z}}_{\kappa }^{\star }.$ Explicit
computations of such values are possible in terms of $\chi $-polarizations
of prime double wormhole solutions. For double 4-d BE configurations of type
(\ref{elipsoid1}) with fixed $\tau _{0},$ such formulas can be also defined
and computed. Nevertheless, this information geometric thermodynamic
paradigm is different form that considered in \cite%
{wormh22,wormh19,wormh21,wormh17,wormh18,wormh13} for the case of wormhole
solutions in GR and higher dimension extensions with diagonalizable metrics.
In our approach, the formulas encode nonassociative geometric flow data and
can be applied for locally anisotropic deformations of wormhole solutions
and $\tau $-evolution on $_{s}^{\star }\mathcal{M}$.

\subsubsection{Entangled nonassociative QGIFs and Bell states for double 4-d
wormhole solutions}

For $\tau $-evolving double wormhole configurations (\ref{doublwnonassocwh})
in phase spaces, we associate two nonassociative QGIF systems $\mathcal{A}%
=\{\ _{s}^{\star}A\}=\mathcal{A}^{\star }$ and $\mathcal{B}=\{\ _{s}^{\star
}B\}=\mathcal{B}^{\star }$ and consider respective spanned Hilbert spaces $%
\mathcal{H}_{\mathcal{A}}$ and $\mathcal{H}_{\mathcal{B}},$ when $\mathcal{H}%
_{\mathcal{A},\mathcal{B}}=\{|\hat{1}>_{\mathcal{A},\mathcal{B}},|\hat{2}>_{%
\mathcal{A},\mathcal{B}}\}.$

Let us consider explicit examples of ground state and reduced density matrix
for a $\mathcal{A}$ obtained by taking the partial trace over $\mathcal{H}_{%
\mathcal{B}}$ as in (\ref{puredm}):%
\begin{eqnarray}
|\widehat{\Psi }_{\check{\eta}}^{\star }> &=&\frac{e^{-(\ _{\check{\eta}%
}^{\shortmid }\widehat{\mathcal{E}}_{\underline{1}}^{\star }+\ _{\check{\eta}%
}^{\shortmid }\widehat{\mathcal{E}}_{\underline{2}}^{\star })/2\tau }}{\sqrt{%
2}\ (\ _{\check{\eta}}^{\shortmid }\widehat{\mathcal{Z}}_{\underline{1}%
\kappa }^{\star })^{1/2}(\ _{\check{\eta}}^{\shortmid }\widehat{\mathcal{Z}}%
_{\underline{2}\kappa }^{\star })^{1/2}}(|\hat{1}\hat{2}>-|\hat{2}\hat{1}>)%
\mbox{ and }  \label{natwowhqs} \\
\ _{\check{\eta}}^{\shortmid }\widehat{\rho }_{\mathcal{A}}^{\star } &=&%
\frac{1}{2}(\frac{e^{-\ _{\check{\eta}}^{\shortmid }\widehat{\mathcal{E}}_{%
\underline{1}}^{\star }/\tau }}{\ _{\check{\eta}}^{\shortmid }\widehat{%
\mathcal{Z}}_{\underline{1}\kappa }^{\star }}|\hat{1}>_{\mathcal{A}}\ _{%
\mathcal{A}}<\hat{1}|+\frac{e^{-\ _{\check{\eta}}^{\shortmid }\widehat{%
\mathcal{E}}_{\underline{2}}^{\star }/\tau }}{\ _{\check{\eta}}^{\shortmid }%
\widehat{\mathcal{Z}}_{\underline{2}\kappa }^{\star }}|\hat{2}>_{\mathcal{A}%
}\ _{\mathcal{A}}<\hat{2}|)=\left( 
\begin{array}{cc}
e^{-\ _{\check{\eta}}^{\shortmid }\widehat{\mathcal{E}}_{\underline{1}%
}^{\star }/\tau }/2\ \ _{\check{\eta}}^{\shortmid }\widehat{\mathcal{Z}}_{%
\underline{1}\kappa }^{\star } & 0 \\ 
0 & e^{-\ _{\check{\eta}}^{\shortmid }\widehat{\mathcal{E}}_{\underline{2}%
}^{\star }/\tau }/2\ \ _{\check{\eta}}^{\shortmid }\widehat{\mathcal{Z}}_{%
\underline{2}\kappa }^{\star }%
\end{array}%
\right).  \notag
\end{eqnarray}%
In these formulas, $\ _{\check{\eta}}^{\shortmid }\widehat{\mathcal{Z}}_{%
\underline{1}\kappa }^{\star }= \ _{s}^{\shortmid }\widehat{Z}[\ _{\check{%
\eta}}^{\shortmid }\mathbf{g}_{\underline{1}}(\tau )]$ and $\ _{\check{\eta}%
}^{\shortmid }\widehat{\mathcal{Z}}_{\underline{2}\kappa }^{\star }=\
_{s}^{\shortmid }\widehat{Z}[ \ _{\check{\eta}}^{\shortmid }\mathbf{g}_{%
\underline{2}}(\tau )]$, as in see formula (\ref{spf}), where $\ _{\check{%
\eta} }^{\shortmid }\mathbf{g}_{\underline{1}}(\tau )$ and $\ _{\check{\eta}%
}^{\shortmid }\mathbf{g}_{\underline{2}}(\tau )$ are quasi-stationary
s-metrics of type (\ref{doublwnonassocwh}). Such parametric solutions may be
chosen to define two different classes of nonassociative double wormhole
d-metrics: We can use the same type of h- and c-metrics but for different
integration functions; or as a nonassociative QGIF model consisting from one
4-d wormhole spacetime $\tau $-running configuration and another 4-d
momentum space $\tau $-running configuration. Using the formula (\ref%
{entanglentrcanwh1}) for $\ _{\check{\eta}}^{\shortmid }\widehat{\rho }_{%
\mathcal{A}}^{\star }$ from (\ref{entanglentrcanwh1}), we can define and
compute 
\begin{equation*}
\ _{\check{\eta}}^{\shortmid }\widehat{\mathcal{S}}_{\mathcal{A}}^{\star }\
=-tr_{\mathcal{A}}\left[ \left( 
\begin{array}{cc}
e^{-\ _{\check{\eta}}^{\shortmid }\widehat{\mathcal{E}}_{\underline{1}%
}^{\star }/\tau }/2\ \ _{\check{\eta}}^{\shortmid }\widehat{\mathcal{Z}}_{%
\underline{1}\kappa }^{\star } & 0 \\ 
0 & e^{-\ _{\check{\eta}}^{\shortmid }\widehat{\mathcal{E}}_{\underline{2}%
}^{\star }/\tau }/2\ \ _{\check{\eta}}^{\shortmid }\widehat{\mathcal{Z}}_{%
\underline{2}\kappa }^{\star }%
\end{array}%
\right) \left( 
\begin{array}{cc}
\log [e^{-\ _{\check{\eta}}^{\shortmid }\widehat{\mathcal{E}}_{\underline{1}%
}^{\star }/\tau }/2\ \ _{\check{\eta}}^{\shortmid }\widehat{\mathcal{Z}}_{%
\underline{1}\kappa }^{\star }] & 0 \\ 
0 & \log [e^{-\ _{\check{\eta}}^{\shortmid }\widehat{\mathcal{E}}_{%
\underline{2}}^{\star }/\tau }/2\ \ _{\check{\eta}}^{\shortmid }\widehat{%
\mathcal{Z}}_{\underline{2}\kappa }^{\star }]%
\end{array}%
\right) \right] .
\end{equation*}

For couples of wormholes, we can study examples of Bell type of QGIF
entanglement and nonassociative canonical qubit systems of type (\ref%
{qgifbell}). Such nonassociative two qubit systems can be defined by totally
four independent maximally entangled quasi-stationary QGIFs:%
\begin{eqnarray}
|\ _{\check{\eta}}^{\shortmid }\widehat{B}_{1}^{\star } > &=& \frac{1}{\sqrt{%
2}\ }(\frac{e^{-\ _{\check{\eta}}^{\shortmid }\widehat{\mathcal{E}}_{%
\underline{1}}^{\star }/\tau }}{\ \ _{\check{\eta}}^{\shortmid }\widehat{%
\mathcal{Z}}_{\underline{1}\kappa }^{\star }}|\hat{1}\hat{1}>+\frac{e^{-\ _{%
\check{\eta}}^{\shortmid }\widehat{\mathcal{E}}_{\underline{2}}^{\star
}/\tau }}{\ \ \ _{\check{\eta}}^{\shortmid }\widehat{\mathcal{Z}}_{%
\underline{2}\kappa }^{\star }}|\hat{2}\hat{2}>),\ |\ _{\check{\eta}%
}^{\shortmid }\widehat{B}_{2}^{\star }>=\frac{1}{\sqrt{2}\ }(\frac{e^{-\ _{%
\check{\eta}}^{\shortmid }\widehat{\mathcal{E}}_{\underline{1}}^{\star
}/\tau }}{\ \ _{\check{\eta}}^{\shortmid }\widehat{\mathcal{Z}}_{\underline{1%
}\kappa }^{\star }}|\hat{1}\hat{1}>-\frac{e^{-\ _{\check{\eta}}^{\shortmid }%
\widehat{\mathcal{E}}_{\underline{2}}^{\star }/\tau }}{\ \ \ _{\check{\eta}%
}^{\shortmid }\widehat{\mathcal{Z}}_{\underline{2}\kappa }^{\star }}|\hat{2}%
\hat{2}>),  \label{qgifbelldwmh} \\
|\ _{\check{\eta}}^{\shortmid }\widehat{B}_{3}^{\star } >&=& \frac{e^{-(\ _{%
\check{\eta}}^{\shortmid }\widehat{\mathcal{E}}_{\underline{1}}^{\star }+\ _{%
\check{\eta}}^{\shortmid }\widehat{\mathcal{E}}_{\underline{2}}^{\star
})/2\tau }}{\sqrt{2}\ (\ _{\check{\eta}}^{\shortmid }\widehat{\mathcal{Z}}_{%
\underline{1}\kappa }^{\star })^{1/2}\ (\ _{\check{\eta}}^{\shortmid }%
\widehat{\mathcal{Z}}_{\underline{2}\kappa }^{\star })^{1/2}}(|\hat{1}\hat{2}%
>+|\hat{2}\hat{1}>),\ |\ _{\check{\eta}}^{\shortmid }\widehat{B}_{4}^{\star
}>=\frac{e^{-(\ _{\check{\eta}}^{\shortmid }\widehat{\mathcal{E}}_{%
\underline{1}}^{\star }+\ _{\check{\eta}}^{\shortmid }\widehat{\mathcal{E}}_{%
\underline{2}}^{\star })/2\tau }}{\sqrt{2}(\ _{\check{\eta}}^{\shortmid }%
\widehat{\mathcal{Z}}_{\underline{1}\kappa }^{\star })^{1/2}\ (\ _{\check{%
\eta}}^{\shortmid }\widehat{\mathcal{Z}}_{\underline{2}\kappa }^{\star
})^{1/2}}(|\hat{1}\hat{2}>-|\hat{2}\hat{1}>).  \notag
\end{eqnarray}%
The nonassociative QGIF systems (\ref{qgifbelldwmh}) provide examples of
wormhole nonholonomic deformations of Bell states (Einstein-Podolsky-Rosen
pairs). As quantum systems such states manifest QM aspects in the sense of
nonassociative geometric flows and star product and R-flux violations of the
Bell's inequalities. We can model on quantum computers and in lab
corresponding conditions for nonassociative modifications of wormhole
configurations generalizing the approaches elaborated in \cite%
{wormh22,wormh19,wormh21,wormh17,wormh18,wormh13}. Using generalized
Perelman quantum thermodynamic variables, we can study nonassociative
geometric flows of wormhole data encoded in $\ _{\check{\eta}}^{\shortmid }%
\widehat{\mathcal{E}}_{\underline{1}}^{\star },\ _{\check{\eta}}^{\shortmid }%
\widehat{\mathcal{E}}_{\underline{2}}^{\star }$ and $\ _{\check{\eta}%
}^{\shortmid }\widehat{\mathcal{Z}}_{\underline{1}\kappa }^{\star },\ _{%
\check{\eta}}^{\shortmid } \widehat{\mathcal{Z}}_{\underline{2}\kappa
}^{\star }.$

\subsubsection{Relative and R\'{e}nyi entropy for nonassociative QGIFs of
double 4-d wormholes}

The \textit{relative entropy} of two nonassociative QGIFs (\ref{relativentr}%
) can be defined if we consider on nonassociative phase space two wormhole
density matrices $\ _{\check{\eta}}^{\shortmid }\widehat{\rho }_{\mathcal{A}%
}^{\star }$ and $\ _{\check{\eta}}^{\shortmid }\widehat{\sigma }_{\mathcal{A}%
}^{\star }$ of type (\ref{dmq1wh2}) and/or (\ref{natwowhqs}). All
computations can be performed in $\kappa $-parametric form for canonical
nonholonomic variables using formula 
\begin{equation*}
\ \ _{\check{\eta}}^{\shortmid }\widehat{\mathcal{S}}^{\star }(\ \ _{\check{%
\eta}}^{\shortmid }\widehat{\rho }_{\mathcal{A}}^{\star }\shortparallel \ _{%
\check{\eta}}^{\shortmid }\widehat{\sigma }_{\mathcal{A}}^{\star })=Tr_{%
\mathcal{H}_{\mathcal{B}}}[\ \ _{\check{\eta}}^{\shortmid }\widehat{\rho }_{%
\mathcal{A}}^{\star }(\log \ \ _{\check{\eta}}^{\shortmid }\widehat{\rho }_{%
\mathcal{A}}^{\star }-\log \ \ _{\check{\eta}}^{\shortmid }\widehat{\sigma }%
_{\mathcal{A}}^{\star })],
\end{equation*}%
where $\ _{\check{\eta}}^{\shortmid }\widehat{\mathcal{S}}^{\star }(\ _{%
\check{\eta}}^{\shortmid }\widehat{\rho }_{\mathcal{A}}^{\star
}\shortparallel \ _{\check{\eta}}^{\shortmid }\widehat{\rho }_{\mathcal{A}%
}^{\star })=0.$ We can define a measure of "distance" between two
nonassociative QGIFs of wormhole configuration, or nonholonomic Ricci
solitons for wormholes, using the relative entropy. This allows us to define
the norm $|| \ _{\check{\eta}}^{\shortmid }\widehat{\rho }_{\mathcal{A}%
}^{\star }||=tr(\sqrt{(\ _{\check{\eta}}^{\shortmid }\widehat{\rho }_{%
\mathcal{A}}^{\star })(\ _{\check{\eta}}^{\shortmid }\widehat{\rho }_{%
\mathcal{A}}^{\star })}).$ All formulas, conditions and inequalities claimed
in subsection \ref{ssrelentrna} can be considered for such quasi-stationary
models. We can also study the mutual information of two wormholes by
introducing and computing $\ _{\check{\eta}}^{\shortmid }\widehat{\mathcal{J}%
}^{\star }(\widehat{\mathcal{A}},\widehat{\mathcal{B}})= \ _{\check{\eta}%
}^{\shortmid }\widehat{\mathcal{S}}^{\star }(\ _{\check{\eta}}^{\shortmid }%
\widehat{\rho }_{\mathcal{A\cup B}}^{\star }\shortparallel \ _{\check{\eta}%
}^{\shortmid }\widehat{\rho }_{\mathcal{A}}^{\star }\otimes \ _{\check{\eta}%
}^{\shortmid }\widehat{\rho }_{\mathcal{B}}^{\star }).$

\vskip5pt

The concept of R\'{e}nyi entropy defined for nonassociative QGIFs in
subsection \ref{ssrenyientr} can be also applied to investigate physical
properties of double wormhole configurations being under nonassociative
geometric flow evolution on phase spaces. For such nonassociative
quasi-stationary configurations, we use a density matrix $\ _{\check{\eta}%
}^{\shortmid }\widehat{\rho }_{\mathcal{A}}^{\star }$ and compute the R\'{e}%
nyi entropy 
\begin{equation}
\ _{r}^{\shortmid }\widehat{\mathcal{S}}_{\check{\eta}}^{\star }(\widehat{%
\mathcal{A}}):=\frac{1}{1-r}\log [tr_{\mathcal{A}}(\ _{\check{\eta}%
}^{\shortmid }\widehat{\rho }_{\mathcal{A}}^{\star })^{r}],
\label{renentr2wmh}
\end{equation}%
with an integer replica parameter $r.$ The replica formalism can be
elaborated for an analytic continuation of $r$ to a real number with a
defined limit $\ _{\check{\eta}}^{\shortmid }\widehat{\mathcal{S}}_{\mathcal{%
A}}^{\star }(\ _{\check{\eta}}^{\shortmid }\widehat{\rho }_{\mathcal{A}%
}^{\star}) =\lim_{r\rightarrow 1}\ _{r}^{\shortmid }\widehat{\mathcal{S}}_{%
\check{\eta}}^{\star }(\widehat{\mathcal{A}})$ and normalization $tr_{%
\mathcal{A}}(\ _{\check{\eta}}^{\shortmid }\widehat{\rho }_{\mathcal{A}%
}^{\star })$ for $r\rightarrow 1.$ We assume that, for such conditions, the
formula (\ref{renentr2wmh}) reduces to the nonasociative entanglement
entropy $\ _{\check{\eta}}^{\shortmid }\widehat{\mathcal{S}}_{\mathcal{A}%
}^{\star }$ (\ref{entanglentrcanwh1}). Similarly to the double wormhole
versions of $\ _{r}^{\shortmid }\widehat{\mathcal{S}}_{\check{\eta}}^{\star
} ( \widehat{\mathcal{A}})$ (\ref{renentr2wmh}), we can consider the concept
of relative R\'{e}nyi entropy for QGIFs (\ref{relatrenyi}) when all
thermodynamic information variables are determined by wormhole data. Proofs
of validity of such formulas, monotonic properties, and inequalities follow
from unifications of constructions point by point along causal curves on
respective nonholonomic Lorentz manifolds/ (co) bundles. This allow us to
select physically allowed nonassociative geometric flow evolution scenarios
and study, for instance, models of nonassociative phase space entanglement
of wormhole configurations.

\subsection{Nonassociative 8-d phase space wormholes and entanglement}

We study how the concept of entanglement and respective models of GIFs and
QGIFs can be applied for understanding physical properties of nonassociative
geometric flow deformations of higher dimension wormholes with warped extra
dimensions \cite{sv14a,warpwormh} and/or nonholonomic structures \cite{vwh3}.

\subsubsection{Parametric solutions for 8-d nonassociative wormholes}

We consider a different class of nonassociative star product R-flux deformed
solutions which are very different form the models with double 4-d wormholes
defined \ respectively by s-metrics (\ref{pm1a}) and (\ref{doublwnonassocwh}%
). In this subsection, both the prime and target s-metrics define higher
dimension quasi-stationary configurations which nonholonomically configured
as 8-d phase space wormholes and/or their $\tau $-parametric evolution as
nonassociative quasi-stationary s-metrics.

\paragraph{Prime 8-d metrics embedding 7-d phase space warped wormholes: 
\newline
}

Let us consider a 8-d phase space $\ _{s}\mathcal{M}$ endowed with local
coordinates $\ ^{\shortmid }\bar{u}^{\alpha _{s}^{\prime }}=(\bar{x}^{1}= 
\bar{r},\bar{x}^{2},\bar{y}^{3},\bar{y}^{4}=t,\bar{p}_{a_{3}},\bar{p}_{7},%
\bar{p}_{8}=E)= (\bar{x}^{i_{2}^{\prime }},\bar{p}_{a_{s}^{\prime }})$ for $%
\alpha _{s}^{\prime }=1,2,...8$ and $\bar{r}=\sqrt{%
(x^{1})^{2}+(x^{2})^{2}+(y^{3})^{2}+(p_{5})^{2}+(p_{6})^{2}}$ used for
defining the quadratic line element 
\begin{eqnarray}
d\bar{s}^{2} &=&\ ^{\shortmid }\bar{g}_{\alpha _{s}^{\prime }}(d\
^{\shortmid }\bar{u}^{\alpha _{s}^{\prime }})^{2}=\ ^{\shortmid }\bar{g}%
_{i_{2}^{\prime }}(d\bar{x}^{i_{2}^{\prime }})^{2}+\ ^{\shortmid }\bar{g}%
^{a_{s}^{\prime }}(d\bar{p}_{a_{s}^{\prime }})^{2}  \label{pm2} \\
&=&-dt^{2}+(1-\frac{\bar{b}_{0}^{2}}{\bar{r}^{2}})^{-1}d\bar{r}^{2}+\bar{r}%
^{2}d\bar{\Omega}_{[4]}^{2}+\bar{P}^{2}(1-\frac{\bar{b}_{0}^{2}}{\bar{r}^{2}}%
)d\bar{p}_{7}-dE^{2},  \notag
\end{eqnarray}%
where $d\bar{\Omega}_{[4]}^{2}$ is the 4-d spherical volume element and the
constants $\bar{b}_{0}$ and $\bar{P}$ are chosen to model a warping
configuration on momentum type coordinate $\bar{p}_{7}.$ For some
hypersurfaces $\bar{p}_{7}=const$ and $E=const$, the metric (\ref{pm2})
defines an Ellis-Bronnikov phase space wormhole with spherical topology $%
S^{4}$ instead of $S^{2}$ as we considered in (\ref{pm1}). We cite \cite%
{warpwormh,sv14a} for details and references on constructing higher
dimension wormhole solutions and corresponding models on phase spaces with
momentum like coordinates.

To apply the AFCDM for deforming prime s-metrics not involving singular
off-diagonal terms, we can consider well-defined coordinate transforms $\
^{\shortmid }\bar{u}^{\alpha _{s}}=\ ^{\shortmid }\bar{u}^{\alpha _{s}}(\
^{\shortmid }\bar{u}^{\alpha _{s}^{\prime }})$ as we explained above for
formulas (\ref{pmwh}), (\ref{pmwc}) and (\ref{pm1a}). The prime metric (\ref%
{pm2}) \ transforms into a s-metric of type 
\begin{eqnarray}
d\bar{s}^{2} &=&\ ^{\shortmid }\bar{g}_{\alpha _{s}}(\ ^{\shortmid }\bar{u}%
^{\beta _{s}})(d\ ^{\shortmid }\bar{u}^{\alpha _{s}})^{2}  \label{pm2a} \\
&=&\ ^{\shortmid }\bar{g}_{i_{1}}(\bar{x}^{k_{1}})(d\ \bar{x}^{i_{1}})^{2}+\
^{\shortmid }\bar{g}_{a_{2}}(\bar{x}^{k_{1}},\bar{y}^{3})(\mathbf{\bar{e}}%
^{a_{2}})^{2}+\ ^{\shortmid }\bar{g}^{a_{3}}(\bar{x}^{k_{1}},\bar{y}^{b_{2}},%
\bar{p}_{5})(\ ^{\shortmid }\mathbf{\bar{e}}_{a_{3}})^{2}+\ ^{\shortmid }%
\bar{g}^{a_{4}}(\bar{x}^{k_{1}},\bar{y}^{b_{2}},\bar{p}_{b_{3}},\bar{p}%
_{7})(\ ^{\shortmid }\mathbf{\bar{e}}_{a_{4}})^{2},  \notag \\
\mathbf{\bar{e}}^{a_{2}} &=&d\bar{y}^{a_{2}}+\bar{N}_{i_{1}}^{a_{2}}(\bar{x}%
^{k_{1}},\bar{y}^{3})d\ \bar{x}^{i_{1}};\ ^{\shortmid }\mathbf{\bar{e}}%
_{a_{3}}=d\bar{p}_{a_{3}}+\ ^{\shortmid }\bar{N}_{a_{3}i_{2}}(\bar{x}%
^{k_{1}},\bar{y}^{b_{2}},\bar{p}_{5})d\ \bar{x}^{i_{2}};  \notag \\
\ ^{\shortmid }\mathbf{\bar{e}}_{a_{4}} &=&d\bar{p}_{a_{4}}+\ ^{\shortmid }%
\bar{N}_{a_{4}i_{3}}(\bar{x}^{k_{1}},\bar{y}^{b_{2}},\bar{p}_{b_{3}},\bar{p}%
_{7})d\ \ ^{\shortmid }\bar{x}^{i_{3}}.  \notag
\end{eqnarray}%
In these formulas, the indices are not primed but the coefficients of
s-metric and N-connection (and the local coordinates) are overlined in order
to distinguish the constructions for the higher dimension wormholes from
those for the 4-d ones. The system of s-adapted frames and coordinates is
chosen in such a form that the coefficients of (\ref{pm2a}) do not depend on
coordinates $\ ^{\shortmid }\bar{u}^{4}=t$ and $\ ^{\shortmid }\bar{u}%
^{8}=E. $ Such conditions allow us to construct $\tau $-running
quasi-stationary nonassociative deformations of the prime metric (\ref{pm2})
using the method described in Appendix \ref{asstables} when corresponding
effective nonassociative generating sources, generating functions and
integration functions are adapted to $\bar{u}$-coordinates preserving the
quasi-stationary character and Killing symmetry on $\partial /\partial E.$
In brief, the coefficients of the prime s-metric (\ref{pm2a}) and respective
N-connection coefficients can be labelled in the form $\ _{s}^{\shortmid }%
\mathbf{\bar{g}}=\{^{\shortmid }\mathbf{\bar{g}}_{\alpha _{s}\beta _{s}}\}$
and $\ _{s}^{\shortmid }\mathbf{\bar{N}}=\{\ ^{\shortmid }\bar{N}%
_{i_{s-1}}^{a_{s}}\}$ for shells $s=1,2,3,4.$ They encode data for a 7-d
wormhole configurations determined by a diagonal metric (\ref{pm2}) and then
subjected to curved phase space coordinate transforms when the coefficients
of s-geometric objects are quasi-stationary and with a fixed energy
parameter $E_{0}.$

\paragraph{Nonassociative quasi-stationary phase space deformed/ warped 7-d
wormholes: \newline
}

Applying the AFCDM outlined in Appendix \ref{asstables},\footnote{%
in symbolic geometric and analytic forms, we can use data from Table 3 and
re-define correspondingly the formulas for $\tau $-families of
quasi-stationary s-metrics (\ref{qst8})} we can construct parametric
solutions describing 8-d nonassociative geometric flow deformations of the
prime 7-d wormhole s-metric (\ref{pm2a}). To distinguish the formulas from
similar ones in previous subsection, we underline respective prime s-metrics
and the data for target s-metric. For instance, the gravitational $\eta $%
-polarizations of higher dimension phase space wormhole s-metrics are
labeled in the form 
\begin{equation}
\ _{s}^{\shortmid }\mathbf{\bar{g}}=[\ ^{\shortmid }\bar{g}_{\alpha _{s}},\
^{\shortmid }\bar{N}_{i_{s-1}}^{a_{s}}]\rightarrow \ _{\eta }^{\shortmid }%
\mathbf{\bar{g}}(\tau )=[\ ^{\shortmid }\bar{\eta}_{\alpha _{s}}(\tau )\
^{\shortmid }\bar{g}_{\alpha _{s}}(\tau ),\ ^{\shortmid }\bar{\eta}%
_{i_{s-1}}^{a_{s}}(\tau )\ ^{\shortmid }\bar{N}_{i_{s-1}}^{a_{s}}(\tau )],
\label{etapolgen2}
\end{equation}%
substituting respective data, $\ _{s}^{\shortmid }\mathbf{\mathring{g}}(\tau
)\rightarrow \ _{s}^{\shortmid }\mathbf{\bar{g}}$ and $\ _{s}^{\shortmid}\Im
\rightarrow \ _{s}^{\shortmid }\bar{\Im}^{\star }(\tau )$ in (\ref%
{offdiagpolfr1}). Here we note that $\ _{s}^{\shortmid }\Im ^{\star }(\tau ) 
$ \ and $\ _{s}^{\shortmid }\bar{\Im}^{\star }(\tau )$ are defined for
deforming prime s-configurations involving different types of wormhole
energy-momentum tensors stated respectively for 4-d and 7-d wormhole
configurations. Generated $\tau $-families of target metrics are determined
by prime associative and commutative wormhole phase s-metrics with
coefficients $\ ^{\shortmid }\mathbf{\bar{g}}_{\alpha _{s}\beta _{s}}$. Such
coefficients do not depend on the flow parameter $\tau $, but general target
coefficients $\ _{\eta }^{\shortmid }\mathbf{\bar{g}}_{\alpha _{s}\beta
_{s}}(\tau )$ are s-metrics solving the nonassociative geometric flow
equations (\ref{nonassocrf}) defined by $\tau $-running canonical
s-connection structure. In this subsection, we shall omit hats on formulas
to simplify the system of notations considering that "overlining" of
necessary symbols and formulas is enough for defining nonassociative
geometric/ GIF / QGIF variables as functionals of certain canonical
s-objects.

Using phase space coordinates adapted to prime wormhole solutions (\ref{pm2a}%
), we parameterize the $\tau $-running generating functions (\ref{etapolgen2}%
) in the form:%
\begin{eqnarray*}
\bar{\psi}(\tau ) &\simeq &\bar{\psi}(\hbar ,\kappa ;\tau ,\bar{r},\bar{x}%
^{2};\ _{1}\bar{\Lambda}^{\star }(\tau );\ \ _{1}^{\shortmid }\bar{\Im}%
^{\star }(\tau )),\bar{\eta}_{4}(\tau )\ \simeq \bar{\eta}_{4}(\tau ,\bar{x}%
^{k_{1}},\bar{y}^{3};\ _{2}\bar{\Lambda}^{\star }(\tau );\ \ _{2}^{\shortmid
}\bar{\Im}^{\star }(\tau )), \\
\ ^{\shortmid }\bar{\eta}^{6}(\tau ) &\simeq &\ ^{\shortmid }\bar{\eta}%
^{6}(\tau ,\bar{x}^{k_{2}},\bar{p}_{5};\ _{3}\bar{\Lambda}^{\star }(\tau );\
\ _{3}^{\shortmid }\bar{\Im}^{\star }(\tau )),\ ^{\shortmid }\bar{\eta}%
^{8}(\tau )\simeq \ ^{\shortmid }\bar{\eta}^{8}(\tau ,\bar{x}^{k_{2}},\bar{p}%
_{a_{3}},\bar{p}_{7};\ _{4}\bar{\Lambda}^{\star }(\tau );\ \ _{4}^{\shortmid
}\bar{\Im}^{\star }(\tau )).
\end{eqnarray*}%
In this subsection, we consider nonlinear symmetries relating $\
_{s}^{\shortmid }\bar{\Im}^{\star }(\tau )$ to $\ _{s}\bar{\Lambda}^{\star
}(\tau )$ for any shell apart, which is different, for instance, from the
formulas (\ref{statistgenf2wh}), (\ref{energy2wh}) and (\ref{perentr2wh}),
where the $\tau $-running cosmological constants were adapted to 4-d
wormhole configurations in respective spacetime and momentum space. We shall
omit overlining of s-geometric objects with $\bar{\eta}(\tau )$-deformations
and use only left or write labels $\eta $ if the "main" symbol already
contain overlining. For instance, we shall write $\ _{\eta }^{\shortmid }%
\mathbf{\bar{g}}_{\alpha _{s}\beta _{s}}(\tau )$ instead of $\ _{\bar{\eta}%
}^{\shortmid }\mathbf{\bar{g}}_{\alpha _{s}\beta _{s}}(\tau )$ and, for
instance, $\ _{\eta }^{\shortmid }\mathbf{\bar{D}}$ instead of $\
_{s}^{\shortmid }\mathbf{\hat{D}}$ determined by $\bar{\eta}$-deformations.
Following such conventions, we can generate $\tau $-families of
nonassociative quasi-stationary s-metrics: 
\begin{eqnarray}
d\ \ _{\eta }^{\shortmid }\bar{s}^{2}(\tau ) &=&\ \ _{\eta }^{\shortmid }%
\bar{g}_{\alpha _{s}\beta _{s}}(\hbar ,\kappa ,\tau ,\bar{x}^{i_{1}},\bar{y}%
^{a_{2}},\bar{p}_{a_{3}},\bar{p}_{7};\ ^{\shortmid }\bar{g}_{\alpha _{s}};%
\bar{\eta}_{4}(\tau ),\ ^{\shortmid }\bar{\eta}^{6}(\tau ),\ ^{\shortmid }%
\bar{\eta}^{8}(\tau ),\ _{s}\bar{\Lambda}^{\star }(\tau );\ _{s}^{\shortmid }%
\bar{\Im}^{\star }(\tau ))d~\ ^{\shortmid }\bar{u}^{\alpha _{s}}d~\
^{\shortmid }\bar{u}^{\beta _{s}}  \notag \\
&=&e^{\bar{\psi}(\tau )}[(d\bar{x}^{1}(\bar{r},\bar{x}^{2^{\prime }}))^{2}+(d%
\bar{x}^{2}(\bar{r},\bar{x}^{2^{\prime }}))^{2}]-  \label{defwh7} \\
&&\frac{[\partial _{3}(\bar{\eta}_{4}(\tau )\ \bar{g}_{4}(\tau ))]^{2}}{%
|\int d\bar{y}^{3}\ _{2}^{\shortmid }\bar{\Im}^{\star }(\tau )\partial _{3}(%
\bar{\eta}_{4}(\tau )\ \bar{g}_{4}(\tau ))|\ (\bar{\eta}_{4}(\tau )\bar{g}%
_{4}(\tau ))}\{d\bar{y}^{3}+\frac{\partial _{i_{1}}[\int d\bar{y}^{3}\ _{2}%
\bar{\Im}(\tau )\ \partial _{3}(\bar{\eta}_{4}(\tau )\bar{g}_{4}(\tau ))]}{\
_{2}^{\shortmid }\bar{\Im}^{\star }(\tau )\partial _{3}(\bar{\eta}_{4}(\tau )%
\bar{g}_{4}(\tau ))}d\bar{x}^{i_{1}}\}^{2}+  \notag \\
&&\bar{\eta}_{4}(\tau )\bar{g}_{4}(\tau ))\{dt+[\ _{1}n_{k_{1}}(\tau )+\
_{2}n_{k_{1}}(\tau )\int \frac{d\bar{y}^{3}[\partial _{3}(\bar{\eta}%
_{4}(\tau )\bar{g}_{4}(\tau ))]^{2}}{|\int d\bar{y}^{3}\ _{2}^{\shortmid }%
\bar{\Im}^{\star }(\tau )\partial _{3}(\bar{\eta}_{4}(\tau )\bar{g}_{4}(\tau
))|\ [\bar{\eta}_{4}(\tau )\bar{g}_{4}(\tau )]^{5/2}}]d\bar{x}^{k_{1}}\}- 
\notag
\end{eqnarray}%
\begin{eqnarray*}
&&\frac{[\ ^{\shortmid }\partial ^{5}(\ ^{\shortmid }\bar{\eta}^{6}(\tau )\
^{\shortmid }\bar{g}^{6}(\tau ))]^{2}}{|\int d\bar{p}_{5}~\ _{3}^{\shortmid }%
\bar{\Im}^{\star }(\tau )\ \ ^{\shortmid }\partial ^{5}(\ ^{\shortmid }\bar{%
\eta}^{6}(\tau )\ \ ^{\shortmid }\bar{g}^{6}(\tau ))\ |\ (\ ^{\shortmid }%
\bar{\eta}^{6}(\tau )\ ^{\shortmid }\bar{g}^{6}(\tau ))}\{d\bar{p}_{5}+\frac{%
\ ^{\shortmid }\partial _{i_{2}}[\int d\bar{p}_{5}\ _{3}^{\shortmid }\bar{\Im%
}(\tau )\ ^{\shortmid }\partial ^{5}(\ ^{\shortmid }\bar{\eta}^{6}(\tau )\
^{\shortmid }\bar{g}^{6}(\tau ))]}{~\ _{3}^{\shortmid }\bar{\Im}^{\star
}(\tau )\ ^{\shortmid }\partial ^{5}(\ ^{\shortmid }\bar{\eta}^{6}(\tau )\
^{\shortmid }\bar{g}^{6}(\tau ))}d\bar{x}^{i_{2}}\}^{2}+ \\
&&(\ ^{\shortmid }\bar{\eta}^{6}(\tau )\ ^{\shortmid }\bar{g}^{6}(\tau ))\{d%
\bar{p}_{6}+[\ _{1}^{\shortmid }n_{k_{2}}(\tau )+\ _{2}^{\shortmid
}n_{k_{2}}(\tau )\int \frac{d\bar{p}_{5}[\ ^{\shortmid }\partial ^{5}(\
^{\shortmid }\bar{\eta}^{6}(\tau )\ ^{\shortmid }\bar{g}^{6}(\tau ))]^{2}}{%
|\int d\bar{p}_{5}\ _{3}^{\shortmid }\bar{\Im}^{\star }(\tau )\ \partial
^{5}(\ ^{\shortmid }\bar{\eta}^{6}(\tau )\ ^{\shortmid }\bar{g}^{6}(\tau
))|\ [\ ^{\shortmid }\bar{\eta}^{6}(\tau )\ ^{\shortmid }\bar{g}^{6}(\tau
)]^{5/2}}]d\bar{x}^{k_{2}}\}-
\end{eqnarray*}%
\begin{eqnarray*}
&&\frac{[\ ^{\shortmid }\partial ^{7}(\ ^{\shortmid }\bar{\eta}^{8}(\tau )\
^{\shortmid }\bar{g}^{8}(\tau ))]^{2}}{|\int d\bar{p}_{7}\ _{4}^{\shortmid }%
\bar{\Im}^{\star }(\tau )\ ^{\shortmid }\partial ^{8}(\ ^{\shortmid }\bar{%
\eta}^{7}(\tau )\ ^{\shortmid }\bar{g}^{7}(\tau ))\ |\ (\ ^{\shortmid }\bar{%
\eta}^{7}(\tau )\ ^{\shortmid }\bar{g}^{7}(\tau ))}\{d\bar{p}_{7}+\frac{\
^{\shortmid }\partial _{i_{3}}[\int d\bar{p}_{7}\ _{4}^{\shortmid }\bar{\eta}%
^{\star }(\tau )\ ^{\shortmid }\partial ^{7}(\ ^{\shortmid }\bar{\eta}%
^{8}(\tau )\ \ ^{\shortmid }\bar{g}^{8}(\tau ))]}{\ _{4}^{\shortmid }\bar{\Im%
}^{\star }(\tau )\ ^{\shortmid }\partial ^{7}(\ ^{\shortmid }\eta ^{8}(\tau
)\ ^{\shortmid }\bar{g}^{8}(\tau ))}d\ ^{\shortmid }\bar{x}^{i_{3}}\}^{2}+ \\
&&(\ ^{\shortmid }\bar{\eta}^{8}(\tau )\ ^{\shortmid }\bar{g}^{8}(\tau
))\{dE+[\ _{1}n_{k_{3}}(\tau )+\ _{2}n_{k_{3}}(\tau )\int \frac{d\bar{p}%
_{7}[\ ^{\shortmid }\partial ^{7}(\ ^{\shortmid }\bar{\eta}^{8}(\tau )\
^{\shortmid }\bar{g}^{8}(\tau ))]^{2}}{|\int d\bar{p}_{7}\ _{4}^{\shortmid }%
\bar{\Im}^{\star }(\tau )[\ ^{\shortmid }\partial ^{7}(\ ^{\shortmid }\bar{%
\eta}^{8}(\tau )\ ^{\shortmid }\bar{g}^{8}(\tau ))]|\ [\ ^{\shortmid }\bar{%
\eta}^{8}(\tau )\ ^{\shortmid }\bar{g}^{8}(\tau )]^{5/2}}]d\ ^{\shortmid }%
\bar{x}^{k_{3}}\}.
\end{eqnarray*}%
For self-similar configurations with $\tau =\tau _{0},$ these formulas
define parametric solutions of vacuum nonassociative gravitational equations
with shell effective cosmological constants $_{s}^{\shortmid }\bar{\Lambda}%
_{0}^{\star }=\ _{s}^{\shortmid }\bar{\Lambda}^{\star }(\tau _{0}).$ The
target s-metrics $\ _{\eta }^{\shortmid }\mathbf{\bar{g}}(\tau )=$ $\
_{s}^{\shortmid }\mathbf{\bar{g}}(\tau )$ (\ref{defwh7}) involves also a $%
\tau $-family $\bar{\psi}(\tau ,\bar{r},\bar{x}^{2})$ of solutions of 2-d
Poisson equations 
\begin{equation*}
\partial _{11}^{2}\bar{\psi}(\tau ,\bar{r},\bar{x}^{2})+\partial _{22}^{2}%
\bar{\psi}(\tau ,\bar{r},\bar{x}^{2})=2 \ _{1}^{\shortmid }\bar{\Im}^{\star
}(\tau ,\bar{r},\bar{x}^{2}).
\end{equation*}%
Involving nonlinear symmetries and 2-d conformal transforms, we can use in
certain equivalent forms the solutions of $\partial _{11}^{2}\bar{\psi}(\tau
)+\partial _{22}^{2}\bar{\psi}(\tau )=2\ _{1}\bar{\Lambda}^{\star }(\tau ).$

Finally, we note that $\kappa$-linear decompositions of s-metrics (\ref%
{defwh7}) can be performed by a corresponding adapting of formulas (\ref%
{whpolf1}).

\subsubsection{Nonassociative QGIF phase space entanglement for 8-d R-flux
deformed wormholes}

We can formulate GIF and QGIF models of higher dimension wormhole warped
solutions and their nonassociative geometric flow evolution as in the case
of double 4-d wormhole configurations but re-defining and computing
fundamental variables and solutions in terms of "overlined" geometric and
quantum information s-objects. This provides also abstract geometric proofs
for physically important inequalities and conditions of relative/ mutual
information / entanglement etc. We omit motivations and details which
technically are similar to those introduced and computed in subsection \ref%
{ssdoublenworm} and outline the main formulas and nonassociative GIF and
QGIF concepts applicable to $\tau $-parametric solutions (\ref{defwh7}) .

\paragraph{GIFs thermodynamic variables for nonassociative 7-d warped
wormholes: \newline
}

Let us consider "overlined" formulas for $\kappa $-linear parametric
decompositions of the nonassociative generalized Perelman variables (\ref%
{nagthermodvalues}) describing $\tau $-evolution of phase space higher
dimension and warped wormhole configurations (\ref{defwh7}). We denote the
respective volume functionals in phase spaces as $\ ^{\shortmid }\delta \
^{\shortmid }\mathcal{\bar{V}}(\tau )$ and change the statistical generating
function $_{s}\widehat{\mathcal{Z}}^{\star }(\tau )$ (\ref{spf}) into $\
_{\eta }^{\shortmid }\mathcal{\bar{Z}}^{\star }(\tau ).$ In abstract
geometric form, the thermodynamic variables (\ref{thvcannpd}) are computed
in terms of overlined variables:%
\begin{eqnarray}
\ _{\eta }^{\shortmid }\mathcal{\bar{Z}}_{\kappa }^{\star }(\tau ) &=&\exp %
\left[ \int\nolimits_{\tau ^{\prime }}^{\tau }\frac{d\tau }{(2\pi \tau )^{4}}%
\frac{1}{\sqrt{|\ _{1}\bar{\Lambda}^{\star }(\tau )\ _{2}\bar{\Lambda}%
^{\star }(\tau )\ _{3}^{\shortmid }\bar{\Lambda}^{\star }(\tau )\
_{4}^{\shortmid }\bar{\Lambda}^{\star }(\tau )|}}\ _{\eta }^{\shortmid }%
\mathcal{\bar{V}}(\tau )\right] ,  \label{statistgenf7wh} \\
\ _{\eta }^{\shortmid }\overline{\mathcal{E}}_{\kappa }^{\star }(\tau )
&=&-\int\nolimits_{\tau ^{\prime }}^{\tau }\frac{d\tau }{(4\pi )^{4}\tau ^{3}%
}\frac{\tau \lbrack \ _{1}\bar{\Lambda}^{\star }(\tau )+\ _{2}\bar{\Lambda}%
^{\star }(\tau )+\ _{3}^{\shortmid }\bar{\Lambda}^{\star }(\tau )+\
_{4}^{\shortmid }\bar{\Lambda}^{\star }(\tau )]-4}{\sqrt{|\ _{1}\bar{\Lambda}%
^{\star }(\tau )\ _{2}\bar{\Lambda}^{\star }(\tau )\ _{3}^{\shortmid }\bar{%
\Lambda}^{\star }(\tau )\ _{4}^{\shortmid }\bar{\Lambda}^{\star }(\tau )|}}\
_{\eta }^{\shortmid }\mathcal{\bar{V}}_{\kappa }(\tau ),  \label{energy7wh}
\\
\ _{\eta }^{\shortmid }\overline{\mathcal{S}}_{\kappa }^{\star }(\tau )
&=&-\ _{s}^{\shortmid }\overline{\mathcal{W}}_{\kappa }^{\star }(\tau
)=-\int\nolimits_{\tau ^{\prime }}^{\tau }\frac{d\tau }{(4\pi \tau )^{4}}%
\frac{\tau \lbrack \ _{1}\bar{\Lambda}^{\star }(\tau )+\ _{2}\bar{\Lambda}%
^{\star }(\tau )+\ _{3}^{\shortmid }\bar{\Lambda}^{\star }(\tau )+\
_{4}^{\shortmid }\bar{\Lambda}^{\star }(\tau )]-8}{\sqrt{|\ _{1}\bar{\Lambda}%
^{\star }(\tau )\ _{2}\bar{\Lambda}^{\star }(\tau )\ _{3}^{\shortmid }\bar{%
\Lambda}^{\star }(\tau )\ _{4}^{\shortmid }\bar{\Lambda}^{\star }(\tau )|}}\
_{\eta }^{\shortmid }\mathcal{\bar{V}}_{\kappa }(\tau ).  \label{pernetr7wh}
\end{eqnarray}%
In these formulas, we use the running phase space volume functional $\
_{\eta }^{\shortmid }\mathcal{\bar{V}}_{\kappa }(\tau )= \int_{\
_{s}^{\shortmid }\widehat{\Xi }}\ ^{\shortmid }\delta \ _{\eta }^{\shortmid }%
\mathcal{\bar{V}}_{\kappa }(\ _{s}^{\shortmid }\bar{\Im}^{\star }(\tau ),\
^{\shortmid }\bar{g}_{\alpha _{s}})$ can be computed explicitly if we
prescribe certain classes of generating $\eta $-functions, effective
generating sources $\ _{s}^{\shortmid }\bar{\Im}(\tau ),$ coefficients of a
prime s-metric $\ ^{\shortmid }\bar{g}_{\alpha _{s}}$ and nonholonomic
distributions defining a closed hyper-surface $\ _{s}^{\shortmid }\widehat{%
\Xi }\subset \ _{s}^{\shortmid }\mathcal{M}^{\star }.$ For the purposes of
this work, it is enough to use general volume functionals $\ _{\eta
}^{\shortmid }\mathcal{\bar{V}}_{\kappa }(\tau )$ when the thermodynamic
information constructions can be normalized for certain classes of
(modified) statistical generating functions.

For nonassociative higher dimension phase space wormhole $\tau $-running
solutions $\ _{\eta }^{\shortmid }\mathbf{\bar{g}}(\tau )$ (\ref{defwh7}),
the Shannon--Perelman entropy (\ref{shanonper}) is computed: 
\begin{eqnarray*}
\ _{\eta }^{\shortmid }\overline{S}_{A}^{\star } &\simeq
&\sum\nolimits_{z=1}^{z=r}\ _{\eta }^{\shortmid }\overline{\mathcal{S}}%
_{\kappa }^{\star }(\tau _{z}),%
\mbox{  nonassociative Ricci flows for a set of discrete values
of }\tau ; \\
&=&\ _{\eta }^{\shortmid }\overline{\mathcal{S}}_{\kappa }^{\star }(\tau
_{0}),\mbox{ for a
solution of nonassociative Ricci soliton/ Einstein eqs. }; \\
&\simeq &\int\nolimits_{\tau _{r}}^{\tau _{r}}d\tau \ \ _{\eta }^{\shortmid }%
\overline{\mathcal{S}}_{\kappa }^{\star }(\tau ),%
\mbox{nonasociative
geometric flows with  continuous running of }\tau .
\end{eqnarray*}%
Such variables are derived for $\ _{\eta }^{\shortmid }\mathcal{\bar{Z}}%
_{\kappa }^{\star }(\tau )$ (\ref{statistgenf7wh}), for respective energy
operators computed as in $\ _{\eta }^{\shortmid }\overline{\mathcal{E}}%
_{\kappa }^{\star }(\tau )$ (\ref{energy7wh}). For explicit applications and
elaborating lab experiments, it is necessary to perform further small
parametric decompositions of the s-metric (\ref{defwh7}) in a form similar
to (\ref{whpolf1}). Such constructions have to be performed in s-adapted
form using overlined geometric objects and adapting to overlined s-frames
and coordinates. We do not provide such cumbersome formulas in this work.
For nonassociative GIF models of higher dimension wormholes, the
thermodynamic constructions are physical if the corresponding $\ ^{\shortmid
}\overline{S}_{A}^{\star }\simeq \sum\nolimits_{z=1}^{z=r} \ _{\eta
}^{\shortmid } \overline{\mathcal{S}}_{\kappa }^{\star }(\tau _{z})\geq 0$
and, correspondingly, $\ _{\eta }^{\shortmid }\overline{\mathcal{S}}%
_{\kappa}^{\star }(\tau _{0}) \geq 0,$ or $\int\nolimits_{\tau _{r}}^{\tau
_{r}}d\tau \ _{\eta }^{\shortmid }\overline{\mathcal{S}}_{\kappa }^{\star
}(\tau )\geq 0.$

In overlined form, we can define and compute continuous analogs of (\ref%
{suadditivgif}) for nonassociative GIFs of nonassociative higher dimension
solutions $\ _{\eta }^{\shortmid }\mathbf{\bar{g}}(\tau )$ (\ref{defwh7}).
We consider a generalized partition function 
\begin{eqnarray*}
\ _{AB}^{\shortmid }\overline{\mathcal{Z}}[\ \ _{\eta }^{\shortmid }\mathbf{%
\bar{g}}(\tau ),\ _{1\eta }^{\shortmid }\mathbf{\bar{g}}(\tau )\ (\tau )]
&=&\ \int_{\ _{s}^{\shortmid }\widehat{\Xi }}\int_{\ _{1s}^{\shortmid }%
\widehat{\Xi }}(4\pi \tau )^{-8}e^{-\ _{AB}\widehat{f}}\sqrt{|\ \ _{\eta
}^{\shortmid }\mathbf{\bar{g}}(\tau )|}\sqrt{|\ \ _{1\eta }^{\shortmid }%
\mathbf{\bar{g}}(\tau )|}d^{8}\ ^{\shortmid }\overline{u}\ d^{8}\
_{1}^{\shortmid }\bar{u}(-\ _{AB}\widehat{f}+8), \\
&&\mbox{ for
}\ _{s}^{\shortmid }\mathcal{M}^{\star }\otimes \ _{s}^{\shortmid }\mathcal{M%
}^{\star }\mathbf{.}
\end{eqnarray*}%
Correspondingly, we compute the canonical thermodynamic entropy function
(similarly to (\ref{pernetr7wh}) ) using formulas 
\begin{eqnarray*}
\ _{AB}^{\shortmid }\overline{\mathcal{S}}_{\eta }^{\star }(\tau ) &=&\
_{\eta }^{\shortmid }\overline{\mathcal{S}}^{\star }\ [\ _{s}^{\star }%
\widehat{A},\ _{s}^{\star }\widehat{B}](\tau )=-\int_{\ _{s}^{\shortmid }%
\widehat{\Xi }}\int_{\ _{s1}^{\shortmid }\widehat{\Xi }}(4\pi \tau
)^{-8}e^{-\ _{AB}\widehat{f}}\sqrt{|\ \ _{\eta }^{\shortmid }\mathbf{\bar{g}}%
(\tau )|}\sqrt{|\ _{1\eta }^{\shortmid }\mathbf{\bar{g}}(\tau )|}d^{8}\
^{\shortmid }\bar{u}\ d^{8}\ _{1}^{\shortmid }\bar{u} \\
&&\left[ \tau (\ _{\eta }^{\shortmid }\overline{R}^{\star }(\tau )+\ _{1\eta
}^{\shortmid }\widehat{R}^{\star }(\tau )+|\ \ _{AB}^{\shortmid }\overline{%
\mathbf{D}}^{\star }\widehat{f}|^{2})+\ _{AB}\widehat{f}-16\right] ,
\end{eqnarray*}%
where (using nonlinear symmetries) the respective scalar s-curvatures can be
transformed into $\tau $-running cosmological constants $\ _{s}\bar{\Lambda}%
^{\star }(\tau ).$

\paragraph{QGIFs entanglement entropy and Bell states for nonassociative 7-d
warped wormholes: \newline
}

We apply the abstract nonassociative geometric and quantum formulas from
subsection \ref{ssthgdsnonassoc} and compute the geometric and thermodynamic
information variables for higher dimension s-metrics $\ _{\eta }^{\shortmid }%
\mathbf{\bar{g}}(\tau )$ (\ref{defwh7}). A discrete model with effective
energy $\ _{\eta }^{\shortmid }\overline{\mathcal{E}}_{\underline{a}}^{\star
}:= \ _{\eta }^{\shortmid }\overline{\mathcal{E}}_{\kappa }^{\star }(\tau _{%
\underline{a}})$ (for a discrete $\tau _{\underline{a}},$ when $\underline{a}%
=1,2,...,\check{r})$ can be constructed for energy functionals (\ref%
{energy7wh}). We can compute the canonical quantum variables following
formulas (\ref{dmq1}) 
\begin{eqnarray}
|\ _{\eta }^{\shortmid }\widehat{\Psi }^{\star } >&=&(\ _{\eta }^{\shortmid }%
\overline{\mathcal{Z}}^{\star })^{-1/2}\sum_{\underline{a}}e^{- \ _{\eta
}^{\shortmid }\overline{\mathcal{E}}_{\underline{a}}^{\star }/2\tau }|\psi _{%
\mathcal{A}}^{\underline{a}}>\otimes |\psi _{\mathcal{B}}^{\underline{a}}>,
\label{dmq1wh7} \\
\ _{\eta }^{\shortmid }\widehat{\rho }_{\mathcal{A}}^{\star } &=&(\ _{\eta
}^{\shortmid }\overline{\mathcal{Z}}^{\star })^{-1}\sum_{\underline{a}}e^{-\
\ _{\eta }^{\shortmid }\overline{\mathcal{E}}_{\underline{a}}^{\star }/\tau
}|\widehat{\psi }_{\mathcal{A}}^{\underline{a}}>\otimes <\widehat{\psi }_{%
\mathcal{A}}^{\underline{a}}|=(\ _{\eta }^{\shortmid }\overline{\mathcal{Z}}%
^{\star })^{-1}e^{-\ \ _{\eta }^{\shortmid }\widehat{\boldsymbol{H}}_{%
\mathcal{A}}^{\star }/\tau },  \notag
\end{eqnarray}%
In such formulas, there are considered any prescribed/computed data $(\
_{\eta }^{\shortmid }\overline{\mathcal{Z}}^{\star },\ \ _{\eta }^{\shortmid
}\overline{\mathcal{E}}_{\underline{a}}^{\star }),$ where $\
_{\eta}^{\shortmid }\widehat{\boldsymbol{H}}_{\mathcal{A}}^{\star }|\widehat{%
\psi } _{\mathcal{A}}^{\underline{a}}>=\ _{\eta }^{\shortmid }\overline{%
\mathcal{E}}_{\underline{a}}^{\star }|\widehat{\psi }_{\mathcal{A}}^{%
\underline{a}}>.$

For higher dimension nonassociative wormholes, the von Neumann--Perelman
entropy (i.e. the entanglement canonical entropy of respective
nonassociative QGIF system $\mathcal{A}=\{\ _{\eta }^{\shortmid }A^{\star}\}=%
\mathcal{A}^{\star }),$ when 
\begin{equation}
\ _{\eta }^{\shortmid }\overline{\mathcal{S}}_{\mathcal{A}}^{\star }\ =-tr_{%
\mathcal{A}}[\ \ _{\eta }^{\shortmid }\widehat{\rho }_{\mathcal{A}}^{\star
}\log \ \ _{\eta }^{\shortmid }\widehat{\rho }_{\mathcal{A}}^{\star }]
\label{entanglentrcanwh2}
\end{equation}%
is computed for the $\ _{\eta }^{\shortmid }\widehat{\rho }_{\mathcal{A}%
}^{\star }$ and/or $|\ _{\eta }^{\shortmid }\widehat{\Psi }^{\star }>$ from (%
\ref{dmq1wh7}).

\paragraph{Entangled nonassociative QGIFs and Bell phase states for 7-d
warped wormholes: \newline
}

The formulas (\ref{natwowhqs}) can be modified to describe double quantum
thermo-geometric states encoding nonassociative data, when 
\begin{eqnarray}
|\ _{\eta }^{\shortmid }\widehat{\Psi }^{\star }> &=&\frac{e^{-(\ _{\eta
}^{\shortmid }\overline{\mathcal{E}}_{\underline{1}}^{\star }+\ _{\eta
}^{\shortmid }\overline{\mathcal{E}}_{\underline{2}}^{\star })/2\tau }}{%
\sqrt{2}\ (\ _{\eta }^{\shortmid }\overline{\mathcal{Z}}_{\underline{1}%
}^{\star })^{1/2}(\ _{\eta }^{\shortmid }\overline{\mathcal{Z}}_{\underline{2%
}}^{\star })^{1/2}}(|\hat{1}\hat{2}>-|\hat{2}\hat{1}>)\mbox{ and }
\label{dmq1wh7a} \\
\ _{\eta }^{\shortmid }\widehat{\rho }_{\mathcal{A}}^{\star } &=&\frac{1}{2}(%
\frac{e^{-\ _{\eta }^{\shortmid }\overline{\mathcal{E}}_{\underline{1}%
}^{\star }/\tau }}{\ _{\eta }^{\shortmid }\overline{\mathcal{Z}}_{\underline{%
1}}^{\star }}|\hat{1}>_{\mathcal{A}}\ _{\mathcal{A}}<\hat{1}|+\frac{e^{-\
_{\eta }^{\shortmid }\overline{\mathcal{E}}_{\underline{2}}^{\star }/\tau }}{%
\ _{\eta }^{\shortmid }\overline{\mathcal{Z}}_{\underline{2}}^{\star }}|\hat{%
2}>_{\mathcal{A}}\ _{\mathcal{A}}<\hat{2}|)=\left( 
\begin{array}{cc}
e^{-\ _{\eta }^{\shortmid }\overline{\mathcal{E}}_{\underline{1}}^{\star
}/\tau }/2\ _{\eta }^{\shortmid }\overline{\mathcal{Z}}_{\underline{1}%
}^{\star } & 0 \\ 
0 & e^{-\ _{\eta }^{\shortmid }\overline{\mathcal{E}}_{\underline{2}}^{\star
}/\tau }/2\ _{\eta }^{\shortmid }\overline{\mathcal{Z}}_{\underline{2}%
}^{\star }%
\end{array}%
\right) .  \notag
\end{eqnarray}%
In these formulas, $\ _{\eta }^{\shortmid }\overline{\mathcal{Z}}_{%
\underline{1}}^{\star }=\ _{s}^{\shortmid }\widehat{Z}[\ _{\eta }^{\shortmid
}\mathbf{\bar{g}}_{_{\underline{1}}}(\tau )]$ and $\ _{\eta }^{\shortmid}%
\overline{\mathcal{Z}}_{_{\underline{2}}}^{\star }= \ _{s}^{\shortmid }%
\widehat{Z}[\ _{\eta }^{\shortmid }\mathbf{\bar{g}}_{_{\underline{2}%
}}(\tau)] $ and respective thermodynamic variables (see formulas (\ref{spf})
and (\ref{statistgenf7wh})) are defined by two nonassociative higher
dimension solutions $\ _{\eta }^{\shortmid }\mathbf{\bar{g}}_{_{\underline{1}%
}}(\tau )$ and $\ _{\eta }^{\shortmid }\mathbf{\bar{g}}_{_{\underline{2}%
}}(\tau )$ for ansatz of form (\ref{defwh7}). We can consider two classes of
different such $\tau $-running quasi-stationary solutions determined by
different effective sources and/or different sets of shell cosmological
constants, or the same class of solutions but with different integration
functions.

We can compute 
\begin{equation*}
\ _{\eta }^{\shortmid }\overline{\mathcal{S}}_{\mathcal{A}}^{\star }\ =-tr_{%
\mathcal{A}}\left[ \left( 
\begin{array}{cc}
e^{-\ _{\eta }^{\shortmid }\overline{\mathcal{E}}_{\underline{1}}^{\star
}/\tau }/2\ _{\eta }^{\shortmid }\overline{\mathcal{Z}}_{\underline{1}%
}^{\star } & 0 \\ 
0 & e^{-\ _{\eta }^{\shortmid }\overline{\mathcal{E}}_{\underline{2}}^{\star
}/\tau }/2\ _{\eta }^{\shortmid }\overline{\mathcal{Z}}_{\underline{2}%
}^{\star }%
\end{array}%
\right) \left( 
\begin{array}{cc}
\log [e^{-\ _{\eta }^{\shortmid }\overline{\mathcal{E}}_{\underline{1}%
}^{\star }/\tau }/2\ \ _{\eta }^{\shortmid }\overline{\mathcal{Z}}_{%
\underline{1}}^{\star }] & 0 \\ 
0 & \log [e^{-\ _{\check{\eta}}^{\shortmid }\widehat{\mathcal{E}}_{%
\underline{2}}^{\star }/\tau }/2\ \ _{\check{\eta}}^{\shortmid }\widehat{%
\mathcal{Z}}_{\underline{2}\kappa }^{\star }]%
\end{array}%
\right) \right]
\end{equation*}%
following the\ definition $\ _{\eta }^{\shortmid }\overline{\mathcal{S}}_{%
\mathcal{A}}^{\star }\ \ $(\ref{entanglentrcanwh2}) but for $\ _{\check{\eta}%
}^{\shortmid }\widehat{\rho }_{\mathcal{A}}^{\star }$ from (\ref{dmq1wh7a}).

Next, we consider how we can define examples of Bell type of QGIF
entanglement and nonassociative canonical qubit systems of type (\ref%
{qgifbell}) constructed in explicit form for quantum thermo-geometric
variables (\ref{dmq1wh7a}). We using the respective quantum statistical
generating functions and effective energy s-operators (\ref{statistgenf7wh})
and (\ref{energy7wh}). Such nonassociative two qubit systems are determined
by totally four independent maximally entangled quasi-stationary QGIFs for
nonassociative higher dimension wormholes (\ref{defwh7}):%
\begin{eqnarray*}
|\ \ _{\eta }^{\shortmid }\widehat{B}_{1}^{\star } &>&=\frac{1}{\sqrt{2}\ }(%
\frac{e^{-\ \ _{\eta }^{\shortmid }\overline{\mathcal{E}}_{\underline{1}%
}^{\star }/\tau }}{\ \ _{\eta }^{\shortmid }\overline{\mathcal{Z}}_{%
\underline{1}}^{\star }}|\hat{1}\hat{1}>+\frac{e^{-\ _{\eta }^{\shortmid }%
\overline{\mathcal{E}}_{\underline{2}}^{\star }/\tau }}{\ \ \ _{\eta
}^{\shortmid }\overline{\mathcal{Z}}_{\underline{2}}^{\star }}|\hat{2}\hat{2}%
>),\ |\ \ _{\eta }^{\shortmid }\widehat{B}_{2}^{\star }>=\frac{1}{\sqrt{2}\ }%
(\frac{e^{-\ \ _{\eta }^{\shortmid }\overline{\mathcal{E}}_{\underline{1}%
}^{\star }/\tau }}{\ \ _{\eta }^{\shortmid }\overline{\mathcal{Z}}_{%
\underline{1}}^{\star }}|\hat{1}\hat{1}>-\frac{e^{-\ _{\check{\eta}%
}^{\shortmid }\widehat{\mathcal{E}}_{\underline{2}}^{\star }/\tau }}{\ \ \ \
_{\eta }^{\shortmid }\overline{\mathcal{Z}}_{\underline{2}}^{\star }}|\hat{2}%
\hat{2}>), \\
|\ \ _{\eta }^{\shortmid }\widehat{B}_{3}^{\star } &>&=\frac{e^{-(\ \ _{\eta
}^{\shortmid }\overline{\mathcal{E}}_{\underline{1}}^{\star }+\ _{\eta
}^{\shortmid }\overline{\mathcal{E}}_{\underline{2}}^{\star })/2\tau }}{%
\sqrt{2}\ (\ _{\eta }^{\shortmid }\overline{\mathcal{Z}}_{\underline{1}%
}^{\star })^{1/2}\ (\ _{\eta }^{\shortmid }\overline{\mathcal{Z}}_{%
\underline{2}}^{\star })^{1/2}}(|\hat{1}\hat{2}>+|\hat{2}\hat{1}>),\ |\ \
_{\eta }^{\shortmid }\widehat{B}_{4}^{\star }>=\frac{e^{-(\ \ _{\eta
}^{\shortmid }\overline{\mathcal{E}}_{\underline{1}}^{\star }+\ _{\eta
}^{\shortmid }\overline{\mathcal{E}}_{\underline{2}}^{\star })/2\tau }}{%
\sqrt{2}(\ _{\eta }^{\shortmid }\overline{\mathcal{Z}}_{\underline{1}%
}^{\star })^{1/2}\ (\ _{\eta }^{\shortmid }\overline{\mathcal{Z}}_{%
\underline{2}}^{\star })^{1/2}}(|\hat{1}\hat{2}>-|\hat{2}\hat{1}>).
\end{eqnarray*}

The nonassociative QGIF systems analyzed in this subsection for higher
dimension phase space wormhole stationary deformations are characterized by
nonholonomic deformations of Bell states, i.e. of Einstein-Podolsky-Rosen
pairs. As quantum systems such states manifest QM aspects but they also
encode nonassociative geometric flows and star product and R-flux violations
of the Bell's inequalities.

\subsection{Comments on nonassociative QGIFs and transversable R-flux
deformed wormholes}

\label{sscomments}Let us discuss and interpret certain important physical
properties (in the context of the theory of relativistic Ricci flows,
modified gravity, and quantum information) of nonassociative deformed
wormhole solutions.

\subsubsection{Nonassocitaive $\protect\tau $-running phase space bridges}

Wormhole solutions in GR are considered as some hypothetical geometric
structures that link two distinct regions of the same spacetime. The first
model of the so-called Einstein-Rosen, ER, bridge was proposed as a vacuum
solution of gravitational field equations \cite{er35}. That solution was
derived from the Schwarzschild metric and such a wormhole is not traversable
because of the presence of a singularity. Morris and Thorne constructed a
static and spherically symmetric wormhole configuration having a traversable
throat at the centre \cite{morris88}. Various classes of wormhole metrics
were found in the framework of GR and MGTs \cite{wormh21a,roy20,souza22} but
their existence demands the presence of exotic matter when the null energy
condition, NEC, is violated in order to achieve a stable and traversable
structure. Due to this fact, many authors excluded them as reliable
astrophysical objects. Other authors concluded that there are possibilities
to realize wormholes without considering exotic matter and/or emphasized
that, at the moment, theories with NEC are used in modern cosmology. Such
solutions are not experimentally prohibited due to no final result on the
existence of other types of particles beyond the Standard Model. It seems
reasonable also to consider the standard observed matter and search for new
classes of theories and solutions where geometric corrections and generic
off-diagonal metrics, nonholonomic structures and generalized connections
play a leading role \cite{sv14a,vwh3}.

\vskip5pt On the other hand, a plethora of alternative theories of gravity
were elaborated with the goal of addressing the phenomenology ranging from
the fundamental Planck scale up to cosmological distances; we cite \cite%
{vbv18} and references therein for reviews of MGTs with generalizations on
(co) tangent Lorentz bundles. Such a geometric formalism is important for
elaborating nonassociative and noncommutative R-flux models on phase spaces.
Various theoretical tests are continuously performed to examine the validity
of GR at various energy regimes. It is found that several shortcomings to
geometric flows and various nonassociative and noncommutative structures
occur both at ultraviolet and infrared scales so that extension or
modifications of GR seem mandatory to address issues like QG, dark matter,
or cosmological accelerated expansion. This paper belongs to a series of
partner works \cite{partner01,partner02,partner03,lbdssv22,lbsvevv22}
devoted to nonassociative geometric flows, gravity and (quantum) information
models determined by star product R-flux deformations in string theory in
the form \cite{blumenhagen16,aschieri17,szabo19}. Nonassociative wormhole
solutions consist of explicit examples which allow one to elaborate on
thought experiments and analogous lab modelling of transversable wormhole
geometric flow evolution scenarios and/or nonholonomic Ricci soliton
dynamics using recent ideas from \cite{wormh22,wormh19,wormh19a,wormh21}.

\vskip5pt

The quadratic line element (\ref{pm1a}) (or equivalently (\ref{pmwc}))
defines two independent configurations of 4-d wormholes in phase space $\
_{s}\mathcal{M}$. One ER bridge is defined on the base spacetime and the
second one is guided by momenta-like coordinates. Nonassociative
quasi-stationary deformations and nonholonomic geometric flows result in
s-metrics $\ _{\check{\eta}}^{\shortmid }\mathbf{g}(\tau )= \
_{s}^{\shortmid }\mathbf{g}(\tau )$ (\ref{doublwnonassocwh}) as generic
off-diagonal solutions of $\tau $- and $\kappa $-parametric system of
nonlinear PDEs (\ref{nonassocrf}). In general, such target solutions
describe $\tau $-evolution (in particular nonholonomic Ricci solitons) of
very sophisticated nonlinear configurations which are different from primary
wormhole configurations. Nevertheless, we can model bridge-like phase space
configurations for some special classes of nonholonomic conditions (\ref%
{elipsoid1}), e. g. generating and connecting two 4-d ellipsoidal type
wormholes in quasi-stationary backgrounds $\ _{s}^{\star }\mathcal{M}$
encoding nonassociative data.

\vskip5pt

The prime 8-d s-metric (\ref{pm2a}) (or equivalently (\ref{pm2})) defines a
7-d warped wormhole embedded in $\ _{s}\mathcal{M}.$ Such a generalized
Einstein-Rosen bridge connects two asymptotic regions in phase spaces, for
instance, defined as certain asymptotic regions (we can model higher
dimension nonassociative BHs). We argue that similarly to the Randall
Sundrum II model and various nonholonomic generalizations \cite%
{warpwormh,sv14a,vwh3} phase spaces allow for traversable wormhole
solutions. From the outside, they resemble an effective mass (with effective
charge and small horizon and anisotropic masses) BHs. They can be generated
to be of big size on the base spacetime manifold when a human traveller can
survive the tidal forces. Such an observer would take a very short proper
time to traverse, with dependencies on momentum like coordinates, but a long
time as seen from the outside by other spacetime observers. A supposed
traveller acquires a very large boost factor depending also on momentum
variables, as it goes through the center of the phase space wormhole. We can
discuss plausible mechanisms for higher dimension/ warped wormhole
formations if we involve nonassociative R-flux deformation and $\tau $%
-evolution scenarios. For general $\eta $-deformations, the s-metrics (\ref%
{elipsoid1}) do not describe wormhole solutions but we can state respective
nonholonomic conditions for small ellipsoidal $\chi $-deformations which
describe nonassociative geometric flow evolution models for locally
anisotropic wormholes and asymptotic BE studied in \cite%
{partner02,partner03,lbdssv22,lbsvevv22}.

\subsubsection{Energy conditions for nonassociative geometric flows and
Perelman's thermodynamics}

Various energy conditions are discussed in GR and MGTs in the context of
possible violations of some energy conditions. Usually, there are considered
models with standard matter fluids when additional contributions to a
stress-energy tensor act as an additional source in the field equations. For
the consistency of certain types of wormhole solutions, this means that
realistic and viable models can be formulated without involving exotic
matter sources. In most gravity theories, in general, with quantum
semi-classical approximations for gravitational and (effective) matter
fields, it is of special interest to search for wormholes as solutions for
the dynamical equations but, in all cases, the problems of stability and
traversability of associated (multi) bridge structures have to be considered.

\vskip5pt Let us explain in brief how traversable wormholes require a
violation of the so-called \textit{average null energy condition,} ANEC. For
a local QFT, the ANEC states that along a complete achronal null geodesic
the energy-momentum tensor for matter fields, $T_{\mu \nu }$ (following our
system of notations from Appendix \ref{ass1}, we consider two shells $s=1,2$%
), satisfies the conditions $\int T_{\mu _{s}\nu\,_{s}}k^{\mu
_{s}}k^{\nu_{s}}d\lambda \geq 0,$ where $k^{\mu _{s}}$ is a tangent d-vector
and $\lambda $ is an affine parameter. For classical theories, violations of
the ANEC are prevented by the null energy conditions, NEC, $T_{\mu _{s}\nu
_{s}}k^{\mu _{s}}k^{\nu _{s}}d\lambda \geq 0,$ implying the integral
variant, i.e. the ANEC. Such conditions must be valid for any physically
reasonable theory, which has to be completed with other important criteria
on causality, topological censorship, absence of singularities etc. However,
for elaborating explicit physical models involving some special classes of
solutions (in modern cosmology and astrophysics), it is admitted that QM
effects may induce negative null energy, leading to violations of some NECs
and/or ANECs. Usually, to sustain a traversable wormhole one introduces
certain negative null energy and various nonlocal / nonachronal
constructions. For instance, we can treat the matter fields as quantum ones,
but the gravitational field is treated classically. In this context, the
goal is to solve some semi-classical Einstein's equations with an effective
source $<T_{\mu _{s}\nu _{s}}>$ taken as the expectation value of the
stress-energy tensor in a given quantum state. For certain models, the
1-loop expectation value of the stress-energy tensor does indeed lead to $%
\int <T_{\mu _{s}\nu _{s}}>k^{\mu _{s}}k^{\nu _{s}}dd\lambda <0$. Such
conditions allow one to construct transversable Einstein-Rosen bridges with
certain interesting physical properties. This provides the possibility to
transfer information between the two asymptotic spacetime boundaries and
such a process can be viewed as a teleportation protocol (see details and
references in \cite{wormh19}). Such conditions are consistent with the fact
that while entangled, various theories involving the thermofield double
states may not interact.

\vskip5pt This work is devoted to a class of nonassociative QGIF theories
related to nonassociative gravity and string theory and considers certain
applications for quasi-stationary $\tau $-evolution and off-diagonal R-flux
deformations of phase space wormhole configurations. Positively, we have to
formulate and analyze (modified) NEC and ANEC conditions in order to
elaborate realistic and physically viable theoretical and quantum
information/ lab models. Here we emphasize specific points characterizing
such nonassociative geometric flow constructions and applications of the
AFCDM:

\begin{enumerate}
\item Nonassociative star product R-flux deformations of GR result in
off-diagonal nonassociative symmetric metrics and nonsymmetric
nonassociative metrics; and other nonholonomic phase space structures
determined by N-connections. Such nonassociative geometric s-objects are
subject to nonholonomic $\tau $-evolution scenarios for nonassociative
geometric flow models, which can be formulated as variants of
thermo-geometric theories for nonassociative QGIFs (see section \ref%
{ssthgeomds}). Corresponding systems of fundamental nonlinear PDEs for
defining such theories, for instance, in $\kappa $-parametric form (\ref%
{nonassocrf}), are very different from the PDEs and (for diagonal ansatz)
ODEs used in GR.

\item Nonassociative quasi-stationary and wormhole solutions considered in
this work are constructed following the AFCDM when prime metrics are $\eta $%
- / $\chi $-deformed into nonassociative quasi-stationary solutions
characterized by respective nonlinear symmetries relating effective
generating sources, $\ _{s}^{\shortmid }\Im ^{\star }(\tau ),$ and effective 
$\tau $-running cosmological constants, $\ _{s}^{\shortmid }\Lambda
^{\star}(\tau ).$ The formulas for generating functions, generating sources,
effective cosmological constants and their nonlinear symmetries are provided
in Table 3 from Appendix \ref{asstables} (see details in partner works \cite%
{partner02,partner03,lbdssv22,lbsvevv22}).

\item We can consider on base spacetime some prime metrics constructed as
solutions of Einstein equations for a well defined $T_{\mu _{s}\nu \,_{s}}$
when ANEC and/or NEC are satisfied for $s=1,2.$ Nonassociative $\tau $- / $%
\eta $- / $\chi $-deformed target quasi-stationary s-metrics are constructed
for nonholonomic deformations on phase space $\ _{s}^{\star }\mathcal{M}$
when $T_{\mu _{s}\nu \,_{s}}\rightarrow \ _{s}^{\shortmid }\Im ^{\star}(\tau
)$ and/or $T_{\mu _{s}\nu \,_{s}}\rightarrow \ _{s}^{\shortmid}\Lambda
^{\star}(\tau ).$ Imposing additional constraints on respective classes of
generating and integration data, we can extract LC-configurations and
certain well-defined causal / transversable / with geometric entanglement
locally anisotropic wormhole solutions as we found in previous subsections.
In this way, we formulate a nonassociative geometric flow protocol for the
chosen prime wormhole solutions. They may be generalized to include certain
quantum corrections to $<T_{\mu _{s}\nu _{s}}>$ and star product R-flux
deformation data.

\item We note that the conservation laws for nonholonomic geometric flows
and related off-diagonal solutions are different from those considered for
holonomic configurations. It is similar to nonholonomic mechanics when
certain integration constants must be introduced additionally and then the
Lagrange/ Hamilton functions have to be redefined in new nonholonomic
variables with more sophisticated conservation laws. Using Perelman
thermodynamic variables (\ref{nagthermodvalues}), we can characterize such
models by an effective energy $\ ^{\shortmid}\widehat{\mathcal{E}}^{\star }$
and entropy, $\ ^{\shortmid }\widehat{\mathcal{S}}^{\star }.$ We argue that
if such variables are well-defined in certain phase space regions and for
certain values of $\tau $-parameters, we can construct respective GIF and/or
QGIF models as we proved in previous sections. This is possible even for the
target s-metrics under geometric evolution and/or off-diagonal deformations
which do not preserve (for general nonassociative deformations) the primary
wormhole configurations.
\end{enumerate}

\subsubsection{Running cosmological constants \& nonassociative wormhole
transversability in the lab}

We can use also the Schwarzschild-AdS metric and construct wormhole
solutions of the Einstein equations with a nontrivial cosmological constant.
However, such Einstein-Rosen bridges in AdS are also not traversable. The
interest in studying such wormhole solutions had increased with the
discovery of the AdS/ CFT correspondence \cite{mald99}. Together with the
holographic principle \cite{thooft93,susskind95} such ideas and theoretical
constructions resulted in an enormous amount of work done as an important
guidepost to understanding QG. In this context, holography means that there
is an exact equivalence (or duality) between two descriptions of some
systems with an effective cosmological constant. Such models use various
higher symmetry effects which are reminiscent of 2-d dilaton gravities and
give teleportation protocols which are more efficient and facilitate the
transfer of information between the members of the dual Bell pairs (see a
recent review of wormholes in holography \cite{wormh21a}).

\vskip5pt The nonassociative QGIF theory formulated in this work involves $%
\tau $-running cosmological constants, $\ _{s}^{\shortmid }\Lambda
^{\star}(\tau)$, which can be prescribed for generalized Schwarzschild-AdS;
black ellipsoids, BE, and BE-AdS; and/or other types of quasi-stationary and
wormhole configurations, and their (nonassociative) geometric flow evolution 
\cite{partner03,lbdssv22,lbsvevv22}. In principle, we can model certain
transforms to locally anisotropic versions of CFT involving, for instance,
higher dimension momentum-like coordinates. Nevertheless, the AFCDM allows
us to decouple and integrate nonassociative GIF equations in certain general
forms when certain high symmetry configurations and duality conditions are
not prescribed. We can construct nonassociative versions of BH and wormhole
solutions and consider models of their $\tau $-evolution in the context of
quantum information theory. For instance, we can investigate important
properties of such solutions by defining QGIFs both in a quantum channel and
encoding nonassociative entanglement witnesses. We can define modifications
of AdS/ CFT protocols that allow either the bounding of the channel's GIF
entanglement capacity or the determination of aspects of the nonassociative
entanglement structure between two classes of solutions of (modified) flow
equations, or nonassociative Ricci solitons. Such generalized protocols and
adapted/ auxiliary connections allow for the use of quantum channel
techniques in the study of nonassociative geometric flow and gravitational
theories and vice versa. More generally, the results of this paper suggest a
purely quantum information-theoretic criterion for recognizing when the
product of two GIF and QGIF phase space theories has a classical bulk
interpretation in the framework of a Perelman-like model encoding
nonassociative data.

\vskip5pt A crucial property of the wormhole configurations studied in GR
and certain MGTs is that energy-violating matter fields are typically needed
to support transversability through the throat. However, recent theoretical
progress has shown that these violations may naturally occur in QG, which is
also supported by our theory of nonassociative QGIFs. This suggests that
real-world transversable wormholes might exist in certain generalized
geometric flow and gravity theories and raises fascinating questions and a
number of problems about nature and phenomenology, for instance, of models
encoding nonassociative data. We argue that elaborating on further theoretic
and technologically advanced methods, such nonholonomic configurations may
be created in a laboratory and used in experiments on quantum teleportation
as proposed in \cite{wormh19a,wormh21}. The authors of such works explored a
subject called `quantum gravity in the lab' when the basic concepts are
motivated from the non-gravitational side of the duality which is physically
realized as a strongly coupled quantum system in a low-energy physics lab.
Then, the emergent holographic gravity dual is treated as `QG in the lab',
which can be treated in the context of size-winding and holographic
teleportation \cite{wormh19,wormh17}. For instance, the paper \cite{wormh21}
contains a more in-depth examination of `teleportation by size' examples. We
emphasize that `QG in the lab' studied in those works, as the authors have
defined it, must involve holography. In this article, we elaborated a
different class of nonassociative QGIF theories which provides an advanced
geometric and quantum information techniques for developing various thought
models and lab experiments. Such explorations involve QM, gravity, and labs,
which fall beyond the scope of this work. Here, we also note that any
experiment that has anything to do with the gravitational field that stops
the lab equipment from floating off into space \cite{goldman86,page81} is
beyond our definition of `QG and QGIFs in the lab'. In the language of
AdS/CFT, the quantum hardware lives in the `boundary', but using our
nonassociative geometric flow and nonholonomic methods, and the AFCDM, we
can obtain not only the gravity in the boundary but GIF and QGIF theories in
the bulk. In particular, we can construct solutions and study emergent
gravity in the bulk \cite{lbdssv22,lbsvevv22}.

\vskip5pt In 4-d gravity theories, the direct wormhole coupling can arise
from the local and causal dynamics. Such solutions have a nontrivial
topology that can be detectable by the observers who are at asymptotical
infinity. In GR, such configurations are prohibited from topological
censorship at the classical level, but quantum effects enable us to
construct such a configuration. GIF and QGIF theories are related to
topologic and analytical constructions for which the corresponding
conjectures and theorems can be proven, for instance, for the geometric flow
evolution of Riemannian metrics. How and if such results can be formulated
and proven for nonassociative and/ or relativistic geometric flow models it
is not clear. Nevertheless, we established in our partner works \cite%
{partner02,partner03,lbdssv22,lbsvevv22} that we can decouple and solve in
certain general forms physically important systems of nonlinear PDEs. This
can be done under very general assumptions on the respective classes of
generating functions, generating sources and integration functions. The
issues of topology/ stability/ causality etc. have to be solved for certain
special classes of primary geometric data and prescribed nonlinear
symmetries of target solutions. In GR and various MGTs, traversable
wormholes are known to be forbidden by the average null energy condition
(ANEC) \cite{morris88}. In a recent work \cite{wormh19}, the authors
proposed that traversable wormholes can be modelled by turning on a coupling
between the two boundaries of the AdS eternal black hole geometry. Such
nonlocal constructions violate the ANEC, correspond to thermal field double
states and may describe two entangled systems. In our approach, we
considered thermo-geometric doubles states when nonassociative traversable
wormholes describe the fact that scrambled QGIFs in one system can be
restored from the other system; we have to make use of the coupling and the
pre-existing nonassociative entanglement between such quasi-stationary
configurations. This is similar to quantum teleportation and is also related
to the Hayden-Preskill protocol of recovering information from the black
hole \cite{hyden07} but generalized for nonassociative geometric flow
theories. This generalize the nonholonimc geometric approaches elaborated in 
\cite{wormh22,wormh19,wormh19a,wormh21,wormh17,wormh18,wormh13,wormh21}. For
such GIF and QGIF theories, we have to extend the thermodynamic
informational paradigm from the Bekenstein-Hawking, AdS/CFT and holography
to a more general Perelman-type paradigm including quantum thermodynamic
variables, nonassociative geometric flows and MGTs modelled on phase spaces
with star product R-flux deformations.

\section{Discussion and conclusions}

\label{sec5} In this paper, we have formulated the nonassociative geometric
and information flow theory (i.e. the respective GIF and QGIFs, for quantum
models). We studied also possible applications of the anholonomic frame and
connection deformation method, AFCDM, for constructing parametric
off-diagonal solutions of nonholonomic geometric flow evolution and
dynamical equations in nonassociative QGIFs, modified gravity theories,
MGTs, and in nonassociative modifications of general relativity, GR. Such
new theoretical results and innovative geometric and quantum information
methods were applied for constructing and investigating
theoretical-information properties of new classes of nonassociative star
product R-flux deformations of wormhole solutions in relativistic phase
spaces involving both Lorentz manifold/spacetime variables and (co) tangent
momentum like variables. We elaborated on double 4-d wormhole configurations
and, for another class, on higher dimension phase space warped wormhole
solutions; concluded that such wormholes are transversable for
nonassociative qubits; and analyzed perspectives for quantum computer
modelling and lab tests of quantum gravity, QG, MGTs and string gravity, and
nonassociative geometric flow and gravity theories.

\vskip5pt Our main idea, involved in the Hypothesis and Claims C1-C6 from
section \ref{sshypothesis}, is that in fundamental classical and quantum
theories the thermodynamic informational paradigm is based on the concept of
Perelman entropy. This approach is more general and provides better methods
for research than the well-known Bekenstein-Hawking paradigm for the BH
physics, related AdS classical and quantum duality of theories and
geometries, and holographic principles. Basic ingredients from the classical
and quantum information theories have to be revised and extended to include
concepts of Shannon--Perelman entropy for nonassociative GIFs;
nonassociative von Neumann-- / R\'{e}nyi--Perelman entropy; respective
relative and subadditive inequalities; and the concepts of nonassociative
entangled QGIF and Bell states.

\vskip5pt We summarize the main results of this work following Objectives (%
\textbf{Obj1 -- Obj4}) stated in section \ref{ssobjectives}:

\begin{enumerate}
\item Although our approach with nonassociative geometric flows and methods
of constructing parametric solutions builds on the detailed constructions
from partner works \cite{partner01,partner02,partner03,lbdssv22,lbsvevv22},
with nonassociative star product R-flux deformations in string theory and
MGTs \cite{blumenhagen16,aschieri17,szabo19}, the results for \textbf{Obj1}
in section \ref{sec2} are largely independent and provide an introduction
into the theory of nonassociative geometric flows and generalized Perelman
thermodynamics. Necessary proofs are sketched following a corresponding
abstract star-symbolic geometric formalism as we explain in detail in
Appendix \ref{asstables}. In particular, we showed that such models include
nonassociative gravity theories as certain examples of nonholonomic Ricci
solitons encoding nonassociative data in effective sources and shell-adapted
cosmological constants. The nonassociative GIF theory was elaborated using a
synthesis of the concepts of the Shannon and Perelman entropy performing a
thermodynamic information study on nonassociative conditional entropy and
relative entropy sketching proofs for important inequalities involving
strong subadditivity and monotonicity of mutual information. We also derived
the main formulas in nonassociative GIF theory and geometric thermodynamics
of nonholonomic Einstein systems encoded in off-diagonal form and with
effective source star product R-flux deformations.

\item In section \ref{sec3}, we had to reformulate the basic ingredients of
quantum information theory (reviewed in \cite{preskill,witten20,nishioka18})
and formulate the theory of nonassociative QGIFs. This provided solutions
for \textbf{Obj2}, which involve definitions and methods of computations of
density matrices and quantum entropy for QGIFs. In such an approach, the
statistical partition (thermodynamic) generating function (for relativistic
nonholonomic generalizations of Perelman theory and GIFs) is generalized as
the density matrix for a corresponding thermo-geometric theory, where the
geometric flow parameter is identified with temperature. We proved that the
constructions can be performed both in nonassociative geometric and quantum
forms for respective QGIF theories. Using thermo-geometric double states, we
introduced the concept of nonassociative von Neumann-- / R\'{e}nyi--Perelman
entropy and showed how to define and compute parametric star product
deformations of separable and entangled QGIFs and Bell states. Basic
inequalities for entropies of nonassociative Ricci solitons / nonholonomic
Einstein spaces and their effective temperature evolutions were formulated.

\item The first goal of section \ref{sec4} was to apply the AFCDM for
constructing two classes of $\tau$-running quasi-stationary nonassociative
parametric solutions. In this work, we analyze star product R-flux
deformations of phase space configurations defined as pairs of 4-d prime
wormholes, respectively, on base and cofiber spaces, or as higher dimension
warped wormholes. Explicit conditions for defining locally anisotropic
configurations were stated. Such constructions can be related to
corresponding models of nonassociative star deformations of BH and BE
solutions studied in section 5 of \cite{lbsvevv22}. We have described how
QGIF models can be derived from GIF thermodynamic variables and computed
important quantum information s-objects for nonassociative entangled
wormholes, and respectively entangled QGIF and Bell states. This solved the
remaining goals of \textbf{Obj3}. More broadly, in section \ref{sec4} we
have presented a quantum information-theoretical description of
nonassociative geometric flow theory and gravity for a class of solutions
describing phase space wormholes and their star product R-flux deformations.

\item In conclusion, we found that wormholes on Lorentz manifold spacetimes,
in co-fiber subspaces with momentum-like variables and for higher dimension
warped wormhole phase space configurations, are transversable for
nonassociative geometric and information flows encoding star product R-flux
deformations in string theory. This opens additional geometric and quantum
channels considered with the AdS/ CFT framework. Specifically, we defined a
nonassociative QGIF protocol in the bulk that can be used to open and/or
maximize the number of qubits that can be sent through the traversable
wormhole. As a result, we solved the goals of \textbf{Obj4}.
\end{enumerate}

The above objectives support the Hypothesis from section \ref{sshypothesis}
and conclude that it is possible to elaborate transversable wormhole
protocols using nonassociative geometric and quantum information flows when
the thermodynamic information variables are derived as in Perelman
thermodynamics. The AFCDM of constructing generic off-diagonal
exact/parametric solutions for such physical and information-theoretic
systems of nonlinear PDEs is of crucial importance for elaborating thought
experiments, in lab modelling and tests of QG, GR and MGTs. In our research
program on nonassociative QGIF theory and applications in modern physics and
cosmology, we follow the Claims C1-C6 stated in the Introduction and
respective updated queries Q1a-Q4a and Q5 from the previous partner work 
\cite{lbsvevv22}. We shall develop the above directions and report on
progress in future works.

\vskip6pt \textbf{Acknowledgments:} This is the third work after \cite%
{lbdssv22,lbsvevv22} partially supported by a Fulbright senior fellowship
for SV and hosted in 2022-2023 by the physics department at California State
University at Fresno, USA. The second part of the paper provides a bridge to
another research program for 2024 at CAS LMU Munich, Germany. Together with
partner works \cite{partner01,partner02,partner03} this paper extends to
nonassociative geometric and quantum information flows and applications in
modern gravity the results of former visiting programs hosted by professors
H. Dehnen, D. L\"{u}st and J. Moffat. The research of DS was supported in
part by National Science Foundation under Grant No. NSF PHY-1748958.

\appendix
\setcounter{equation}{0} \renewcommand{\theequation}
{A.\arabic{equation}} \setcounter{subsection}{0} 
\renewcommand{\thesubsection}
{A.\arabic{subsection}}

\label{appendixa}

\section{Glossary and tables for generating nonassociative quasi-stationary
solutions \& QGIFs}

In this appendix, we summarize the main concepts, conventions and formulas
for generating exact and parametric quasi-stationary solutions of
nonassociative geometric flow/ Ricci soliton equations and, in particular,
nonholonomic Einstein equations. Details and proofs can be found in the main
text and partner works \cite%
{partner01,partner02,partner03,lbdssv22,lbsvevv22}, see also relevant
previous results and methods on relativistic and modified geometric flows 
\cite{sv20,vbv18}.

\subsection{Concepts, conventions, and notations}

\label{ass1}

We summarize the notations used in this article and partner works and
explain in brief the meaning and scopes of such conventions. There are
provided respective formulas and subsections in the main part of the text
containing details and references. Unfortunately, it is not possible to
formulate a simple system of notations for a series of partner works with
general goals to develop geometric methods for decoupling and integrating
physically important systems of nonlinear PDEs in nonassociative geometric
flow and gravity theories. We have to consider special (2+2)+(2+2)
nonholonomic splitting with canonical decompositions with respect to certain
classes of adapted bases and auxiliary connections which allow a general
decoupling of geometric evolution/ gravitational / matter field equations.
The constructions are performed on (co) tangent bundles and then subjected
to star product R-flux deformations to nonassociative models. So, we have to
introduce different types of indices encoding shell by shell decompositions,
parametric decompositions etc. To elaborate on statistical/ geometric
thermodynamic information models we need to consider (3+1)+(3+1) phase space
splitting; when, for certain purposes with decoupling of equations, it is
convenient to consider adapted index coefficients, but for other cases (for
instance, to define nonlinear symmetries), it is important to develop an
abstract geometric formalism with distinguishing labels. In GR, such a
geometric techniques was elaborated in \cite{misner}. In a more general
form, which allows us to formulate and develop the AFCDM in nonassociative
geometric flow and nonassociative gravity theories, such abstract geometric
and N-adapted constructions are provided in our partner works \cite%
{partner01,partner02,partner03,lbdssv22,lbsvevv22}, see therein references
on associative and, for commutative constructions, \cite{vbv18}.

\subsubsection{Associative and commutative (co) tangent bundles with
nonholonomic 2-d shells}

\begin{enumerate}
\item A base 4-d spacetime $\mathbf{V}$ is modelled as a nonholonomic
Lorentz manifold with (dyadic) 2+2 splitting by a conventional N-connection $%
\mathbf{N}:\ T\mathbf{V}=\ ^{1}hV\oplus \ ^{2}vV,$ defined by a Whitney sum $%
\oplus$ for $\dim (\ ^{1}hV)=\dim (\ ^{1}hV)=2;$ local coordinates $%
(x,y)=(u^{\alpha _{2}})=(x^{i_{1}},y^{a_{2}}),$ for $i_{1},j_{1},...=1,2$
and $a_{2},b_{2},...=3,4;$ convention for indices: $\alpha
_{2}=(i_{1},a_{2});$ enabled with a pseudo-Riemannian metric $%
g(x,y)=\{g_{a_{2}\beta _{2}}(u^{\alpha _{2}})\}$ with local pseudo-Euclidean
signature $(+++-),$ when $u^{4}=y^{4}=t$ is considered as a time like
coordinate and $(x^{i_{1}},y^{3})$ are space like coordinates; the
Levi-Civita, LC, connection $\nabla =\{\Gamma _{\ \beta
_{2}\gamma_{2}}^{\alpha _{2}}\}$ is uniquely defined by $g$ if and only if
the zero torsion and metric compatibility, $\nabla g=0$, conditions; $%
Ric[\nabla ]=\{R_{a_{2}\beta _{2}}(u^{\gamma _{2}})\}$ and $\bigtriangleup
=\nabla ^{2}$ are respectively the Ricci tensor and Laplace operator of a
chosen $\nabla $ (we may write this as a functional dependence $[\nabla ]$);
where $Rs=R$ is the curvatures scalar determined by geometric data $(\mathbf{%
g,}\nabla ).$ The term nonholonomic (equivalently, anholonomic) is used
because the $\oplus $ is defined in general as a non-integrable distribution
characterised by respective nonholonomic (co) frames (in literature being
used respectively terms like tetrads, dual frames, co-frames), $\mathbf{e}%
_{\alpha _{2}}$ and $\mathbf{e}^{\alpha _{2}}.$ The constructions with
coefficients of geometric objects determined with respect to local
coordinate frames, $e_{\alpha _{2}}=\partial _{\alpha _{2}}=\partial
/\partial u^{\alpha _{2}}$ and $e^{\alpha _{2}}=du^{\alpha _{2}}$, are not
N-adapted.

\item We use boldface symbols and write, for instance, $\mathbf{N,g},\mathbf{%
D}$ (for an arbitrary/affine linear connection, called distinguished
connection, d-connection) and a corresponding Ricci tensor, called d-tensor, 
$\mathbf{R}ic[\mathbf{D}]$ etc., if a spacetime/ phase space is enabled with
a N-connection structure and the geometric objects, d-objects, are adapted
(i.e. preserve under parallel transports) the N-connection structure.

\item For relativistic flows on $\mathbf{V}$ parameterized by a real $\tau $%
-parameter, $0\leq \tau \leq \tau _{0}$ (it can be a temperature like one
and take continuous or discrete values), there are consider families of
geometric objects like $g(\tau ,x,y)=\{g_{a_{2}\beta _{2}}(\tau ,u^{\alpha
_{2}})\};$ in brief, we write respectively $g(\tau )=\{g_{a_{2}\beta
_{2}}(\tau )\},\nabla (\tau )=\{\Gamma _{\ \beta _{2}\gamma _{2}}^{\alpha
_{2}}(\tau )\}, Ric(\tau ),$ $\mathbf{N}(\tau ),\mathbf{D}(\tau )$ etc.,
where $\nabla\lbrack \mathbf{g}](\tau )$ denotes a $\tau $-family of
Levi-Civita, LC, connections. In linear approximations for Riemannian
metrics, the geometric flow equations (\ref{heq}) \ are diffusion type ones
being determined by the operator $\bigtriangleup $. For Lorentzian metrics
and linearized equations, we obtain the wave type d' Alambert operator $%
\square $; see explanations for formulas and (\ref{heqps}).

\item A 8-d phase space $\mathcal{M\simeq }T\mathbf{V}$ with velocity type
fiber coordinates $v=\{v^{a_{3}},v^{a_{4}}\}$ is modelled on a tangent
Lorentz bundle, $T\mathbf{V},$ on a base spacetime manifold $\mathbf{V.}$
Local coordinates and abstract/coordinate indices are adapted to a
conventional dyadic splitting of type (2+2)+(2+2), when $u^{%
\alpha_{4}}=(x^{i_{1}},y^{a_{2}},v^{a_{3}},v^{a_{4}})=(x^{i_{3}},v^{a_{4}})=(x^{i_{2}},v^{a_{3}},v^{a_{4}}) 
$ and $u^{\alpha
_{3}}=(x^{i_{1}},y^{a_{2}},v^{a_{3}})=(x^{i_{3}})=(x^{i_{2}},v^{a_{3}});$
where indices run respective values $a_{4},b_{4},...=7,8,$ for $%
a_{3},b_{3},...=5,6,$ with $i_{2}=(i_{1},a_{2})=1,2,3,4$ and $%
i_{3}=(i_{2},a_{a})=1,2,3,4,5,6$ etc. This defines a dyadic shell structure,
s-structure, of shells $s=1,2,3,4.$ It can be defined geometrically as a
nonholonomic dyadic distribution determined by a respective N-connection
structure $\ _{s}\mathbf{N}=\{\ N_{\ i_{s}a_{s}}(u)\}$ (\ref{ncon}) for any
point $u=(x,v)=\ (\ _{1}x,\ _{2}y,\ _{3}v,\ _{4}v)\in \mathbf{TV.}$ Such a
dyadic N-connection splitting allows to construct s-adapted decompositions
of geometric objects, called s-objects and fundamental geometric evolution/
field equations on $\mathcal{M}.$ We emphasize that a phase space is with a
nonholonomic s-shell splitting if we write $\ _{s}\mathcal{M}.$

\item For elaborating nonassociative geometric and (quantum) information
models in this work, we shall work mainly on a phase space $\ ^{\shortmid }%
\mathcal{M\simeq }T^{\ast }\mathbf{V}$ with momentum like coordinates on
co-fibers, $p=\{p_{a_{3}},p_{a_{4}}\},$ where $p_{8}=E$ is an energy type
phase space coordinate. We use a label "$\ ^{\shortmid }$" to distinguish if
necessary the geometric objects on $\ ^{\shortmid }\mathcal{M}$ from those
on $\mathcal{M}.$ For instance, the local coordinates on shell $s=4$ of a $\
_{s}^{\shortmid }\mathcal{M}$ are labeled in the form $\ ^{\shortmid
}u^{\alpha _{4}}=(x^{i_{1}}, y^{a_{2}},p_{a_{3}}, p_{a_{4}})= (\
^{\shortmid}x^{i_{3}},p_{a_{4}})=(x^{i_{2}},p_{a_{3}},p_{a_{4}}),$ or in
abstract form for any point $u=(x,p)=\ ^{\shortmid }u=(\ _{1}x,\ _{2}y,\
_{3}p,\ _{4}p)\in \mathbf{T}^{\ast }\mathbf{V.}$ The nonholonomic
s-structure is defined by a N-connection $\ _{s}^{\shortmid }\mathbf{N}=\{\
^{\shortmid }N_{\ i_{s}a_{s}}(\ ^{\shortmid }u)\}$ (\ref{ncon}) and
respective N-linear (co) frame structures $\ ^{\shortmid }\mathbf{e}_{\alpha
_{s}}[\ ^{\shortmid }N_{\ i_{s}a_{s}}]$ (\ref{nadapbdsc}) and $\ ^{\shortmid}%
\mathbf{e}^{\alpha _{s}}[\ ^{\shortmid }N_{\ i_{s}a_{s}}]$ (\ref{nadapbdss}%
), see also footnote \ref{fnshells}.

\item Metric s-structures (\ref{sdm}), s-metrics, on $\ _{s}^{\shortmid }%
\mathcal{M}$, are written for dyadic shell decompositions, 
\begin{equation}
\ _{s}^{\shortmid }\mathbf{g}=(h_{1}\ ^{\shortmid }\mathbf{g}, v_{2}\
^{\shortmid }\mathbf{g},\ c_{3}\ ^{\shortmid }\mathbf{g}, c_{4}\ ^{\shortmid
}\mathbf{g})=\ ^{\shortmid }\mathbf{g}_{\alpha _{s}\beta _{s}}(\
_{s}^{\shortmid }u)\ \ ^{\shortmid }\mathbf{e}^{\alpha _{s}}\otimes _{s}\
^{\shortmid }\mathbf{e}^{\beta _{s}}= \{\ ^{\shortmid }\mathbf{g}_{\alpha
_{s}\beta _{s}}=(\ ^{\shortmid }\mathbf{g}_{i_{1}j_{1}}, \ ^{\shortmid }%
\mathbf{g}_{a_{2}b_{2}},\ \ ^{\shortmid }\mathbf{g}^{a_{3}b_{3}},\ \
^{\shortmid }\mathbf{g}^{a_{4}b_{4}})\}.  \notag
\end{equation}%
In general, such a s-metric is generic off-diagonal if the coefficients are
re-defined with respect to coordinate s-bases.

\item A s--connection $\ _{s}^{\shortmid }\mathbf{D}=(h_{1}\ ^{\shortmid
}D,v_{2}\ ^{\shortmid }D,c_{3}\ ^{\shortmid }D,c_{4}\ ^{\shortmid }D)$ is a
linear connection preserving under parallelism the N--connection splitting $%
\ _{s}^{\shortmid }\mathbf{N}$ (\ref{ncon}). It defines a covariant
N--adapted derivative $\ _{s}^{\shortmid }\mathbf{D}_{\mathbf{X}}\
_{s}^{\shortmid }\mathbf{Y}$ of a s--vector field $\ _{s}^{\shortmid }%
\mathbf{Y}$ in the direction of a d--vector $\ _{s}^{\shortmid }\mathbf{X}.$
With respect to N-adapted frames (\ref{nadapbdsc}) and (\ref{nadapbdss}),
any $\ _{s}^{\shortmid }\mathbf{D}$ can be computed as in GR and/or metric
affine gravity but with the coefficients defined by h-, v- and c-indices, 
\begin{eqnarray*}
\ _{s}^{\shortmid }\mathbf{D} &=&\{\ ^{\shortmid }\mathbf{\Gamma }_{\ \alpha
_{s}\beta _{s}}^{\gamma _{s}}=(\ ^{\shortmid }L_{j_{s}k_{s}}^{i_{s-1}},\
^{\shortmid }\acute{L}_{b_{s}k_{s-1}}^{a_{s}};\ ^{\shortmid }\acute{C}%
_{j_{s-1}c_{s}}^{i_{s-1}},\ ^{\shortmid }C_{b_{s}c_{s}}^{a_{s}})\}, 
\mbox{
where } \\
&& h\ ^{\shortmid }D = (\ ^{\shortmid}L_{j_{s-1}k_{s-1}}^{i_{s-1}}, \
^{\shortmid }\acute{L} _{b_{s}k_{s-1}}^{a_{s}})\mbox{ and } v\ ^{\shortmid
}D=(\ ^{\shortmid }\acute{C}_{j_{s-1}c_{s}}^{i_{s-1}},\ ^{\shortmid
}C_{b_{s}c_{s}}^{a_{s}}),\mbox{ for } s=2,3,4,
\end{eqnarray*}%
which additionally split into (co) vertical components and with inverted
indices on respective shells. By definition, any s--connection is
characterized by three fundamental geometric s-objects, which (by
corresponding definitions in abstract form) are: 
\begin{eqnarray*}
\ _{s}^{\shortmid }\mathcal{T}(\ _{s}^{\shortmid }\mathbf{X,}\
_{s}^{\shortmid }\mathbf{Y})&:=& \ _{s}^{\shortmid } \mathbf{D}_{\
_{s}^{\shortmid }\mathbf{X}}\ _{s}^{\shortmid }\mathbf{Y} - \
_{s}^{\shortmid }\mathbf{D}_{\ _{s}^{\shortmid }\mathbf{Y}}\ _{s}^{\shortmid
}\mathbf{X}- [\ _{s}^{\shortmid }\mathbf{X,\ _{s}^{\shortmid }Y}], 
\mbox{
torsion s-tensor, s-torsion}; \\
\ _{s}^{\shortmid }\mathcal{R}(\ _{s}^{\shortmid }\mathbf{X,}\
_{s}^{\shortmid }\mathbf{Y})&:=&\ _{s}^{\shortmid }\mathbf{D}_{\
_{s}^{\shortmid }\mathbf{X}}\ _{s}^{\shortmid }\mathbf{D}_{\ _{s}^{\shortmid}%
\mathbf{Y}} - \ _{s}^{\shortmid }\mathbf{D}_{\ _{s}^{\shortmid }\mathbf{Y}}\
_{s}^{\shortmid }\mathbf{D}_{\ _{s}^{\shortmid }\mathbf{X}}- \
_{s}^{\shortmid }\mathbf{D}_{\mathbf{[\ _{s}^{\shortmid }X,\
_{s}^{\shortmid}Y]}}, \mbox{ curvature s-tensor, s-curvature}; \\
\mathcal{Q}(\ _{s}^{\shortmid }\mathbf{X})&:=&\ _{s}^{\shortmid }\mathbf{D}%
_{\ _{s}^{\shortmid }\mathbf{X}}\ _{s}^{\shortmid }\mathbf{g,} %
\mbox{nonmetricity s-fiels, s-nonmetricity},
\end{eqnarray*}%
which for (2+2)-splitting are called respectively distinguished, d, tensors.
The N-adapted coefficients of such geometric s-objects are computed by
considering $\ _{s}^{\shortmid }\mathbf{X}=\ ^{\shortmid }\mathbf{e}%
_{\alpha_{s}}$ and $\ _{s}^{\shortmid }\mathbf{Y}=\ ^{\shortmid }\mathbf{e}%
_{\beta _{s}},$ defined by (\ref{nadapbdsc}), and respective h-v-c-splitting
for $\ _{s}^{\shortmid }\mathbf{D}=\{\ ^{\shortmid }\mathbf{\Gamma }_{\
\alpha _{s}\beta _{s}}^{\gamma _{s}}\},$ see details in \cite{vbv18}, 
\begin{eqnarray*}
\ _{s}^{\shortmid }\mathcal{T} &=&\{\ ^{\shortmid }\mathbf{T}_{\ \alpha
_{s}\beta _{s}}^{\gamma _{s}}=\left( \ ^{\shortmid }T_{\
j_{s-1}k_{s-1}}^{i_{s-1}},\ ^{\shortmid }T_{\ j_{s-1}a_{s}}^{i_{s-1}},\
^{\shortmid }T_{\ j_{s-1}i_{s-1}}^{a_{s}},\ ^{\shortmid }T_{\
b_{s}i_{s-1}}^{a_{s}},\ ^{\shortmid }T_{\ b_{s}c_{s}}^{a_{s}}\right) \}; \\
\ _{s}^{\shortmid }\mathcal{R} &\mathbf{=}&\mathbf{\{\ ^{\shortmid }R}_{\
\beta _{s}\gamma _{s}\delta _{s}}^{\alpha _{s}}\mathbf{=}\left( \
^{\shortmid }R_{\ h_{s-1}j_{s-1}k_{s-1}}^{i_{s-1}}\mathbf{,}\ ^{\shortmid
}R_{\ b_{s}j_{s-1}k_{s-1}}^{a_{s}}\mathbf{,}\ ^{\shortmid }R_{\
h_{s-1}j_{s-1}a_{s}}^{i_{s-1}}\mathbf{,}\ ^{\shortmid }R_{\
b_{s}j_{s-1}a_{s}}^{c_{s}}\mathbf{,}\ ^{\shortmid }R_{\
h_{s-1}b_{s}a_{s}}^{i_{s-1}},\ ^{\shortmid }R_{\
b_{s}e_{s}a_{s}}^{c_{s}}\right) \mathbf{\};} \\
\ \ _{s}^{\shortmid }\mathcal{Q} &=&\mathbf{\{\ ^{\shortmid }Q}_{\ \alpha
_{s}\beta _{s}}^{\gamma _{s}}=\ ^{\shortmid }\mathbf{D}^{\gamma _{s}}\
^{\shortmid }\mathbf{g}_{\alpha _{s}\beta _{s}}=(\ ^{\shortmid }Q_{\
i_{s-1}j_{s-1}}^{k_{s-1}},\ ^{\shortmid }Q_{\ i_{s-1}j_{s-1}}^{c_{s}},\
^{\shortmid }Q_{\ a_{s}b_{s}}^{k_{s-1}},\ ^{\shortmid }Q_{\
a_{s}b_{s}}^{c_{s}})\},
\end{eqnarray*}%
for $s=2,3,4.$ Omitting labels "$\ ^{\shortmid }$" and considering
dependencies of geometric s-objects on coordinates $u^{\alpha
_{4}}=(x^{i_{1}},y^{a_{2}},v^{a_{3}},v^{a_{4}}),$ we can define similar
s-metric, s-connection structures and respective fundamental geometric
objects on $\ _{s}\mathcal{M}.$

\item Using a s-metric $\ _{s}^{\shortmid }\mathbf{g}$ (\ref{sdm}), we can
define two important linear connection structures : 
\begin{equation*}
(\ _{s}^{\shortmid }\mathbf{g,\ _{s}^{\shortmid }N})\rightarrow \left\{ 
\begin{array}{cc}
\ _{s}^{\shortmid }\mathbf{\nabla :} & \ _{s}^{\shortmid }\mathbf{\nabla }\
_{s}^{\shortmid }\mathbf{g}=0;\ _{\nabla }^{\shortmid }\mathcal{T}=0,\ %
\mbox{\ the LC--connection }; \\ 
\ _{s}^{\shortmid }\widehat{\mathbf{D}}: & 
\begin{array}{c}
\ _{s}^{\shortmid }\widehat{\mathbf{Q}}=0;\ h_{1}\ ^{\shortmid }\widehat{%
\mathcal{T}}=0,v_{2}\ ^{\shortmid }\widehat{\mathcal{T}}=0,\ h_{1}v_{2}\
^{\shortmid }\widehat{\mathcal{T}}\neq 0,h_{1}c_{3}\ ^{\shortmid }\widehat{%
\mathcal{T}}\neq 0,h_{1}c_{4}\ ^{\shortmid }\widehat{\mathcal{T}}\neq 0, \\ 
v_{2}c_{3}\ ^{\shortmid }\widehat{\mathcal{T}}\neq 0,v_{2}c_{4}\ ^{\shortmid
}\widehat{\mathcal{T}}\neq 0,c_{2}c_{3}\ ^{\shortmid }\widehat{\mathcal{T}}%
\neq 0\mbox{ the canonical
s-connection}.%
\end{array}%
\end{array}%
\right.
\end{equation*}%
We use "hat" labels for geometric s-objects written in such a canonical
form. For any $\ _{s}^{\shortmid }\widehat{\mathbf{D}},$ we can define and
compute the canonical fundamental geometric s-objects, for instance, $\
_{s}^{\shortmid }\widehat{\mathcal{R}}=\{\ ^{\shortmid }\widehat{\mathbf{R}}%
_{\ \beta _{s}\gamma _{s}\delta _{s}}^{\alpha _{s}}\}$ and $\
_{s}^{\shortmid }\widehat{\mathcal{T}}=\{\ ^{\shortmid }\widehat{\mathbf{T}}%
_{\ \alpha _{s}\beta _{s}}^{\gamma _{s}}\}$. In a similar form, we can
compute the fundamental geometric objects defined by $\ _{s}^{\shortmid }%
\mathbf{\nabla ,}$ for instance, $\ _{\nabla }^{\shortmid }\mathcal{R}=\{\
_{\nabla }^{\shortmid }R_{\ \beta \gamma \delta }^{\alpha }\}$ and $\
^{\shortmid }\mathcal{R}ic[\ ^{\shortmid }\nabla ]=\{\
^{\shortmid}R_{a_{s}\beta _{s}}(\ ^{\shortmid }u^{\gamma _{s}})\}$ (in such
cases, boldface indices are not used). Because both linear connections $\
_{s}^{\shortmid }\mathbf{\nabla }$ and $\ _{s}^{\shortmid }\widehat{\mathbf{D%
}}$ are defined by the same s-metric $\ _{s}^{\shortmid }\mathbf{g,}$ we can
define a canonical distortion relation $\ _{s}^{\shortmid }\widehat{\mathbf{D%
}}=\ _{s}^{\shortmid }\nabla \lbrack \ _{s}^{\shortmid }\mathbf{g]}+ \
_{s}^{\shortmid }\widehat{\mathbf{Z}}[\ _{s}^{\shortmid }\mathbf{g,}\
_{s}^{\shortmid }\mathbf{N],}$ where the canonical distortion s-tensor $\
_{s}^{\shortmid }\widehat{\mathbf{Z}}[\mathbf{g]}$ is an algebraic
combination of s-indices of the nonholonomic induced canonical s-torsion
structure $\ _{s}^{\shortmid}\widehat{\mathcal{T}}[\ _{s}^{\shortmid }%
\mathbf{g,}\ _{s}^{\shortmid }\mathbf{N].}$ Using such a distortion
relation, we can compute respective canonical distortions of fundamental
geometric s-objects relating, for instance, $\ _{\nabla }^{\shortmid }R_{\
\beta \gamma \delta }^{\alpha }$ and $\ ^{\shortmid }\widehat{\mathbf{R}}_{\
\beta _{s}\gamma _{s}\delta _{s}}^{\alpha _{s}}.$ In our works, we prefer to
work with the canonical s-connection $\ _{s}^{\shortmid }\widehat{\mathbf{D}}
$ because it allows a general decoupling and integration of (modified)
Einstein equations in terms of generating functions, generating sources and
integration functions. If necessary, from such general classes of solutions
determined by generic off-diagonal metrics (depending in principle on all
spacetime and phase space coordinates) and generalized (non) linear
connections, we can extract LC-configurations by restricting the class of
generating and integration functions to configurations when $\
_{s}^{\shortmid }\widehat{\mathbf{Z}}=0$ and $\ _{s}^{\shortmid }\widehat{%
\mathbf{D}}\rightarrow \ _{s}^{\shortmid}\nabla .$ This is the main claim of
the AFCDM for constructing exact and parametric solutions in MGTs and GR 
\cite{vbv18}. It works also in nonassociative gravity \cite%
{partner01,partner02,partner03,lbdssv22}. In brief, this geometric method of
generating off-diagonal quasi-stationary solutions is summarized in appendix %
\ref{asstables}.

\item Nonholonomic geometric flows of s-metrics $\ _{s}^{\shortmid }\mathbf{g%
}(\tau )=\{\ ^{\shortmid }\mathbf{g}_{\alpha _{s}\beta _{s}}(\tau )\}$ on a
phase space $_{s}\mathcal{M}$ can be described in canonical s-adapted form
for geometric data $[\ _{s}^{\shortmid }\mathbf{g}(\tau ),\ _{s}^{\shortmid }%
\mathbf{N(\tau ),}\ _{s}^{\shortmid }\widehat{\mathbf{D}}(\tau )]$ as
solutions of generalized R. Hamilton equations, $\frac{\partial \
_{s}^{\shortmid }\mathbf{g}(\tau )}{\partial \tau }= -2\ _{s}^{\shortmid }%
\widehat{\mathcal{R}}ic[\ _{s}^{\shortmid }\mathbf{g},\ _{s}^{\shortmid }%
\widehat{\mathbf{D}}](\tau )+\ldots$, where dots denote possible additional
terms determined by quadratic curvature terms, normalization functions etc.
In a series of papers devoted respectively to (nonholonomic) relativistic
geometric/ generalized Finsler-Lagrange-Hamilton flows, emergent gravity,
and quantum methods in physics and information theory \cite%
{sv20,lbsvevv22,partner03,lbdssv22}, we prove that such equations can be
derived following s-adapted variational and/or abstract nonholonomic
geometric principles from generalized Perelman functionals $\
_{s}^{\shortmid }\widehat{\mathcal{F}}(\tau )=\ _{s}^{\shortmid }\widehat{%
\mathcal{F}}(\tau , \ _{s}^{\shortmid }\mathbf{g}(\tau ),\ _{s}^{\shortmid }%
\widehat{\mathbf{D}}(\tau ),\ _{s}^{\shortmid }\widehat{\mathbf{R}}s(\tau),
\ _{s}^{\shortmid }\widehat{f}(\tau ))$ and $\ _{s}^{\shortmid }\widehat{%
\mathcal{W}}(\tau )=\ _{s}^{\shortmid }\widehat{\mathcal{W}}(\tau, \
_{s}^{\shortmid }\mathbf{g}(\tau ),\ _{s}^{\shortmid }\widehat{\mathbf{D}}%
(\tau ),\ _{s}^{\shortmid }\widehat{\mathbf{R}}s(\tau ), \ _{s}^{\shortmid }%
\widehat{f}(\tau )),$ where $\ _{s}^{\shortmid }\widehat{f}(\tau )=\
_{s}^{\shortmid }\widehat{f}(\tau ,\ ^{\shortmid }u)$ are shell
normalization functions determining together with $\ _{s}^{\shortmid }%
\mathbf{g}(\tau )$ certain $\tau $-families of integration measures on $_{s}%
\mathcal{M}$.

\item Above mentioned system of nonholonomic geometric flow evolution
equations can be written for certain types of nonholonomic constraints on $%
\tau $-running distortions s-tensors and possible $\tau $-families of matter
fields can be written in the forms $\ ^{\shortmid }\widehat{\mathbf{R}}_{\ \
\gamma _{s}}^{\beta _{s}}(\tau )= {\delta }_{\ \ \gamma _{s}}^{\beta _{s}}\
_{s}^{\shortmid }\Im (\tau )$, where $\ _{s}^{\shortmid }\Im (\tau )$ are
effective generating sources for respective shell $s$. For some general
assumptions and re-definitions of generating functions and generating
sources, there are well-defined nonlinear symmetries transforming $\
_{s}^{\shortmid }\Im (\tau )\rightarrow \ _{s}^{\shortmid }\Lambda (\tau ),$
where $\ _{s}^{\shortmid }\Lambda (\tau)$ are effective $\tau $-running
cosmological constants not depending on $\ ^{\shortmid }u$. In result, we
can consider equivalent systems of type $\widehat{\mathcal{R}}ic[\tau ,\
_{s}^{\shortmid }\mathbf{g}(\tau ),\ _{s}^{\shortmid }\widehat{\mathbf{D}}%
(\tau )]= \ _{s}^{\shortmid }\Lambda (\tau )\ _{s}^{\shortmid }\mathbf{g}%
(\tau )$. Other important geometric and physical examples are defined for
self-similar configurations for $\ _{s}^{\shortmid }\mathbf{g}(\tau )$ and a
fixed parameter $\tau =\tau _{0}$, we obtain nonholonomic Ricci soliton
equations $\ ^{\shortmid }\widehat{\mathbf{R}}_{\ \alpha _{s}\beta _{s}}+ \
^{\shortmid }\widehat{\mathbf{D}}_{\alpha _{s}}\ ^{\shortmid }\widehat{%
\mathbf{D}}_{\beta _{s}} \ _{s}^{\shortmid }\varpi (\ _{s}^{\shortmid }u)= \
_{s}^{\shortmid}\lambda \ _{s}^{\shortmid }\mathbf{g},$ where $%
_{s}^{\shortmid }\varpi $ are smooth potential functions on every shell $%
s=1,2,3,4$ and $\ \ _{s}^{\shortmid }\lambda =const.$ Above systems of
nonlinear PDEs can be decoupled and integrated in various general forms
using the AFCDM, see details, proofs and reviews of results in \cite{vbv18}.
Various classed of solutions are determined by generic off-diagonal
s-metrics and generalized s-connections in MGTs and nonholonomic geometric
flow theories. For additional nonholonomic constraints, we can extract
LC-configurations and generate new classes of solutions in GR or MGTs. There
were constructed various quasi-stationary configurations (for instance,
black ellipsoid, toroid BHs, locally anisotropic wormholes etc.) and
elaborated local anisotropic cosmological scenarios.

\item It should be noted that the W-functional $\ _{s}^{\shortmid }\widehat{%
\mathcal{W}}(\tau )$ has properties of "minus" entropy. Generalizing in
relativistic form on phase spaces the Perelman constructions \cite{perelman1}
we can characterize nonholonomic s-adapted geometric flows by corresponding
models of statistical/ geometric thermodynamics. For using the method
described with respect formulas (21) and (22) in \cite{lbdssv22} for
canonical s-flows with $\ ^{\shortmid }\widehat{\mathbf{D}}_{\alpha _{s}}$
determined by a solution of (nonholonomic) geometric flows, we introduce a
partition function (statistical thermodynamic generating function) 
\begin{equation*}
\ _{s}\widehat{\mathcal{Z}}(\tau )=\exp [\int_{\ _{s}^{\shortmid }\widehat{%
\Xi }}[-\ _{s}^{\shortmid }\widehat{f}+4]\ \left( 4\pi \tau \right)
^{-4}e^{-\ _{s}^{\shortmid }\widehat{f}}\ ^{\shortmid }\delta \ ^{\shortmid }%
\mathcal{V}(\tau ),
\end{equation*}%
where the volume form $d\ \ ^{\shortmid }\mathcal{V}ol(\tau )$ is defined by
formula (\ref{volform}), $_{s}^{\shortmid }\widehat{\Xi }$ labels a closed
phase space hypersurface defined for an additional (3+1)+(3+1) splitting,
together with a (2+2)+(2+2) one. Such a double nonholonomic fibration on
phase space $\ _{s}^{\shortmid }\mathcal{M}$ allows us to provide proofs and
computations\footnote{%
variational and abstract geometric procedures from \cite%
{perelman1,kleiner06,morgan06,cao06}, were generalized in nonholonomic form
in \cite{sv20} using $\ _{s}^{\shortmid }\widehat{\mathcal{W}}(\tau )$ and $%
\ _{s}\widehat{\mathcal{Z}}(\tau )$} of such relativistic thermodynamic
variables: 
\begin{eqnarray*}
\mbox{ average energy },\ _{s}^{\shortmid }\widehat{\mathcal{E}}(\tau )\
&=&-\tau ^{2}\int_{\ _{s}^{\shortmid }\widehat{\Xi }}\ \left( 4\pi \tau
\right) ^{-4}\left( \ _{s}^{\shortmid }\widehat{\mathbf{R}}sc+|\
_{s}^{\shortmid }\widehat{\mathbf{D}}\ _{s}^{\shortmid }\widehat{f}|^{2}-%
\frac{4}{\tau }\right) e^{-\ _{s}^{\shortmid }\widehat{f}}\ ^{\shortmid
}\delta \ ^{\shortmid }\mathcal{V}(\tau ); \\
\mbox{ entropy },\ \ _{s}^{\shortmid }\widehat{S}(\tau ) &=&-\int_{\
_{s}^{\shortmid }\widehat{\Xi }}\left( 4\pi \tau \right) ^{-4}\left( \tau (\
_{s}^{\shortmid }\widehat{\mathbf{R}}sc+|\ _{s}^{\shortmid }\widehat{\mathbf{%
D}}\ _{s}^{\shortmid }\widehat{f}|^{2})+\ _{s}^{\shortmid }\tilde{f}%
-8\right) e^{-\ _{s}^{\shortmid }\widehat{f}}\ ^{\shortmid }\delta \
^{\shortmid }\mathcal{V}(\tau ); \\
\mbox{ fluctuation },\ _{s}^{\shortmid }\widehat{\sigma }(\tau ) &=&2\ \tau
^{4}\int_{\ _{s}^{\shortmid }\widehat{\Xi }}\left( 4\pi \tau \right) ^{-4}|\
\ ^{\shortmid }\widehat{\mathbf{R}}_{\alpha _{s}\beta _{s}}+\ ^{\shortmid }%
\widehat{\mathbf{D}}_{\alpha _{s}}\ ^{\shortmid }\widehat{\mathbf{D}}_{\beta
_{s}}\ \ _{s}^{\shortmid }\widehat{f}-\frac{1}{2\tau }\mathbf{g}_{\alpha
_{s}\beta _{s}}|^{2}e^{-\ _{s}^{\shortmid }\widehat{f}}\ \ ^{\shortmid
}\delta \ ^{\shortmid }\mathcal{V}(\tau ).
\end{eqnarray*}%
The data $\left[ \ _{s}\widehat{\mathcal{Z}}(\tau ),\ _{s}^{\shortmid }%
\widehat{\mathcal{E}}(\tau ),\ _{s}^{\shortmid }\widehat{S}(\tau ), \
_{s}^{\shortmid }\widehat{\sigma }(\tau )\right] $ define a relativistic
thermodynamic paradigm for nonholonomic geometric flows on phase spaces,
which is more general than that stated in GR by Bekenstein and Hawking \cite%
{bek1,bek2,haw1,haw2} for the BH solutions and certain generalizations, when
there are involved conventional horizons/ duality conditions/ holographic
configurations. The generalized Perelman thermodynamic variables can be
computed even for generic off-diagonal solutions in GR which do not involve
hypersurface configurations and the Bekenstein-Hawking paradigm is not valid.

\item For discrete values of geometric flow/ temperature parameter ($\tau _{%
\underline{a}},$ for $\underline{a}=1,2,...,\check{r}$) , we can consider
geometric information flow systems (GIFs, $A_{_{\underline{a}}}\simeq \
_{s}^{\shortmid }A_{_{\underline{a}}})$ defined by nonholonomic phase space
Ricci solitons characterized by respective canonical partition functions $\
_{s}^{\shortmid }\widehat{Z}_{\underline{a}}=\ _{s}^{\shortmid }\widehat{Z}%
[\ ^{\shortmid }\mathbf{g}(\tau _{\underline{a}})]$ (\ref{spf}), with $\
_{s}^{\shortmid }\widehat{Z}=\sum_{\underline{a}}\ _{s}^{\shortmid }\widehat{%
Z}_{\underline{a}}.$ Respectively we can compute the thermodynamic values $\
^{\shortmid }\widehat{\mathcal{E}}_{\underline{a}}^{\star }:=\
_{s}^{\shortmid }\widehat{\mathcal{E}}(\tau _{\underline{a}})$ and $\
^{\shortmid }\widehat{\mathcal{S}}_{\underline{a}}:=\ _{s}^{\shortmid }%
\widehat{\mathcal{S}}(\tau _{\underline{a}}).$ We can elaborate on models of
associative and commutative nonholonomic QGIF systems $\mathcal{A}_{_{%
\underline{a}}}\simeq \ _{s}^{\shortmid }\mathcal{A}_{_{\underline{a}}}$
considering respective quantum states determined by canonical vector states
and/or canonical density matrices, 
\begin{eqnarray*}
|\widehat{\Psi } > &=&\ _{s}^{\shortmid }\widehat{Z}^{-1/2}\sum_{\underline{a%
}}e^{-\ ^{\shortmid }\widehat{\mathcal{E}}_{\underline{a}}/2\tau }|\psi _{%
\mathcal{A}}^{\underline{a}}>\otimes |\psi _{\mathcal{B}}^{\underline{a}}>,
\\
\ ^{\shortmid }\widehat{\rho }_{\mathcal{A}} &=&\ _{s}^{\shortmid }\widehat{%
\rho }=\ _{s}^{\shortmid }\widehat{Z}^{-1}\sum_{\underline{a}}e^{-\
^{\shortmid }\widehat{\mathcal{E}}_{\underline{a}}/\tau }|\widehat{\psi }_{%
\mathcal{A}}^{\underline{a}}>\otimes <\widehat{\psi }_{\mathcal{A}}^{%
\underline{a}}|=\ _{s}^{\shortmid }\widehat{Z}^{-1}e^{-\widehat{\boldsymbol{H%
}}_{\mathcal{A}}/\tau },
\end{eqnarray*}%
where $\widehat{\boldsymbol{H}}_{\mathcal{A}}|\widehat{\psi }_{\mathcal{A}}^{%
\underline{a}}>=\ ^{\shortmid }\widehat{\mathcal{E}}_{\underline{a}}^{\star
}|\widehat{\psi }_{\mathcal{A}}^{\underline{a}}>.$ Nonholonomic canonical
thermo-field, thermo-geometric and QGIFs can be formulated for continuous
value of $\tau $ using $\ ^{\shortmid }\widehat{\rho }_{\mathcal{A}}=\
_{s}^{\shortmid }\widehat{\rho }$ and defining $\ ^{\shortmid }\widehat{%
\mathcal{S}}_{\mathcal{A}}=-tr_{\mathcal{A}}[\ ^{\shortmid }\widehat{\rho }_{%
\mathcal{A}}\log \ ^{\shortmid }\widehat{\rho }_{\mathcal{A}}]$ as in (\ref%
{entanglentrcan}). Such associative and commutative GIF and QGIF theories
and applications were studied in \cite{sv20,lbsvevv22}. They allow to
introduce, compute and exploits new concepts of nonholonomic classical and
quantum conditional and relative entropy, nonholonomic geometric and quantum
entanglement etc. Certain associative and commutative geometric evolution
and dynamical field theories consist the $\kappa ^{0}$-part of
nonassociative GIF and QGIF models elaborated and studied in this work.
\end{enumerate}

\subsubsection{Nonassociative star product R-flux deformed phase spaces and
nonholonomic splitting}

The nonholonomic geometric and quantum constructions outlined in previous
subsection can be generalized for nonassociative and noncommutative
geometric and physical theories with star product R-flux deformations
studied in \cite{blumenhagen16,aschieri17} and, in nonholonomic s-adapted
forms, in partner works \cite{partner01,partner02,partner03,lbsvevv22}.

\begin{enumerate}
\item The noholonomic s-adapted \textbf{star product} $\star _{s}$ (\ref%
{starpn}) involves in definition s-adapted N-elongated s-derivati\-ves $\
^{\shortmid }\mathbf{e}_{i_{s}}$ (\ref{nadapbdsc}) when the original
definitions for star products $\star $ where in terms of coordinate bases 
\cite{blumenhagen16,aschieri17}. Our approach developed nonholonomic
s-adapted 2+2+... splitting allows to decouple and integrate in certain
general forms physically important systems of nonlinear PDEs and develop the
AFCDM \cite{vbv18}. We use a label $"s"$ on a spacetime/ phase space and/or
a geometric s-object (even writing $\otimes _{s}$ and $\star _{s}$ instead
of "standard" $\otimes _{s}$ and $\star _{s}$) in order to emphasize that
the geometric/ physical constructions involve a shell dyadic decomposition
with corresponding s-adapting. For nonassociative deformations with a $\star
_{s},$ we put star labels (they can be left/ right and/or up/low) for
respective symbols. A star product transform a standard phase space into a
respective nonassociative one, which is written as $_{s}\mathcal{M}%
\rightarrow \ _{s}^{\star }\mathcal{M},$ $_{s}\mathcal{M}\rightarrow \ _{s}%
\mathcal{M}^{\star }.$ To emphasize that the constructions are performed on
dual phase spaces with real momentum coordinates $p_{a}$ we write, for
instance, $_{s}\mathcal{M}\rightarrow \ _{s}^{\shortmid }\mathcal{M}^{\star
}.$ The definition of star product involves also the complex unity $i,$ when 
$i^{2}=-1,$ and, in certain geometric and physical models, there are used
complex momentum coordinates $\ ^{\shortparallel }p_{a}:=ip_{a},$ for
instance, see \cite{blumenhagen16,aschieri17,partner01,partner02}. This is
natural for elaborating quantum mechanical and QGT theories. In such cases,
in our works on nonassociative geometry and gravity, we write $_{s}\mathcal{M%
}\rightarrow \ _{s}^{\shortparallel }\mathcal{M}^{\star }.$ Here we
emphasize that in theories elaborated on (co) tangent bundles we can always
work with so-called almost complex/ symplectic structures when the geometric
models are real with a matrix operator $I^{2}=-1$ substituting the complex
unity, see details and references in \cite{vbv18,partner01,partner02}. In
this work, we shall work only with real momentum variables and use only the
labels "$\ ^{\shortmid }"$ and $\ "_{s}^{\shortmid }".$ The labels/symbols $%
\star $ and $\star _{s}$ are used if it is necessary to state that
respective spaces and/or geometric/physical are nonassociative ones. This
way we can elaborate in abstract form nonassociative and/or nonholonomic
s-adapted geometric and physical theories and derive geometric
evolutions/dynamical nonlinear and linear systems of PDEs. To decouple and
solve such systems in certain general/ explicit forms we have to introduce
respective systems of nonanholonomic s-adapted frames and coordinates with
additional labels of indices $\ ^{\shortmid }\mathbf{e}_{\alpha _{s}}[\
^{\shortmid }N_{\ i_{s}a_{s}}]$ (\ref{nadapbdsc}) and $\ ^{\shortmid }%
\mathbf{e}^{\alpha _{s}}[\ ^{\shortmid }N_{\ i_{s}a_{s}}] $ (\ref{nadapbdss}%
), see also footnote \ref{fnshells} and paragraphs 4 and 5 in previous
subsection.

\item Nonassociative nonholonomic s-adapted star product R-flux deformations 
$\star _{s}:\ _{s}\mathcal{M}\rightarrow \ _{s}^{\shortmid }\mathcal{M}%
^{\star }$ are defined and computed for respective deformations of metric
affine structures, $\ (\ _{s}^{\shortmid }\mathbf{N,}\ _{s}^{\shortmid }%
\mathbf{g,}\ _{s}^{\shortmid }\mathbf{D})\rightarrow (\ _{s}^{\shortmid }%
\mathbf{N},\ _{\star }^{\shortmid }\mathfrak{g}= (\ _{\star }^{\shortmid }%
\mathfrak{\check{g}}, \ _{\star }^{\shortmid }\mathfrak{a}), \
_{s}^{\shortmid }\mathbf{D}^{\star }),$ where $\ _{\star }^{\shortmid }%
\mathfrak{\check{g}}$ and $\ _{\star }^{\shortmid }\mathfrak{a}$ are
respective symmetric and nonsymmetric components of the nonassociative
metric, and $\ _{\star }^{\shortmid }\mathfrak{g}$, see formulas (\ref%
{aux40a}) - (\ref{aux40aa}). The nonassociative s-connection $\
_{s}^{\shortmid }\mathbf{D}^{\star }$ is the corresponding analog for a
R-flux deformation of a s-connection $\ _{s}^{\shortmid }\mathbf{D}$. The
Convention 2 (\ref{conv2s}) is formulated in such a form that allows to
define and compute in abstract/ coefficient formulas of corresponding star
deformations of geometric s-objects, for instance, of torsion s-tensors, $\
_{s}^{\shortmid }\mathcal{T}\rightarrow \ _{s}^{\shortmid }\widehat{\mathcal{%
T}}^{\star },$ and Riemann curvature s-tensors, $\ _{s}^{\shortmid }\mathcal{%
R}\rightarrow \ _{s}^{\shortmid }\widehat{\mathcal{R}}^{\star },$ etc., see
explanations for formulas (\ref{mafgeomobn}). Such nonassociative geometric
constructions can be performed in canonical s-forms and/or using star
deformed LC-connections, see definitions (\ref{twoconsstar}).

\item The abstract s-adapted nonassociative geometric star deformation
formalism allows to define in abstract geometric form on $\ _{s}^{\shortmid }%
\mathcal{M}^{\star }$ the nonassociative and noncommutative modified
Einstein equations (\ref{nonassocdeinst1}); the star product deformed
Perelman functionals, $\ _{s}^{\shortmid }\widehat{\mathcal{F}}^{\star
}(\tau )$ (\ref{naffunct}) and/or $\ _{s}^{\shortmid }\widehat{\mathcal{W}}%
^{\star }(\tau )$ (\ref{nawfunct}); nonassociative geometric flow equations (%
\ref{nonassocgeomfl}); nonassociative Ricci soliton equations (\ref{naricsol}%
); and respective thermodynamic variables $_{s}\widehat{\mathcal{Z}}^{\star
}(\tau )$ (\ref{spf}) and $[\ \ _{s}^{\shortmid }\widehat{\mathcal{E}}%
^{\star }(\tau ),\ _{s}^{\shortmid }\widehat{\mathcal{S}}^{\star }(\tau ),\
_{s}^{\shortmid }\widehat{\sigma }^{\star }(\tau )]$ (\ref{nagthermodvalues}%
).

\item The definition of star products of type (\ref{starpn}) involve also
the Planck constant, $\hbar ,$ and the string constant $\mathit{\ell }_{s}$
(here $s$ is for string), which allows to define a small parameter $\kappa =%
\mathit{\ell }_{s}^{3}/6\hbar .$ For explicit computations of star product
deformations of nonassociative geometric objects in \cite%
{blumenhagen16,aschieri17}, there were elaborated abstract and coordinate
base decompositions on $\hbar $ and $\kappa $ of geometric/physical objects
in nonassociative geometry. In our partner/this works \cite%
{partner01,partner02}, the $\kappa $-parametric decompositions of necessary
nonassocitaive geometric s-objects, physically important systems of
nonlinear PDEs etc. were formulated and computed in nonholonomic s-adapted
forms. For instance, we provide formulas (\ref{nadriemhopfcan}) - (\ref%
{driccicanonstar1}) for so-called $\kappa $-linear forms.

\item For $\kappa $-parametric linear s-adapted decompositions, we obtain
respective parametric deformations of the Perelman nonassociative
functionals $\ _{s}^{\shortmid }\widehat{\mathcal{F}}_{\kappa }^{\star
}(\tau )$ (\ref{naffunctp}) and $\ _{s}^{\shortmid }\widehat{\mathcal{W}}%
_{\kappa }^{\star }(\tau )$ (\ref{nawfunctp}). Corresponding, we can define
and compute the $\kappa $-parametric nonassociative geometric flow equations
(\ref{nonassocgeomflp}) which can be represented in the form $\ ^{\shortmid }%
\widehat{\mathbf{R}}_{\ \ \gamma _{s}}^{\beta _{s}}(\tau )={\delta }_{\ \
\gamma _{s}}^{\beta _{s}}\ _{s}^{\shortmid }\Im ^{\star }(\tau )$ (\ref%
{nonassocrf}). The effective sources $\ _{s}^{\shortmid }\Im ^{\star }(\tau
) $ (\ref{cannonsymparamc2b}) encode data on nonassociative star product
R-flux deformations of geometric s-objects and associative and commutative
generating sources $\ _{s}^{\shortmid }\Im (\tau ).$ Such systems of
nonlinear PDEs can be decoupled and integrated in certain general forms by
applying the AFCDM as we explain in subsection \ref{ansatzss} and appendix %
\ref{asstables}. In principle, we use the same method as for constructing
exact/parametric solutions in the theory of associative and commutative
nonholonomic geometric flows and gravitational equations in MGTs but for
another type of effective sources $\ _{s}^{\shortmid }\Im ^{\star }(\tau )$
and nonassociative generalized nonlinear symmetries, when the solutions are $%
\tau $-parametric and encoding nonassociative data.

\item In this work, we consider explicit examples and applications for
nonassociative quasi-stationary solutions of $\tau $-parametric canonical
Einstein equations (\ref{nonassocrf}), which define $\tau $-linear
nonassociative star product R-flux deformations of wormhole solutions on
phase spaces $\ _{s}^{\shortmid }\mathcal{M}^{\star }.$

\item Two classes of nonassociative wormhole solutions studied in section %
\ref{sec4} can't be characterized, in general, in terms of the thermodynamic
Bekenstein-Hawking paradigm \cite{bek1,bek2,haw1,haw2}. Nevertheless, they
admit a Perelman like statistical thermodynamic description with $\kappa $%
-linear thermodynamic variables of type (\ref{spf}) and (\ref%
{nagthermodvalues}), with symbolic data $[\ _{s}^{\shortmid }\widehat{%
\mathcal{Z}}_{\kappa }^{\star }(\tau ),\ _{s}^{\shortmid }\widehat{\mathcal{E%
}}_{\kappa }^{\star }(\tau ),\ _{s}^{\shortmid }\widehat{\mathcal{S}}%
_{\kappa }^{\star }(\tau ),\ _{s}^{\shortmid }\widehat{\sigma }_{\kappa
}^{\star }(\tau )].$ Such thermodynamic variables were computed in our
partner works for certain classes of nonassociative BH and BE solutions \cite%
{partner03,lbdssv22, lbsvevv22}.
\end{enumerate}

\subsubsection{Conventions on nonassociative GIFs and QGIFs and information
thermodynamics}

Nonassociative GIF and QGIF model can be elaborated for a generating
(statistical) partition function $\ _{s}^{\shortmid }\widehat{Z}(\tau )$ (%
\ref{spf}) when the "star" energy $\ _{s}^{\shortmid }\widehat{\mathcal{E}}%
_{\kappa }^{\star }(\tau )$ and star-entropy $\ _{s}^{\shortmid }\widehat{S}%
^{\star }(\tau )$ are computed as in (\ref{nagthermodvalues}), for a
s-metric $\ ^{\shortmid }\mathbf{g}_{\alpha _{s}\beta _{s}}(\tau )$ taken as
a solution of nonassociative geometric flow equations (\ref{nonassocrf}),
parameterized by an ansatz (\ref{ssolutions}). In brief, we say that we
consider nonassociative canonical geometric flow systems $\ _{s}^{\star }%
\widehat{A}(\tau )=\{\ _{s}^{\shortmid }\mathfrak{g}^{\star }(\tau ),\
_{s}^{\shortmid }\widehat{\mathbf{D}}^{\star }(\tau )\},$ when "hats" for
symbols of some geometric/ information systems can be omitted (writing $\
_{s}^{\star }A(\tau ),\ _{s}A(\tau ),\ _{s}^{\star }B,\ _{s}B$ etc.) in
order to simplify the system of notations. For explicit computations, we
shall use $\kappa $- parametric decompositions with $\ _{s}^{\shortmid }%
\widehat{Z}(\tau )$ and $\ _{s}^{\shortmid }\widehat{S}_{\kappa }^{\star
}(\tau ).$

\paragraph{Discrete and continuous nonassociative GIFs: \newline
}

\begin{enumerate}
\item Variants of the Shannon--Perelman entropies (\ref{shanonper}) are
defined: $\ ^{\shortmid }\widehat{S}_{A}^{\star }=\sum\nolimits_{z=1}^{z=r}\
_{s}^{\shortmid }\widehat{S}^{\star }(\tau _{z})\simeq
\sum\nolimits_{z=1}^{z=r}\ _{s}^{\shortmid }\widehat{S}_{\kappa
}^{\star}(\tau _{z}),$ for a set of discrete $\tau _{z};$ and $\ ^{\shortmid
}\widehat{S}_{A}^{\star }= \int\nolimits_{\tau _{r}}^{\tau _{r}}d\tau \
_{s}^{\shortmid }\widehat{S}^{\star }(\tau )\simeq \int\nolimits_{\tau
_{r}}^{\tau _{r}}d\tau \ _{s}^{\shortmid }\widehat{S}_{\kappa }^{\star}(\tau
),$ with continuous $\tau .$

\item We can introduce, respectively, the nonassociative canonical
conditional entropy $\ ^{\shortmid }\widehat{S}_{A|B}^{\star }$ and
nonassociative canonical mutual information $\ ^{\shortmid }I^{\star }(\
_{s}^{\star }A;\ _{s}^{\star }B)$, see (\ref{condes}) and (\ref{minfs}).

\item Conditions of subadditivity in nonassociative GIF theories, can be
proved for $\kappa $-parametric Ricci solitonic configurations (\ref%
{suadditivgif}) and then claimed for physical applications:\newline
$\ ^{\shortmid }\widehat{S}^{\star }(\ _{P}^{\shortmid }\widehat{Z}%
_{A,B}\parallel \ _{Q}^{\shortmid }\widehat{Z}_{A,B})= \ ^{\shortmid }%
\widehat{S}_{A}^{\star }+\ ^{\shortmid }\widehat{S}_{B}^{\star }-\
^{\shortmid }\widehat{S}_{AB}^{\star }= \ ^{\shortmid }I^{\star }(\
_{s}^{\star }A;\ _{s}^{\star }B)\geq 0$.

\item Thermodynamic generating functions (with generalizations of partition
function (\ref{spf})) and canonical entropies for two, ($\ _{s}^{\star }A,\
_{s}^{\star }B$), and three ($\ _{s}^{\star }A,\ _{s}^{\star }B,\
_{s}^{\star }C$), GIFs of NESs: 
\begin{eqnarray*}
\ _{AB}^{\shortmid }\widehat{\mathcal{Z}}[\mathbf{g}(\tau ),\ _{1}\mathbf{g}%
(\tau )]\mbox{ and }\ _{AB}^{\shortmid }\widehat{\mathcal{S}}^{\star }(\tau
) &=&\ ^{\shortmid }\widehat{\mathcal{S}}^{\star }\ [\ _{s}^{\star }\widehat{%
A},\ _{s}^{\star }\widehat{B}](\tau ),\mbox{ for
}\ _{s}^{\shortmid }\mathcal{M}^{\star }\otimes \ _{s}^{\shortmid }\mathcal{M%
}^{\star }\mathbf{;} \\
\ _{ABC}\widehat{\mathcal{Z}}[\mathbf{g}(\tau ),\ _{1}\mathbf{g}(\tau ),\
_{2}\mathbf{g}(\tau )]\mbox{ and }\ _{ABC}\widehat{\mathcal{S}}^{\star
}(\tau ) &=&\ \widehat{\mathcal{S}}^{\star }[\widehat{A},\widehat{B},%
\widehat{C}](\tau ),\mbox{ for }\ _{s}^{\shortmid }\mathcal{M}^{\star
}\otimes \ _{s}^{\shortmid }\mathcal{M}^{\star }\otimes \ _{s}^{\shortmid }%
\mathcal{M}^{\star };
\end{eqnarray*}%
see respective formulas (\ref{twogenf}) and (\ref{twoentr}); and (\ref%
{threegenf}) and (\ref{threeentr}).
\end{enumerate}

\paragraph{Discrete and continuous nonassociative QGIFs: \newline
}

\begin{enumerate}
\item The \textit{von Neumann--Perelman entropy} for $\ ^{\shortmid }%
\widehat{\rho }_{\mathcal{A}}$ (\ref{candm}), for instance, when $\
^{\shortmid }\widehat{\rho }_{\mathcal{A}}=\ _{s}^{\shortmid }\widehat{\rho }
$ (\ref{densmcan}), is defined as the \textit{entanglement canonical entropy}
of a nonassociative QGIF system $\mathcal{A}=\{\ _{s}^{\star }A\}= \mathcal{A%
}^{\star },$ \newline
$\ ^{\shortmid }\widehat{\mathcal{S}}_{\mathcal{A}}^{\star }\ =-tr_{\mathcal{%
A}}[\ ^{\shortmid }\widehat{\rho }_{\mathcal{A}}\log \ ^{\shortmid }\widehat{%
\rho }_{\mathcal{A}}]$ (\ref{entanglentrcan}). Such an entropy is determined
by a canonical partition function $\ _{s}^{\shortmid }\widehat{Z}[\
^{\shortmid }\mathbf{g}(\tau )]$ (\ref{spf}) for a solution $\ ^{\shortmid }%
\mathbf{g}(\tau )$ of $\kappa $--parametric nonassociative geometric flows
equations (\ref{nonassocrf}). It can be considered as the quantum variant of
the thermodynamic entropy $\ _{s}^{\shortmid }\widehat{\mathcal{S}}^{\star
}(\tau )$ from (\ref{nagthermodvalues}). For a pure state (\ref{puredm}), $\
^{\shortmid }\widehat{\mathcal{S}}_{\mathcal{A}}^{\star }$ (\ref%
{entanglentrcan}) vanishes.

\item Canonical \textit{nonassociative qubit systems} and examples of Bell
type nonassociative entanglement for double quantum thermo-geometric \
nonassociative QGIF systems $|\widehat{B}_{1}^{\star }>, |\widehat{B}%
_{2}^{\star }>, |\widehat{B}_{3}^{\star }>, |\widehat{B}_{4}^{\star }>$ (\ref%
{qgifbell}), are defined by nonassociative geometric flow data via $\
^{\shortmid }\widehat{\mathcal{E}}_{\underline{1}}^{\star }, \ ^{\shortmid }%
\widehat{\mathcal{E}}_{\underline{2}}^{\star }$ and $\ ^{\shortmid }\widehat{%
Z}_{\underline{1}}^{1/2},\ ^{\shortmid }\widehat{Z}_{\underline{2}}^{1/2}.$

\item The \textit{relative entropy} $\ ^{\shortmid }\widehat{\mathcal{S}}%
^{\star }(\ ^{\shortmid }\widehat{\rho }_{\mathcal{A}}\shortparallel \
^{\shortmid }\widehat{\sigma }_{\mathcal{A}})$ (\ref{relativentr}) is
defined for two nonassociative QGIFs with density matrices $\ ^{\shortmid }%
\widehat{\rho }_{\mathcal{A}}$ and $\ ^{\shortmid }\widehat{\sigma }_{%
\mathcal{A}}$.

\item The R\'{e}nyi entropy $\ _{r}^{\shortmid }\widehat{\mathcal{S}}^{\star
}(\widehat{\mathcal{A}}):=(1-r)^{-1}\log [tr_{\mathcal{A}}(\ ^{\shortmid }%
\widehat{\rho }_{\mathcal{A}})^{r}]$ (\ref{renentr}) can be defined and
computed for a nonassociative QGIF density matrix $\ ^{\shortmid }\widehat{%
\rho }_{\mathcal{A}}$ and developing a corresponding replica formalism.
\end{enumerate}

\subsection{Nonassociative off-diagonal quasi-stationary solutions with
fixed energy parameter}

\label{asstables}The anholonomic frame and connection deformation method,
AFCDM, allows to construct exact and parametric solutions in various classes
of MGTs including nonassociative gravity and certain extensions for
nonassociative geometric flows, see proofs and details in \cite%
{partner02,partner03,lbdssv22,lbsvevv22}. For associative and commutative
8-d solutions (they may include as particular examples 4-d configurations in
GR, or higher dimension configurations in higher dimension gravity,
relativistic Finsler-Lagrange-Hamilton gravity, phase space gravity theories
etc.); on AFCDM, see also review \cite{vbv18}. Similar classes of solutions
can be generated for $\kappa $-parametric and $\tau $-dependence of
nonassociative geometric flow equations (\ref{nonassocrf}). In a formal
abstract symbolic geometric form, we can take any class of exact/parametric
solution and consider it with additional $\tau $-dependence for
parameterizations of type (\ref{ans1rf}). There are necessary additional
assumptions on adapting nonassociative effective sources $\ _{s}^{\shortmid
}\Im ^{\star }(\tau )=\{\ ^{\shortmid }\Im _{\star \ \beta _{s}}^{\alpha
_{s}}~(\tau ,\ ^{\shortmid }u^{\gamma _{s}})\}$ (\ref{cannonsymparamc2b}) to
the nonholonomic s-frame structure in order to generate physically important
solutions in certain explicit forms. In this Appendix, we generalize for $%
\tau $-dependent ansatz with coefficients determined by $\ _{s}^{\shortmid
}\Im ^{\star }(\tau )$ and explain such generalizations and modifications in
the next subsections of the Appendix to this work. The Tables 1, 2 and 3 are
constructed for quasi-stationary generic off-diagonal $\tau $-running
solutions in such a way that they will allow to generate nonassociative star
product R-flux deformations of 4-d and extra dimensional wormhole solutions
in GR and MGTs.

\subsubsection{Metric ansatz and systems of nonlinear ODEs and PDEs}

Parameterizations of frames/coordinates for $\tau $-families of 4-d base
spaces modelled by nonholonomic Lorentz manifolds and canonical geometric
d-objects $[\mathbf{g}(\tau ),\mathbf{N}(\tau ),\widehat{\mathbf{D}}(\tau )]$
and generating of (effective) sources $\ _{s}^{\shortmid }\Im ^{\star }(\tau
),$ for shells $s=1$ and $2,$ are provided in Table 1. There are considered
two types of $\tau $-running generic off-diagonal metric ansatz. The first
class is for modelling nonassociative $\kappa $-parametric geometric flows
of quasi-stationary metrics with dependence only on space coordinates. In
particular, the 4-d components of such solutions describe nonassociative
deformations of BH and/or wormhole solutions in GR. The second class can be
considered for constructing nonassociative deformations of various classes
of locally anisotropic cosmological solutions with dependence on a time-like
coordinate and possible additional dependencies on two other space like
coordinates. In associative and commutative forms, generic off-diagonal
cosmological scenarios were studied in various classes of MGTs, GIF and QGIF
theories. We consider 4-d cosmological ansatz with underlined symbols in
Table 1 with the aim to state the conventions of distinguishing
quasi-stationary solutions from locally cosmological solutions. It is also
emphasized that for applications of the AFCDM such $\tau $-evolving
configurations can be such way parameterized that various classes of locally
anisotropic cosmological solutions can be generated a being $t$-dual to
those for quasi-stationary configurations. Respective coefficients for
ansatz (\ref{ssolutions}) depend on different types of coordinates with
changing of local signature, various types of (non) linear symmetries and
physical interpretation. In this work, we study possible physical
implications of generic off-diagonal ansatz with Killing symmetry on $%
\partial _{4},$ for quasi-stationary configurations (\ref{ans1rf}).

Similar parameterizations and ansatz from Table 1 can be lifted on (co)
fibers of respective (co) tangent bundles to Lorentz manifolds. This allows
to construct (co) fiber BH, BE, wormhole solutions depending, for instance,
on momentum type coordinates, when $u^{\alpha}=(x^{1},x^{2},y^{3},y^{4}=t)%
\longrightarrow p_{a_{s}},$ with $s=3,4$ and respective $p_{5,}p_{6},p_{7},$
when (for the purposes of this work) $p_{8}=E_{0}=const.$ Quasi-stationary
configurations can be generated for "rainbow" s-metrics when, for instance, $%
p_{8}=E$ is variable, but $p_{7}=const.$ Such phase space solutions model $%
\tau $-running of base spacetime quasi-stationary solutions being
nonassociatively correlated with certain locally anisotropic co-fiber
cosmological models when $E$ plays the role of "time" like coordinate. Other
classes of (locally anisotropic cosmological) solutions can be generated for
d-metrics with coefficients depending on time like coordinate $y^{4}=t$ on
the shell $s=2.$ Typically, we underline such d-metric coefficients and
respective generating sources, for instance, $\underline{\mathbf{g}}%
_{a}(\tau ,x^{k},t)$ and $\ _{2}^{\shortmid }\underline{\Im }_{\ }^{\star
}(\tau ,x^{i},t).$ We do not consider such examples (i.e. solutions with
explicit dependencies on $t$ and $E,$ for d-metrics) in this work.
Nevertheless, it should be noted that such solutions can be generated in
abstract geometric forms using respective nonlinear symmetries and duality
properties of quasi-stationary configurations and locally anisotropic
cosmological d-metrics. This reflects both the priorities of the AFCDM and
specific geometric and physical properties of nonassociative phase space
geometric evolution of generic off-diagonal gravitational systems.


{\scriptsize 
\begin{eqnarray*}
&&%
\begin{tabular}{l}
\hline\hline
\begin{tabular}{lll}
& {\ \textsf{Table 1:\ Diagonal and off-diagonal ansatz resulting in systems
of nonlinear ODEs and PDEs} } &  \\ 
& applying the Anholonomic Frame and Connection Deformation Method, \textbf{%
AFCDM}, &  \\ 
& \textit{for constructing 4-d generic off-diagonal exact and parametric
solutions of \ }(\ref{nonassocrf}) & 
\end{tabular}%
\end{tabular}
\\
&&{%
\begin{tabular}{lll}
\hline
diagonal ansatz: PDEs $\rightarrow $ \textbf{ODE}s &  & AFCDM: \textbf{PDE}s 
\textbf{with decoupling; \ generating functions} \\ 
radial coordinates $u^{\alpha }=(r,\theta ,\varphi ,t)$ & $u=(x,y),\tau :$ & 
\mbox{ nonholonomic 2+2
splitting, } (\ref{ans1rf})$~u^{\alpha }=(x^{1},x^{2},y^{3},y^{4}=t),\tau $
\\ 
LC-connection $\mathring{\nabla}(\tau )\simeq \mathring{\nabla}(\tau ,u)$ & 
[connections] & $%
\begin{array}{c}
\mathbf{N}(\tau ):T\mathbf{V}=hT\mathbf{V}\oplus vT\mathbf{V,}%
\mbox{ locally
}\mathbf{N}(\tau )=\{N_{i}^{a}(\tau ,x,y)\} \\ 
\mbox{ canonical connection distortion }\widehat{\mathbf{D}}(\tau )=\nabla
(\tau )+\widehat{\mathbf{Z}}(\tau );\widehat{\mathbf{D}}\mathbf{g=0,} \\ 
\widehat{\mathcal{T}}[\mathbf{g}(\tau )\mathbf{,N}(\tau ),\widehat{\mathbf{D}%
}(\tau )]\mbox{ canonical
d-torsion}%
\end{array}%
$ \\ 
$%
\begin{array}{c}
\mbox{ diagonal ansatz  }g_{\alpha \beta }(\tau ,u) \\ 
\simeq \left( 
\begin{array}{cccc}
\mathring{g}_{1}(\tau ) &  &  &  \\ 
& \mathring{g}_{2}(\tau ) &  &  \\ 
&  & \mathring{g}_{3}(\tau ) &  \\ 
&  &  & \mathring{g}_{4}(\tau )%
\end{array}%
\right)%
\end{array}%
$ & $\mathbf{g}(\tau )\Leftrightarrow $ & $%
\begin{array}{c}
g_{\alpha \beta }(\tau )=%
\begin{array}{c}
g_{\alpha \beta }(\tau ,x^{i},y^{a})\mbox{ general frames / coordinates} \\ 
\left[ 
\begin{array}{cc}
g_{ij}(\tau )+N_{i}^{a}(\tau )N_{j}^{b}(\tau )h_{ab}(\tau ) & N_{i}^{b}(\tau
)h_{cb}(\tau ) \\ 
N_{j}^{a}(\tau )h_{ab}(\tau ) & h_{ac}(\tau )%
\end{array}%
\right] ,%
\end{array}
\\ 
\mathbf{g}_{\alpha \beta }(\tau )=[g_{ij}(\tau ),h_{ab}(\tau )], \\ 
\mathbf{g}(\tau )=\mathbf{g}_{i}(\tau ,x^{k})dx^{i}\otimes dx^{i}+\mathbf{g}%
_{a}(\tau ,x^{k},y^{b})\mathbf{e}^{a}(\tau )\otimes \mathbf{e}^{b}(\tau )%
\end{array}%
$ \\ 
$\mathring{g}_{\alpha \beta }(\tau )=\left\{ 
\begin{array}{cc}
\mathring{g}_{\alpha }(\tau ,r) & \mbox{ for BHs} \\ 
\mathring{g}_{\alpha }(\tau ,t) & \mbox{ for FLRW }%
\end{array}%
\right. $ & [coord.frames] & $g_{\alpha \beta }(\tau )=\left\{ 
\begin{array}{cc}
g_{\alpha \beta }(\tau ,x^{i},y^{3}) & 
\mbox{ quasi-stationary
configurations} \\ 
\underline{g}_{\alpha \beta }(\tau ,x^{i},y^{4}=t) & 
\mbox{locally anisotropic
cosmology}%
\end{array}%
\right. $ \\ 
&  &  \\ 
$%
\begin{array}{c}
\mbox{coord. transf. }e_{\alpha }(\tau )=e_{\ \alpha }^{\alpha ^{\prime
}}(\tau )\partial _{\alpha ^{\prime }}, \\ 
e^{\beta }(\tau )=e_{\beta ^{\prime }}^{\ \beta }(\tau )du^{\beta ^{\prime
}}, \\ 
\mathring{g}_{\alpha \beta }(\tau )=\mathring{g}_{\alpha ^{\prime }\beta
^{\prime }}(\tau )e_{\ \alpha }^{\alpha ^{\prime }}(\tau )e_{\ \beta
}^{\beta ^{\prime }}(\tau ) \\ 
\begin{array}{c}
\mathbf{\mathring{g}}_{\alpha }(\tau ,x^{k},y^{a})\rightarrow \mathring{g}%
_{\alpha }(\tau ,r), \\ 
\mbox{ or }\mathring{g}_{\alpha }(\tau ,t), \\ 
\mathring{N}_{i}^{a}(\tau ,x^{k},y^{a})\rightarrow 0.%
\end{array}%
\end{array}%
$ & [N-adapt. fr.] & $\left\{ 
\begin{array}{cc}
\begin{array}{c}
\mathbf{g}_{i}(\tau ,x^{k}),\mathbf{g}_{a}(\tau ,x^{k},y^{3}), \\ 
\mbox{ or }\mathbf{g}_{i}(\tau ,x^{k}),\underline{\mathbf{g}}_{a}(\tau
,x^{k},t),%
\end{array}
& \mbox{ d-metrics } \\ 
\begin{array}{c}
N_{i}^{3}(\tau )=w_{i}(\tau ,x^{k},y^{3}),N_{i}^{4}(\tau )=n_{i}(\tau
,x^{k},y^{3}), \\ 
\mbox{ or }\underline{N}_{i}^{3}(\tau )=\underline{n}_{i}(\tau ,x^{k},t),%
\underline{N}_{i}^{4}(\tau )=\underline{w}_{i}(\tau ,x^{k},t),%
\end{array}
& 
\end{array}%
\right. $ \\ 
$\mathring{\nabla}(\tau ),$ $Ric(\tau )=\{\mathring{R}_{\ \beta \gamma
}(\tau )\}$ & Ricci d-tensors & $\widehat{\mathbf{D}}(\tau ),\ \widehat{%
\mathcal{R}}ic(\tau )=\{\widehat{\mathbf{R}}_{\ \beta \gamma }(\tau )\}$ \\ 
$~^{m}\mathcal{L}[\phi (\tau )]\mathcal{\rightarrow }\ ^{m}\mathbf{T}%
_{\alpha \beta }[\phi (\tau )]$ & 
\begin{tabular}{l}
generating \\ 
sources%
\end{tabular}
& $%
\begin{array}{c}
\ _{2}^{\shortmid }\Im _{\ \nu }^{\star \mu }(\tau )=\mathbf{e}_{\ \mu
^{\prime }}^{\mu }\mathbf{e}_{\nu }^{\ \nu ^{\prime }}\ _{2}^{\shortmid }\Im
_{\ \nu ^{\prime }}^{\star \mu ^{\prime }}[\ ^{m}\mathcal{L}(\mathbf{\varphi
),}T_{\mu \nu },\ _{2}^{\shortmid }\Lambda ^{\star }] \\ 
\begin{array}{c}
=diag[\ \ _{1}^{\shortmid }\Im _{\ }^{\star }(\tau ,x^{i})\delta _{j}^{i},\
\ _{2}^{\shortmid }\Im _{\ }^{\star }(\tau ,x^{i},y^{3})\delta _{b}^{a}], \\ 
\mbox{ quasi-stationary configurations};%
\end{array}
\\ 
\begin{array}{c}
=diag[\ \ _{1}^{\shortmid }\Im _{\ }^{\star }(\tau ,x^{i})\delta _{j}^{i},\
\ _{2}^{\shortmid }\underline{\Im }_{\ }^{\star }(\tau ,x^{i},t)\delta
_{b}^{a}], \\ 
\mbox{ locally anisotropic cosmology};%
\end{array}%
\end{array}%
$ \\ 
trivial equations for $\mathring{\nabla}(\tau )$-torsions & LC-conditions & $%
\widehat{\mathbf{D}}_{\mid \widehat{\mathcal{T}}(\tau )\rightarrow 0}(\tau )=%
\mathbf{\nabla }(\tau )\mbox{ extracting new classes of solutions in GR}$ \\ 
\hline\hline
\end{tabular}%
}
\end{eqnarray*}%
}

\subsubsection{Decoupling and quasi-stationary solutions of nonassociative
geometric flow equations}

The key steps for generating stationary off-diagonal exact solutions of $%
\kappa $-parametric nonassociative geometric flow equations are outlined in
Table 2. Such $\tau $-families of generic off-diagonal solutions are, in
general, with nontrivial nonholonomically induced canonical d-torsions \ $\
_{s}^{\shortmid }\widehat{\mathcal{T}}^{\star }(\tau )=\{\ ^{\shortmid }%
\widehat{\mathbf{T}}_{\ \star \beta _{s}\gamma _{s}}^{\alpha _{s}}(\tau )\}$
(\ref{mafgeomobn}). The s-metrics can be expressed and re-defined
equivalently in terms of respective families of generating functions $\Psi
(\tau ,x^{k},y^{3})$ or $\Phi (\tau ,x^{k},y^{3})$ using nonlinear
symmetries, see details in \cite{partner02,partner03,lbdssv22,lbsvevv22}.
For $\kappa $-linear decompositions, considering $\eta $--polarization
functions, respective $\tau $-families of d-metrics and N-connections can be
parameterized to describe nonholonomic deformations of a family of primary
d-metrics $\mathbf{\mathring{g}}(\tau )$ (for instance, a BH or wormhole
solution) \ into target generic off diagonal stationary solutions $\widehat{%
\mathbf{g}}(\tau )$ of type (\ref{ans1rf}), when 
\begin{equation}
\mathbf{\mathring{g}}(\tau )\rightarrow \widehat{\mathbf{g}}(\tau
,x^{k},y^{3})=[g_{\alpha }(\tau ,x^{k},y^{3})=\eta _{\alpha }(\tau
,x^{k},y^{3})\mathring{g}_{\alpha }(\tau ),\ \eta _{i}^{a}(\tau ,x^{k},y^{3})%
\mathring{N}_{i}^{a}(\tau )].  \label{etapolarizations}
\end{equation}%
%
%
{\scriptsize 
\begin{eqnarray*}
&&%
\begin{tabular}{l}
\hline\hline
\begin{tabular}{l}
{\large \textsf{Table 2:\ Off-diagonal quasi-stationary nonassociative
geometric evolution}} \\ 
Exact/parametric solutions for $\ ^{\shortmid }\widehat{\mathbf{R}}_{\ \
\gamma _{2}}^{\beta _{2}}(\tau )={\delta }_{\ \ \gamma _{2}}^{\beta _{2}}\
_{s}^{\shortmid }\Im ^{\star }(\tau )$ (\ref{nonassocrf}) ($s=1,2)$ and
systems of nonlinear PDEs with decoupling%
\end{tabular}
\\ 
\end{tabular}
\\
&&%
\begin{tabular}{lll}
\hline\hline
&  &  \\ 
$%
\begin{array}{c}
\mbox{d-metric ansatz with} \\ 
\mbox{Killing symmetry }\partial _{4}=\partial _{t}%
\end{array}%
$ &  & $%
\begin{array}{c}
ds^{2}=g_{i_{1}}(\tau ,x^{k_{1}})(dx^{i_{1}})^{2}+g_{a_{2}}(\tau
,x^{k_{1}},y^{3})(dy^{a_{2}}+N_{i_{1}}^{a_{2}}(\tau
,x^{k_{1}},y^{3})dx^{i_{1}})^{2},\mbox{ for } \\ 
g_{i_{1}}(\tau )=e^{\psi (\tau ,x^{k_{1}})},g_{a_{2}}(\tau )=h_{a_{2}}(\tau
,x^{k_{1}},y^{3}), \\ 
N_{i_{1}}^{3}(\tau )=w_{i_{1}}(\tau ,x^{k},y^{3}),N_{i_{1}}^{4}(\tau
)=n_{i_{1}}(\tau ,x^{k},y^{3});%
\end{array}%
$ \\ 
&  &  \\ 
Effective matter sources &  & $\ ^{\shortmid }\Im _{\ \nu _{2}}^{\star \mu
_{2}}=[\ \ _{1}^{\shortmid }\Im ^{\star }(\tau ,x^{k_{1}})\delta
_{j_{1}}^{i_{1}},\ _{2}^{\shortmid }\Im ^{\star }(\tau ,x^{k},y^{3})\delta
_{b_{2}}^{a_{2}}],\mbox{ for }i_{1},j_{1},..=1,2;a_{2},b_{2},...=3,4$ \\ 
\hline
Nonlinear PDEs with decoupling: &  & $%
\begin{array}{c}
\psi ^{\bullet \bullet }+\psi ^{\prime \prime }=2\ \ _{1}^{\shortmid }\Im
^{\star }(\tau ); \\ 
\varpi ^{\ast }\ h_{4}^{\ast }=2h_{3}h_{4}\ \ _{2}^{\shortmid }\Im ^{\star
}(\tau ); \\ 
\beta w_{i}-\alpha _{i}=0; \\ 
n_{k}^{\ast \ast }+\gamma n_{k}^{\ast }=0;%
\end{array}%
$ for $%
\begin{array}{c}
\varpi (\tau ){=\ln |\partial _{3}h_{4}/\sqrt{|h_{3}h_{4}|}|,} \\ 
\alpha _{i}(\tau )=(\partial _{3}h_{4})\ (\partial _{i}\varpi ),\beta (\tau
)=(\partial _{3}h_{4})\ (\partial _{3}\varpi ),\  \\ 
\ \gamma (\tau )=\partial _{3}\left( \ln |h_{4}|^{3/2}/|h_{3}|\right) , \\ 
\partial _{1}q=q^{\bullet },\partial _{2}q=q^{\prime },\partial
_{3}q=q^{\ast }%
\end{array}%
$ \\ \hline
$%
\begin{array}{c}
\mbox{ Generating functions:}\ h_{3}(\tau ,x^{k},y^{3}), \\ 
\Psi (\tau ,x^{k},y^{3})=e^{\varpi },\Phi (\tau ,x^{k},y^{3}); \\ 
\mbox{integration functions:}\ h_{4}^{[0]}(\tau ,x^{k}),\  \\ 
_{1}n_{k}(\tau ,x^{i}),\ _{2}n_{k}(\tau ,x^{i}); \\ 
\mbox{\& nonlinear symmetries}%
\end{array}%
$ &  & $%
\begin{array}{c}
\ (\Psi ^{2}(\tau ))^{\ast }=-\int dy^{3}\ \ _{2}^{\shortmid }\Im ^{\star
}(\tau )h_{4}^{\ \ast }, \\ 
\Phi ^{2}(\tau )=-4\ _{2}\Lambda ^{\star }(\tau )h_{4}; \\ 
h_{4}=h_{4}^{[0]}-\Phi ^{2}/4\ \ _{2}\Lambda ^{\star }(\tau ),h_{4}^{\ast
}\neq 0,\ \ _{2}\Lambda ^{\star }\neq 0=const%
\end{array}%
$ \\ \hline
Off-diag. solutions, $%
\begin{array}{c}
\mbox{d--metric} \\ 
\mbox{N-connec.}%
\end{array}%
$ &  & $%
\begin{array}{c}
\ g_{i_{1}}(\tau )=e^{\ \psi (\tau ,x^{k})}%
\mbox{ as a solution of 2-d
Poisson eqs. }\psi ^{\bullet \bullet }+\psi ^{\prime \prime }=2\
_{1}^{\shortmid }\Im ^{\star }(\tau ); \\ 
h_{3}(\tau )=-(\Psi ^{\ast })^{2}/4\ (\ \ _{2}^{\shortmid }\Im ^{\star
}(\tau ))^{2}h_{4}; \\ 
h_{4}(\tau )=h_{4}^{[0]}(\tau )-\int dy^{3}(\Psi ^{2})^{\ast }/4\ \
_{2}^{\shortmid }\Im ^{\star }=h_{4}^{[0]}(\tau )-\Phi ^{2}/4\ _{2}\Lambda
^{\star }(\tau ); \\ 
\\ 
w_{i}(\tau )=\partial _{i}\ \Psi /\ \partial _{3}\Psi =\partial _{i}\ \Psi
^{2}/\ \partial _{3}\Psi ^{2}|; \\ 
n_{k}(\tau )=\ _{1}n_{k}+\ _{2}n_{k}\int dy^{3}(\Psi ^{\ast })^{2}/\ \
_{2}^{\shortmid }\Im ^{\star }|h_{4}^{[0]}-\int dy^{3}\frac{(\Psi
^{2})^{\ast }}{4\ \ _{2}^{\shortmid }\Im ^{\star }}|^{5/2}. \\ 
\\ 
\end{array}%
$ \\ \hline
LC-configurations (\ref{lccondnonass}) &  & $%
\begin{array}{c}
\partial _{3}w_{i}=(\partial _{i}-w_{i}\partial _{3})\ln \sqrt{|h_{3}|}%
,(\partial _{i}-w_{i}\partial _{3})\ln \sqrt{|h_{4}|}=0, \\ 
\partial _{k}w_{i}=\partial _{i}w_{k},\partial _{3}n_{i}=0,\partial
_{i}n_{k}=\partial _{k}n_{i}; \\ 
\mbox{ see d-metric }(\ref{ans1rf})\mbox{ for } \\ 
\Psi (\tau )=\check{\Psi}(\tau ,x^{i},y^{3}),(\partial _{i}\check{\Psi}%
)^{\ast }=\partial _{i}(\check{\Psi}^{\ast })\mbox{ and } \\ 
\ \ _{2}^{\shortmid }\Im ^{\star }(\tau ,x^{i},y^{3})=\ \ _{2}^{\shortmid
}\Im ^{\star }[\check{\Psi}]=\ \ _{2}^{\shortmid }\check{\Im}^{\star },%
\mbox{ or }\ _{2}^{\shortmid }\check{\Im}^{\star }=\ _{2}^{\shortmid }\check{%
\Im}^{\star }(\tau ). \\ 
\end{array}%
$ \\ \hline
N-connections, zero torsion &  & $%
\begin{array}{c}
w_{i}(\tau )\simeq \partial _{i}\check{A}(\tau ,x^{i},y^{3})=\left\{ 
\begin{array}{c}
\partial _{i}(\int dy^{3}\ \ \ _{2}^{\shortmid }\check{\Im}^{\star }\ \check{%
h}_{4}{}^{\ast }])/\ \ _{2}^{\shortmid }\check{\Im}^{\star }\ \check{h}%
_{4}{}^{\ast }; \\ 
\partial _{i}\check{\Psi}/\check{\Psi}^{\ast }; \\ 
\partial _{i}(\int dy^{3}\ \ \ _{2}^{\shortmid }\check{\Im}^{\star }(\check{%
\Phi}^{2})^{\ast })/(\check{\Phi})^{\ast }\ \ _{2}^{\shortmid }\check{\Im}%
^{\star };%
\end{array}%
\right. \\ 
\mbox{ and }n_{k}(\tau )\simeq \check{n}_{k}=\partial _{k}n((\tau ),x^{i}).%
\end{array}%
$ \\ \hline
$%
\begin{array}{c}
\mbox{polarization functions} \\ 
\mathbf{\mathring{g}}\rightarrow \widehat{\mathbf{g}}\mathbf{=}[g_{\alpha
}=\eta _{\alpha }\mathring{g}_{\alpha },\ \eta _{i}^{a}\mathring{N}_{i}^{a}]%
\end{array}%
$ &  & $%
\begin{array}{c}
\\ 
ds^{2}=\eta _{1}(\tau ,r,\theta )\mathring{g}_{1}(\tau ,r,\theta
)[dx^{1}(r,\theta )]^{2}+\eta _{2}(\tau ,r,\theta )\mathring{g}_{2}(r,\theta
)[dx^{2}(r,\theta )]^{2}+ \\ 
\eta _{3}(\tau ,r,\theta ,\varphi )\mathring{g}_{3}(\tau ,r,\theta
)[d\varphi +\eta _{i}^{3}(\tau ,r,\theta ,\varphi )\mathring{N}_{i}^{3}(\tau
,r,\theta )dx^{i}(r,\theta )]^{2}+ \\ 
\eta _{4}(\tau ,r,\theta ,\varphi )\mathring{g}_{4}(r,\theta )[dt+\eta
_{i}^{4}(\tau ,r,\theta ,\varphi )\mathring{N}_{i}^{4}(\tau ,r,\theta
)dx^{i}(r,\theta )]^{2}, \\ 
\end{array}%
$ \\ \hline
Solutions for polarization funct. &  & $%
\begin{array}{c}
\eta _{i}(\tau )\simeq e^{\ \psi (\tau ,x^{k})}/\mathring{g}_{i};\eta _{3}%
\mathring{h}_{3}=-\frac{4[(|\eta _{4}\mathring{h}_{4}|^{1/2})^{\ast }]^{2}}{%
|\int dy^{3}\ \ _{2}^{\shortmid }\check{\Im}^{\star }[(\eta _{4}\mathring{h}%
_{4})]^{\ast }|\ }; \\ 
\eta _{4}(\tau )\simeq \eta _{4}(\tau ,x^{k},y^{3})%
\mbox{ as a generating
function}; \\ 
\ \eta _{i}^{3}\ \mathring{N}_{i}^{3}=\frac{\partial _{i}\ \int dy^{3}\
_{2}^{\shortmid }\check{\Im}^{\star }\ (\eta _{4}\ \mathring{h}_{4})^{\ast }%
}{\ _{2}^{\shortmid }\check{\Im}^{\star }\ (\eta _{4}\ \mathring{h}%
_{4})^{\ast }}; \\ 
\eta _{k}^{4}\ \mathring{N}_{k}^{4}=\ _{1}n_{k}+16\ \ _{2}n_{k}\int dy^{3}%
\frac{\left( [(\eta _{4}\mathring{h}_{4})^{-1/4}]^{\ast }\right) ^{2}}{|\int
dy^{3}\ _{2}^{\shortmid }\check{\Im}^{\star }\ [(\eta _{4}\ \mathring{h}%
_{4})]^{\ast }|\ }%
\end{array}%
$ \\ \hline
Polariz. funct. with zero torsion &  & $%
\begin{array}{c}
\eta _{i}(\tau )\simeq e^{\ \psi (\tau ,x^{k})}/\mathring{g}_{i};\eta _{4}=%
\check{\eta}_{4}(\tau ,x^{k},y^{3})\mbox{ as a generating function}; \\ 
\eta _{3}(\tau )=-\frac{4[(|\eta _{4}\mathring{h}_{4}|^{1/2})^{\ast }]^{2}}{%
\mathring{g}_{3}|\int dy^{3}\ _{2}^{\shortmid }\check{\Im}^{\star }\ [(%
\check{\eta}_{4}\mathring{h}_{4})]^{\ast }|\ };\eta _{i}^{3}(\tau )=\partial
_{i}\check{A}/\mathring{w}_{k},\eta _{k}^{4}(\tau )=\frac{\ \partial _{k}n}{%
\mathring{n}_{k}}. \\ 
\end{array}%
$ \\ \hline\hline
\end{tabular}%
\end{eqnarray*}%
} 
We can extract torsionless LC-configurations (in general, encoding
nonassociative data via off-diagonal terms and effective sources) by
imposing additional nonholonomic conditions (\ref{lccondnonass})\ for
respective $\tau $-family. Such conditions can be satisfied for a more
special class of "integrable" generating functions $(\check{h}_{4}(\tau
,x^{k},y^{3}),$ or $\check{\Psi}(\tau ,x^{k},y^{3})$ and/or $\check{\Phi}%
(\tau ,x^{k},y^{3}))$ for respective effective sources $\ _{2}\check{\Upsilon%
}(\tau ,x^{k},y^{3})$ related via nonlinear symmetries to a $\tau $-family \
of effective cosmological constants $\ _{2}\Lambda (\tau ).$

\vskip5pt The assumption that (effective) generating sources $\ ^{\shortmid
}\Im _{\alpha _{s}\beta _{s}}^{\star }(\tau )$ for a $\tau $-family of $%
\kappa $-parametric nonassociative geometric flow equations (\ref%
{nonassocgeomflef}) can be parameterized in N-adapted form as $\ _{\star
}^{\shortmid }\Im _{\ \beta _{2}}^{\alpha _{2}}=[\ \ _{\star }^{\shortmid
}\Im _{1}(\tau )\delta _{\ \ j_{1}}^{i_{1}},\ \ _{\star }^{\shortmid }\Im
_{2}(\tau )\delta _{\ \ b_{2}}^{a_{2}}]$ (\ref{cannonsymparamc2b}) is
important because it allows to integrate the system of nonlinear PDEs (\ref%
{ans1rf}) in explcit form. The shell $s=1,2$ data $\ _{\star }^{\shortmid
}\Im _{2}(\tau )$ are related via certain algebraic relations (choosing
respectively the coefficients of frame transforms) to certain $\tau $%
-families of energy-momentum tensors for matter fields, star product R-flux $%
\kappa $-parametric deformations, and $\partial _{\tau }\ ^{\shortmid }%
\mathbf{g}_{\alpha _{s}\beta _{s}}(\tau ).$ Such algebraic relations involve
partial derivatives $\partial _{\tau }$, on geometric flow parameter and can
be prescribed for explicit symmetries of solutions, additional deformation
parameters (like eccentricity, small interactions with matter fields). They
are stated as some nonholonomic constraints for some classes of nonholonomic
transforms which can be solved explicitly for additional assumptions on
nonassociative/ nonholonomic data, symmetries of interactions, prescriptions
of some effective constants, or trivial ansatz for extra dimension
components. This refers to certain nonholonomic frame and special coordinate
conditions. Nevertheless, the effective/ $\tau $-parametric geometric flow
equations (\ref{ans1rf}) can be decoupled and integrated in certain general
forms applying the standard AFCDM re-defined by $\ _{\star }^{\shortmid }\Im
_{\ \beta _{2}}^{\alpha _{2}}(\tau )$ and certain effective $\ _{2}\Lambda
^{\star }(\tau ).$ We use a star label even for such effective cosmological
(running) cosmological constants because they encode in direct holonomic, or
in nonholonomic forms, certain nonassociative data from extra dimensional
velocity (momentum) like (co) fiber coordinates. This allows us to prove
general decoupling and integration properties of nonassociative geometric
flow and/or gravitational equations as in \cite{partner02} by considering
additional dependencies of s-metrics and effective sources on $\tau .$ To
model $\tau $-evolution of 4-d quasi-stationary configurations, we can
consider two effective generating sources, for instance, when that $\
_{\star }^{\shortmid }\Im _{1}(\tau )=\ _{\star }^{\shortmid }\Im _{1}(\tau
,x^{i})$ and $\ _{\star }^{\shortmid }\Im _{2}(\tau )=\ _{\star }^{\shortmid
}\Im _{2}(\tau ,x^{i},y^{3}).$

\vskip5pt We generate generic off-diagonal solutions describing
quasi-stationary geometric evolution of nonholonomic Einstein systems when
the effective sources $\ _{\star }^{\shortmid }\Im _{\ \beta _{2}}^{\alpha
_{2}}(\tau )$ are prescribed from certain physical considerations. They can
be related to effective $\tau $-running cosmological constants $\
_{2}\Lambda ^{\star }(\tau ),$ or considered as certain symbols encoding
nonassociative data and/or encoding nonholonomic deformations and
distortions. To elaborate on explicit computations such $\ _{\star
}^{\shortmid }\Im _{\ \beta _{2}}^{\alpha _{2}}(\tau )$ may be considered as
generating sources determined by contributions, for instance, from some
effective classical and/or quantum interactions, extra dimension etc. For
small parametric decompositions and using recurrent formulas, we can
generate new classes of parametric solutions when their physical
interpretation is similar to "not-deformed" prime models. In such cases,
nonassociative geometric flow data are encoded into certain locally
anisotropic polarized physical constants, star product deformed horizons,
additional off-diagonal terms, induced nonsymmetric components of metrics
etc. In general, for certain parameterizations of generating sources, it is
not clear if a corresponding class of solutions may have, or not, importance
for physical theories. Nevertheless, using the AFCDM we are able to
investigate to construct explicit classes of exact/parametric solutions and
analyze properties of off-diagonal nonlinear gravitational and (effective)
matter field interactions subjected to nonassociative star product R-flux
deformations and their $\tau $-parametric geometric flow equations. This is
more general than in the case when the (modified) Einstein equations are
transformed into systems of nonlinear ODEs for diagonal metrics and
LC-configurations. Examples of physically important quasi-stationary
solutions are studied in section \ref{sec4} for modified wormhole solutions
encoding nonassociative data.

\subsubsection{Nonassociative quasi-stationary geometric flow solutions with
fixed energy parameter}

Such quasi-stationary $\tau $-running solutions are determined by
nonholonomic flows of momentum type phase configurations modeled on
cotangent Lorentz bundles with $p_{8}=E_{0}=const.$ The Table 3 summarize
the data on nonassociative star product R-flux deformations when the
effective sources $\ _{s}^{\shortmid }\widehat{\Upsilon }$ are changed into
nonassociative and $\tau $-running ones $\ _{s}^{\shortmid }\Im ^{\star
}(\tau ).$ It also extends the results of Table 2 above for $s=3,4$ on $%
T^{\ast }\mathbf{V}$ in a form with $p_{8}=E_{0}$ keeping the
quasi-stationary character of solutions.

{\scriptsize 
\begin{eqnarray*}
&&%
\begin{tabular}{l}
\hline\hline
\begin{tabular}{l}
{\large \textsf{Table 3:\ Quasi-stationary phase space nonassociative
geometric flows, fixed energy}} \\ 
Exact solutions of $\ ^{\shortmid }\widehat{\mathbf{R}}_{\ \ \gamma
_{s}}^{\beta _{s}}(\tau )={\delta }_{\ \ \gamma _{s}}^{\beta _{s}}\
_{s}^{\shortmid }\Im ^{\star }(\tau )$ (\ref{nonassocrf}) ($s=1,2,3,t)$ on $%
T_{s}^{\ast }V$ decoupled into nonlinear PDEs%
\end{tabular}
\\ 
\end{tabular}
\\
&&%
\begin{tabular}{lll}
\hline\hline
&  &  \\ 
$%
\begin{array}{c}
\mbox{ s-metric ansat with} \\ 
\mbox{Killing symmetries on } \\ 
\partial _{4}=\partial _{t},\ ^{\shortmid}\partial ^{8}%
\end{array}%
$ &  & $%
\begin{array}{c}
ds^{2}=g_{i_{1}}(\tau ,x^{k_{1}})(dx^{i_{1}})^{2}+g_{a_{2}}(\tau
,x^{k_{1}},y^{3})(dy^{a_{2}}+N_{i_{1}}^{a_{2}}(\tau
,x^{k_{1}},y^{3})dx^{i_{1}})^{2} \\ 
+\ ^{\shortmid}g^{a_{3}}(\tau ,x^{k_{2}},p_{5})(dp_{a_{3}}+ \
^{\shortmid}N_{i_{2}a_{3}}(\tau ,x^{k_{2}},p_{5})dx^{i_{2}})^{2} \\ 
+\ ^{\shortmid}g^{a_{4}}(\tau ,\ ^{\shortmid}x^{k_{3}},p_{7})(dp_{a_{4}}+\
^{\shortmid}N_{i_{3}a_{4}}(\tau ,\ ^{\shortmid}x^{k_{3}},p_{7})d\
^{\shortmid}x^{i_{3}})^{2},\mbox{
for } \\ 
g_{i_{1}}(\tau )=e^{\psi {(\tau ,x}^{k_{1}}{)}},\ g_{a_{2}}(\tau
)=h_{a_{2}}(\tau ,x^{k_{1}},y^{3}), \\ 
N_{i_{1}}^{3}(\tau )=\ ^{2}w_{i_{1}}(\tau )=w_{i_{1}}(\tau
,x^{k_{1}},y^{3}),N_{i_{1}}^{4}(\tau )=\ ^{2}n_{i_{1}}(\tau )=n_{i_{1}}(\tau
,x^{k_{1}},y^{3}), \\ 
\ ^{\shortmid}g^{a_{3}}(\tau )=\ ^{\shortmid}h^{a_{3}}(\tau
,x^{k_{2}},p_{5}),\ ^{\shortmid}N_{i_{2}5}(\tau )=\ _{\shortmid
}^{3}w_{i_{2}}(\tau )=\ ^{\shortmid}w_{i_{2}}(\tau ,x^{k_{2}},p_{5}), \\ 
\ ^{\shortmid}N_{i_{2}6}(\tau )=\ _{\shortmid }^{3}n_{i_{2}}(\tau )=\
^{\shortmid}n_{i_{2}}(\tau ,x^{k_{2}},p_{5}), \\ 
\ ^{\shortmid}g^{a_{4}}(\tau )=\ ^{\shortmid}h^{a_{4}}(\tau ,\
^{\shortmid}x^{k_{3}},p_{7}),\ ^{\shortmid}N_{i_{3}7}(\tau )= \ _{\shortmid
}^{4}w_{i_{3}}(\tau )=\ ^{\shortmid}w_{i_{3}}(\tau ,x^{k_{3}},p_{7}), \\ 
\ ^{\shortmid}N_{i_{3}8}(\tau )=\ _{\shortmid }^{4}n_{i_{3}}(\tau )=\
^{\shortmid}n_{i_{3}}(\tau ,x^{k_{3}},p_{7}), \ _{s}^{\shortmid }\Im _{\ \nu
_{s}}^{\star \mu _{s}}(\tau )\simeq%
\end{array}%
$ \\ 
Effective matter sources &  & $\lbrack \ _{1}^{\shortmid }\Im ^{\star }(\tau
,{x}^{k_{1}})\delta _{j_{1}}^{i_{1}}, \ _{2}^{\shortmid }\Im ^{\star }(\tau ,%
{x}^{k_{1}},y^{3})\delta _{b_{2}}^{a_{2}},\ _{3}^{\shortmid }\Im ^{\star
}(\tau ,{x}^{k_{2}},p_{5})\delta _{b_{3}}^{a_{3}}, \ _{4}^{\shortmid }\Im
^{\star }(\tau ,{x}^{k_{3}},p_{7})\delta _{b_{4}}^{a_{4}}]$ \\ \hline
Decoupled nonlinear PDEs &  & $%
\begin{tabular}{lll}
$%
\begin{array}{c}
\psi ^{\bullet \bullet }+\psi ^{\prime \prime }=2\ \ _{1}^{\shortmid }\Im
^{\star }(\tau ); \\ 
\ ^{2}\varpi ^{\ast }\ h_{4}^{\ast }=2h_{3}h_{4}\ _{2}^{\shortmid }\Im
^{\star }(\tau ); \\ 
\ ^{2}\beta \ ^{2}w_{i_{1}}-\ ^{2}\alpha _{i_{1}}=0; \\ 
\ ^{2}n_{k_{1}}^{\ast \ast }+\ ^{2}\gamma \ ^{2}n_{k_{1}}^{\ast }=0;%
\end{array}%
$ &  & $%
\begin{array}{c}
\ ^{2}\varpi {=\ln |\partial _{3}h_{4}/\sqrt{|h_{3}h_{4}|}|,} \\ 
\ ^{2}\alpha _{i_{1}}=(\partial _{3}h_{4})\ (\partial _{i_{1}}\ ^{2}\varpi ),
\\ 
\ ^{2}\beta =(\partial _{3}h_{4})\ (\partial _{3}\ ^{2}\varpi ),\  \\ 
\ \ ^{2}\gamma =\partial _{3}\left( \ln |h_{4}|^{3/2}/|h_{3}|\right) , \\ 
\partial _{1}q=q^{\bullet },\partial _{2}q=q^{\prime },\partial
_{3}q=q^{\ast }%
\end{array}%
$ \\ 
$%
\begin{array}{c}
\ ^{\shortmid}\partial ^{5}(\ _{\shortmid}^{3}\varpi)\ ^{\shortmid}\partial
^{5}\ ^{\shortmid}h^{6}= 2\ ^{\shortmid}h^{5}\ ^{\shortmid}h^{6} \
_{3}^{\shortmid }\Im ^{\star}(\tau); \\ 
\ _{\shortmid }^{3}\beta \ _{\shortmid }^{3}w_{i_{2}}-\
_{\shortmid}^{3}\alpha _{i_2}=0; \\ 
\ ^{\shortmid}\partial ^{5}(\ ^{\shortmid}\partial ^{5}\
_{\shortmid}^{3}n_{k_2})+ \ _{\shortmid}^{3}\gamma \ ^{\shortmid}\partial
^{5}(\ _{\shortmid }^{3}n_{k_2})=0;%
\end{array}%
$ &  & $%
\begin{array}{c}
\\ 
\ _{\shortmid }^{3}\varpi {=\ln |\ ^{\shortmid}\partial ^{5}\
^{\shortmid}h^{6}/\sqrt{|\ ^{\shortmid}h^{5}\ ^{\shortmid}h^{6}|}|,} \\ 
\ _{\shortmid}^{3}\alpha _{i_{2}}=(\ ^{\shortmid}\partial ^{5}\
^{\shortmid}h^{6})\ (\partial _{i_{2}}\ _{\shortmid }^{3}\varpi ), \\ 
\ _{\shortmid }^{3}\beta =(\mathbf{\ ^{\shortmid }}\partial ^{5}\mathbf{\
^{\shortmid }}h^{6})\ (\mathbf{\ ^{\shortmid }}\partial ^{5}\ _{\shortmid
}^{3}\varpi ),\  \\ 
\ \ _{\shortmid }^{3}\gamma =\mathbf{\ ^{\shortmid }}\partial ^{5}\left( \ln
|\mathbf{\ ^{\shortmid }}h^{6}|^{3/2}/|\mathbf{\ ^{\shortmid }}h^{5}|\right)
,%
\end{array}%
$ \\ 
$%
\begin{array}{c}
\mathbf{\ ^{\shortmid }}\partial ^{7}(\ _{\shortmid }^{4}\varpi )\ \mathbf{\
^{\shortmid }}\partial ^{7}\ \mathbf{\ ^{\shortmid }}h^{8}=2\mathbf{\
^{\shortmid }}h^{7}\mathbf{\ ^{\shortmid }}h^{8}\ \ _{4}^{\shortmid }\Im
^{\star }(\tau ); \\ 
\ _{\shortmid }^{4}\beta \ _{\shortmid }^{4}w_{i_{3}}-\ _{\shortmid
}^{4}\alpha _{i_{3}}=0; \\ 
\mathbf{\ ^{\shortmid }}\partial ^{7}(\mathbf{\ ^{\shortmid }}\partial ^{7}\
_{\shortmid }^{4}n_{k_{3}})+\ _{\shortmid }^{4}\gamma \mathbf{\ ^{\shortmid }%
}\partial ^{7}(\ _{\shortmid }^{4}n_{k_{3}})=0;%
\end{array}%
$ &  & $%
\begin{array}{c}
\\ 
\ _{\shortmid }^{4}\varpi {=\ln |\mathbf{\ ^{\shortmid }}\partial ^{7}%
\mathbf{\ ^{\shortmid }}h^{8}/\sqrt{|\mathbf{\ ^{\shortmid }}h^{7}\mathbf{\
^{\shortmid }}h^{8}|}|,} \\ 
\ _{\shortmid }^{4}\alpha _{i_{3}}=(\mathbf{\ ^{\shortmid }}\partial ^{7}%
\mathbf{\ ^{\shortmid }}h^{8})\ (\mathbf{\ ^{\shortmid }}\partial _{i_{3}}\
_{\shortmid }^{4}\varpi ), \\ 
\ _{\shortmid }^{4}\beta =(\mathbf{\ ^{\shortmid }}\partial ^{7}\mathbf{\
^{\shortmid }}h^{8})\ (\mathbf{\ ^{\shortmid }}\partial ^{7}\ _{\shortmid
}^{4}\varpi ),\  \\ 
\ \ _{\shortmid }^{4}\gamma =\mathbf{\ ^{\shortmid }}\partial ^{7}\left( \ln
|\mathbf{\ ^{\shortmid }}h^{8}|^{3/2}/|\mathbf{\ ^{\shortmid }}h^{7}|\right)
,%
\end{array}%
$%
\end{tabular}%
$ \\ \hline
$%
\begin{array}{c}
\mbox{ Gener.  functs:}\ h_{3}(\tau ,x^{k_{1}},y^{3}), \\ 
\ ^{2}\Psi (\tau ,x^{k_{1}},y^{3})\simeq e^{\ ^{2}\varpi (\tau )}, \\ 
\ ^{2}\Phi (\tau ,x^{k_{1}},y^{3}), \\ 
\mbox{integr. functs:}\ h_{4}^{[0]}(\tau ,x^{k_{1}}),\  \\ 
_{1}n_{k_{1}}(\tau ,x^{i_{1}}),\ _{2}n_{k_{1}}(\tau ,x^{i_{1}}); \\ 
\mbox{Gener. functs:}\mathbf{\ ^{\shortmid }}h^{5}(\tau ,x^{k_{2}},p_{5}),
\\ 
\ \ _{\shortmid }^{3}\Psi (\tau ,x^{k_{2}},p_{5})\simeq e^{\ \ _{\shortmid
}^{3}\varpi (\tau )}, \\ 
\ \ \ _{\shortmid }^{3}\Phi (\tau ,x^{k_{2}},p_{5}), \\ 
\mbox{integr. functs:}\ h_{6}^{[0]}(\tau ,x^{k_{2}}),\  \\ 
_{1}^{3}n_{k_{2}}(\tau ,x^{i_{2}}),\ _{2}^{3}n_{k_{2}}(\tau ,x^{i_{2}}); \\ 
\mbox{Gener. functs:}\ ^{\shortmid}h^{7}(\tau ,\
^{\shortmid}x^{k_{3}},p_{7}), \\ 
\ \ _{\shortmid }^{4}\Psi (\tau ,x^{k_{2}},p_{7})\simeq e^{ \ _{\shortmid
}^{4}\varpi (\tau )}, \\ 
\ \ \ _{\shortmid }^{4}\Phi (\tau ,\ ^{\shortmid}x^{k_{3}},p_{7}), \\ 
\mbox{integr. functs:}\ h_{8}^{[0]}(\tau ,\ ^{\shortmid}x^{k_{3}}), \\ 
_{1}^{4}n_{k_{3}}(\tau,\ ^{\shortmid}x^{i_{3}}),\ _{2}^{4}n_{k_{3}}(\tau ,\
^{\shortmid}x^{i_{3}}); \\ 
\mbox{\& nonlinear symmetries}%
\end{array}%
$ &  & $%
\begin{array}{c}
\ ((\ ^{2}\Psi )^{2})^{\ast }=-\int dy^{3} \ _{2}^{\shortmid }\Im ^{\star
}(\tau )h_{4}^{\ \ast }, \\ 
(\ ^{2}\Phi )^{2}=-4\ \ _{2}\Lambda ^{\star }(\tau )\ h_{4}, \\ 
h_{4}=h_{4}^{[0]}-(\ ^{2}\Phi )^{2}/4\ \ _{2}\Lambda ^{\star }(\tau
),h_{4}^{\ast }\neq 0,\ \ _{2}\Lambda ^{\star }(\tau )\neq 0; \\ 
\\ 
\mathbf{\ ^{\shortmid }}\partial ^{5}((\ \ _{\shortmid }^{3}\Psi
)^{2})=-\int dp_{5}\ \ _{3}^{\shortmid }\Im ^{\star }(\tau )\mathbf{\
^{\shortmid }}\partial ^{5}\mathbf{\ ^{\shortmid }}h^{6}, \\ 
(\ \ _{\shortmid }^{3}\Phi )^{2}=-4\ \ _{3}\Lambda ^{\star }(\tau )\mathbf{\
^{\shortmid }}h^{6}, \\ 
\mathbf{\ ^{\shortmid }}h^{6}=\mathbf{\ ^{\shortmid }}h_{[0]}^{6}-(\ \
_{\shortmid }^{3}\Phi )^{2}/4\ _{3}\Lambda ^{\star }(\tau ),\mathbf{\
^{\shortmid }}\partial ^{5}\mathbf{\ ^{\shortmid }}h^{6}\neq 0,\ _{3}\Lambda
^{\star }(\tau )\neq 0; \\ 
\\ 
\mathbf{\ ^{\shortmid }}\partial ^{7}((\ _{\shortmid }^{4}\Psi )^{2})=-\int
dp_{7}\ \ _{4}^{\shortmid }\Im ^{\star }(\tau )\mathbf{\ ^{\shortmid }}%
\partial ^{7}\mathbf{\ ^{\shortmid }}h^{8}, \\ 
(\ _{\shortmid }^{4}\Phi )^{2}=-4\ _{4}\Lambda ^{\star }(\tau )\mathbf{\
^{\shortmid }}h^{8}, \\ 
\mathbf{\ ^{\shortmid }}h^{8}=\mathbf{\ ^{\shortmid }}h_{[0]}^{8}-(\
_{\shortmid }^{4}\Phi )^{2}/4\ _{4}\Lambda ^{\star }(\tau ),\mathbf{\
^{\shortmid }}\partial ^{7}\mathbf{\ ^{\shortmid }}h^{8}\neq 0,_{4}\Lambda
^{\star }(\tau )\neq 0;%
\end{array}%
$ \\ \hline
$%
\begin{array}{c}
\mbox{Off-diag. solutions} \\ 
\mbox{ s-metrics} \\ 
\mbox{N-connections}%
\end{array}%
$ &  & $%
\begin{tabular}{l}
$%
\begin{array}{c}
\ g_{i}=e^{\ \psi (x^{k})}\mbox{ as a solution of 2-d Poisson eqs. }\psi
^{\bullet \bullet }+\psi ^{\prime \prime }=2~\ _{1}^{\shortmid }\Im ^{\star
}(\tau ); \\ 
h_{3}=-(\Psi ^{\ast })^{2}/4\ \ _{2}^{\shortmid }\Im ^{\star }(\tau )h_{4};
\\ 
h_{4}=h_{4}^{[0]}-\int dy^{3}(\Psi ^{2})^{\ast }/4\ \ _{2}^{\shortmid }\Im
^{\star }(\tau )=h_{4}^{[0]}-\Phi ^{2}/4\ _{2}\Lambda ^{\star }(\tau ); \\ 
w_{i}=\partial _{i}\ \Psi /\ \partial _{3}\Psi =\partial _{i}\ \Psi ^{2}/\
\partial _{3}\Psi ^{2}|; \\ 
n_{k}=\ _{1}n_{k}+\ _{2}n_{k}\int dy^{3}(\Psi ^{\ast })^{2}/\ \
_{2}^{\shortmid }\Im ^{\star }(\tau )|h_{4}^{[0]}-\int dy^{3}(\Psi
^{2})^{\ast }/4\ \ _{2}^{\shortmid }\Im ^{\star }(\tau )|^{5/2};%
\end{array}%
$ \\ 
$%
\begin{array}{c}
\mathbf{\ ^{\shortmid }}h^{5}=-(\mathbf{\ ^{\shortmid }}\partial ^{5}\
_{\shortmid }^{3}\Psi )^{2}/4\ \ _{3}^{\shortmid }\Im ^{\star }(\tau )%
\mathbf{\ ^{\shortmid }}h^{6}; \\ 
\mathbf{\ ^{\shortmid }}h^{6}=\mathbf{\ ^{\shortmid }}h_{[0]}^{6}-\int dp_{5}%
\mathbf{\ ^{\shortmid }}\partial ^{5}((\ \ _{\shortmid }^{3}\Psi )^{2})/4\ \
_{3}^{\shortmid }\Im ^{\star }(\tau )=\mathbf{\ ^{\shortmid }}h_{[0]}^{6}-(\
\ _{\shortmid }^{3}\Phi )^{2}/4\ _{3}\Lambda ^{\star }(\tau ); \\ 
w_{i_{2}}=\partial _{i_{2}}(\ _{\shortmid }^{3}\Psi )/\mathbf{\ ^{\shortmid }%
}\partial ^{5}(\ _{\shortmid }^{3}\Psi )=\partial _{i_{2}}(\ _{\shortmid
}^{3}\Psi )^{2}/\ \mathbf{\ ^{\shortmid }}\partial ^{5}(\ _{\shortmid
}^{3}\Psi )^{2}|; \\ 
n_{k_2}=\ _{1}n_{k_2}+\ _{2}n_{k_2}\int dp_{5}(\ ^{\shortmid}\partial ^{5}\
_{\shortmid}^{3}\Psi )^{2}/ \ _{3}^{\shortmid}\Im ^{\star }(\tau)|\
^{\shortmid}h_{[0]}^{6}-\int dp_{5}\ ^{\shortmid}\partial ^{5}((\
_{\shortmid }^{3}\Psi )^{2})/4\ _{3}^{\shortmid}\Im ^{\star}(\tau)|^{5/2}%
\end{array}%
$ \\ 
$%
\begin{array}{c}
\mathbf{\ ^{\shortmid }}h^{7}=-(\mathbf{\ ^{\shortmid }}\partial ^{7}\
_{\shortmid }^{4}\Psi )^{2}/4\ \ _{4}^{\shortmid }\Im ^{\star }(\tau )%
\mathbf{\ ^{\shortmid }}h^{8}; \\ 
\mathbf{\ ^{\shortmid }}h^{8}=\mathbf{\ ^{\shortmid }}h_{[0]}^{8}-\int dp_{7}%
\mathbf{\ ^{\shortmid }}\partial ^{7}((\ _{\shortmid }^{4}\Psi )^{2})/4\ \
_{4}^{\shortmid }\Im ^{\star }(\tau )=h_{8}^{[0]}-(\ \ _{\shortmid }^{4}\Phi
)^{2}/4\ \ _{4}\Lambda ^{\star }(\tau ); \\ 
\mathbf{\ ^{\shortmid }}w_{i_{3}}=\mathbf{\ ^{\shortmid }}\partial
_{i_{3}}(\ \ _{\shortmid }^{4}\Psi )/\ \mathbf{\ ^{\shortmid }}\partial
^{7}(\ _{\shortmid }^{4}\Psi )=\mathbf{\ ^{\shortmid }}\partial _{i_{3}}(\ \
_{\shortmid }^{4}\Psi )^{2}/\ \mathbf{\ ^{\shortmid }}\partial ^{7}(\
_{\shortmid }^{4}\Psi )^{2}|; \\ 
\mathbf{\ ^{\shortmid }}n_{k_{3}}=\ _{1}^{\shortmid }n_{k_{3}}+\
_{2}^{\shortmid }n_{k_{3}}\int dp_{7}(\ _{\shortmid }^{4}\Psi )^{2}/\ \
_{4}^{\shortmid }\Im ^{\star }(\tau )|h_{8}^{[0]}-\int dp_{7}\mathbf{\
^{\shortmid }}\partial ^{7}((\ \ _{\shortmid }^{4}\Psi )^{2})/4\
_{4}^{\shortmid }\Im ^{\star }(\tau )|^{5/2}%
\end{array}%
$%
\end{tabular}%
$ \\ \hline\hline
\end{tabular}%
\end{eqnarray*}%
} In functional form, such nonassociative phase space s-metrics define a
class of nonlinear quadratic elements:

\begin{eqnarray}
d\widehat{s}_{[8d]}^{2} &=&\widehat{g}_{\alpha _{s}\beta _{s}}(\tau
,x^{k},y^{3},p_{5},p_{7};h_{4}(\tau ),\mathbf{\ ^{\shortmid }}h^{6}(\tau ),%
\mathbf{\ ^{\shortmid }}h^{8}(\tau );\ \ _{2}^{\shortmid }\Im ^{\star }(\tau
);\ \ _{s}\Lambda ^{\star }(\tau ))d\mathbf{\ ^{\shortmid }}u^{\alpha _{s}}d%
\mathbf{\ ^{\shortmid }}u^{\beta _{s}}  \label{qst8} \\
&=&e^{\psi (\tau ,x^{k},\ \ _{1}^{\shortmid }\Im ^{\star
})}[(dx^{1})^{2}+(dx^{2})^{2}]-\frac{(h_{4}^{\ast })^{2}}{|\int dy^{3}[\
_{2}^{\shortmid }\Im ^{\star }h_{4}]^{\ast }|\ h_{4}}\{dy^{3}+\frac{\partial
_{i_{1}}[\int dy^{3}(\ _{2}^{\shortmid }\Im ^{\star })\ h_{4}^{\ast }]}{\ \
_{2}^{\shortmid }\Im ^{\star }\ h_{4}^{\ast }}dx^{i_{1}}\}^{2}+  \notag \\
&&h_{4}\{dt+[\ _{1}n_{k_{1}}+\ _{2}n_{k_{1}}\int dy^{3}\frac{(h_{4}^{\ast
})^{2}}{|\int dy^{3}[\ _{2}^{\shortmid }\Im ^{\star }h_{4}]^{\ast }|\
(h_{4})^{5/2}}]dx_{1}^{k}\}+  \notag \\
&&\frac{(\mathbf{\ ^{\shortmid }}\partial ^{5}\mathbf{\ ^{\shortmid }}%
h^{6})^{2}}{|\int dp_{5}\mathbf{\ ^{\shortmid }}\partial ^{5}[\
_{3}^{\shortmid }\Im ^{\star }\mathbf{\ ^{\shortmid }}h^{6}]|\ \mathbf{\
^{\shortmid }}h^{6}}\{dp_{5}+\frac{\partial _{i_{2}}[\int dp_{5}(\
_{3}^{\shortmid }\Im ^{\star })\mathbf{\ ^{\shortmid }}\partial ^{5}\mathbf{%
\ ^{\shortmid }}h^{6}]}{\ \ _{3}^{\shortmid }\Im ^{\star }\mathbf{\
^{\shortmid }}\partial ^{5}\ \mathbf{^{\shortmid }}h^{6}}dx^{i_{2}}\}^{2}+ 
\notag \\
&&\mathbf{\ ^{\shortmid }}h^{6}\{dp_{5}+[\ _{1}n_{k_{2}}+\ _{2}n_{k_{2}}\int
dp_{5}\frac{(\mathbf{\ ^{\shortmid }}\partial ^{5}\mathbf{\ ^{\shortmid }}%
h^{6})^{2}}{|\int dp_{5}\mathbf{\ ^{\shortmid }}\partial ^{5}[\
_{3}^{\shortmid }\Im ^{\star }\mathbf{\ ^{\shortmid }}h^{6}]|\ (\mathbf{\
^{\shortmid }}h^{6})^{5/2}}]dx^{k_{2}}\}+  \notag \\
&&\frac{(\mathbf{\ ^{\shortmid }}\partial ^{7}\mathbf{\ ^{\shortmid }}%
h^{8})^{2}}{|\int dp_{7}\mathbf{\ ^{\shortmid }}\partial ^{7}[\
_{4}^{\shortmid }\Im ^{\star }\mathbf{\ ^{\shortmid }}h^{8}]|\ \mathbf{\
^{\shortmid }}h^{8}}\{dp_{7}+\frac{\partial _{i_{3}}[\int dp_{7}(\
_{4}^{\shortmid }\Im ^{\star })\ \mathbf{\ ^{\shortmid }}\partial ^{7}%
\mathbf{\ ^{\shortmid }}h^{8}]}{\ _{4}^{\shortmid }\Im ^{\star }\ \mathbf{\
^{\shortmid }}\partial ^{7}\mathbf{\ ^{\shortmid }}h^{8}}d\mathbf{\
^{\shortmid }}x^{i_{3}}\}^{2}+  \notag \\
&&\mathbf{\ ^{\shortmid }}h^{8}\{dE+[\ _{1}^{\shortmid }n_{k_{3}}+\
_{2}^{\shortmid }n_{k_{3}}\int dp_{7}\frac{(\mathbf{\ ^{\shortmid }}\partial
^{7}\mathbf{\ ^{\shortmid }}h^{8})^{2}}{|\int dp_{7}\mathbf{\ ^{\shortmid }}%
\partial ^{7}[\ \ _{3}^{\shortmid }\Im ^{\star }\mathbf{\ ^{\shortmid }}%
h^{8}]|\ (\mathbf{\ ^{\shortmid }}h^{8})^{5/2}}]d\mathbf{\ ^{\shortmid }}%
x^{k_{3}}\}.  \notag
\end{eqnarray}%
The nonlinear symmetries of such s-metrics for generic off-diagonal flows on
nonassociative phase spaces are included in Table 3. They allow to allow to
re-define the generating functions and generating sources and relate them to
conventional $\tau $-running cosmological constants $\ _{s}\Lambda ^{\star
}(\tau ).$

\subsubsection{Nonassociative quasi-stationary evolution and gravitational
polarizations}

\label{ssqsgrpol}

Quasi-stationary $\tau $-running solutions can be re-defined in terms of
gravitational $\eta $-polarizations (\ref{etapolarizations}). We can study
off-diagonal deformations of a family of prescribed prime s-metrics $\
_{s}^{\shortmid }\mathbf{\mathring{g}}(\tau )$ into other families of target
ones, 
\begin{equation*}
\ _{s}^{\shortmid }\mathbf{\mathring{g}}(\tau )=[\ ^{\shortmid }\mathring{g}%
_{\alpha _{s}}(\tau ),\ ^{\shortmid }\mathring{N}_{i_{s-1}}^{a_{s}}(\tau
)]\rightarrow \ _{s}^{\shortmid }\mathbf{g}(\tau )=[\ ^{\shortmid }\eta
_{\alpha _{s}}(\tau )\ ^{\shortmid }\mathring{g}_{\alpha _{s}}(\tau ),\
^{\shortmid }\eta _{i_{s-1}}^{a_{s}}(\tau )\ ^{\shortmid }\mathring{N}%
_{i_{s-1}}^{a_{s}}(\tau )].
\end{equation*}%
The s-adapted coefficients of $\ _{s}^{\shortmid }\mathbf{\mathring{g}}%
(\tau) $ can be arbitrary ones or chosen to define certain physically
important solutions of some nonholonomic geometric flows, Ricci solitons, or
(modified) Einstein equations, for instance of BH or BE types. The $\tau $%
-running generating functions can be parameterize in the form:%
\begin{equation}
\psi (\tau )\simeq \psi (\hbar ,\kappa ;\tau ,x^{k_{1}}),\eta _{4}(\tau )\
\simeq \eta _{4}(\tau ,x^{k_{1}},y^{3}),\ ^{\shortmid }\eta ^{6}(\tau
)\simeq \ ^{\shortmid }\eta ^{6}(\tau ,x^{i_{2}},p_{5}),\ ^{\shortmid }\eta
^{8}(\tau )\simeq \ ^{\shortmid }\eta ^{8}(\tau ,x^{i_{2}},p_{5},p_{7}),
\label{etapolgen}
\end{equation}%
when the target s-metrics $\ _{s}^{\shortmid }\mathbf{g}(\tau )$ are
solutions of type (\ref{qst8}). Such quasi-stationary $\tau $-parametric
s-metrics define nonlinear quadratic elements\footnote{%
proofs are similar to those presented in \cite{partner02,lbsvevv22} but for
different types of effective generating sources encoding nonassociative star
product R-flux data} 
\begin{eqnarray}
d\ \ ^{\shortmid }\widehat{s}^{2}(\tau ) &=&\ \ ^{\shortmid }g_{\alpha
_{s}\beta _{s}}(\hbar ,\kappa ,\tau ,x^{k},y^{3},p_{a_{3}},p_{a_{4}};\
^{\shortmid }\mathring{g}_{\alpha _{s}}(\tau );\eta _{4}(\tau ),\
^{\shortmid }\eta ^{6}(\tau ),\ ^{\shortmid }\eta ^{8}(\tau ),\ _{s}\Lambda
^{\star }(\tau );\ \ _{s}^{\shortmid }\Im ^{\star }(\tau ))d~\ ^{\shortmid
}u^{\alpha _{s}}d~\ ^{\shortmid }u^{\beta _{s}}  \notag \\
&=&e^{\psi (\tau ,\ _{1}\Lambda ^{\star }(\tau
))}[(dx^{1})^{2}+(dx^{2})^{2}]-  \label{offdiagpolfr1} \\
&&\frac{[\partial _{3}(\eta _{4}(\tau )\ \mathring{g}_{4}(\tau ))]^{2}}{%
|\int dy^{3}\ _{2}^{\shortmid }\Im ^{\star }(\tau )\partial _{3}(\eta
_{4}(\tau )\ \mathring{g}_{4}(\tau ))|\ (\eta _{4}(\tau )\mathring{g}%
_{4}(\tau ))}\{dy^{3}+\frac{\partial _{i_{1}}[\int dy^{3}\ _{2}\Im (\tau )\
\partial _{3}(\eta _{4}(\tau )\mathring{g}_{4}(\tau ))]}{\ _{2}^{\shortmid
}\Im ^{\star }(\tau )\partial _{3}(\eta _{4}(\tau )\mathring{g}_{4}(\tau ))}%
dx^{i_{1}}\}^{2}+  \notag \\
&&\eta _{4}(\tau )\mathring{g}_{4}(\tau ))\{dt+[\ _{1}n_{k_{1}}(\tau )+\
_{2}n_{k_{1}}(\tau )\int \frac{dy^{3}[\partial _{3}(\eta _{4}(\tau )%
\mathring{g}_{4}(\tau ))]^{2}}{|\int dy^{3}\ _{2}^{\shortmid }\Im ^{\star
}(\tau )\partial _{3}(\eta _{4}(\tau )\mathring{g}_{4}(\tau ))|\ [\eta
_{4}(\tau )\mathring{g}_{4}(\tau )]^{5/2}}]dx^{k_{1}}\}-  \notag
\end{eqnarray}%
\begin{eqnarray*}
&&\frac{[\ ^{\shortmid }\partial ^{5}(\ ^{\shortmid }\eta ^{6}(\tau )\
^{\shortmid }\mathring{g}^{6}(\tau ))]^{2}}{|\int dp_{5}~\ _{3}^{\shortmid
}\Im ^{\star }(\tau )\ \ ^{\shortmid }\partial ^{5}(\ ^{\shortmid }\eta
^{6}(\tau )\ \ ^{\shortmid }\mathring{g}^{6}(\tau ))\ |\ (\ ^{\shortmid
}\eta ^{6}(\tau )\ ^{\shortmid }\mathring{g}^{6}(\tau ))}\{dp_{5}+\frac{\
^{\shortmid }\partial _{i_{2}}[\int dp_{5}\ _{3}^{\shortmid }\Im (\tau )\
^{\shortmid }\partial ^{5}(\ ^{\shortmid }\eta ^{6}(\tau )\ ^{\shortmid }%
\mathring{g}^{6}(\tau ))]}{~\ _{3}^{\shortmid }\Im ^{\star }(\tau )\
^{\shortmid }\partial ^{5}(\ ^{\shortmid }\eta ^{6}(\tau )\ ^{\shortmid }%
\mathring{g}^{6}(\tau ))}dx^{i_{2}}\}^{2}+ \\
&&(\ ^{\shortmid }\eta ^{6}(\tau )\ ^{\shortmid }\mathring{g}^{6}(\tau
))\{dp_{6}+[\ _{1}^{\shortmid }n_{k_{2}}(\tau )+\ _{2}^{\shortmid
}n_{k_{2}}(\tau )\int \frac{dp_{5}[\ ^{\shortmid }\partial ^{5}(\
^{\shortmid }\eta ^{6}(\tau )\ ^{\shortmid }\mathring{g}^{6}(\tau ))]^{2}}{%
|\int dp_{5}\ _{3}^{\shortmid }\Im ^{\star }(\tau )\ \partial ^{5}(\
^{\shortmid }\eta ^{6}(\tau )\ ^{\shortmid }\mathring{g}^{6}(\tau ))|\ [\
^{\shortmid }\eta ^{6}(\tau )\ ^{\shortmid }\mathring{g}^{6}(\tau )]^{5/2}}%
]dx^{k_{2}}\}-
\end{eqnarray*}%
\begin{eqnarray*}
&&\frac{[\ ^{\shortmid }\partial ^{7}(\ ^{\shortmid }\eta ^{8}(\tau )\
^{\shortmid }\mathring{g}^{8}(\tau ))]^{2}}{|\int dp_{7}\ _{4}^{\shortmid
}\Im ^{\star }(\tau )\ ^{\shortmid }\partial ^{8}(\ ^{\shortmid }\eta
^{7}(\tau )\ ^{\shortmid }\mathring{g}^{7}(\tau ))\ |\ (\ ^{\shortmid }\eta
^{7}(\tau )\ ^{\shortmid }\mathring{g}^{7}(\tau ))}\{dp_{7}+\frac{\
^{\shortmid }\partial _{i_{3}}[\int dp_{7}\ _{4}^{\shortmid }\Im ^{\star
}(\tau )\ ^{\shortmid }\partial ^{7}(\ ^{\shortmid }\eta ^{8}(\tau )\ \
^{\shortmid }\mathring{g}^{8}(\tau ))]}{\ _{4}^{\shortmid }\Im ^{\star
}(\tau )\ ^{\shortmid }\partial ^{7}(\ ^{\shortmid }\eta ^{8}(\tau )\
^{\shortmid }\mathring{g}^{8}(\tau ))}d\ ^{\shortmid }x^{i_{3}}\}^{2}+ \\
&&(\ ^{\shortmid }\eta ^{8}(\tau )\ ^{\shortmid }\mathring{g}^{8}(\tau
))\{dE+[\ _{1}n_{k_{3}}(\tau )+\ _{2}n_{k_{3}}(\tau )\int \frac{dp_{7}[\
^{\shortmid }\partial ^{7}(\ ^{\shortmid }\eta ^{8}(\tau )\ ^{\shortmid }%
\mathring{g}^{8}(\tau ))]^{2}}{|\int dp_{7}\ _{4}^{\shortmid }\Im ^{\star
}(\tau )[\ ^{\shortmid }\partial ^{7}(\ ^{\shortmid }\eta ^{8}(\tau )\
^{\shortmid }\mathring{g}^{8}(\tau ))]|\ [\ ^{\shortmid }\eta ^{8}(\tau )\
^{\shortmid }\mathring{g}^{8}(\tau )]^{5/2}}]d\ ^{\shortmid }x^{k_{3}}\}.
\end{eqnarray*}

String parametric deformations are modeled for $\ _{s}^{\shortmid }\eta
(\tau )\ ^{\shortmid }\mathring{g}_{\alpha _{s}}(\tau )\sim \ ^{\shortmid
}\zeta _{\alpha _{s}}(\tau )(1+\kappa \ ^{\shortmid }\chi _{\alpha
_{s}}(\tau ))\ \ ^{\shortmid }\mathring{g}_{\alpha _{s}}(\tau ).$

\subsubsection{Nonassociative s-metrics with small parametric
quasi-stationary evolution of double phase space wormholes}

We provide explicit formulas for explicit small parametric $\tau $- and $%
\kappa $-parametric decompositions of solutions (\ref{doublwnonassocwh})
using \ $\chi $-polarizations as we explain in subsection (\ref{ssdwhsmall}%
): 
\begin{eqnarray*}
d\ \widehat{s}^{2}(\tau ) &=&\widehat{g}_{\alpha \beta }(\hbar ,\kappa ,\tau
,l,\theta ,\varphi ,p_{l},p_{\varphi };\ ^{\shortmid }\mathring{g}_{\alpha
_{s}};\eta _{4}(\tau ),\ ^{\shortmid }\eta ^{6}(\tau ),\ ^{\shortmid }\eta
^{8}(\tau ),\Lambda _{\lbrack h]}^{\star }(\tau );\ _{s}^{\shortmid }\Im
^{\star }(\tau ),\ ^{\shortmid }\Lambda _{\lbrack c]}^{\star }(\tau
))du^{\alpha }du^{\beta } \\
&=&e^{\psi _{0}(\hbar ,\tau ,l,\theta )}[1+\kappa \ ^{\psi (\hbar ,\tau
,l,\theta )}\chi (l,\theta )][(dx^{1}(l,\theta ))^{2}+(dx^{2}(l,\theta
))^{2}]
\end{eqnarray*}%
\begin{eqnarray*}
&&-\{\frac{4[\partial _{\varphi }(|\zeta _{4}\ \breve{g}_{4}|^{1/2})]^{2}}{%
\breve{g}_{3}|\int d\varphi \{\Lambda _{\lbrack h]}^{\star }(\tau )\partial
_{\varphi }(\zeta _{4}\ \breve{g}_{4})\}|}-\kappa \lbrack \frac{\partial
_{\varphi }(\chi _{4}|\zeta _{4}\breve{g}_{4}|^{1/2})}{4\partial _{\varphi
}(|\zeta _{4}\ \breve{g}_{4}|^{1/2})}-\frac{\int d\varphi \{\ \Lambda
_{\lbrack h]}^{\star }(\tau )\partial _{\varphi }[(\zeta _{4}\ \breve{g}%
_{4})\chi _{4}]\}}{\int d\varphi \{\Lambda _{\lbrack h]}^{\star }(\tau
)\partial _{\varphi }(\zeta _{4}\ \breve{g}_{4})\}}]\}\ \breve{g}_{3} \\
&&\{d\varphi +[\frac{\partial _{i_{1}}\ \int d\varphi \ \Lambda _{\lbrack
h]}^{\star }(\tau )\ \partial _{\varphi }\zeta _{4}}{(\check{N}%
_{i_{1}}^{3})\ \Lambda _{\lbrack h]}^{\star }(\tau )\partial _{\varphi
}\zeta _{4}}+\kappa (\frac{\partial _{i_{1}}[\int d\varphi \ \Lambda
_{\lbrack h]}^{\star }(\tau )\ \partial _{\varphi }(\zeta _{4}\chi _{4})]}{%
\partial _{i_{1}}\ [\int d\varphi \ \Lambda _{\lbrack h]}^{\star }(\tau
)\partial _{\varphi }\zeta _{4}]}-\frac{\partial _{\varphi }(\zeta _{4}\chi
_{4})}{\partial _{\varphi }\zeta _{4}})]\check{N}_{i_{1}}^{3}dx^{i_{1}}\}^{2}
\end{eqnarray*}%
\begin{eqnarray}
&&+\zeta _{4}(1+\kappa \ \chi _{4})\ \breve{g}_{4}\{dt+[(\check{N}%
_{k_{1}}^{4})^{-1}[\ _{1}n_{k}+16\ _{2}n_{k_{1}}[\int d\varphi \frac{\left(
\partial _{\varphi }[(\zeta _{4}\ \breve{g}_{4})^{-1/4}]\right) ^{2}}{|\int
d\varphi \partial _{\varphi }[\Lambda _{\lbrack h]}^{\star }(\tau )(\zeta
_{4}\ \breve{g}_{4})]|}]  \notag \\
&&+\kappa \frac{16\ _{2}n_{k_{1}}\int d\varphi \frac{\left( \partial
_{\varphi }[(\zeta _{4}\ \breve{g}_{4})^{-1/4}]\right) ^{2}}{|\int d\varphi
\partial _{\varphi }[\ \Lambda _{\lbrack h]}^{\star }(\tau )(\zeta _{4}\ 
\breve{g}_{4})]|}(\frac{\partial _{\varphi }[(\zeta _{4}\ \breve{g}%
_{4})^{-1/4}\chi _{4})]}{2\partial _{\varphi }[(\zeta _{4}\ ^{cy}g)^{-1/4}]}+%
\frac{\int d\varphi \partial _{\varphi }[\Lambda _{\lbrack h]}^{\star }(\tau
)(\zeta _{4}\chi _{4}\ \breve{g}_{4})]}{\int d\varphi \partial _{\varphi
}[(\zeta _{4}\ \breve{g}_{4})]})}{\ _{1}n_{k_{1}}+16\ _{2}n_{k_{1}}[\int
d\varphi \frac{\left( \partial _{\varphi }[(\zeta _{4}\ \breve{g}%
_{4})^{-1/4}]\right) ^{2}}{|\int d\varphi \partial _{\varphi }[\Lambda
_{\lbrack h]}^{\star }(\tau )(\zeta _{4}\ \breve{g}_{4})]|}]}]\check{N}%
_{k_{1}}^{4}dx^{k_{1}}\}^{2}  \label{whpolf1}
\end{eqnarray}%
\begin{eqnarray*}
&&-\{\frac{4[\ ^{\shortmid }\partial ^{5}(|\ ^{\shortmid }\zeta ^{6}\ \
^{\shortmid }\breve{g}^{6}|^{1/2})]^{2}}{\ ^{\shortmid }\breve{g}^{5}|\int
dp_{5}\{\ ^{\shortmid }\Lambda _{\lbrack c]}^{\star }(\tau )\ ^{\shortmid
}\partial ^{5}(\ ^{\shortmid }\zeta ^{6}\ \ ^{\shortmid }\breve{g}^{6})\}|}%
-\kappa \lbrack \frac{\ ^{\shortmid }\partial ^{5}(\ ^{\shortmid }\chi
^{6}|\ ^{\shortmid }\zeta ^{6}\ \ ^{\shortmid }\breve{g}^{6}|^{1/2})}{4\
^{\shortmid }\partial ^{5}(|\ ^{\shortmid }\zeta ^{6}\ \ ^{\shortmid }\breve{%
g}^{6}|^{1/2})}-\frac{\int dp_{5}\{\ ^{\shortmid }\Lambda _{\lbrack
c]}^{\star }(\tau )\ ^{\shortmid }\partial ^{5}[(\ ^{\shortmid }\zeta ^{6}\
\ ^{\shortmid }\breve{g}^{6})\ ^{\shortmid }\chi ^{6}]\}}{\int dp_{5}\{\
^{\shortmid }\Lambda _{\lbrack c]}^{\star }(\tau )\ ^{\shortmid }\partial
^{5}(\ ^{\shortmid }\zeta ^{6}\ \ ^{\shortmid }\breve{g}^{6})\}}]\}\
^{\shortmid }\breve{g}^{5} \\
&&\{dp_{5}+[\frac{\ ^{\shortmid }\partial _{i_{2}}\ \int dp_{5}\ \
^{\shortmid }\Lambda _{\lbrack c]}^{\star }(\tau )\ ^{\shortmid }\partial
^{5}\ ^{\shortmid }\zeta ^{6}}{(\ ^{\shortmid }\check{N}_{5i_{2}})\ \
^{\shortmid }\Lambda _{\lbrack c]}^{\star }(\tau )\ ^{\shortmid }\partial
^{5}\ ^{\shortmid }\zeta ^{6}}+\kappa (\frac{\ ^{\shortmid }\partial
_{i_{2}}[\int dp_{5}\ \ ^{\shortmid }\Lambda _{\lbrack c]}^{\star }(\tau )\
\ ^{\shortmid }\partial ^{5}(\ ^{\shortmid }\zeta ^{6}\ ^{\shortmid }\chi
^{6})]}{\ ^{\shortmid }\partial _{i_{2}}\ [\int dp_{5}\ \ ^{\shortmid
}\Lambda _{\lbrack c]}^{\star }(\tau )\ ^{\shortmid }\partial ^{5}\
^{\shortmid }\zeta ^{6}]}-\frac{\ ^{\shortmid }\partial ^{5}(\ ^{\shortmid
}\zeta ^{6}\ ^{\shortmid }\chi ^{6})}{\ ^{\shortmid }\partial ^{5}\
^{\shortmid }\zeta ^{6}})]\ ^{\shortmid }\check{N}_{5i_{2}}dx^{i_{2}}\}^{2}
\end{eqnarray*}%
\begin{eqnarray*}
&&+\ ^{\shortmid }\zeta ^{6}(1+\kappa \ \ ^{\shortmid }\chi ^{6})\ \
^{\shortmid }\breve{g}^{6}\{dt+[(\ ^{\shortmid }\check{N}_{5i_{2}})^{-1}[\
_{1}n_{k_{2}}+16\ _{2}n_{k_{2}}[\int dp_{5}\frac{\left( \ ^{\shortmid
}\partial ^{5}[(\ ^{\shortmid }\zeta ^{6}\ \ ^{\shortmid }\breve{g}%
^{6})^{-1/4}]\right) ^{2}}{|\int dp_{5}\ ^{\shortmid }\partial ^{5}[\
^{\shortmid }\Lambda _{\lbrack c]}^{\star }(\tau )(\ ^{\shortmid }\zeta
^{6}\ \ ^{\shortmid }\breve{g}^{6})]|}] \\
&&+\kappa \frac{16\ _{2}n_{k_{2}}\int dp_{5}\frac{\left( \ ^{\shortmid
}\partial ^{5}[(\ ^{\shortmid }\zeta ^{6}\ \ ^{\shortmid }\breve{g}%
^{6})^{-1/4}]\right) ^{2}}{|\int dp_{5}\ ^{\shortmid }\partial ^{5}[\ \
^{\shortmid }\Lambda _{\lbrack c]}^{\star }(\tau )(\ ^{\shortmid }\zeta
^{6}\ \ ^{\shortmid }\breve{g}^{6})]|}(\frac{\ ^{\shortmid }\partial ^{5}[(\
^{\shortmid }\zeta ^{6}\ \ ^{\shortmid }\breve{g}^{6})^{-1/4}\ ^{\shortmid
}\chi ^{6})]}{2\ ^{\shortmid }\partial ^{5}[(\ ^{\shortmid }\zeta ^{6}\ \
^{\shortmid }\breve{g}^{6})^{-1/4}]}+\frac{\int dp_{5}\ ^{\shortmid
}\partial ^{5}[\ ^{\shortmid }\Lambda _{\lbrack c]}^{\star }(\tau )(\ \
^{\shortmid }\chi ^{6}\ \ ^{\shortmid }\zeta ^{6}\ \ ^{\shortmid }\breve{g}%
^{6})]}{\int dp_{5}\ ^{\shortmid }\partial ^{5}[(\ ^{\shortmid }\zeta ^{6}\
\ ^{\shortmid }\breve{g}^{6})]})}{\ _{1}n_{k_{2}}+16\ _{2}n_{k_{2}}[\int
dp_{5}\frac{\left( \ ^{\shortmid }\partial ^{5}[(\ ^{\shortmid }\zeta ^{6}\
\ ^{\shortmid }\breve{g}^{6})^{-1/4}]\right) ^{2}}{|\int dp_{5}\ ^{\shortmid
}\partial ^{5}[\ ^{\shortmid }\Lambda _{\lbrack c]}^{\star }(\tau )(\
^{\shortmid }\zeta ^{6}\ \ ^{\shortmid }\breve{g}^{6})]|}]}] \\
&&\ ^{\shortmid }\check{N}_{6k_{2}}dx^{k_{2}}\}^{2}
\end{eqnarray*}%
\begin{eqnarray*}
&&-\{\frac{4[\ ^{\shortmid }\partial ^{7}(|\ ^{\shortmid }\zeta ^{8}\ \
^{\shortmid }\breve{g}^{8}|^{1/2})]^{2}}{\ ^{\shortmid }\breve{g}^{7}|\int
dp_{7}\{\ ^{\shortmid }\Lambda _{\lbrack c]}^{\star }(\tau )\ ^{\shortmid
}\partial ^{7}(\ ^{\shortmid }\zeta ^{8}\ \ ^{\shortmid }\breve{g}^{8})\}|}%
-\kappa \lbrack \frac{\ ^{\shortmid }\partial ^{7}(\ ^{\shortmid }\chi
^{8}|\ ^{\shortmid }\zeta ^{8}\ \ ^{\shortmid }\breve{g}^{8}|^{1/2})}{4\
^{\shortmid }\partial ^{7}(|\ ^{\shortmid }\zeta ^{8}\ \ ^{\shortmid }\breve{%
g}^{8}|^{1/2})}-\frac{\int dp_{7}\{\ ^{\shortmid }\Lambda _{\lbrack
c]}^{\star }(\tau )\ ^{\shortmid }\partial ^{7}[(\ ^{\shortmid }\zeta ^{8}\
\ ^{\shortmid }\breve{g}^{8})\ ^{\shortmid }\chi ^{8}]\}}{\int dp_{7}\{\
^{\shortmid }\Lambda _{\lbrack c]}^{\star }(\tau )\ ^{\shortmid }\partial
^{7}(\ ^{\shortmid }\zeta ^{8}\ \ ^{\shortmid }\breve{g}^{8})\}}]\}\
^{\shortmid }\breve{g}^{7} \\
&&\{dp_{7}+[\frac{\ ^{\shortmid }\partial _{i_{3}}\ \int dp_{7}\ \
^{\shortmid }\Lambda _{\lbrack c]}^{\star }(\tau )\ ^{\shortmid }\partial
^{7}\ ^{\shortmid }\zeta ^{8}}{(\ ^{\shortmid }\check{N}_{7i_{2}})\ \
^{\shortmid }\Lambda _{\lbrack c]}^{\star }(\tau )\ ^{\shortmid }\partial
^{7}\ ^{\shortmid }\zeta ^{8}}+\kappa (\frac{\ ^{\shortmid }\partial
_{i_{3}}[\int dp_{7}\ \ ^{\shortmid }\Lambda _{\lbrack c]}^{\star }(\tau )\
\ ^{\shortmid }\partial ^{7}(\ ^{\shortmid }\zeta ^{8}\ ^{\shortmid }\chi
^{8})]}{\ ^{\shortmid }\partial _{i_{3}}\ [\int dp_{7}\ \ ^{\shortmid
}\Lambda _{\lbrack c]}^{\star }(\tau )\ ^{\shortmid }\partial ^{7}\
^{\shortmid }\zeta ^{8}]}-\frac{\ ^{\shortmid }\partial ^{7}(\ ^{\shortmid
}\zeta ^{8}\ ^{\shortmid }\chi ^{8})}{\ ^{\shortmid }\partial ^{7}\
^{\shortmid }\zeta ^{8}})]\ ^{\shortmid }\check{N}_{8i_{3}}dx^{i_{3}}\}^{2}
\end{eqnarray*}%
\begin{eqnarray*}
&&+\ ^{\shortmid }\zeta ^{8}(1+\kappa \ \ ^{\shortmid }\chi ^{8})\ \
^{\shortmid }\breve{g}^{8}\{dE+[(\ ^{\shortmid }\check{N}_{7i_{3}})^{-1}[\
_{1}n_{k_{3}}+16\ _{2}n_{k_{3}}[\int dp_{7}\frac{\left( \ ^{\shortmid
}\partial ^{7}[(\ ^{\shortmid }\zeta ^{8}\ \ ^{\shortmid }\breve{g}%
^{8})^{-1/4}]\right) ^{2}}{|\int dp_{7}\ ^{\shortmid }\partial ^{7}[\
^{\shortmid }\Lambda _{\lbrack c]}^{\star }(\tau )(\ ^{\shortmid }\zeta
^{8}\ \ ^{\shortmid }\breve{g}^{8})]|}] \\
&&+\kappa \frac{16\ _{2}n_{k_{3}}\int dp_{7}\frac{\left( \ ^{\shortmid
}\partial ^{7}[(\ ^{\shortmid }\zeta ^{8}\ \ ^{\shortmid }\breve{g}%
^{8})^{-1/4}]\right) ^{2}}{|\int dp_{7}\ ^{\shortmid }\partial ^{7}[\ \
^{\shortmid }\Lambda _{\lbrack c]}^{\star }(\tau )(\ ^{\shortmid }\zeta
^{8}\ \ ^{\shortmid }\breve{g}^{8})]|}(\frac{\ ^{\shortmid }\partial ^{7}[(\
^{\shortmid }\zeta ^{8}\ \ ^{\shortmid }\breve{g}^{8})^{-1/4}\ ^{\shortmid
}\chi ^{8})]}{2\ ^{\shortmid }\partial ^{7}[(\ ^{\shortmid }\zeta ^{8}\ \
^{\shortmid }\breve{g}^{8})^{-1/4}]}+\frac{\int dp_{7}\ ^{\shortmid
}\partial ^{7}[\ ^{\shortmid }\Lambda _{\lbrack c]}^{\star }(\tau )(\ \
^{\shortmid }\chi ^{8}\ \ ^{\shortmid }\zeta ^{8}\ \ ^{\shortmid }\breve{g}%
^{8})]}{\int dp_{7}\ ^{\shortmid }\partial ^{7}[(\ ^{\shortmid }\zeta ^{8}\
\ ^{\shortmid }\breve{g}^{8})]})}{\ _{1}n_{k_{3}}+16\ _{2}n_{k_{3}}[\int
dp_{7}\frac{\left( \ ^{\shortmid }\partial ^{7}[(\ ^{\shortmid }\zeta ^{8}\
\ ^{\shortmid }\breve{g}^{8})^{-1/4}]\right) ^{2}}{|\int dp_{7}\ ^{\shortmid
}\partial ^{7}[\ ^{\shortmid }\Lambda _{\lbrack c]}^{\star }(\tau )(\
^{\shortmid }\zeta ^{8}\ \ ^{\shortmid }\breve{g}^{8})]|}]}]\times \\
&&\ ^{\shortmid }\check{N}_{8k_{8}}dx^{k_{8}}\}^{2}.
\end{eqnarray*}%
Details on proofs for $\tau $-families of such nonassociative s-metrics can
be found, for instance, for BH and BE configurations in partner works \cite%
{partner02,partner03,lbdssv22,lbsvevv22}. In abstract symbolic geometric
forms, we can change BH coordinates into respective wormhole ones, then the
prime s-metrics are taken for wormholes.


\begin{thebibliography}{999}
\bibitem{preskill} J. Preskill, lecture notes, available at\newline
http://www.theory.caltech.edu/\symbol{126}preskill/ph219/index.html\#lecture

\bibitem{witten20} E. Witten, A mini-introduction to information theory,
Riv. Nuovo Cim. 43 (2020) 187-227, arXiv: 1805.11965

\bibitem{ryu16} S. Ryu and T. Takayanagi, Holographic derivation of
entanglement entropy from AdS/CFT, Phys. Rev. Lett. 96 (2006) 181602

\bibitem{vanraam10} M. Van Raamsdonk, Building up spacetime with quantum
entanglement, Gen. Rel. Grav. 42 (2010) 2323 [Int. J. Mod. Phys. D 19 (2010)
2429]

\bibitem{jacobson16} T. Jacobson, Entanglement equilibrium and the Einstein
equation, Phys. Rev. Lett. 116 (2016) 201101

\bibitem{pastavski15} F. Pastawski, B. Yoshida, D. Harlow and J. Preskill,
Holographic quantum error-correcting codes: Toy models for the bulk/boundary
correspondence, JHEP, 1506 (2015) 149

\bibitem{cover91} T. M. Cover and J. A. Thomas, Elements of Information
Theory (John Wiley \& Sons, 1991)

\bibitem{nielsen10} M. A. Nielsen and I. L. Chuang, Quantum computation and
quantum information (Cambridge University Press, 2010)

\bibitem{wilde13} M. M. Wilde, Quantum Information Theory (Cambridge
University Press, 2013)

\bibitem{hayashi17} M. Hayashi, Quantum Information Theory (Springer, 2017)

\bibitem{watrous18} J. Watrous, The theory of quantum information (Cambridge
University Press, 2018)

\bibitem{aolita15} L. Aolita, F. de Melo, L. Davidovich, Opens-system
dynamics of entanglement, Rep. Progr. Phys. 78 \ (2015) 042001

\bibitem{nishioka18} T. Nishioka, Entanglement entropy: holography and
renormalization group, Rev. Mod. Phys. 90 (2018) 03500

\bibitem{svnc00} S. Vacaru, Gauge and Einstein gravity from non-Abelian
gauge models on noncommutative spaces, Phys. Lett. B 498 (2001) 74-82;
arXiv: hep-th/0009163



\bibitem{svnonh08} S. Vacaru, Nonholonomic Ricci flows: II. Evolution
equations and dynamics, J. Math. Phys. 49 (2008) 043504 (27 pages); arXiv:
math.DG/0702598

\bibitem{svmpnc09} S. Vacaru, Spectral functionals, nonholonomic Dirac
operators, and noncommutative Ricci flows, J. Math. Phys. 50 (2009) 073503;
arXiv: 0806.3814 [math-ph]






\bibitem{sv20} S. Vacaru, Geometric information flows and G. Perelman
entropy for relativistic classical and quantum mechanical systems, Eur.
Phys. J. C 80 (2020) 639; arXiv: 1905.12399






\bibitem{lbdssv22} L. Bubuianu, D. Singleton, and S. Vacaru, Nonassociative
black holes in R-flux deformed phase spaces and relativistic models of G.
Perelman thermodynamics, JHEP 05 (2023) 057; arXiv: 2207.0515

\bibitem{lbsvevv22} L. Bubuianu, S. Vacaru and Elsen Veli Veliev,
Nonassociative Ricci flows, swampland conjectures and G. Perelman
thermodynamics for star product and R-flux deformed black holes, Fortschr.
Phys. 71 (2023) 2200140

\bibitem{hamilton82} R. S. Hamilton, Three-manifolds with positive Ricci
curvature, J. Diff. Geom. 17 (1982) 255-306

\bibitem{friedan80} D. Friedan, Nonlinear models in 2 + $\varepsilon $
dimensions, Phys. Rev. Lett. 45 (1980) 1057-1060

\bibitem{perelman1} G. Perelman, The entropy formula for the Ricci flow and
its geometric applications, arXiv: math. DG/0211159



\bibitem{kleiner06} B. Kleiner and J. Lott, Notes on Perelman's papers,
Geometry \& Topology 12 (2008) 2587-2855; arXiv: math/0605667

\bibitem{morgan06} J. W. Morgan and G. Tian, Ricci flow and the Poincar\'{e}
conjecture, AMS, Clay Mathematics Monogaphs, vol. 3 (2007); arXiv: math/
0607607

\bibitem{cao06} H. -D. Cao and H. -P. Zhu, A complete proof of the Poincar%
\'{e} and geometrization conjectures - application of the Hamilton--Perelman
theory of the Ricci flow, Asian J. Math. 10 (2006) 165-495; see also a
preprint version: H. -D. Cao and H. -P. Zhu, Hamilton-Perelman's proof of
the Poincar\'{e} conjecture and the geometrization conjectures, arXiv:
math/0612069

\bibitem{kehagias19} A. Kehagias, D. L\"{u}st and S. L\"{u}st, Swampland,
gradient flow and infinite distance, JHEP 04 (2020) 170; arXiv: 1910.00453

\bibitem{biasio20} D. De Basio and D. L\"{u}st, Geometric flow equations for
Schwarzschild-AdS space-time and Hawking page phase transition, Fortsch.
Phys. 68 (2020) 8, 2000053; arXiv: 2006.03076

\bibitem{biasio21} D. De Biasio, J. Freigang and D. L\"{u}st, Geometric flow
equations for the number of space-time dimensions, arXiv: 2104.05261

\bibitem{lueben21} M. L\"{u}ben, D. L\"{u}st and Ariadna Ribes, The black
hole entropy distance conjecture and black hole evaporation, Fortsch. Phys.
69 (2021) 3, 2000130; arXiv: 2011.12331

\bibitem{biasio22} D. De Basio and D. L\"{u}st, Geometric flow of bubbles,
Nucl. Phys. B 980 (2022) 115812;\ arXiv: 2201.01679

\bibitem{sv14a} S. Vacaru, Exact solutions in modified massive gravity and
off-diagonal wormhole deformations, Eur. Phys. J. C 74 (2014) 2781;\ arXiv:
1403.1815








\bibitem{blumenhagen16} R. Blumenhagen and M. Fuchs, Towards a theory of
nonassociative gravity, JHEP 1601 (2016) 039; arXiv: 1604.03253

\bibitem{aschieri17} P. Aschieri, M. Dimitrijevi\'{c} \'{C}iri\'{c}, R. J.
Szabo, Nonassociative differential geometry and gravity with non-geometric
fluxes, JHEP 02 (2018) 036; arXiv: 1710.11467

\bibitem{szabo19} R. J. Szabo, An introduction to nonassociative physics,
Published in: PoS CORFU2018 (2019) 100; arXiv: 1903.05673

\bibitem{partner01} S. Vacaru, Elsen Veli Veliev, and Laurentiu Bubuianu,
Nonassociative geometry of nonholonomic phase spaces with star R-flux string
deformations and (non) symmetric metrics, Fortschr. Phys. 69 (2021) 2100029;
arXiv: 2106.01320

\bibitem{partner02} Elsen Veli Veliev, Laurentiu Bubuianu, and Sergiu I.
Vacaru, Decoupling and integrability of nonassociative vacuum phase space
gravitational equations with star and R-flux parametric deformations,
Fortschr. Phys. 69 (2021) 2100030; arXiv: 2106.01869

\bibitem{partner03} L. Bubuianu, S. Vacaru and E. V. Veliev, Nonassociative
black ellipsoids distorted by R-fluxes and four dimensional thin locally
anisotropic accretion disks, Eur. Phys. J. C 81 (2021) 1145; arXiv:
2108.04689

\bibitem{alvarez06} L. Alvarez-Gaume, F. Meyer, and M. A. Vazquez-Mozo,
Comments on noncommutative gravity, Nucl. Phys. B753 (2006) 92-127; arXiv:
hep-th/0605113

\bibitem{luest10} D. L\"{u}st, T-duality and closed string non-commutative
(doubled) geometry, JHEP 12 (2010) 084; arXiv: 1010.1361

\bibitem{blumenhagen10} R. Blumenhagen and E. Plauschinn, Nonassociative
gravity in string theory? J. Phys. A44 (2011) 015401; arXiv: 1010.1263

\bibitem{condeescu13} C. Condeescu, I. Florakis, C.\ Kounnas, and D. L\"{u}%
st, Gauged supergravities and non-geometric Q/R-fluxes from asymmetric
orbifold CFT's, JHEP 10 (2012) 057; arXiv: 1307.0999

\bibitem{blumenhagen13} R. Blumenhagen, M. Fuchs, F. Ha\ss ler, D. L\"{u}st,
and R.\ Sun, Non-associative deformations of geometry in duble field theory,
JHEP 04 (2014) 141; arXiv: 1312.0719

\bibitem{kupriyanov15} V. G. Kupriyanov and D. V. Vassilevich,
Nonassociative Weyl star products, JHEP 1509 (2015) 102; arXiv: 1506.02329

\bibitem{gunaydin} M. G\"{u}naydin, D. L\"{u}st and E. Malek,
Non-associativity in non-geometric string and M-theory background, the
algebra of octonions, and missing momentum models, JHEP 1611 (2016) 027,
arXiv: 1607.06474

\bibitem{kupriyanov19a} V. G. Kupriyanov, Non-commutative deformations of
Chern-Simons theory, Eur. Phys. J. C80 (2020) 42; arXiv: 1905.08753

\bibitem{shafer95} R. D. Schafer, An Introduction to Nonassociative Algebras
(Dover Publications, New York, 1995)

\bibitem{baez02} J. C. Baez, The octonions, Bull. Ammer. Math. Soc. 39
(2002) 145-202 [Erratum: 42 (2005) 2013] arXiv: math-ra/0105155

\bibitem{jordan32} P. Jordan, \"{U}ber Eine Klasse Nichassociativer
Hyperkomplexer Algebre, Nachr. Ges. Wiss. G\"{o}ttingen (1932) 569-575

\bibitem{jordan34} P. Jordan, J. von Neumann and E. Wigner, On algebraic
generalization of the quantum mechanical formalism, Ann. Math. 35 (1934)
29-64

\bibitem{okubo} S. Okubo, Introduction to Octonion and other Non-associative
Algebras in Physics (Cambridge Univ. Press, 1995)


\bibitem{mylonas13} D. Mylonas, P. Schupp, and R. J. Szabo, Non-geometric
fluxes, quasi-Hopf twist deformations and nonassociative quantum mechanics,
J. Math. Phys. 55 (2014) 122301; arXiv: 1312.1621


\bibitem{weinberg95} S. Weinberg, The Quantum Theory of Fields (three
volumes: I Foundations 1995, II Modern Applications, 1996, III
Supersymmetry, 2000,\ Cambridge University Press

\bibitem{becker06} K. Becker, M. Becker and J. H. Schwarz, String Theory and
M-Theory: A Modern Introduction (Cambridge University Press, 2006)

\bibitem{blumenhagen12} R. Bumenhagen, D. L\"{u}st and S. Theisen, Basic
Concepts of String theory (Springer, 2012)

\bibitem{kramer03} D. Kramer, H. Stephani, E. Herdlt, and M. A. H.
MacCallum, Exact Solutions of Einstein's Field Equations, 2d edition
(Cambridge University Press, 2003)

\bibitem{misner} C. W. Misner, K. S. Thorne and J. A. Wheeler, Gravitation
(Freeman, 1973)

\bibitem{hawking73} S. W. Hawking and C.F. R. Ellis, The Large Scale
Structure of Spacetime (Cambridge University Press, 1973)

\bibitem{wald82} R. W. Wald, General Relativity (Universtiy of Chicago
Press, Chicago, IL, 1984)

\bibitem{gomez19} C. G\'{o}mez, Gravity as universal UV completion: towards
a unified view of swampland conjectures, Fortsch. Phys. 69 (2021) 2,
2000096; arXiv: 1907.13386

\bibitem{bek1} J. D. Bekenstein, Black holes and entropy, Phys. Rev. D 7
(1973) 2333-2346

\bibitem{bek2} J. D. Bekenstein, Generalized second law of thermodynamics in
black hole physics, Phys. Rev. D 9 (1974) 3292-3300

\bibitem{haw1} S. W. Hawking, Particle creation by black holes, Commun.
Math. Phys. 43 (1975) 199-220, Erratum: 46 (1976) 2006

\bibitem{haw2} S. W. Hawking, Black holes and thermodynamics, Phys. Rev. D
13 (1976) 191-197

\bibitem{drinf89} V. G. Drinfeld, Quasi-Hopf algebras, Alg. Anal. 1 N6
(1989) 114-148


\bibitem{vbv18} L. Bubuianu and S. Vacaru, Axiomatic formulations of
modified gravity theories with nonlinear dispersion relations and
Finsler-Lagrange-Hamilton geometry, Eur. Phys. J. C 78 (2018) 969 (this is
the published part of a preprint containing a status report with historical
remarks in Introduction, Conclusions and Appendix B: S. Vacaru, On axiomatic
formulation of gravity and matter field theories with MDRs and
Finsler-Lagrange-Hamilton geometry on (co) tangent Lorentz bundles, arXiv:
1801.06444)



\bibitem{birrell82} N. D. Birrell and P. C. W. Davies, Quantum Fields in
Curved Space (Cambridge University Press, 1982)

\bibitem{cottrell19} W. Cottrell, B. Breivogel, D. M. Hofman and S. F.
Lokhande, How to build the thermofield double state, JHEP 02 (2019) 058;
arXiv: 1811.11528

\bibitem{greenberger} D. M. Greenberger, M. A. Horne, A. Shimony and A.
Zeilenger, Bells theorem without inequalities, Am.\ J.\ Phys. 58 (1990)
1131-1143

\bibitem{renyi61} A. R\'{e}nyi, On measures of entropy and information, in:
Fourth Berkeley Symposium on Mathematical Statistics and Probablility
(1961), pp. 547-561

\bibitem{bao19} N. Bao, M. Moosa and I. Shehzad, The holographyc dual of R%
\'{e}nyi relative entropy, arXiv: 1904.08433

\bibitem{zycz03} K. Zyczkowski, R\'{e}nyi extrapolation of Shannon entropy,
Open Systems \& Information Dynamics 10 (2003) 297-310



\bibitem{vwh3} S. Vacaru and D. Singleton, Warped, anisotropic wormhole /
soliton configurations in vacuum 5D gravity, Class. Quant. Grav. 19 (2002)
2793-2811; arXiv: hep-th / 0111045


\bibitem{warpwormh} S. Kar, Wormholes with a warped extra dimension? Gen. Rel. Gravit. 54 (2022) 66; arXiv:
2203.1463

\bibitem{wormh21a} A. Kundu, Wormholes and holography: an introduction, Eur.
Phys. J. C 82 (2022) 447; arXiv: 2110.14958

\bibitem{morris88} M. S. Morris and K. S. Thorne, Wormhole in spacetime and
their use for interstellar travel: a tool for teaching general relativity,
Am. J. Phys. 56 (1988) 395-412

\bibitem{kar94} S. Kar, S. N. Minwalla, D. Mishra and D. Sahdev, Resonances
in the transmission of massless scalar waves in a class of wormholes, Phys.
Rev. D 51 (1994) 1632-1638

\bibitem{roy20} P. D. Roy, S. Aneesh, and S. Kar, Revisiting a family of
wormholes: geometry, matter, scalar quasinormal modes and echoes, Eur. Phys.
J. 80 (2020) 850, arXiv: 1910.08746

\bibitem{souza22} T. F. de Souza, A. C. A Ramos, R. N. Costa Filho and J.
Furtado, Generalized Ellis-Bronnikov graphene wormhole, arXiv: 2208.06869

\bibitem{wormh22} D. Jafferis, A. Zlokapa, J. D. Lykken, D. K. Kolchmeyer,
S. I. Davis, N. Lauk, H. Neven and M. Spiropulu, Transversable wormhole
dynamics on a quantum processor, Nature 612 (2022) 51-55

\bibitem{wormh19} P. Gao and D. L. Jafferis, A transversable wormhole
teleportation protocol in the SYK model, JHEP 07 (2021) 097, arXiv:
1911.07416

\bibitem{wormh19a} A. R. Brown, H. Gharibyan, S. Leichenauer, H. W. Lin, S.
Neami, G. Satlon, L. Sussking, B. Swingle and M. Walter, Quantum gravity in
the lab: teleportation by size and transversable wormholes,  PRX Quantum  4 (2023) 010320; arxiv: 1911.06314

\bibitem{wormh21} S. Nezami, H. W. Lin, A. R. Brown, H. Gharibyan, S.
Leichenauer, G. Salton, L. Susskind, B. Swingle, and M. M. Walter, Quantum
gravity in the lab: teleportation by size and transversable wormholes, part
II, PRX Quantum  4 (2023) 010321; arxiv: 2102.01064

\bibitem{thooft93} G. 't Hooft, Dimensional reduction in quantum gravity,
Conference on Highlights of Particle and Condensed Matter Physics
(SALAMFEST) Trieste, Italy, March 8-12, 1993, Conf. Proc. C930308, 284
(1993), arXiv:gr-qc/9310026

\bibitem{susskind95} L. Susskind, The World as a Hologram, J. Math. Phys.
36, 6377 (1995), arXiv:hep-th/9409089

\bibitem{wormh17} J. Maldacena, D. Stanford, and Z. Yang, Diving into
traversable wormholes, Fortschr. Phys. 65 (2017) 1700034, arXiv: 1704.05333

\bibitem{wormh18} L. Susskind and Y. Zhao, Teleportation through the
wormhole, Phys. Rev. D 98 (2018) 046016, arXiv: 1707.04354

\bibitem{er35} A. Einstein and N.\ Rosen, The particle problem in the
general theory of relativity, Phys. Rev. 48 (1935) 73-77

\bibitem{epr35} A. Einsten, B.\ Podolsky and N. Rosen, Can
quantum-mechanical description of physical reality be considered complete?
Phys. Rev. 47 (1935) 777-780

\bibitem{wormh13} J. Maldacena and L. Susskind, Cool horizons for entangled
black holes, Fortschr. Phys. 61 (2013) 781-811, arXiv: 1306.0533

\bibitem{mald99} J. M. Maldacena, The Large N limit of superconformal field
theories and supergravity, Int. J. Theor. Phys. 38, 1113 (1999), [Adv.
Theor. Math. Phys.2,231(1998)], arXiv:hep-th/9711200

\bibitem{goldman86} T. Goldman, R. J. Hughes, and M. M. Nieto, Experimental
evidence for quantum gravity? Physics Letters B 171 (1986) 217

\bibitem{page81} D. N. Page and C. Geilker, Indirect evidence for quantum
gravity, Phys. Rev. Letters 47, 9 (1981) 79

\bibitem{hyden07} P. Hayden and J. Preskill, Black holes as mirrors: quantum
information in random subsystems, J. High. Energy Physics 0709 (2007) 120;
arXiv: 0708.4025
\end{thebibliography}
\end{document}